\documentclass[namedreferences]{solarphysics}
\usepackage[hyperref,optionalrh,solaromanenum]{spr-sola-addons} 
\usepackage{graphicx}                    
\usepackage{color}                       
\usepackage{breakurl}                         
\usepackage{pdflscape}

\newcommand{\ie}{{\it i.e.~}}
\newcommand{\eg}{{\it e.g.}}

\chardef\us=`\_

\begin{document}

\begin{article}

\begin{opening}

\title{Spatio-temporal dynamics of sources of hard X-ray pulsations in solar flares}

%
\author[addressref={aff1,aff2,aff3},email={}]{\inits{S.A.~}\fnm{S.A.~}\lnm{Kuznetsov}}
\author[addressref={aff3,aff4,aff5,aff2},corref,email={ivanzim@iki.rssi.ru}]{\inits{I.V.~}\fnm{I.V.~}\lnm{Zimovets}}
\author[addressref={aff1,aff2},email={}]{\inits{A.S.~}\fnm{A.S.~}\lnm{Morgachev}}
\author[addressref={aff3,aff6},email={}]{\inits{A.B.~}\fnm{A.B.~}\lnm{Struminsky}}
%
\runningauthor{Dynamics of sources of hard X-ray pulsations in solar flares}
\runningtitle{Kuznetsov et al.}

\address[id={aff1}]{Radiophysical Research Institute (NIRFI), Bolshaya Pecherskaya str. 25/12a, Nizhny Novgorod 603950, Russia}
\address[id={aff2}]{Central Astronomical Observatory at Pulkovo of the Russian Academy of Sciences, Pulkovskoye chaussee 65/1, Saint-Petersburg 196140, Russia }
\address[id={aff3}]{Space Research Institute (IKI) of the Russian Academy of Sciences, Profsoyuznaya str. 84/32, Moscow 117997, Russia   }
\address[id={aff4}]{State Key Laboratory of Space Weather (SKSW), National Space Science Center (NSSC) of the Chinese Academy of Sciences, No.1 Nanertiao, Zhongguancun, Haidian District, Beijing 100190, China  }
\address[id={aff5}]{International Space Science Institute -- Beijing (ISSI-BJ), No.1 Nanertiao, Zhongguancun, Haidian District, Beijing 100190, China }
\address[id={aff6}]{Moscow Institute of Physics and Technology (State University), Institutskyi per. 9, Dolgoprudny, Moscow Region 141700, Russia   }

\begin{abstract}
We present systematic analysis of the spatio-temporal evolution of sources of hard X-ray (HXR) pulsations in solar flares. We concentrate on disk flares whose impulsive phase are accompanied by a series of more than three successive peaks (pulsations) of HXR emission detected in the RHESSI 50--100 keV energy channel with 4--second time cadence. 29 such flares observed from February 2002 to June 2015 with characteristic time differences between successive peaks $P \approx 8-270$ s are studied. The main observational result of the analysis is that sources of HXR pulsations in all flares are not stationary, they demonstrate apparent movements/displacements in parental active regions from pulsation to pulsation. The flares can be subdivided into two main groups depending on the character of dynamics of HXR sources. The group-1 consists of 16 flares ($55\%$) with the systematic dynamics of the HXR sources from pulsation to pulsation with respect to a magnetic polarity inversion line (MPIL), which has simple extended trace on the photosphere. The group-2 consists of 13 flares ($45\%$) with more chaotic displacements of the HXR sources with respect to an MPIL having more complicated structure, and sometimes several MPILs are presented in parental active regions of such flares. Based on the observations we conclude that the mechanism of the flare HXR pulsations (at least with time differences of the considered range) is related to successive triggering of flare energy release process in different magnetic loops (or bundles of loops) of parental active regions. Group-1 flare regions consist of loops stacked into magnetic arcades extended along MPILs. Group-2 flare regions have more complicated magnetic structures and loops are arranged more chaotically and randomly there. We also found that at least 14 ($88\%$) group-1 flares and 11 ($85\%$) group-2 flares are accompanied by coronal mass ejections (CMEs), \ie the absolute majority of the flares studied are eruptive events. This gives a strong indication that eruptive processes play important role in generation of HXR pulsations in flares. We suggest that an erupting flux rope can act as a trigger of flare energy release. Its successive interaction with different loops of a parental active region can lead to apparent motion of HXR sources and to a series of HXR pulsations. However, the exact mechanism responsible for the generation of pulsations remains unclear and requires more detailed investigation.    
\end{abstract}

%

\end{opening}

%

\section{Introduction}\label{S-Intro} 

One of the major properties of solar flares is their non-stationarity and the irregularity (intermittency) of their energy release processes in time. In particular, the light curves of hard X-ray (HXR) and microwave emission of many flares are composed of correlated sequences of peaks (bursts, pulses) with durations ranging from tens of milliseconds to a few minutes \citep{1988SoPh..118...49D,2002SSRv..101....1A,2011SSRv..159..225W}. This indicates that electrons are accelerated in a series of many individual episodes of short duration with respect to total duration of a flare, which can reach several hours. Perhaps an even more interesting phenomenon related to flares is the quasi-periodic character of HXR and microwave emission observed in some events (see, \eg, \citealp{1987SoPh..111..113A}; \citealp{1988SoPh..118...49D}; \citealp{2009SSRv..149..119N}; and references therein). In these cases, the time intervals between successive peaks (pulses or pulsations) of flare emission do not differ much from each other, and by extension the Fourier or wavelet spectra of the emission time profiles contain significant peaks. We would like to emphasize here that this definition of quasi-periodicity of pulsations is quite intuitive and not strict. To the best of our knowledge, there is no strict, generally accepted definition of quasi-periodic pulsations (QPPs) of solar flare emission (see discussions, \eg, in \citealp{1998ApJ...505..941A, 2007ApJ...662..691M,2010PPCF...52l4009N,2011A&A...533A..61G,2015ApJ...798..108I}).  

QPPs in HXR and microwave emissions can represent special properties of flare energy release, in particular, some specific regimes of magnetic reconnection (\eg, \citealp{1987ApJ...321.1031T,2000A&A...360..715K,2006ApJ...644L.149O,2012SoPh..277..283A,2012A&A...548A..98M,2015ApJ...807...72L,2015ApJ...804L...8L}). QPPs can also be a result of modulation of flare energy release by some external quasi-periodic sources, such as oscillating coronal loops or sunspots located in a flare region or in close proximity (\eg, \citealp{1975ApJ...200..734B,1982SvAL....8..132Z,2005A&A...440L..59F,2006A&A...452..343N,2009A&A...505..791S,2013SoPh..286..427J,2014SoPh..289.2073J}). The scenario similar to the ``domino effect'' is also possible, when one episode of flare energy release causes another episode of energy release, which in turn causes another one, and so on \citep{1981ApL....22..171E,2011ApJ...730L..27N,2016ApJ...817....5H}. However, the physical mechanisms responsible for QPPs are an open question and a matter of active debate. A better understanding of these mechanisms is important for our general understanding of solar and stellar flares. Moreover, any good and reliable flare model must be able to explain the QPP-phenomenon in a natural way.            

A number of models were proposed to explain flare QPPs (see, \eg, \citealp{1987SoPh..111..113A,2008PhyU...51.1123Z,2009SSRv..149..119N,2016SSRv..XXX..YYYZ}, and references therein). The basis of the majority of these models is magnetohydrodynamic (MHD) oscillations of coronal magnetic (current-carrying) flux tubes (loops, in particular). Typical observed periods of the HXR and microwave QPPs in solar flares are of the order of $\sim 1-100$ seconds. This range of QPPs' periods corresponds well to the range of periods of coronal loop oscillations observed in the extreme-ultraviolet (EUV) waveband (\eg, \citealp{2005LRSP....2....3N}, \citealp{2016A&A...585A.137G}). This is the main argument in favor of the hypothesis that the flare QPPs can be associated with MHD oscillations of coronal loops. An additional indirect argument in favor of this hypothesis is a close correlation of HXR and microwave pulsations with EUV and soft X-ray (SXR) pulsations \citep{2012ApJ...749L..16D,2013ApJ...777..152S,2015SoPh..290.3625S} and sunspot oscillations \citep{2009A&A...505..791S} observed in some events. However, despite the fact that oscillations of coronal magnetic structures (and waves in them) have been well observed for more than fifteen years, to our knowledge, no direct, unambiguous and indisputable observational evidence has yet been presented which supports the link between the flare HXR QPPs and MHD oscillations and waves. We need to emphasize here that the aforementioned observational works \citep{2013ApJ...777..152S,2015SoPh..290.3625S} lead to the conclusion that it is not necessarily to explain the observed pulsations in terms of MHD oscillations or waves in coronal loops. Other mechanisms are also  possible. Moreover, the pulsations observed across several wavelengths ranges by \inlinecite{2012ApJ...749L..16D} impose strong constraints to the applicability of MHD oscillations of coronal loops as the drivers of pulsations.  

Spatially-resolved observations made with \textit{Yohkoh}/HXT \citep{1991SoPh..136...17K} and RHESSI \citep{2002SoPh..210....3L} with high angular resolution up to $\approx 2.2^{\prime\prime}$ have clearly demonstrated that the HXR ($\gtrsim 30$ keV) footpoint sources are usually not stationary. Instead, they move (change their position) in parental active regions during flares \citep{1998ASSL..229..273S,2002SoPh..210..307F,2003ApJ...595L.103K,2005ApJ...630..561B,2008AdSpR..41..908G,2009ApJ...693..132Y}. Based on the detailed analysis of a few flares it was shown that the sources of HXR QPPs can also move significantly during flares \citep{2005ApJ...625L.143G,2009SoPh..258...69Z,2012ApJ...748..139I}. In the studied events the footpoint sources of the HXR QPPs moved predominantly along the magnetic polarity inversion line (MPIL) in parental active regions of the studied flares. This indicates that the flare QPPs, at least those observed in the HXR range, are not necessarily related to MHD oscillations of coronal loops, as usually assumed. Rather, different pulsations are emitted successively from different magnetic flux tubes (loops) of flare arcades extended along the MPIL. This supports older ideas about ``elementary flare bursts'' put forward many years ago by \inlinecite{1974spre.conf..447V,1978SoPh...58..127D,1979SoPh...64..135D}. Moreover, there is evidence (\eg, \citealp{2013AstL...39..267Z}) that similar scenarios can also take place in flares with microwave QPPs emitted from sources with apparent ``single'' loop structure \citep{2010SoPh..267..329K}. Most probably, the angular resolution of modern solar radio telescopes is insufficient to resolve thin magnetic flux tubes (loops or threads), successive triggering of which leads to the QPPs. 

To our knowledge, until now, there have been no investigations of QPPs on the basis of a systematic analysis of a large number of solar flares observed in the HXR range with high spatial resolution. All previous investigations are case studies \citep{2005ApJ...625L.143G,2008SoPh..247...77L,2009SoPh..258...69Z,2010SoPh..263..163Z,2012ApJ...748..139I,2013ApJ...777...30I}. Only a few of the most interesting flares accompanied by remarkable and obvious HXR QPPs were investigated in these works. Due to this, it is not yet possible to draw general conclusions about the physical mechanisms responsible for HXR QPPs in flares. It is still not known whether the movement of HXR sources of the QPPs is a common phenomenon, or whether only some specific flares are accompanied by apparent displacements of the HXR sources. It is also not clear whether the HXR sources of the QPPs manifest some special patterns (characters) of movement or not. The goal of our work is to investigate spatio-temporal dynamics (if present) of sources of solar HXR pulsations on the base of systematic analysis of a large sample of flares observed with the RHESSI from the beginning of its regular operation on February 2002 till the end of June 2015 (the start of this work), \ie for the fourteen-year time interval covering the decay phase of the 23rd solar cycle and the rise phase of the 24th solar cycle. In our opinion, such a systematic approach, in contrast to case studies, can help us to understand the general picture of the sources of HXR pulsations in flares and to place some restrictions on their models. 

The paper is organized as follows. The methodology of the work is briefly explained in Section~\ref{S-Methodology}. The method of selection of solar flares for further analysis is described in Section~\ref{S-Select}. Analysis of the selected flares is presented in Section~\ref{S-Method}. The main results of analysis of the selected flares are summarized and discussed in Section~\ref{S-sumres}. Conclusions are given in Section~\ref{S-conclu}.            

\section{Methodology}\label{S-Methodology} 

As was stated in Section~\ref{S-Intro}, the main goal of this work is to perform a systematic study of spatio-temporal evolution of sources of HXR pulsations in solar flares, using spatially resolved RHESSI observations. The first step of the work is the selection of flares from the RHESSI events catalog for further detailed analysis. The flare selection procedure is described in Section~\ref{S-Select}. The idea is to select and analyze as many flares as possible accompanied by multiple (more than three) HXR peaks/pulsations (regardless of whether they are considered quasi-periodic or not) in the energy range $E_{\gamma} \gtrsim 50$ keV to find some general properties of the sources of HXR pulsations. 

For each event, the times of all significant HXR peaks (pulsations) above the background level in RHESSI's 50--100 keV (and 25--50 keV as well) count rates are found and tabulated. This is necessary to decompose the analyzed flares into sequences of ``elementary bursts/pulsations'' (to the extent allowed by RHESSI's 4-second time resolution). This decomposition later (Sections~\ref{Ss-imgrec}--\ref{SS-hxrcha}) will allow us to reconstruct images of the sources of individual HXR pulsations and to analyze the spatio-temporal evolution of these sources in parental active regions over the course of the flares. Prior to these steps, we investigate some temporal properties of the identified pulsations. In particular, in Section~\ref{SS-calper} we analyze the time differences between successive HXR pulsations to uncover important information about the temporal structure of energy release processes in flaring regions. In Section~\ref{SS-periodicities} we also make an attempt to find signatures of quasi-periodicities in the time profiles of HXR emission of the selected flares using standard techniques based on wavelet (Section~\ref{SSS-wavelet}) and Fourier (Section~\ref{SSS-periodogram}) analysis. However, we emphasize that the present work is not specifically focused on QPPs. Here we perform analysis of all HXR pulsations detected with RHESSI --- no matter whether they are quasi-periodic or not. More focused and detailed search for the presence of quasi-periodicities in the selected flares will be presented elsewhere. 

Further, we reconstruct sequences of images of the sources of the identified HXR pulsations implementing the standard reconstruction techniques on RHESSI's spatially-resolved data (Section~\ref{Ss-imgrec}). After we synthesize cubes of the HXR images for individual pulsations in the selected flares, we superpose HXR images with the photospheric line-of-sight magnetograms (Section~\ref{SS-hxrpos}). After the superposition is done, we find the positions of the HXR sources of all pulsations relative to the magnetic polarity inversion lines (MPILs) for each flare region. This allows us to investigate the spatio-temporal evolution of the HXR sources from pulsation to pulsation in relation to the magnetic structure of the flare regions. Based on qualitative analysis, in Section~\ref{SS-twogroups} we attempt to categorize the selected flares (if possible) into main groups in accordance with the spatio-temporal evolution of their sources of the HXR pulsations. The last step of our investigation is the analysis of some quantitative characteristics of the spatio-temporal evolution of the sources of the pulsations relative to the MPILs. This is done (in Section~\ref{SS-hxrcha}) only for one group of flares. For the other group of flares such quantitative analysis is not possible due to the specific peculiarity of the spatio-temporal evolution of the HXR sources in these flares.

\section{Flare selection}\label{S-Select} 

We considered all events collected in the RHESSI flare catalog for the time interval from 12 February 2002 until 21 June 2015. This catalog contains around $1.1 \times 10^{5}$ events. The majority of these events are quite weak flares without significant fluxes of HXR emission with energies $\gtrsim 50$ keV.   

For the second step, we used the RHESSI browser (\url{http://sprg.ssl.berkeley.edu/~tohban/browser/}) to select manually only those flares, whose 4-second light curves (time profiles) contain a series of at least four successive bursts (peaks/pulsations) in the 50--100 keV energy channel observed with the naked eye. It makes no sense to discuss phenomenon of pulsations (or quasi-periodic pulsations --- QPPs) in the case when only three or fewer peaks are presented in a flare light curve. We need to emphasize, since we looked at the 4-second light curves to pre-select flares with multiple peaks, we could miss some flares with short characteristic times of $\approx 8-12$ between neighboring peaks, because they can not be resolved sometimes with the naked eye on the quick-look images. This means that we mainly concentrate on flares with characteristic times between neighboring HXR peaks larger than $\approx 8-12$ s, though it will be shown later that the studied flares are accompanied with plenty of such peaks (see Section~\ref{SS-calper}).    

We restrict selection of flares by the 50--100 keV channel for two reasons. First, it is well known that this channel contains mainly bremsstrahlung of non-thermal (accelerated) electrons and is rarely contaminated by emission of hot ``thermal'' plasma occurring in flare regions. Sources of 50--100 keV HXR emission are almost always located in the chromospheric footpoints of flare loops. This allows our investigation to focus on processes directly related to particle acceleration. Second, we found 154 flares with at least four pulsations in the 50--100 keV RHESSI's channel. This number is reasonable for analysis in comparison with the total amount of flares with pulsations observed in the less energetic 25--50 keV channel. On the other hand, the list of flares with pulsations in higher energy channels (\eg, 100--300 keV) is obviously a sub-list of the flares observed in the 50--100 keV channel, because of the inverse power-law character of HXR spectra at the energies $\gtrsim 30$ keV.    

For the third step, we sifted flares from the preliminary list of 154 events, which did not satisfy the following criteria: 
\begin{itemize}
 \item A flare is on the disk and no further than $800^{\prime\prime}$ from the disk center, \ie its location is within $\pm 57^{\circ}$ of the heliographic longitudes and latitudes. This is necessary to relate reliably positions of the HXR sources with magnetic field in parental active regions (see Section~\ref{SS-hxrpos});  
 \item An impulsive phase of a flare is almost fully observed by the RHESSI;  
 \item A signal-to-noise ratio in the RHESSI's 50--100 keV channel is sufficiently high. This is necessary to reliably identify HXR peaks in light curves and to synthesize HXR images of good quality for the analysis; 
 \item A level of high-energy charged particles at the RHESSI detectors is low; 
 \item The RHESSI attenuators do not change too many times during a flare impulsive phase and allow us to reconstruct images for many of the HXR peaks of a flare.       
\end{itemize}

After sifting of events, we obtained the final list of 29 flares suitable for further analysis. It should be emphasized that the criteria mentioned above are quite objective. We specifically tried to avoid subjective criteria to select flares as is often done when choosing flares with QPPs for case studies. We did not explicitly select flares with remarkable QPPs in light curves of HXR emission. All selected flares are just accompanied by at least four peaks of HXR emission in the 50--100 keV range, regardless of their temporal arrangement --- periodic, quasi-periodic or not periodic. 

Some useful general information about the selected flares is summarized in Table~\ref{T-1}: number, date, start, peak and end times, X-ray (GOES) class, National Oceanic and Atmospheric Administration (NOAA) number and Hale class of a parental active region (AR), and the solar coordinates of a flare X-ray source given in the RHESSI catalog. As an additional information we also indicated a time of the first appearance of a coronal mass ejection (CME) in the SOHO/LASCO/C2 field of view. This information was taken from the SOHO/LASCO CME Catalog (\url{http://cdaw.gsfc.nasa.gov/CME_list/}). Also, we noted the standard types of solar decimeter/meter radio bursts which were detected during each event. We took into account only those CME/radio bursts whose appearance was within 120/30 minutes since a peak/start of a flare according to the GOES soft X-ray data.           

\section{Flare analysis}\label{S-Method} 

\subsection{Selection of significant peaks/pulsations}\label{SS-selpea} 

For each selected flare we automatically identified all local minima and maxima in the 4-second RHESSI background-subtracted corrected count rates in the 25--50 and 50--100 keV channels. The obtained maxima/minima correspond to local peaks/dips of the count rates. We considered only those peaks as significant ones, whose amplitudes --- the largest value of differences between a local maximum and preceding and succeeding minima --- exceed threefold standard deviation ($3 \sigma_{b}$) of a pre-flare background. We considered peaks, identified by this way, as real peaks (pulsations) of flare HXR emission, while discarded ones --- as some noise fluctuations not directly related to flare processes, \eg, caused by cosmic rays, radiation belt particles, astrophysical HXR and gamma-ray sources, self-noise of the detectors, and so on. 

\begin{figure} 
\centerline{\includegraphics[width=0.90\textwidth,clip=]{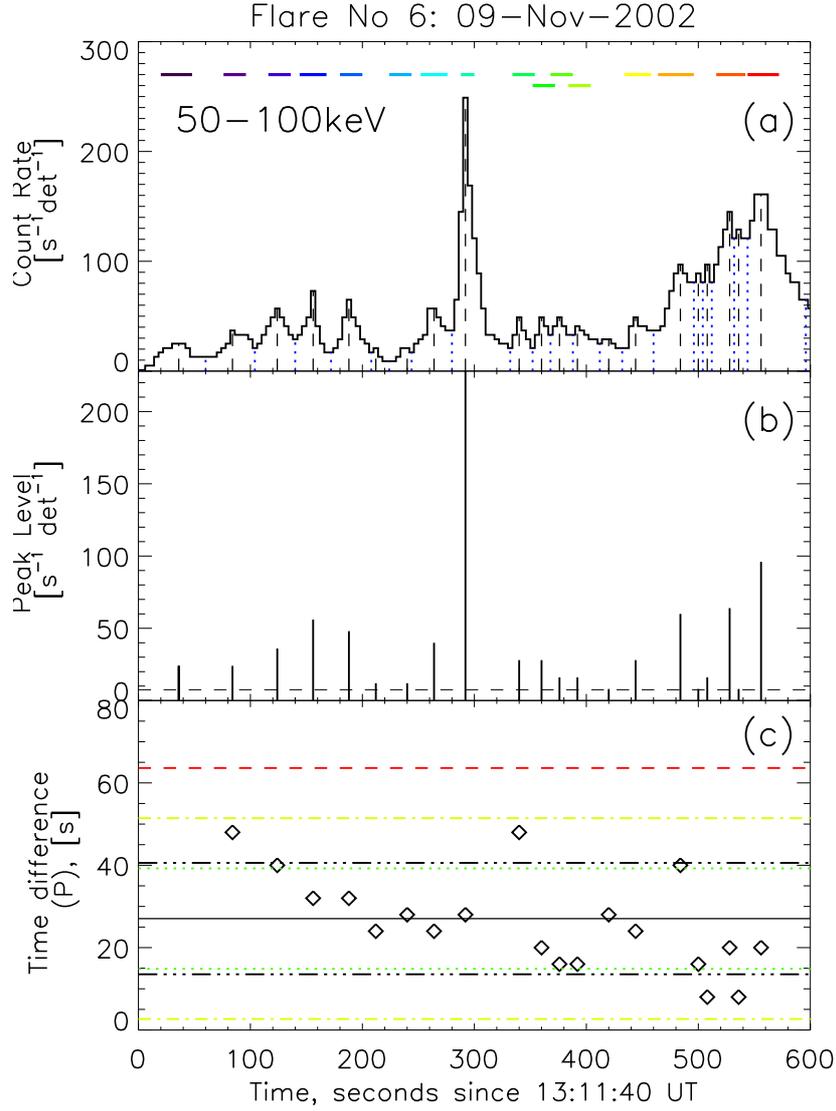}}

\caption{Illustration of selection of significant HXR peaks/pulsations and determination of their time differences on the base of the event No 6 analysis. (a) RHESSI's background-subtracted corrected count rate in the 50-100 keV channel is shown by black curve. Black dashed vertical lines indicate significant local maxima. Blue dotted vertical lines indicate all local minima. Color horizontal segments at the top mark time intervals for which HXR images were synthesized for further analysis (see Figure~\ref{F-5}(b) and Figure~\ref{F-6}, as well as Figures~\ref{AF-5}--\ref{AF-8} in the Appendix~\ref{S-appendix-B}). (b) Amplitudes of significant peaks are shown by black vertical lines. Horizontal dashed line indicates the threefold standard deviation ($3 \sigma_{b}$) value of the pre-flare background. (c) Time differences ($P$) between successive significant peaks are shown by diamonds. Average value of all significant periods (\ie mean time differences --- $\left\langle P \right\rangle$) is shown by black horizontal solid line. Black, green, yellow and red horizontal dashed and dotted lines indicate levels: $\left\langle P \right\rangle \pm 0.5 \left\langle P \right\rangle$, $\left\langle P \right\rangle \pm 1.0 \sigma_{\left\langle P \right\rangle}$, $\left\langle P \right\rangle \pm 2.0 \sigma_{\left\langle P \right\rangle}$ and $\left\langle P \right\rangle \pm 3.0 \sigma_{\left\langle P \right\rangle}$ respectively, where $\sigma_{\left\langle P \right\rangle}$ is the standard deviation of the time differences.} 
\label{F-1}
\end{figure}

An example of the automatic selection of significant peaks for the well known SOL2002-11-09T13:23 flare, \ie{} the event No 6 according to Table~\ref{T-1} (\eg, \citealp{2005ApJ...625L.143G,2012ApJ...748..139I}), is shown in Figure~\ref{F-1}(a,b). The automatic routine found in total 21 peaks in the 50--100 kev channel for this event and the amplitudes of all of them are above $3 \sigma_{b}$-level (see Figure~\ref{F-1}(b), where the horizontal dashed line indicates the threefold standard deviation $3\sigma_{b}$ value of the pre-flare background). Due to this we considered all these peaks to be significant. 

Background-subtracted corrected count rates in RHESSI's 50--100 keV channel with all identified significant and insignificant peaks for all selected flares (except flare No 6) are shown in the top panels of Figures~\ref{AF-1}--\ref{AF-4} in Appendix~\ref{S-appendix-A}. Total numbers of significant peaks ($n_{p}$) in the 25--50 and 50--100 keV channels are shown in columns 2 and 7 of Table~\ref{T-2} respectively.        

We need to note that for the sake of convenience we did not use the original RHESSI count rates but rather the corrected count rates. We found that the use of the corrected count rates did not significantly (within $10 \%$) influence the results of our specific analysis in comparison with the use of the not-corrected count rates, especially in the 50--100 keV band, which is of the principal importance for this work.

\subsection{Time differences between pulsations}\label{SS-calper}

Knowing the times of significant peaks for each event, we calculated time differences ($P$) between successive peaks (see bottom panels of Figure~\ref{F-1} and Figures~\ref{AF-1}--\ref{AF-4} in Appendix~\ref{S-appendix-A}. The total number of time differences in a flare is $N_{P}=n_{p}-1$. Average value of all time differences of a particular flare was calculated and called formally as the mean time difference of this flare ($\left\langle P \right\rangle$; see columns 3 and 8 in Table~\ref{T-2}). For each flare we also calculated standard deviations of time differences ($\sigma_{P}$; see columns 4 and 9 in Table~\ref{T-2}). This value can be considered as a measure of the spread of time differences from the average value, \ie from the mean time difference $\left\langle P \right\rangle$. Intuitively, if the spread is not broad, one could speculate about the phenomenon of quasi-periodic pulsations (QPPs). However, as it was mentioned in Section~\ref{S-Intro}, there is no strict definition of QPPs, and it is not clear what to consider broad or narrow spread around the average value. Due to this, for each flare we calculated several different levels relative to its own mean time difference: $\left\langle P \right\rangle \pm 0.5 \left\langle P \right\rangle$, $\left\langle P \right\rangle \pm 1.0 \sigma_{P}$, $\left\langle P \right\rangle \pm 2.0 \sigma_{P}$ and $\left\langle P \right\rangle \pm 3.0 \sigma_{P}$. They are shown by different horizontal lines on bottom panels of Figure~\ref{F-1} and Figures~\ref{AF-1}--\ref{AF-4} in Appendix~\ref{S-appendix-A}. Numbers (percentages) of significant peaks within the $\left\langle P \right\rangle \pm 0.5 \left\langle P \right\rangle$ and $\left\langle P \right\rangle \pm 3.0 \sigma_{P}$ ranges for all selected flares in the 25--50 and 50--100 keV channels are presented in columns 5--6 and 10--11 of Table~\ref{T-2} respectively. 

In the majority of flares (except for 7 events: No 1, 12, 14, 17, 21, 24 and 26) all time differences in the 50--100 keV channel are within the $\left\langle P \right\rangle \pm 3.0 \sigma_{P}$ range (the same is also for the 25--50 keV channel). This indicates indirectly that $P$ is a random value, which may be distributed normally around some average value (not necessary $\left\langle P \right\rangle$). We tried to check this by plotting distributions of time differences in the form of histograms. However, the number of found significant pulsations is less than 16 in 19 out of 29 studied flares. It does not make sense to build distributions on such a small number of data points. 

\begin{figure}
\centerline{\includegraphics[width=0.32\textwidth,clip=]{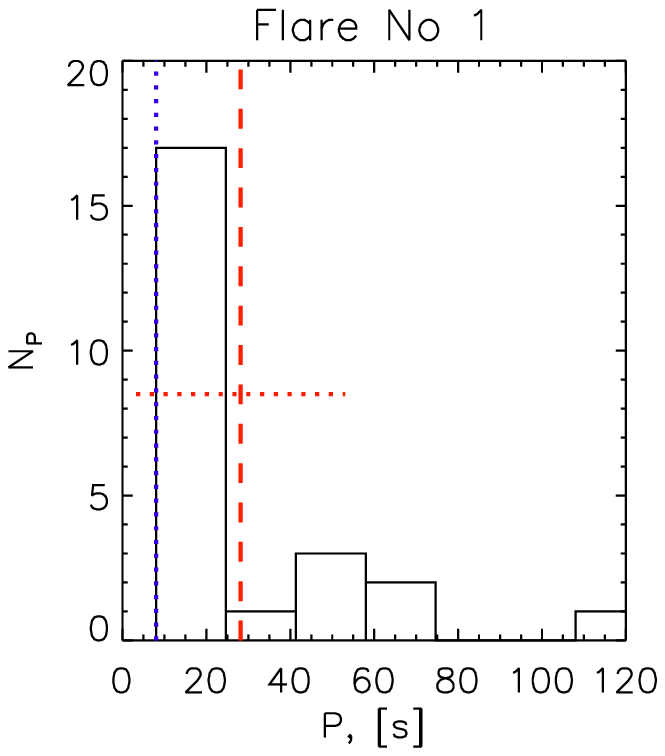}
\includegraphics[width=0.32\textwidth,clip=]{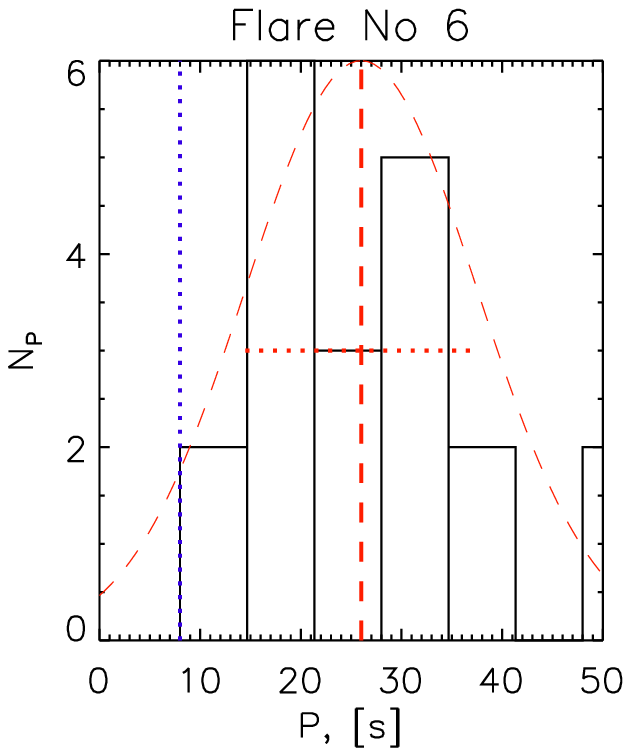}
\includegraphics[width=0.32\textwidth,clip=]{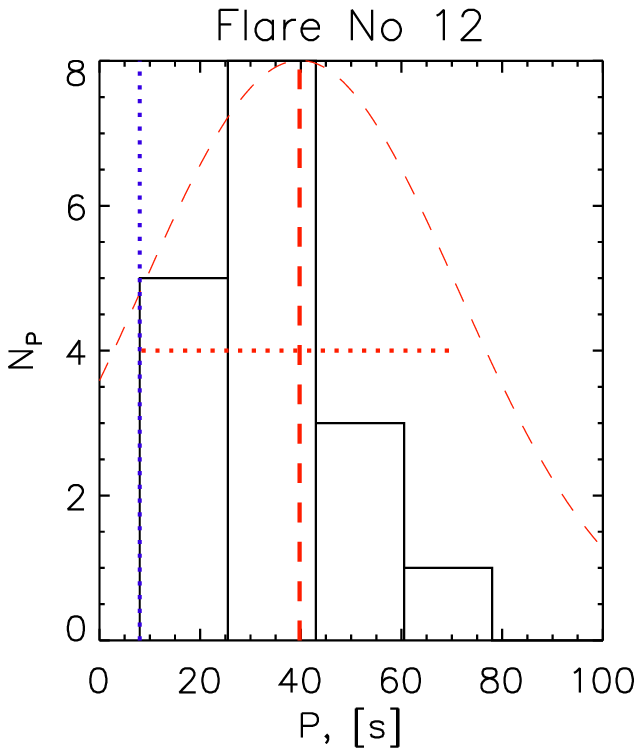}}
\vspace{-0.29\textwidth}   
\centerline{\small \bf     
\hspace{0.24\textwidth}  \color{black}{(a)}
\hspace{0.27\textwidth}  \color{black}{(b)}
\hspace{0.27\textwidth}  \color{black}{(c)}
\hfill}
\vspace{0.25\textwidth}    

\centerline{\includegraphics[width=0.32\textwidth,clip=]{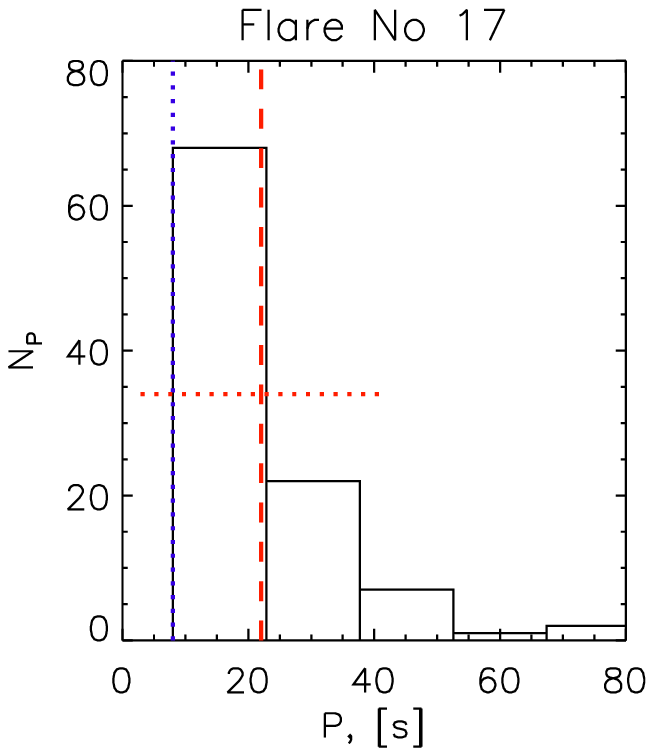}
\includegraphics[width=0.32\textwidth,clip=]{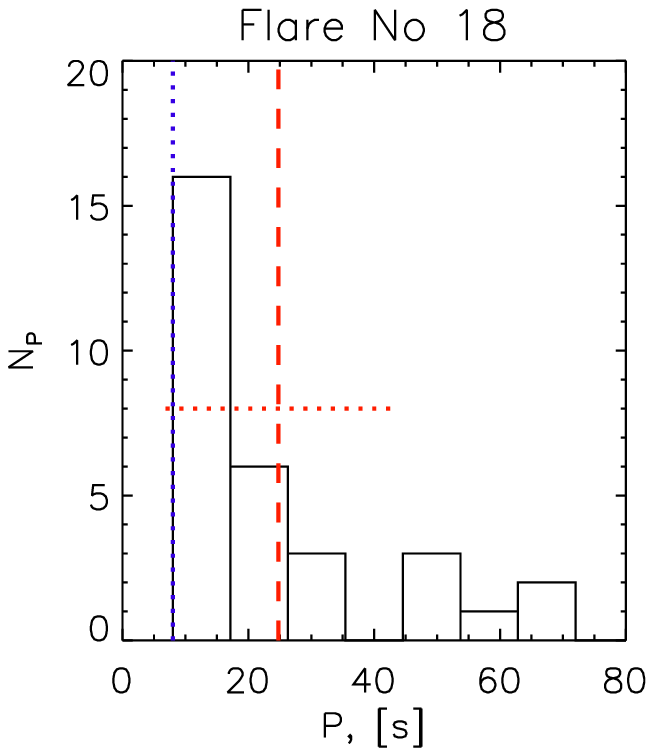}
\includegraphics[width=0.32\textwidth,clip=]{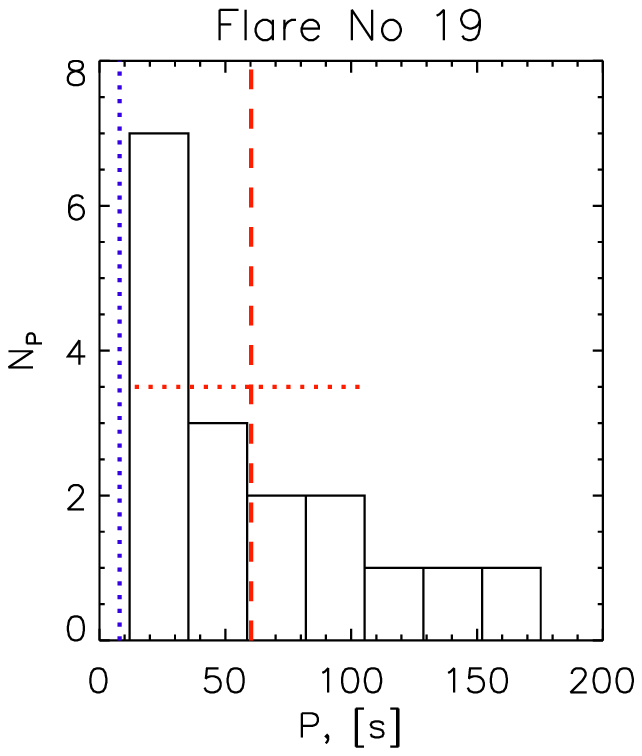}}
\vspace{-0.29\textwidth}   
\centerline{\small \bf     
\hspace{0.24\textwidth}  \color{black}{(d)}
\hspace{0.27\textwidth}  \color{black}{(e)}
\hspace{0.27\textwidth}  \color{black}{(f)}
\hfill}
\vspace{0.25\textwidth}    
		
\centerline{\includegraphics[width=0.32\textwidth,clip=]{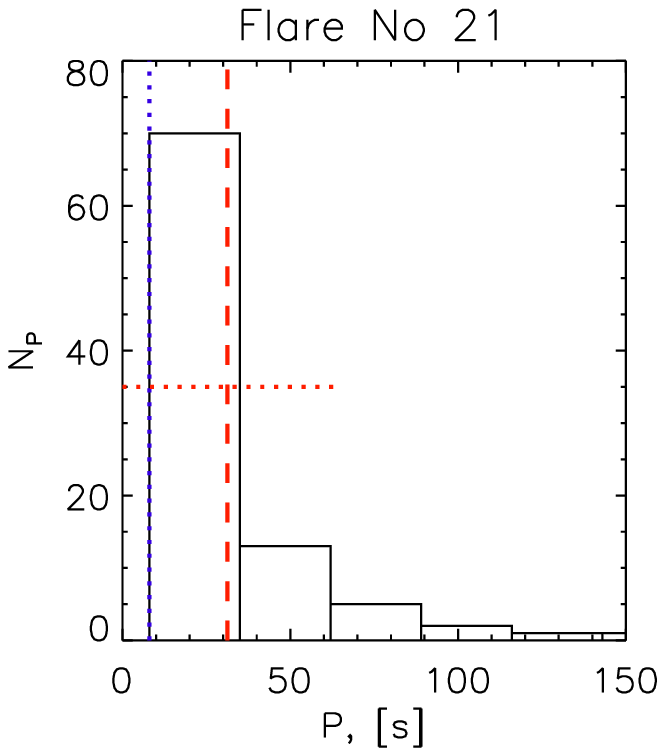}
\includegraphics[width=0.32\textwidth,clip=]{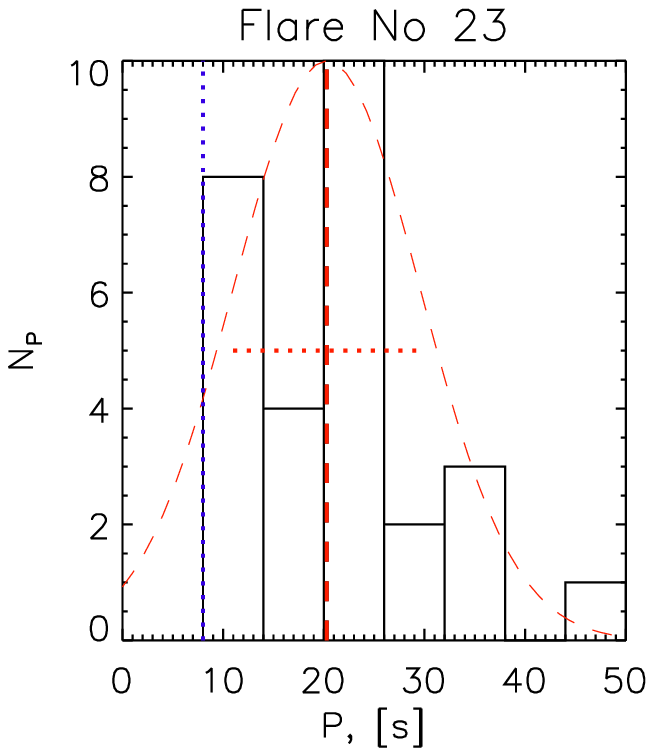}
\includegraphics[width=0.32\textwidth,clip=]{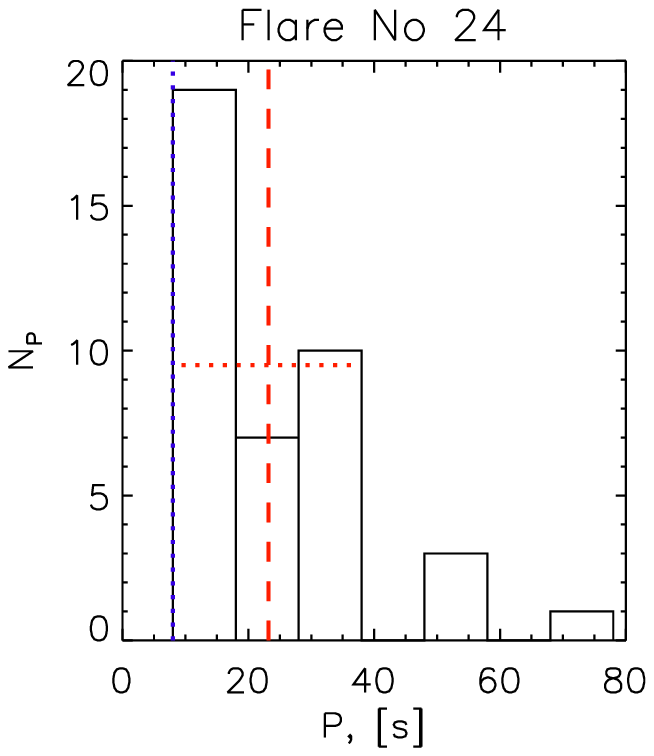}}
\vspace{-0.29\textwidth}   
\centerline{\small \bf     
\hspace{0.24\textwidth}  \color{black}{(g)}
\hspace{0.27\textwidth}  \color{black}{(h)}
\hspace{0.27\textwidth}  \color{black}{(i)}
\hfill}
\vspace{0.25\textwidth}    

\centerline{\includegraphics[width=0.32\textwidth,clip=]{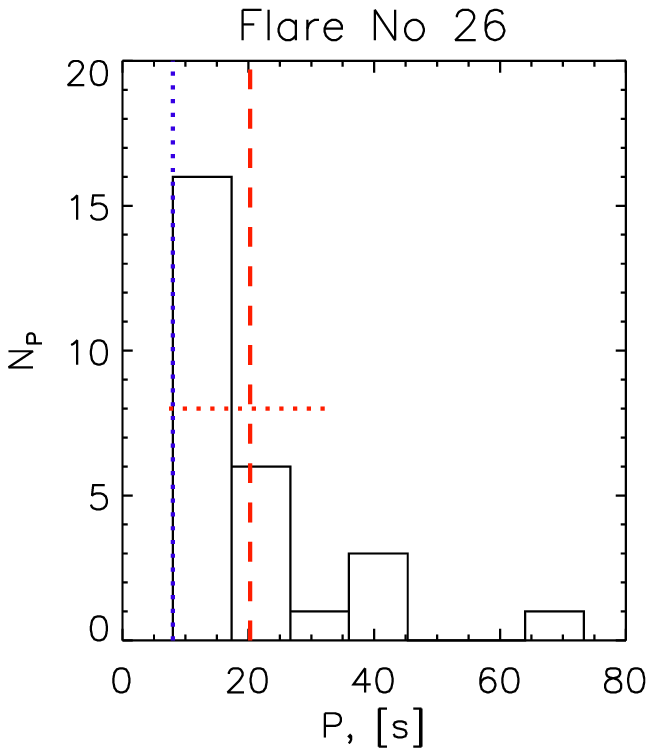}}
\vspace{-0.29\textwidth}   
\centerline{\small \bf     
\hspace{0.57 \textwidth}  \color{black}{(j)}
\hfill}
\vspace{0.25\textwidth}    
						
\caption{Distributions of time differences ($P$) found in the 4-second RHESSI background-subtracted corrected count rates in the 50--100 keV channel for the flares No 1, 6, 12, 17, 18, 19, 21, 23, 24 and 26 respectively (see Table~\ref{T-1}). Red vertical dashed line shows mean time difference $\left\langle P \right\rangle$ of a given flare. Red horizontal dotted line shows standard deviation ($\sigma_{P}$) of time differences at half of maximum level. Blue vertical dotted line shows the lower threshold ($P_{thr}=8$ s) of the time differences, which can be found in the 4-second count rates. Red thin dashed curves on (b), (c) and (h) show Gaussian functions with corresponding $\left\langle P \right\rangle$ and $\sigma_{P}$ parameters for flares No 6, 12, and 23 respectively, whose distributions are similar to bell-like shape.}
\label{F-2}
\end{figure}

Nonetheless, 10 flares (No 1, 6, 12, 17, 18, 19, 21, 23, 24 and 26) have greater than 16 pulsations each. For these flares we plotted distributions of time differences in Figure~\ref{F-2}. One can see that distributions of seven of these flares (No 1, 17, 18, 19, 21, 24 and 26) have sharply decreasing shapes with maxima around $P \approx 8-20$ s, \ie around the lower threshold of time differences ($P_{thr}=8$ s) that can be, in principle, identified in data sets with 4-second cadence. This indicates that in the majority of the studied flares RHESSI does not allow us to resolve the fine time structure of HXR emission and that maxima of real distributions of time differences, if they exist, are somewhere below $P_{thr}=8$ s. 

However, in 3 of these 10 flares (No 6, 12 and 23) the distribution of time differences has a bell-like shape (though not ideal, in particular, because of poor statistics) with maxima in the range of $P \approx 20-40$ s. The positions of these maxima coincide well with mean time differences $\left\langle P \right\rangle$ and widths of these distributions at half of their maxima are close to the standard deviations of time differences in these flares. This indicates that time differences between successive HXR pulsations in these events are random variables distributed normally around their average values --- $\left\langle P\right\rangle$, and that the 4-second RHESSI cadence can be enough to resolve time variations of HXR emission in these particular flares.  

We also plotted collective distributions of time differences combined for all 29 studied flares (see Figure~\ref{F-3}) in order to explore the general picture. The picture is as expected --- distributions have exponentially decreasing shapes with maxima around $P \approx 8-30$ s, \ie around the lower threshold of time differences ($P_{thr} = 8$ s) that is possible for the data with 4-second cadence. The best-fit exponential functions $f=a \exp\left(-P/b\right)$ have the following coefficients for the 50--100 keV and 25--50 keV channels: $a_{50}=1013.09 \pm 7.57$ and $b_{50} = 17.78 \pm 0.11$ s, and $a_{25}=1979.97 \pm 14.18$ and $b_{25} = 15.62 \pm 0.08$ s respectively.

\subsection{Search for quasi-periodicities in flare light curves}\label{SS-periodicities}

In order to search for the presence of possible quasi-periodicities in the light curves of HXR emission of the studied flares we used standard methods of spectral analysis. We would like to emphasize, once again, that it is not the main goal of this work. We apply these standard methods to make our analysis more comprehensive, since these methods are often used by various authors when analyzing pulsations of flare emissions.
 
Firstly (Section~\ref{SSS-wavelet}), we used the wavelet analysis procedure by \inlinecite{1998BAMS...79...61T} for getting the wavelet power spectra and deriving frequency-temporal structures of the investigated signals. Secondly (Section~\ref{SSS-periodogram}), we used the Fast Fourier Transform (the ``\textit{FFT.pro}'' function within IDL) to calculate the normalized periodograms in order to search for significant spectral peaks in the signals. We used the pre-flare background-subtracted RHESSI corrected count rates, normalized to its maxima, without any detrending. This approach ensures that we analyze the original flare signal without introducing any significant distortions due to empirical detrending, which could affect the results of spectral analysis (see, \eg, \citealp{2005A&A...431..391V,2011A&A...533A..61G}). In order to take into account the red-noise influence we searched spectral peaks above the red-noise level at significance levels of 95.0\% and 99.7\%.

\subsubsection{Wavelet analysis}\label{SSS-wavelet}

\begin{figure}
\centerline{\includegraphics[width=0.99\textwidth,clip=]{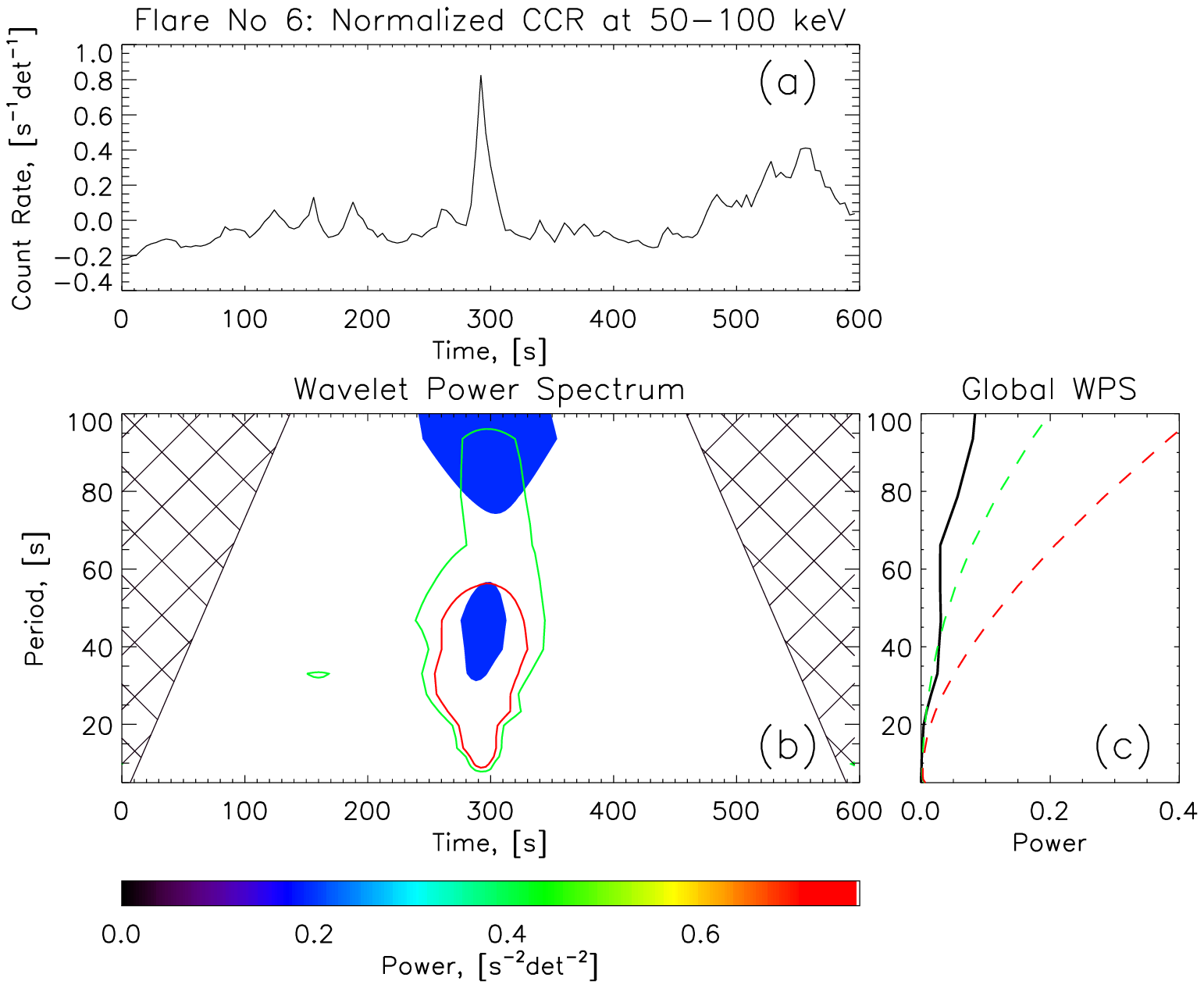}}
\centerline{\includegraphics[width=0.99\textwidth,clip=]{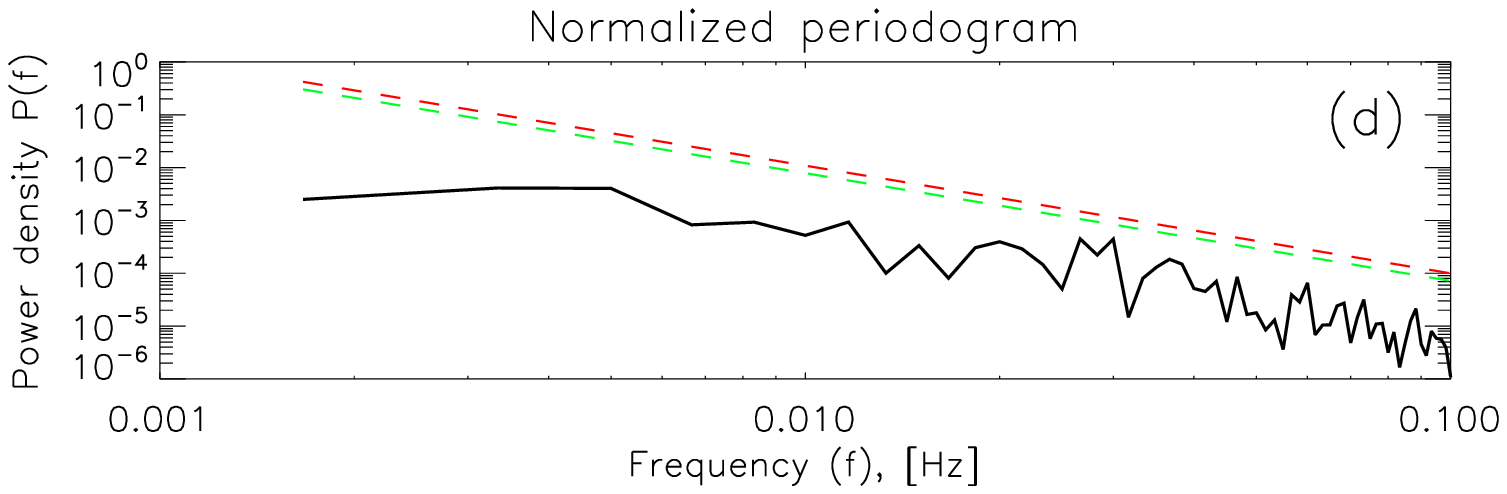}}

\caption{The results of the wavelet and periodogram analysis for the event No 6. (a) RHESSI's background-subtracted, normalized to its maximum, corrected count rate in the 50--100 keV channel. (b) Wavelet power spectrum with significance levels of 95.0\% (green contour) and 99.7\% (red contour) above the red-noise. The cross-hatched region is the cone of influence, where anything above is dubious. (c) Global wavelet power spectrum. Colored lines indicate significance levels of 95.0\% (dashed green line) and 99.7\% (dashed red line) above the red-noise. (d) Normalized periodogram with significance levels above the red-noise of 95.0\% (dashed green line) and 99.7\% (dashed red line).}
\label{F-4}
\end{figure}

The results of the wavelet analysis for the representative event No 6 are shown in Figure~\ref{F-4}(a--c). The characteristic time scale of HXR quasi-periodic pulsations (QPPs) at a significance level of 95.0\% varies from 25 s to 40 s. However, no QPPs are detected at a significance level of 99.7\% for this event. 

The results of the wavelet analysis for all studied events are summarized in Table \ref{T-3} at significance levels of 95.0\% and 99.7\% (columns 4 and 5 respectively). One can see that quasi-periodicities of HXR pulsations have been found in five events at a significance level of 95.0\%. And only three of these events have periodicities at a significance level of 99.7\%.

\subsubsection{Periodogram analysis}\label{SSS-periodogram}

The normalized frequency spectrum for the event No 6 is shown in Figure~\ref{F-4}(d). It is clearly seen that there are no spectral peaks at significance levels of 95.0\% and 99.7\% in this event.

The found spectral peaks (periods) of HXR pulsations for all selected events using FFT are shown in Table~\ref{T-3}. Four events have quasi-periodicities at a significance level of 95.0\%. Only two of these events have quasi-periodicities at a significance level of 99.7\%. These periods correspond to near-the-edge frequencies of the calculated spectra. Moreover, these two events differ from those three events for which quasi-periodicities were found using the wavelet analysis. This gives a reason to seriously doubt in the real significance of quasi-periodicities found. 

To combine the results of the performed wavelet and periodogram analysis we can conclude that quasi-periodicities in solar flare HXR emission are, probably, quite a rare phenomenon, if the red-noise character of the signals is taken into account. This is an expected result (see \citealp{2011A&A...533A..61G,2015ApJ...798..108I}).

\subsection{HXR image reconstruction}\label{Ss-imgrec}

After we identified significant peaks in the RHESSI 50--100 keV (and 25--50 keV) corrected count rates for all selected flares we reconstructed (synthesized) images for some of them using RHESSI data. We used the CLEAN and PIXON algorithms \citep{2002SoPh..210...61H} within the standard RHESSI Solar SoftWare (SSW) package (\url{http://www.lmsal.com/solarsoft/}). The CLEAN algorithm was used for preliminary analysis of flare regions and the PIXON algorithm for detailed analysis of spatio-temporal evolution of the HXR sources of pulsations. We usually used detectors 2--8 with the CLEAN algorithm and all detectors 1--9 with the PIXON. The size of the reconstructed images is $64 \times 64$ pixels with the angular pixel size of $1^{\prime\prime}$ or $2^{\prime\prime}$, depending on the size of a flare region observed in the HXR range. 

To synthesize images we selected time intervals of 12 to 40 s duration around maxima of significant peaks. For some peaks we could not synthesize images because of changing of the attenuators. However, the main problem was small signal-to-noise ratio. This was especially pronounced in the 50--100 keV range, in the 25--50 keV range situation was, in general, better. Unfortunately, we could not synthesize images of reliable quality for some significant peaks of short duration ($\approx 8-12$ s), especially in the 50--100 keV range. Due to this, we restricted our work by reconstructing images mainly in the 25--50 keV range. The effective area of the RHESSI detectors, even in the 25--50 keV range, is not enough to reconstruct images of reliable quality for some peaks of low amplitude and duration. In some cases, to increase the signal-to-noise ratio we needed to synthesize images over time intervals containing several peaks (from 2 to 4). All time intervals selected for image reconstruction are marked by colored horizontal segments in the top panels of Figure~\ref{F-1} and Figures~\ref{AF-1}--\ref{AF-4} in Appendix~\ref{S-appendix-A}. The colors of these segments correspond to colors of contours used to show the HXR sources in Figures~\ref{F-5}--\ref{F-6} and Figures~\ref{AF-5}--\ref{AF-8} in Appendix~\ref{S-appendix-B}. 

It is known (\eg, \citealp{2004ApJ...603L.117V}) that, in some flares, HXRs in the 25--50 keV range are emitted from coronal thick-target sources. However, our experience shows that this situation is rare and in the majority of cases HXR emission in the 25--50 keV range is predominantly bremsstrahlung of non-thermal electrons emitted mainly from the chromospheric (or low coronal) footpoints of flare loops. As an example, Figure~\ref{F-5} shows the positions of HXR sources in the 25--50 keV (a) and 50--100 keV (b) energy ranges for the significant peaks found in flare No 6. One can see that the spatio-temporal evolution of HXR sources for those peaks, for which it was possible to synthesize images of reliable quality, are, in general, similar for both energy ranges. The difference is only in some minor details. This is typical for the majority of the analyzed flares.  

We found that only 5 flares from our list (No 3, 9, 16, 21, and 28)  manifest signatures of coronal emission in the 25--50 keV channel --- in some individual peaks the 25-50 keV sources situated above the magnetic polarity inversion lines (MPIL; see Section~\ref{SS-hxrpos}) and/or had a loop-like shape. However, this did not strongly complicate our analysis of spatio-temporal dynamics of the HXR sources and did not strongly influence the main results.

\subsection{HXR source location}\label{SS-hxrpos}	

Since the magnetic field plays a crucial role in solar flares and because accelerated electrons are magnetized, it is of principal importance to investigate spatio-temporal evolution of flare HXR sources with respect to the magnetic field of parental active regions. For this purpose we used the photospheric line-of-sight magnetograms obtained with the SOHO/MDI \citep{1995SoPh..162..129S} and SDO/HMI \citep{2012SoPh..275..207S} instruments in 2002--2010 and 2011--2015 respectively. For each flare we selected a magnetogram which is closest in time to the beginning of a flare impulsive phase. Using ``\textit{drot\_map.pro}'' routine within SSW we rotated magnetograms to an epoch at the middle of impulsive phase of corresponding flares. For each magnetogram we found magnetic polarity inversion lines (MPILs) as curves separating positive and negative magnetic polarities in a flare region. An MPIL is a natural object, always present in active regions, relative to which it makes sense to investigate dynamics of flare HXR sources (\eg, \citealp{2005ApJ...630..561B,2005ApJ...625L.143G,2008AdSpR..41..908G,2009SoPh..258...69Z}).

For each significant HXR peak, for which it was possible to synthesize RHESSI images of good quality, we found solar coordinates of all pixels within $15\%$, $20\%$, $30\%$ or $40\%$ levels from the brightest pixel at a given image. The percentage level was fixed for all peaks of the same flare (say, $30\%$ or $40\%$), but it varied from flare to flare. Selection of levels is quite subjective. We tried to choose levels so that for each image there are at least one HXR contour on both sides of an MPIL. For the majority of flares we chose the $30\%$ level. 

We automatically separated pixels located in different magnetic polarities. For each image we determined separately the average solar coordinates of all pixels located in positive and negative polarities. These coordinates are used further for obtaining quantitative characteristics of dynamics of HXR sources (see Section~\ref{SS-hxrcha}). HXR sources located in positive and negative magnetic polarities were denoted as $S^{+}$ and $S^{-}$ respectively. For each flare $S^{+}$ and $S^{-}$ sources were overplotted on a magnetogram by different colors for different peaks. This allows to visualize clearly spatio-temporal evolution (if present) of HXR sources from peak to peak relatively to MPILs in parental active regions.

\subsection{Separation of flares into two groups}\label{SS-twogroups}	

After careful visual inspection of the maps of spatio-temporal evolution of HXR sources from peak to peak and of morphology of parental active regions of all flares under study, we found that all events can be subdivided (at least) into two groups. 

Group-1 (16 flares, \ie $55\%$ of all the studied flares) consists mainly of two-ribbon flares with systematic dynamics of HXR sources relative to MPILs, which had simple elongated shapes in these flare regions. For many of the significant HXR peaks in these flares it was possible to identify paired HXR sources ($S^{+}$ and $S^{-}$) located in regions of different magnetic polarities. Most probably, these paired HXR sources are situated in opposite footpoints of flare magnetic flux tubes (loops). By systematic dynamics we mean that the paired HXR sources show one of the four basic types of motions or their combinations (\eg, \citealp{2005ApJ...630..561B,2006ApJ...636L.173J,2008AdSpR..41..908G,2009ApJ...693..132Y}): 
\begin{itemize}
 \item[(Type-A)]{$S^{+}$ and $S^{-}$ move in the same direction parallel to an MPIL;}
 \item[(Type-B)]{$S^{+}$ and $S^{-}$ move in the opposite direction parallel to an MPIL;} 
 \item[(Type-C)]{$S^{+}$ and $S^{-}$ move away from and perpendicular to an MPIL;}
 \item[(Type-D)]{$S^{+}$ and $S^{-}$ move towards and perpendicular to an MPIL.}
\end{itemize}

In group-2 (13 events, \ie $45\%$ of all the studied flares) HXR sources appeared and moved more chaotically than in the group-1 events. It was difficult to unambiguously identify paired HXR sources and to determine their type of movement in these cases. In the majority of group-2 flare regions several separated MPILs of complicated shape were present simultaneously. 

The group classifiation of each analyzed flare is shown in column 11 of Table~\ref{T-1}. Types of paired HXR source motions identified in the group-1 flares are shown in column 16 of Table~\ref{T-4}. Maps of the spatio-temporal evolution of HXR sources for all studied events are shown in Figures~\ref{AF-5}--\ref{AF-8} in Appendix~\ref{S-appendix-B}. 

Mean time differences averaged over the group-1 and group-2 flares and their standard deviations are: $\left\langle P^{g_{1}} \right\rangle \approx 27.9$ s, $\sigma_{P}^{g1}\approx 10.5$ s and $\left\langle P^{g_{2}} \right\rangle \approx 26.4$ s, $\sigma_{P}^{g2}\approx 7.1$ s respectively, \ie they are almost the same. The mean number of significant peaks found in group-1 and group-2 flares and their standard deviations are also almost the same: $n_{p}^{g1} \approx 21.8$, $\sigma_{p}^{g1}\approx 23.7$ and $n_{p}^{g2} \approx 19.3$, $\sigma_{p}^{g2}\approx 22.3$ respectively. Thus, the temporal characteristics of both groups of flares are very similar to each other. This indicates that separation of flares into two groups is rather arbitrary and subjective and, possibly, has no very deep physical meaning. We discuss this in Section~\ref{S-sumres}. Nonetheless, for group-1 flares it is possible to estimate quantitatively some important physical characteristics of dynamics of paired sources of HXR pulsations (see Section~\ref{SS-hxrcha}), whereas it is hard to do for the group-2 flares.

\subsection{Quantitative characterization of dynamics of the HXR sources}\label{SS-hxrcha}	

As was mentioned in Section~\ref{SS-hxrpos}, it is possible to estimate some useful quantitative characteristics of the spatio-temporal evolution of paired $S^{+}$ and $S^{-}$ sources of HXR pulsations in group-1 flares. For the selected peaks of each group-1 flare, for which it was possible to reconstruct images of good quality, we calculated the following characteristics (see Figure~\ref{F-6}(b) for their visual explanation):   
\begin{itemize}
	\item Distance ($D_{S}$) between paired $S^{+}$ and $S^{-}$ sources, which gives an estimate of the length of a flaring loop (with some coefficient, \eg, $\pi/2$, if loop has a semi-circular shape) at a given HXR peak; 
	\item Shear angle ($\alpha_{sh}$) between an MPIL and a segment connecting $S^{+}$ and $S^{-}$ sources. This angle gives estimation of the shear angle of a flaring loop and, to some extent, information about deviation of magnetic field of a loop from potential state, which is a measure of its free magnetic energy; 
	\item Coordinates of an intersection point (IP) between a line segment connecting $S^{+}$ and $S^{-}$ sources and an MPIL; 
	\item Distance ($D_{IP}$) between IPs at two successive HXR peaks. This distance gives an estimate of spacing (in the image plane) between two flaring loops (two bundles of loops), from which two successive HXR peaks were emitted;
	\item Velocity ($v_{IP}$) of apparent IP displacement between two successive HXR peaks. It is calculated as the ratio between $D_{IP}$ and a time difference between two corresponding peaks. This velocity gives an estimate of the velocity of possible triggers of energy release along an MPIL direction; 
	\item Parallel ($v^{+}_{\parallel}$ and $v^{-}_{\parallel}$), perpendicular ($v^{+}_{\bot}$ and $v^{-}_{\bot}$) and total ($v^{+}$ and $v^{-}$) velocities of apparent displacement of $S^{+}$ and $S^{-}$ sources relative to an MPIL; 
	\item Dimensionless parameters $R^{+}=v^{+}_{\parallel}/v^{+}_{\bot}$ and $R^{-}=v^{-}_{\parallel}/v^{-}_{\bot}$. These parameters give information on which type of HXR source movement --- along an MPIL or perpendicular to it --- is more pronounced between two HXR peaks.  
	\end{itemize}

The Time profiles of the parameters calculated for flare No 6 are shown in Figure~\ref{F-7}. One can see the non-stationary pattern of all the calculated characteristics. However, no obvious peak-to-peak correlation between HXR pulsations and characteristics of HXR source dynamics is seen. This indicates that these parameters of HXR source dynamics are not linearly correlated with HXR intensity or that the accuracy of measurements of these parameters is not sufficiently high. 	

We also calculated mean values of these characteristics averaged over the significant peaks of each group-1 flare, for which we had images of good quality. These mean values (placed in $\left\langle \right\rangle$ parentheses) and their standard deviations ($\sigma$) are summarized in Table~\ref{T-4}. They provide a general view of the quantitative characteristics of dynamics of sources of HXR pulsations (with time difference in the range of $P \approx 8-270$ s) in flares.   

\begin{figure} 
 \centerline{\includegraphics[width=0.98\textwidth,clip=]{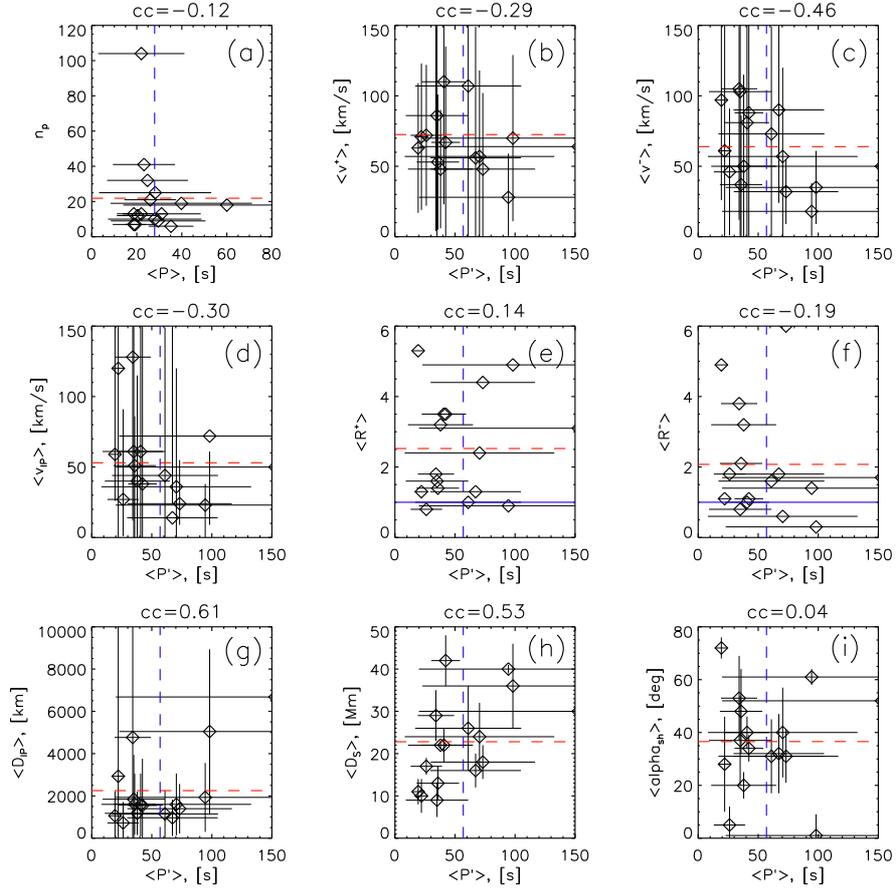}}
 \caption{Mean quantitative characteristics of dynamics of paired HXR sources of pulsations calculated for the group-1 flares (diamonds). Errors (standard deviations; see Table~\ref{T-4}) are shown by black thin solid horizontal and vertical segments. Mean values of the characteristics, averaged over all flares, are shown by red horizontal and blue vertical dashed lines. Blue horizontal solid lines on (e) and (f) show level of 1. The linear Pearson correlation coefficient ($cc$) calculated for each pair of the variables is shown at the top.} 
 \label{F-8}
 \end{figure}

To explore the general picture of the calculated characteristics we plotted them (Figure~\ref{F-8}) \textit{versus} mean time differences, corrected by the special coefficient $r_{p}$, which is different for different flares, \ie \textit{versus} $ \left\langle P^{\prime} \right\rangle = \left\langle P\right\rangle r_{p}$. For a given flare this coefficient (column 15 in Table~\ref{T-2}) is calculated as $r_{p}=n_{p}/k_{p}$, where $n_{p}$ is the number of all significant HXR peaks found in this flare (in the 50--100 keV band) and $k_{p}$ is the number of HXR pulsations of this flare, for which we could reconstruct images of good quality used to calculate the discussed characteristics. This coefficient is used to take into account that we were able to reconstruct images for only part of the found significant peaks, and that times between successive images are not real times between successive peaks. Note, that in Figure~\ref{F-8}(a) we plotted numbers of all significant peaks ($n_{p}$) \textit{versus} $\left\langle P \right\rangle$ (but not \textit{versus} $\left\langle P^{\prime}\right\rangle$ as on all other panels), because $n_{p}$ is calculated directly from RHESSI's corrected count rates as well as $\left\langle P \right\rangle$. For each pair of physical variables, shown in Figure~\ref{F-8}, we  calculated the linear Pearson correlation coefficient ($cc$). Its values are shown above corresponding panels of Figure~\ref{F-8}. One can see that, in general, there is no very strong linear correlation between the shown physical variables. Only $\left\langle D_{IP} \right\rangle$ and $\left\langle D_{S} \right\rangle$ manifest significant positive correlation ($cc \approx 0.61$ and $cc \approx 0.53$ respectively) with $\left\langle P^{\prime} \right\rangle$, indicating some links between distances between neighboring flaring loops (as well as their lengths) and mean time differences of flare HXR pulsations.  

\begin{figure} 
\centerline{\includegraphics[width=0.7\textwidth,clip=]{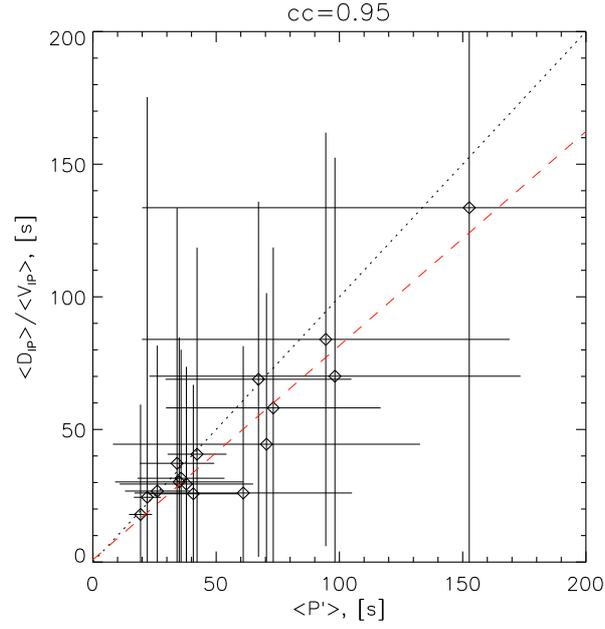}}
\caption{Relation between the ratio $\left\langle D_{IP} \right\rangle / \left\langle v_{IP} \right\rangle$ and corrected mean time difference ($\left\langle\ P^{\prime} \right\rangle $) for all group-1 flares (diamonds). Errors (standard deviations) are shown by black thin solid horizontal and vertical segments. Black thin dotted line shows the line through the center of the coordinate system with the slope $k=1$. Red dashed line shows the linear ($y=ax+b$) best-fit of the data points having the slope $a=0.81 \pm 0.18$ and $b=0.89 \pm 11.64$. The linear Pearson correlation coefficient ($cc$) between $\left\langle D_{IP} \right\rangle / \left\langle v_{IP} \right\rangle$ and $\left\langle P^{\prime} \right\rangle $ variables is shown at the top.} 
\label{F-9}
\end{figure}

However, we believe that more important finding is the high correlation (the linear Pearson correlation coefficient is $cc \approx 0.95$) between the ratio $\left\langle D_{IP} \right\rangle / \left\langle v_{IP} \right\rangle$ and the corrected mean time difference $\left\langle P^{\prime} \right\rangle $ (see Figure~\ref{F-9}). Best fitting of these data points with a linear function 
\begin{equation}
 \left\langle D_{IP} \right\rangle / \left\langle v_{IP} \right\rangle = a \left\langle P^{\prime} \right\rangle + b
\label{E-1}
\end{equation}
gave $b=0.89 \pm 11.64$ and the slope $a=0.81 \pm 0.18$, which is close to 1 within the measurement error. Very high correlation between $\left\langle D_{IP} \right\rangle / \left\langle v_{IP} \right\rangle$ and $\left\langle P^{\prime} \right\rangle $ may be caused by a combination of two reasons. The first reason is methodological, since $v_{IP}$ is calculated as the ratio between $D_{IP}$ and a time difference between two corresponding peaks, \ie $ P $. However, this can not be the only reason, because we also found the high correlation between two independently determined variables $\left\langle D_{IP} \right\rangle$ and $\left\langle P^{\prime} \right\rangle $ (see Figure~\ref{F-8}(g)). This indicates that there must be also a physical reason. We will discuss one possibility in Section~\ref{SS-ppqp}.

\section{Summary of the results and their discussion}\label{S-sumres}	

Here we summarize and discuss the main observational results of the performed analysis of HXR sources in 29 investigated solar flares (see Table~\ref{T-1}) accompanied by series of at least 4 successive HXR peaks (pulses or pulsations) in the 4-second RHESSI corrected count rates in the 50--100 keV channel. We would like to emphasize that, since the criteria of flare selection were objective enough (see Section~\ref{S-Select}), the obtained results, most probably, can be generalized to all solar flares accompanied by series of HXR pulsations, at least with characteristic time differences in the $P \approx 8-270$ s range.

\subsection{Results of the time-sequence analysis}\label{SS-pqpp}

We automatically identified significant peaks in the 4-second RHESSI 50--100 keV (as well as 25--50 keV) background-subtracted corrected count rate data in the studied events. Their general characteristics are summarized in Table~\ref{T-2}. 

The total number of significant peaks in the studied flares varied from $n_{p}^{min}=5$ to $n_{p}^{min}=103$, with the mean values of $\bar{n}_{p}^{50}=22$ and $\bar{n}_{p}^{25}=24$ in the 50--100 keV and 25--50 keV energy channels respectively.  

Time differences between successive peaks in the 50--100 keV channel varied between $P_{min}^{50}=8$ s and $P_{max}^{50}=224$ s, with the mean value of $\bar{P}^{50}=27.1$ s averaged over all flares. In the 25--50 keV channel the situation is similar: $P_{min}^{25}=8$ s, $P_{max}^{25}=272$ s, and $\bar{P}^{25}=23.0$ s respectively. 

\begin{figure}
\centerline{\includegraphics[width=0.45\textwidth,clip=]{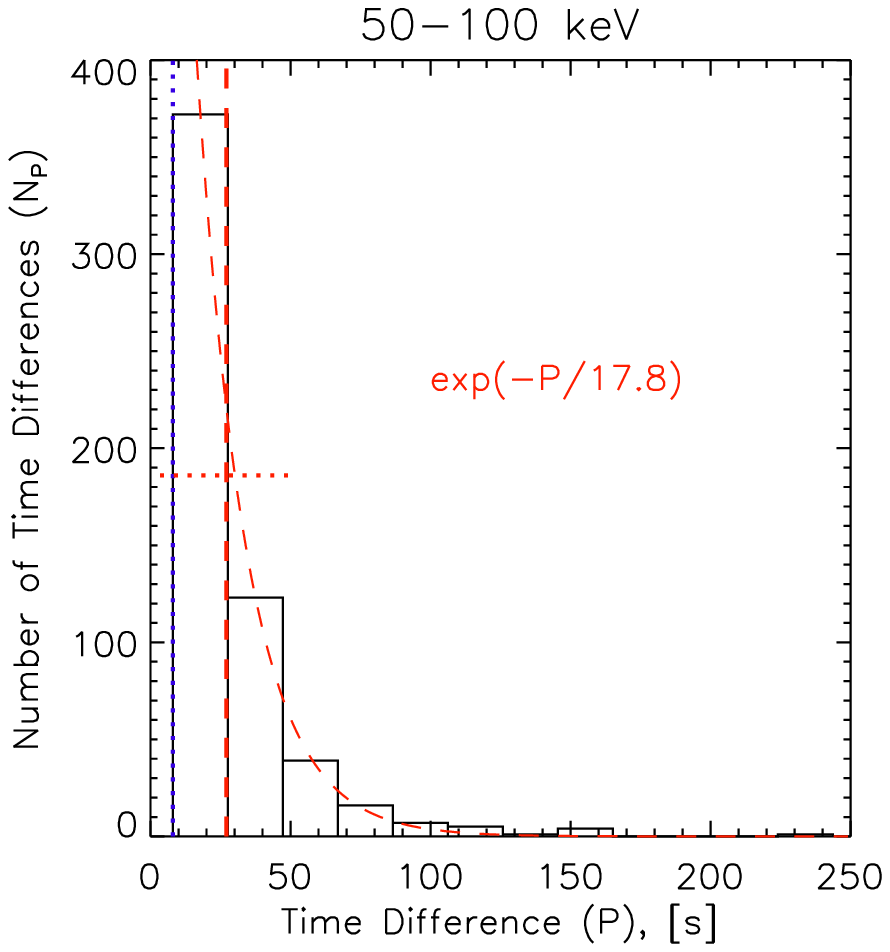}
\includegraphics[width=0.45\textwidth,clip=]{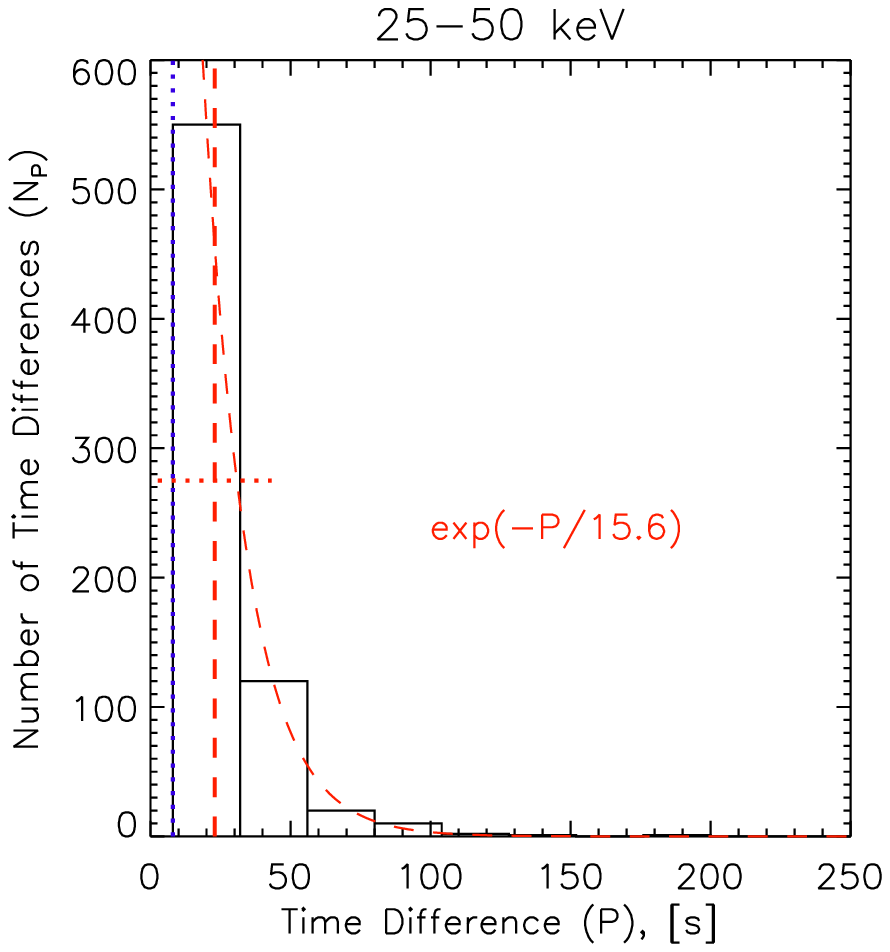}}
\vspace{-0.41\textwidth}   
\centerline{\small \bf     
\hspace{0.4\textwidth}  \color{black}{(a)}
\hspace{0.4\textwidth}  \color{black}{(b)}
\hfill}
\vspace{0.37\textwidth}    
			
\caption{Combined distributions of the time differences ($P$) found in the 4-second RHESSI's background-subtracted corrected count rates in the 50--100 keV (a) and 25--50 keV (b) channels for all 29 studied flares. Red vertical dashed line shows the averaged time difference. Red horizontal dotted line shows standard deviation of time differences at half of maxima of the distributions. Blue vertical dotted line shows the lower threshold ($P_{thr}=8$ s) of time differences, which can be found in the 4-second count rates. Red thin dashed curves show Gaussian functions found by the best-fitting of the distributions.}
\label{F-3}
\end{figure}

Distributions of time differences combined for all 29 flares have rapidly decaying exponential form ($N\left(P\right) \sim \exp \left[-P/a_{p} \right]$) with $a_{p}^{50}=17.78 \pm 0.11$ s in the 50--100 keV channel and $a_{p}^{25}=15.62 \pm 0.08$ s in the 25--50 keV channel (Figure~\ref{F-3}). Maxima of these exponential distributions are close to the lower threshold level ($P_{thr}=8$ s) of time differences that can be identified, in principle, in time profiles with the 4-second step. This indicates clearly that the 4-second sampling is not enough to resolve the majority of peaks, whose time differences are below 8 s. This is an expected result. It is well known from higher cadence observations, made in particular with the CGRO/BATSE instrument, that distributions of widths (and, consequently, of time differences) of HXR pulses have decaying exponential form in the 0.3--3.0 s range \cite{1995ApJ...447..923A}. Thus, in the present work we are restricted to physical processes occurring in flare regions on time scales greater than $P_{min}\approx P_{thr} = 8$ s. This restriction is not due to the physics of flare processes. It is caused by the limited time resolution ($dt=4$ s) of the available data sets.        

Individual distributions of time differences in the majority of inspected flares are similar in shape to the combined distributions (\textit{cf} Figures~\ref{F-2} and \ref{F-3}). They also have a decaying exponential form with maxima close to $P_{thr}=8$ s. However, we found that at least in three flares (No 6, 12, and 23) distributions of time differences have bell-like shapes. They can be approximated (though not ideally) with Gaussian functions with positions of maxima and half-widths at half-maxima corresponding to average values and standard deviations of time differences. Moreover, we found that all time differences in these flares are within the $\left\langle P \right\rangle \pm 3 \sigma_{P}$ range. This indicates that at least in these three flares time differences between successive HXR peaks are random variables distributed normally around their own average values $\left\langle P \right\rangle$. The average value of all time differences in a given flare we refer to as the mean time difference. Physically, this is the most probable time interval between successive HXR peaks of a given flare. As in the majority of flares we observed only exponential tails of distributions of time differences and could not see real maxima of these distributions (if they exist) due to the limited time resolution, we could not find real most probable time differences (\ie true mean time differences) for the majority of flares. Thus, mean time differences ($\left\langle P \right\rangle$), which we estimated, are more correctly treated as artificial (not true) characteristic time intervals between HXR peaks available for observations in the particular time scales between $P_{min} \approx P_{thr} = 8$ s and $P_{max} \approx 270$ s. 

It is worth mentioning that in the majority ($\approx 75 \%$) of the studied flares more than half of found time differences are within the $\left\langle P \right\rangle \pm 0.5 \left\langle P \right\rangle$ range (columns 5 and 10 in Table~\ref{T-2}). Intuitively, this means that in these flares spreading of time differences around their (artificial) mean values is not very high. This may be a signature of quasi-periodicity of pulsations (QPPs). However, this signature is not devoid of subjectivity and shortcomings --- as discussed in Section~\ref{S-Intro}, there is no precise definition of what constitutes quasi-periodiciy. In particular, it is not clear what exact percentage of pulsations in a given flare must be within this $\left\langle P \right\rangle \pm 0.5 \left\langle P \right\rangle$ range to be treated as quasi-periodic ones. Also, the range $\left\langle P \right\rangle \pm 0.5 \left\langle P \right\rangle$ is chosen subjectively. We can not strongly argue why the coefficient in the right hand side should be equal to $0.5$ but not to some other values (see, \eg, \citealp{2013SoPh..286..427J}, for comparison).

On the other hand, we found from wavelet and periodogram analysis, taking into account the power-law (``red noise'') properties of the data \citep{2005A&A...431..391V,2011A&A...533A..61G,2015ApJ...798..108I}, that HXR pulsations in the majority of the studied flares do not show the quasi-periodicity (see Table~\ref{T-3}). This finding is not well consistent with the result discussed in the previous paragraph. It also differs from the recent results by \inlinecite{2015SoPh..290.3625S}. However, our approach to data analysis is not equal to the approach of  \inlinecite{2015SoPh..290.3625S}, who used the time-derivatives of GOES soft X-ray data for the wavelet analysis. We performed the wavelet and periodogram analysis of almost the original RHESSI corrected count rates in the 50--100 keV channel. We only subtracted pre-flare background levels and normalized corrected count rates to its maxima in the time interval studied. We used this approach to modify the original data as little as possible. Also we need to mention that our sample of flares differs from the sample of flares studied by \inlinecite{2015SoPh..290.3625S}. It follows from the aforesaid that the result of finding of quasi-periodicity in the flare HXR data strongly depends on the chosen approach. This problem is non-trivial, and requires further and deeper study, which is out of the scope of the present paper (see, \eg, \citealp{2010PPCF...52l4009N,2011A&A...533A..61G,2015ApJ...798..108I}, for further discussions). The investigation of quasi-periodicity in the studied flares is not the main goal of the present work. Instead, our aim is to investigate dynamics of sources of HXR pulsations in flaring regions, regardless of whether pulsations are quasi-periodic or not.

 \subsection{Results of analysis of dynamics of the HXR sources}\label{SS-ppqp}
  
We investigated the spatio-temporal evolution of HXR sources relative to MPILs in parental active regions of all 29 studied flares. Unfortunately, due to low signal-to-noise ratio and some other instrumental limitations we could not identify the positions of HXR sources of all significant peaks found in time profiles of HXR emission. Nevertheless, we did this for almost half peaks of each flare (see column 12 of Table 2). The average ratio of all significant HXR peaks of each flare ($n_{p}$) to HXR peaks, for which we could find positions of HXR sources ($k_{p}$), is $\left\langle r_{p} \right\rangle = \left\langle n_{p} / k_{p} \right\rangle \approx 2.2$. This is, in principle, enough to make general conclusions about dynamics of sources of HXR pulsations (with time differences $P \approx 8-270$ s) in flares. 

First of all, we found that in all the studied flares HXR sources were not stationary. The positions of HXR sources of each subsequent pulsation differed from positions of HXR sources of previous pulsations, \ie sources of HXR emission showed apparent movement (displacement) in parental active regions from pulsation to pulsation. Generally speaking, this is an expected result, because motion of solar flare HXR sources is a well known and common phenomenon (\eg, \citealp{1998ASSL..229..273S,2005ApJ...630..561B,2006ApJ...636L.173J,2008AdSpR..41..908G,2009ApJ...693..132Y}). However, most previous work was not specifically concerned with the spatio-temporal evolution of sources of HXR pulsations. As far as we know, our study is the first such systematic (based on a large number of flares) investigation focused on dynamics of sources of HXR pulsations. Since the sample of the studied flares is large and almost unbiased, we can conclude with high probability that sources of HXR pulsations, at least with time differences $P \approx 8-270$ s, change their location from pulsation to pulsation virtually in all solar flares.     

Second, we found that all flares can be subdivided into two nominal groups by the character of the dynamics of HXR pulsation sources. Group-1 flares ($55\%$ of all events; see column 11 of Table~\ref{T-1}) showed systematic dynamics of HXR sources from pulsation to pulsation with respect to an single MPIL having relatively simple extended shape on the photosphere. By systematic dynamics we mean one of the standard types of HXR source motion or their combinations (\eg, \citealp{2005ApJ...630..561B,2009ApJ...693..132Y}). We found (see columns 14--16 of Table~\ref{T-4}) that the most common type of motion, presented almost in all flares, is parallel motion of paired HXR sources along a MPIL in the same direction. Also, we found that different types of motion were manifested simultaneously or in different time intervals during the majority of events. This is also a known phenomenon (\citealp{2008AdSpR..41..908G,2009ApJ...693..132Y}). The group-2 flares ($45\%$ of all events) manifested more chaotic displacements of HXR sources with respect to MPILs having more complicated structure. Sometimes several MPILs were presented in parental active regions of these flares. 

We need to emphasize here that subdivision of flares into these two groups is quite subjective and conditional. In some cases, it was difficult to decide unambiguously to which group we must attribute a flare. Most probably, there is no principal difference between physical mechanisms underlying the pulsatory energy release processes in flares of both groups. We found that the temporal characteristics of both groups of flares are very similar to each other. Probably, the only difference that may exist between the flares of group-1 and group-2 is the complexity of the magnetic geometries in their parental active regions. 

\begin{figure} 
\centerline{\includegraphics[width=0.96\textwidth,clip=]{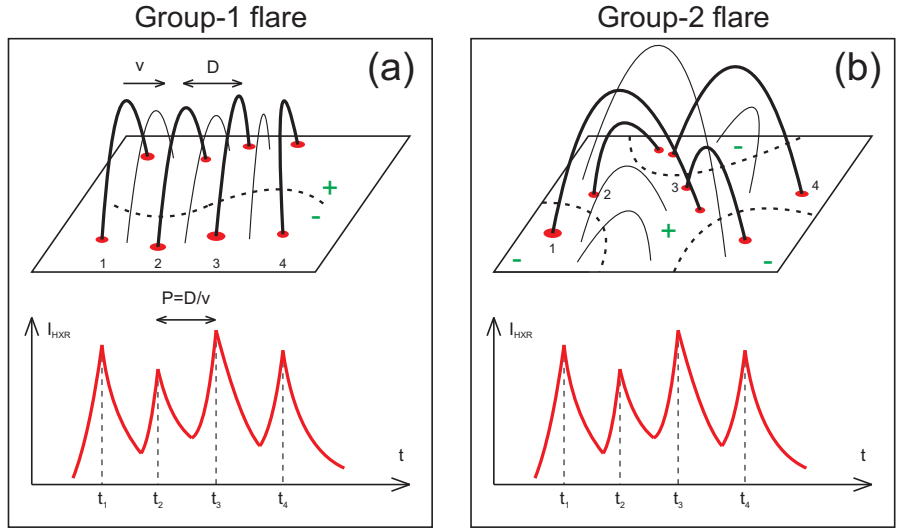}}
\caption{Cartoons generalizing observations of the group-1 (a) and group-2 (b) flares. (\textit{Top}) Morphology of flaring regions. Thick solid arc-shape curves --- flaring loops with HXR sources in their footpoints (red ellipses). Thin solid arc-shape curves --- non-flaring loops, \ie loops without detectable HXR emission. MPILs are shown by dashed curves. Signs of magnetic polarity are marked with green $+$ and $-$ symbols. (\textit{Bottom}) Time profiles of detected flare HXR emission.} 
\label{F-10}
\end{figure}

Based on these observations we suggest that the mechanism of HXR pulsations (at least with time differences of the considered range $P \approx 8-270$ s) in both groups of flares is related to successive ignition/triggering of energy release process and acceleration (injection) of electrons in different magnetic loops (or bundles of loops). The group-1 flare regions consist of loops stacked into magnetic arcades extended along an MPIL. The group-2 flare regions have more complicated magnetic structures and loops are arranged more chaotically and randomly relative to each other and to multiple MPILs. Cartoons summarizing observations of group-1 and group-2 flares are shown in Figure~\ref{F-10}. These ideas are not new. They have roots in the early works of de Jager and his colleagues (\eg, \citealp{1974spre.conf..447V,1974ASSL...42..533V,1978SoPh...58..127D,1979SoPh...64..135D}).
    
Because of the relative simplicity of magnetic geometry, it was possible to investigate the dynamics of paired HXR sources of pulsations in group-1 flares in more detail than in group-2 flares. In particular, for each group-1 flare we estimated distances ($D_{IP}$) between successively flaring loops of magnetic arcades and velocities ($v_{IP}$) of a hypothetical trigger, which could cause initiation of energy release and acceleration (injection) of electrons in the loops. In each flare $D_{IP}$ and $v_{IP}$ have different values for different pulsations (see columns 6--7 and 8--9 of Table~\ref{T-4}). Mean values of these physical parameters, averaged over all flares, are $\bar{D}_{IP} \approx 2250$ km and $\bar{v}_{IP} \approx 53$ km s$^{-1}$. These values are consistent with values estimated in some previous case studies (\eg, \citealp{2005ApJ...625L.143G,2009SoPh..258...69Z,2012ApJ...748..139I}). 

However, in our opinion, the more interesting and physically important finding is the high linear correlation between the ratio of mean values $\left\langle D_{IP} \right\rangle$ and $\left\langle v_{IP} \right\rangle$ (for each flare) and the corrected mean time difference $\left\langle P^{\prime} \right\rangle = \left\langle P \right\rangle r_{p}$ (see Figure~\ref{F-9} and Equation~\ref{E-1}). The linear slope ($a=0.81 \pm 0.18$) between $\left\langle D_{IP} \right\rangle / \left\langle v_{IP} \right\rangle$ and $\left\langle P^{\prime} \right\rangle$ variables is close to unity within the measurement errors. This means that characteristic time between successive pulsations, identified in light curves of flare HXR emission (\ie its mean time difference), is equal to characteristic time required for a trigger of energy release to propagate from one flaring loop (bundle of loops) to the next one (see Figure~\ref{F-10}). The main question that remains --- what is the trigger of energy release?

 \subsection{Erupting flux-rope as a possible trigger of energy release}\label{SS-efrtr}	

It is generally accepted that energy release in flares is related to the process of magnetic reconnection (\eg, \citealp{2002A&ARv..10..313P,2007plas.book.....S,2011LRSP....8....6S}). However, the trigger of magnetic reconnection and flaring energy release is not yet well known and is a matter of active debates. Many models were proposed to explain the pulsatory (bursty) character of energy release processes in flares (see, \eg, \citealp{2008PhyU...51.1123Z,2009SSRv..149..119N,2016SSRv..XXX..YYYZ}; and references therein). It is not our goal to discuss applicability of all these models to the observational results we found in the current study. Here we discuss briefly only one possible interpretation, which, we believe, is well consistent with the observations discussed above.     

We found that at least $88\%$ of group-1 and $85\%$ of group-2 flares were accompanied by CMEs (see column 12 of Table~\ref{T-1}). The apparent absence of CMEs in the remaining $12\%-15\%$ of flares can, in principle, be due to the location of parental active regions close to the solar disk center, according to our selection of the flares (see column 10 of Table~\ref{T-1}). It could be difficult to detect some weak and narrow CMEs, which started from parental active regions located close to the solar disk center and propagated toward the observer. Also, one needs to remember that $\approx 20\%$ of eruptions do not lead to CMEs (\eg, \citealp{2013AdSpR..51.1967S}). Due to this, it is possible that at least some of the flares, in which a CME was not detected, could also be eruptive events. Thus, the majority of the studied flares were eruptive events. Eruptive flares are related to destabilization and ejection of magnetic flux ropes from parental active regions (\eg, \citealp{2009AdSpR..43..739S,2013AdSpR..51.1967S}). Regardless of the magnetic geometry of parental active regions --- quasi-bipolar or multi-polar --- the process of flux rope eruption is associated with the interaction of the magnetic field of a flux rope with surrounding magnetic field. Since a flux rope has an elongated shape it can interact with different surrounding magnetic flux tubes (loops) of a parental active region at different times when erupting, because the surrounding magnetic field is non-homogeneous and, consequently, the process of eruption can be non-uniform along the axis of a flux rope. Interaction of different part of a flux rope with different surrounding magnetic flux tubes (loops) at different times can lead to a sequence of episodes of magnetic reconnection and acceleration of electrons in different places and, as a result, to a sequence of HXR peaks (pulsations) and apparent displacement of HXR sources from pulsation to pulsation. If a parental active region has, in general, a quite simple quasi-bipolar magnetic configuration with an elongated MPIL (see Figure~\ref{F-10}(a) as a cartoon of this situation), then the eruption of a flux rope can be of whipping-like or zipping-like types (\eg, \citealp{2005ApJ...625L.143G,2009ApJ...691.1079L}). This can explain group-1 flares. If a parental active region has more complicated, multi-polar magnetic structure with several MPILs, then an erupting flux rope will interact at different times with different loops oriented more randomly, than in the case of a quasi-bipolar active region. Thus, apparent displacements of HXR sources of pulsations will have more chaotic character in this case (see Figure~\ref{F-10}(b) as a cartoon of this situation). This can explain group-2 flares. 

The natural question is why some loops of parental active regions produce significant HXR peaks, while others do not? We do not know the precise answer to this question. We suggest that: (a) different loops of an active region initially contain different amounts of energy, which can be released during interaction with a flux rope; (b) physical conditions in some loops are more favorable for realization of more energy and/or for more efficient electron acceleration during interaction with a flux rope, than in other loops. This is quite reasonable, because we found from the observations that the physical characteristics of different flaring loops (such as their shear angle and length) differ quite significantly from each other. Also, an interesting question is what determines the observed characteristic spatial scale ($\bar{D}_{IP} \approx 2250$ km; see Figure~\ref{F-8}(g)) between the observed neighborhood flaring loops? Whether this is related to some characteristic spatial scale of magnetic field on the photosphere, \eg, the scale of solar granules (\eg, \cite{2009LRSP....6....2N}, and references therein), or if it is just a consequence of limited angular resolution of RHESSI ($d \approx 2.3^{\prime\prime}$ that corresponds to $\approx 1670$ km)? These questions require further study, which is beyond the scope of the present paper.     

It is worth noting that the slipping reconnection \citep{1995JGR...10023443P,2006SoPh..238..347A} could be a natural part of the scenario discussed above. It was shown by 3D MHD simulations \citep{2013A&A...555A..77J}, as well as with the observations of a real X-class flare \citep{2014ApJ...784..144D}, that slipping reconnection, associated with an eruption of a flux-rope, can be responsible for an apparent motion of hot flare loops' footpoints along flare ribbons with velocities of several tens of km s$^{-1}$. These velocities are consistent with the characteristic velocities of an apparent motion of the sources of the HXR pulsations (see Table~\ref{T-4}). It is also interesting to note here the recent observational indication of the possibility of the quasi-periodic regime of the slipping reconnection during an X-class flare \citep{2015ApJ...804L...8L}. In principle, this regime of slipping reconnection could be responsible for the flare QPPs (see also Section~\ref{SS-ppqp}). However, since there is no direct observational evidence of the link between slipping reconnection and the HXR pulsations we do not want to discuss this subject in more detail here. It requires further more detailed observations and simulations.

 \subsection{On the quasi-periodicity of HXR pulsations}\label{SS-ppqp}	

Finally, we touch on the problem of quasi-periodicity of flare pulsations. \\Though there is no generally accepted definition of the quasi-periodicity of flare emission, intuitively it means that times between emission peaks are not very different from each other. In the terminology of our paper this means that for a given flare all significant time differences (P) are almost the same and, thus, they are approximately equal to the mean time difference $\left\langle P \right\rangle$. In turn, we found from the observations (see Section~\ref{SS-hxrcha}) that $\left\langle D \right\rangle / \left\langle v \right\rangle \approx \left\langle P \right\rangle$, where $\left\langle D \right\rangle$ and $\left\langle v \right\rangle$ are respectively mean values of distances between neighboring flaring loops and speed of a flare trigger, averaged over all HXR peaks of a flare. Consequently, the quasi-periodicity means that the ratio $D/v \approx \left\langle P \right\rangle \approx const$ for all peaks, \ie the ratio $D/v$ can be considered as a kind of quasi-invariant for a given flare. In particular, this can be realized in the case when distances between neighboring flaring loops are almost the same and a trigger (\ie a flux-rope) moves with almost constant speed along an MPIL. Obviously, in natural magnetic systems such as a real flare region with an erupting flux-rope this situation is quite specific and probably not very common. Due to this the quasi-periodicity of HXR pulsations are not found in all flares (see Table~\ref{T-3}).

\section{Conclusion}\label{S-conclu}	

We performed systematic analysis of the spatio-temporal evolution of sources of hard X-ray (HXR) pulsations in solar flares. Our study was focused on the disk flares whose impulsive phases were accompanied by a series of more than three successive peaks (pulsations) of HXR emission detected in RHESSI's 4-second background-subtracted corrected count rates in the 50--100 keV channel. Out of 154 such pre-selected flares detected by RHESSI from February 2002 to June 2015 with characteristic time differences between successive peaks $P \approx 8-270$ s we studied 29 events with the best quality of available data sets. 

The main observational result of our analysis is that the sources of HXR pulsations in all studied flares were not stationary; they demonstrated apparent movements in parental active regions from pulsation to pulsation. We found that flares can be subdivided into two nominal groups depending on the character of dynamics of the HXR sources. Group-1 consists of 16 flares ($55\%$) with the systematic dynamics of HXR sources from pulsation to pulsation with respect to an MPIL having simple extended shape at the photosphere. Group-2 consists of 13 flares ($45\%$) with more chaotic displacements of HXR sources with respect to an MPIL having more complicated structure, and sometimes several MPILs were presented in parental active regions of such flares. 

Based on the observations we came to the conclusion that the mechanism of flare HXR pulsations (at least with time differences of the considered range, \ie $P \approx 8-270$ s) is related with successive triggering of the flare energy release process in different magnetic loops (or bundles of loops) of parental active regions. The group-1 flare regions consist of loops stacked into magnetic arcades extended along an MPIL. The group-2 flares have more complicated magnetic structures and loops are arranged more chaotically and randomly there. We would like to emphasize that the models of flare pulsations based on oscillations of magnetic loops do not explain the observational results.   

We also found that at least 14 ($88\%$) group-1 flares and 11 ($85\%$) group-2 flares were accompanied by CMEs, \ie the majority of the flares studied were eruptive events. This gives a strong indication that eruptive processes play a significant role in generation of HXR pulsations in flares. We discussed the possibility that an erupting flux rope can act as a trigger of flare energy release. If it interacts successively with different loops of a parental active region, this can lead to apparent motion of HXR sources and to a series of HXR pulsations. However, the exact mechanisms responsible for the pulsations (and their quasi-periodic character in some flares) remain unclear yet and require further more detailed investigations.    

%

%

\begin{acks}
We thank Drs. V.M.~Nakariakov, V.F.~Melnikov, S.A.~Anfinogentov, E.P.~Kontar for helpful discussions; Dr. E.G.~Kupriyanova for help with the wavelet analysis; Dr. A.R.~Inglis for help in English editing. We would like to acknowledge the anonymous referee for the useful comments that helped to improve the manuscript. We are grateful to the spacecraft teams and consortia (RHESSI, SOHO/MDI, SOHO/LASCO, SDO/HMI, GOES) and ground-based observatories (RSTN, e-Callisto), whose data were used in this study. This study was supported by the Russian Foundation for Basic Research (grants No. 16-02-00328, 16-32-00535, 15-32-50998, 15-32-21078, 14-02-00924) and by the Marie Curie FP7 PIRSES-GA-2011-295272 ``RadioSun'' Project. We are also grateful to the Specialized Research Fund for State Key Laboratories of China.  
\end{acks} 

\textbf{Disclosure of Potential Conflicts of Interest}  The authors declare that they have no conflicts of interest.

%
%
%
%
%
%



\bibliographystyle{spr-mp-sola}
\bibliography{zimovets_bib_file}  

\begin{thebibliography}{73}
\ifx\bisbn     \undefined \def\bisbn  #1{ISBN #1}\fi
\ifx\binits    \undefined \def\binits#1{#1}\fi
\ifx\bauthor   \undefined \def\bauthor#1{#1}\fi
\ifx\batitle   \undefined \def\batitle#1{#1}\fi
\ifx\bjtitle   \undefined \def\bjtitle#1{\textit{#1}}\fi
\ifx\bvolume   \undefined \def\bvolume#1{\textbf{#1}}\fi
\ifx\byear     \undefined \def\byear#1{#1}\fi
\ifx\bissue    \undefined \def\bissue#1{#1}\fi
\ifx\bfpage    \undefined \def\bfpage#1{#1}\fi
\ifx\blpage    \undefined \def\blpage #1{#1}\fi
\ifx\burl      \undefined \def\burl#1{\textsf{#1}}\fi
\ifx\href      \undefined \def\href#1#2{\textsf{#2}}\fi
\ifx\betal     \undefined \def\betal{\textit{et al.}}\fi
\ifx\bctitle   \undefined \def\bctitle#1{#1}\fi
\ifx\beditor   \undefined \def\beditor#1{#1}\fi
\ifx\bbtitle   \undefined \def\bbtitle#1{\textit{#1}}\fi
\ifx\bedition  \undefined \def\bedition#1{#1}\fi
\ifx\bseriesno \undefined \def\bseriesno#1{\textbf{#1}}\fi
\ifx\blocation \undefined \def\blocation#1{#1}\fi
\ifx\bsertitle \undefined \def\bsertitle#1{\textit{#1}}\fi
\ifx\bsnm      \undefined \def\bsnm#1{#1}\fi
\ifx\bsuffix   \undefined \def\bsuffix#1{#1}\fi
\ifx\bparticle \undefined \def\bparticle#1{#1}\fi
\ifx\barticle  \undefined \def\barticle#1{}\fi
\ifx\binstitute  \undefined \def\binstitute#1{#1}\fi
\ifx\bpublisher  \undefined \def\bpublisher#1{#1}\fi
\ifx\doiurl    \undefined
  \def\doiurl#1{\href{http://dx.doi.org/#1}{\textsf{DOI}}}\fi
\ifx\arxivurl  \undefined
  \def\arxivurl#1{\href{http://arxiv.org/abs/#1}{\textsf{arXiv}}}\fi
\ifx\adsurl    \undefined
  \def\adsurl#1{\href{http://adsabs.harvard.edu/abs/#1}{\textsf{ADS}}}\fi
\ifx\botherref \undefined \def\botherref#1{}\fi
\ifx\url       \undefined \def\url#1{\textsf{#1}}\fi
\ifx\bchapter  \undefined \def\bchapter#1{}\fi
\ifx\bbook     \undefined \def\bbook#1{}\fi
\ifx\bcomment  \undefined \def\bcomment#1{#1}\fi
\ifx\oauthor   \undefined \def\oauthor#1{#1}\fi
\ifx\citeauthoryear \undefined\def \citeauthoryear#1{#1}\fi
\ifx\endbibitem\undefined \def\endbibitem{}\fi
\ifx\bconflocation  \undefined \def\bconflocation#1{#1} \fi

\bibitem[\protect\citeauthoryear{{Artemyev} and
  {Zimovets}}{2012}]{2012SoPh..277..283A}
\begin{barticle}
\bauthor{\bsnm{{Artemyev}}, \binits{A.}},
\bauthor{\bsnm{{Zimovets}}, \binits{I.}}:
\byear{2012},
\batitle{{Stability of Current Sheets in the Solar Corona}}.
\bjtitle{\solphys}
\bvolume{277},
\bfpage{283}.
\doiurl{10.1007/s11207-011-9908-1}.
\adsurl{2012SoPh..277..283A}.
\end{barticle}
\endbibitem

\bibitem[\protect\citeauthoryear{{Aschwanden}}{1987}]{1987SoPh..111..113A}
\begin{barticle}
\bauthor{\bsnm{{Aschwanden}}, \binits{M.J.}}:
\byear{1987},
\batitle{{Theory of radio pulsations in coronal loops}}.
\bjtitle{\solphys}
\bvolume{111},
\bfpage{113}.
\doiurl{10.1007/BF00145445}.
\adsurl{1987SoPh..111..113A}.
\end{barticle}
\endbibitem

\bibitem[\protect\citeauthoryear{{Aschwanden}}{2002}]{2002SSRv..101....1A}
\begin{barticle}
\bauthor{\bsnm{{Aschwanden}}, \binits{M.J.}}:
\byear{2002},
\batitle{{Particle acceleration and kinematics in solar flares - A Synthesis of
  Recent Observations and Theoretical Concepts (Invited Review)}}.
\bjtitle{\ssr}
\bvolume{101},
\bfpage{1}.
\doiurl{10.1023/A:1019712124366}.
\adsurl{2002SSRv..101....1A}.
\end{barticle}
\endbibitem

\bibitem[\protect\citeauthoryear{{Aschwanden}, {Schwartz}, and
  {Alt}}{1995}]{1995ApJ...447..923A}
\begin{barticle}
\bauthor{\bsnm{{Aschwanden}}, \binits{M.J.}},
\bauthor{\bsnm{{Schwartz}}, \binits{R.A.}},
\bauthor{\bsnm{{Alt}}, \binits{D.M.}}:
\byear{1995},
\batitle{{Electron Time-of-Flight Differences in Solar Flares}}.
\bjtitle{\apj}
\bvolume{447},
\bfpage{923}.
\doiurl{10.1086/175930}.
\adsurl{1995ApJ...447..923A}.
\end{barticle}
\endbibitem

\bibitem[\protect\citeauthoryear{{Aschwanden}
  \textit{et~al.}}{1998}]{1998ApJ...505..941A}
\begin{barticle}
\bauthor{\bsnm{{Aschwanden}}, \binits{M.J.}},
\bauthor{\bsnm{{Kliem}}, \binits{B.}},
\bauthor{\bsnm{{Schwarz}}, \binits{U.}},
\bauthor{\bsnm{{Kurths}}, \binits{J.}},
\bauthor{\bsnm{{Dennis}}, \binits{B.R.}},
\bauthor{\bsnm{{Schwartz}}, \binits{R.A.}}:
\byear{1998},
\batitle{{Wavelet Analysis of Solar Flare Hard X-Rays}}.
\bjtitle{\apj}
\bvolume{505},
\bfpage{941}.
\doiurl{10.1086/306200}.
\adsurl{1998ApJ...505..941A}.
\end{barticle}
\endbibitem

\bibitem[\protect\citeauthoryear{{Aulanier}
  \textit{et~al.}}{2006}]{2006SoPh..238..347A}
\begin{barticle}
\bauthor{\bsnm{{Aulanier}}, \binits{G.}},
\bauthor{\bsnm{{Pariat}}, \binits{E.}},
\bauthor{\bsnm{{D{\'e}moulin}}, \binits{P.}},
\bauthor{\bsnm{{DeVore}}, \binits{C.R.}}:
\byear{2006},
\batitle{{Slip-Running Reconnection in Quasi-Separatrix Layers}}.
\bjtitle{\solphys}
\bvolume{238},
\bfpage{347}.
\doiurl{10.1007/s11207-006-0230-2}.
\adsurl{2006SoPh..238..347A}.
\end{barticle}
\endbibitem

\bibitem[\protect\citeauthoryear{{Bogachev}
  \textit{et~al.}}{2005}]{2005ApJ...630..561B}
\begin{barticle}
\bauthor{\bsnm{{Bogachev}}, \binits{S.A.}},
\bauthor{\bsnm{{Somov}}, \binits{B.V.}},
\bauthor{\bsnm{{Kosugi}}, \binits{T.}},
\bauthor{\bsnm{{Sakao}}, \binits{T.}}:
\byear{2005},
\batitle{{The Motions of the Hard X-Ray Sources in Solar Flares: Images and
  Statistics}}.
\bjtitle{\apj}
\bvolume{630},
\bfpage{561}.
\doiurl{10.1086/431918}.
\adsurl{2005ApJ...630..561B}.
\end{barticle}
\endbibitem

\bibitem[\protect\citeauthoryear{{Brown} and
  {Hoyng}}{1975}]{1975ApJ...200..734B}
\begin{barticle}
\bauthor{\bsnm{{Brown}}, \binits{J.C.}},
\bauthor{\bsnm{{Hoyng}}, \binits{P.}}:
\byear{1975},
\batitle{{Betatron acceleration in a large solar hard X-ray burst}}.
\bjtitle{\apj}
\bvolume{200},
\bfpage{734}.
\doiurl{10.1086/153845}.
\adsurl{1975ApJ...200..734B}.
\end{barticle}
\endbibitem

\bibitem[\protect\citeauthoryear{{de Jager}}{1979}]{1979SoPh...64..135D}
\begin{barticle}
\bauthor{\bsnm{{de Jager}}, \binits{C.}}:
\byear{1979},
\batitle{{On the seats of elementary flare bursts}}.
\bjtitle{\solphys}
\bvolume{64},
\bfpage{135}.
\doiurl{10.1007/BF00151122}.
\adsurl{1979SoPh...64..135D}.
\end{barticle}
\endbibitem

\bibitem[\protect\citeauthoryear{{de Jager} and {de
  Jonge}}{1978}]{1978SoPh...58..127D}
\begin{barticle}
\bauthor{\bsnm{{de Jager}}, \binits{C.}},
\bauthor{\bsnm{{de Jonge}}, \binits{G.}}:
\byear{1978},
\batitle{{Properties of elementary flare bursts}}.
\bjtitle{\solphys}
\bvolume{58},
\bfpage{127}.
\doiurl{10.1007/BF00152559}.
\adsurl{1978SoPh...58..127D}.
\end{barticle}
\endbibitem

\bibitem[\protect\citeauthoryear{{Dennis}}{1988}]{1988SoPh..118...49D}
\begin{barticle}
\bauthor{\bsnm{{Dennis}}, \binits{B.R.}}:
\byear{1988},
\batitle{{Solar flare hard X-ray observations}}.
\bjtitle{\solphys}
\bvolume{118},
\bfpage{49}.
\doiurl{10.1007/BF00148588}.
\adsurl{1988SoPh..118...49D}.
\end{barticle}
\endbibitem

\bibitem[\protect\citeauthoryear{{Dolla}
  \textit{et~al.}}{2012}]{2012ApJ...749L..16D}
\begin{barticle}
\bauthor{\bsnm{{Dolla}}, \binits{L.}},
\bauthor{\bsnm{{Marqu{\'e}}}, \binits{C.}},
\bauthor{\bsnm{{Seaton}}, \binits{D.B.}},
\bauthor{\bsnm{{Van Doorsselaere}}, \binits{T.}},
\bauthor{\bsnm{{Dominique}}, \binits{M.}},
\bauthor{\bsnm{{Berghmans}}, \binits{D.}},
\bauthor{\bsnm{{Cabanas}}, \binits{C.}},
\bauthor{\bsnm{{De Groof}}, \binits{A.}},
\bauthor{\bsnm{{Schmutz}}, \binits{W.}},
\bauthor{\bsnm{{Verdini}}, \binits{A.}},
\bauthor{\bsnm{{West}}, \binits{M.J.}},
\bauthor{\bsnm{{Zender}}, \binits{J.}},
\bauthor{\bsnm{{Zhukov}}, \binits{A.N.}}:
\byear{2012},
\batitle{{Time Delays in Quasi-periodic Pulsations Observed during the X2.2
  Solar Flare on 2011 February 15}}.
\bjtitle{\apjl}
\bvolume{749},
\bfpage{L16}.
\doiurl{10.1088/2041-8205/749/1/L16}.
\adsurl{2012ApJ...749L..16D}.
\end{barticle}
\endbibitem

\bibitem[\protect\citeauthoryear{{Dud{\'{\i}}k}
  \textit{et~al.}}{2014}]{2014ApJ...784..144D}
\begin{barticle}
\bauthor{\bsnm{{Dud{\'{\i}}k}}, \binits{J.}},
\bauthor{\bsnm{{Janvier}}, \binits{M.}},
\bauthor{\bsnm{{Aulanier}}, \binits{G.}},
\bauthor{\bsnm{{Del Zanna}}, \binits{G.}},
\bauthor{\bsnm{{Karlick{\'y}}}, \binits{M.}},
\bauthor{\bsnm{{Mason}}, \binits{H.E.}},
\bauthor{\bsnm{{Schmieder}}, \binits{B.}}:
\byear{2014},
\batitle{{Slipping Magnetic Reconnection during an X-class Solar Flare Observed
  by SDO/AIA}}.
\bjtitle{\apj}
\bvolume{784},
\bfpage{144}.
\doiurl{10.1088/0004-637X/784/2/144}.
\adsurl{2014ApJ...784..144D}.
\end{barticle}
\endbibitem

\bibitem[\protect\citeauthoryear{{Emslie}}{1981}]{1981ApL....22..171E}
\begin{barticle}
\bauthor{\bsnm{{Emslie}}, \binits{A.G.}}:
\byear{1981},
\batitle{{An interacting loop model for solar flare bursts.}}
\bjtitle{\aplett}
\bvolume{22},
\bfpage{171}.
\adsurl{1981ApL....22..171E}.
\end{barticle}
\endbibitem

\bibitem[\protect\citeauthoryear{{Fletcher} and
  {Hudson}}{2002}]{2002SoPh..210..307F}
\begin{barticle}
\bauthor{\bsnm{{Fletcher}}, \binits{L.}},
\bauthor{\bsnm{{Hudson}}, \binits{H.S.}}:
\byear{2002},
\batitle{{Spectral and Spatial Variations of Flare Hard X-ray Footpoints}}.
\bjtitle{\solphys}
\bvolume{210},
\bfpage{307}.
\doiurl{10.1023/A:1022479610710}.
\adsurl{2002SoPh..210..307F}.
\end{barticle}
\endbibitem

\bibitem[\protect\citeauthoryear{{Foullon}
  \textit{et~al.}}{2005}]{2005A&A...440L..59F}
\begin{barticle}
\bauthor{\bsnm{{Foullon}}, \binits{C.}},
\bauthor{\bsnm{{Verwichte}}, \binits{E.}},
\bauthor{\bsnm{{Nakariakov}}, \binits{V.M.}},
\bauthor{\bsnm{{Fletcher}}, \binits{L.}}:
\byear{2005},
\batitle{{X-ray quasi-periodic pulsations in solar flares as
  magnetohydrodynamic oscillations}}.
\bjtitle{\aap}
\bvolume{440},
\bfpage{L59}.
\doiurl{10.1051/0004-6361:200500169}.
\adsurl{2005A\%26A...440L..59F}.
\end{barticle}
\endbibitem

\bibitem[\protect\citeauthoryear{{Gan}, {Li}, and
  {Miroshnichenko}}{2008}]{2008AdSpR..41..908G}
\begin{barticle}
\bauthor{\bsnm{{Gan}}, \binits{W.Q.}},
\bauthor{\bsnm{{Li}}, \binits{Y.P.}},
\bauthor{\bsnm{{Miroshnichenko}}, \binits{L.I.}}:
\byear{2008},
\batitle{{On the motions of RHESSI flare footpoints}}.
\bjtitle{Advances in Space Research}
\bvolume{41},
\bfpage{908}.
\doiurl{10.1016/j.asr.2007.05.001}.
\adsurl{2008AdSpR..41..908G}.
\end{barticle}
\endbibitem

\bibitem[\protect\citeauthoryear{{Goddard}
  \textit{et~al.}}{2016}]{2016A&A...585A.137G}
\begin{barticle}
\bauthor{\bsnm{{Goddard}}, \binits{C.R.}},
\bauthor{\bsnm{{Nistic{\`o}}}, \binits{G.}},
\bauthor{\bsnm{{Nakariakov}}, \binits{V.M.}},
\bauthor{\bsnm{{Zimovets}}, \binits{I.V.}}:
\byear{2016},
\batitle{{A statistical study of decaying kink oscillations detected using
  SDO/AIA}}.
\bjtitle{\aap}
\bvolume{585},
\bfpage{A137}.
\doiurl{10.1051/0004-6361/201527341}.
\adsurl{2016A\%26A...585A.137G}.
\end{barticle}
\endbibitem

\bibitem[\protect\citeauthoryear{{Grigis} and
  {Benz}}{2005}]{2005ApJ...625L.143G}
\begin{barticle}
\bauthor{\bsnm{{Grigis}}, \binits{P.C.}},
\bauthor{\bsnm{{Benz}}, \binits{A.O.}}:
\byear{2005},
\batitle{{The Evolution of Reconnection along an Arcade of Magnetic Loops}}.
\bjtitle{\apjl}
\bvolume{625},
\bfpage{L143}.
\doiurl{10.1086/431147}.
\adsurl{2005ApJ...625L.143G}.
\end{barticle}
\endbibitem

\bibitem[\protect\citeauthoryear{{Gruber}
  \textit{et~al.}}{2011}]{2011A&A...533A..61G}
\begin{barticle}
\bauthor{\bsnm{{Gruber}}, \binits{D.}},
\bauthor{\bsnm{{Lachowicz}}, \binits{P.}},
\bauthor{\bsnm{{Bissaldi}}, \binits{E.}},
\bauthor{\bsnm{{Briggs}}, \binits{M.S.}},
\bauthor{\bsnm{{Connaughton}}, \binits{V.}},
\bauthor{\bsnm{{Greiner}}, \binits{J.}},
\bauthor{\bsnm{{van der Horst}}, \binits{A.J.}},
\bauthor{\bsnm{{Kanbach}}, \binits{G.}},
\bauthor{\bsnm{{Rau}}, \binits{A.}},
\bauthor{\bsnm{{Bhat}}, \binits{P.N.}},
\bauthor{\bsnm{{Diehl}}, \binits{R.}},
\bauthor{\bsnm{{von Kienlin}}, \binits{A.}},
\bauthor{\bsnm{{Kippen}}, \binits{R.M.}},
\bauthor{\bsnm{{Meegan}}, \binits{C.A.}},
\bauthor{\bsnm{{Paciesas}}, \binits{W.S.}},
\bauthor{\bsnm{{Preece}}, \binits{R.D.}},
\bauthor{\bsnm{{Wilson-Hodge}}, \binits{C.}}:
\byear{2011},
\batitle{{Quasi-periodic pulsations in solar flares: new clues from the Fermi
  Gamma-Ray Burst Monitor}}.
\bjtitle{\aap}
\bvolume{533},
\bfpage{A61}.
\doiurl{10.1051/0004-6361/201117077}.
\adsurl{2011A\%26A...533A..61G}.
\end{barticle}
\endbibitem

\bibitem[\protect\citeauthoryear{{Hood}
  \textit{et~al.}}{2016}]{2016ApJ...817....5H}
\begin{barticle}
\bauthor{\bsnm{{Hood}}, \binits{A.W.}},
\bauthor{\bsnm{{Cargill}}, \binits{P.J.}},
\bauthor{\bsnm{{Browning}}, \binits{P.K.}},
\bauthor{\bsnm{{Tam}}, \binits{K.V.}}:
\byear{2016},
\batitle{{An MHD Avalanche in a Multi-threaded Coronal Loop.}}
\bjtitle{\apj}
\bvolume{817},
\bfpage{5}.
\doiurl{10.3847/0004-637X/817/1/5}.
\adsurl{2016ApJ...817....5H}.
\end{barticle}
\endbibitem

\bibitem[\protect\citeauthoryear{{Hurford}
  \textit{et~al.}}{2002}]{2002SoPh..210...61H}
\begin{barticle}
\bauthor{\bsnm{{Hurford}}, \binits{G.J.}},
\bauthor{\bsnm{{Schmahl}}, \binits{E.J.}},
\bauthor{\bsnm{{Schwartz}}, \binits{R.A.}},
\bauthor{\bsnm{{Conway}}, \binits{A.J.}},
\bauthor{\bsnm{{Aschwanden}}, \binits{M.J.}},
\bauthor{\bsnm{{Csillaghy}}, \binits{A.}},
\bauthor{\bsnm{{Dennis}}, \binits{B.R.}},
\bauthor{\bsnm{{Johns-Krull}}, \binits{C.}},
\bauthor{\bsnm{{Krucker}}, \binits{S.}},
\bauthor{\bsnm{{Lin}}, \binits{R.P.}},
\bauthor{\bsnm{{McTiernan}}, \binits{J.}},
\bauthor{\bsnm{{Metcalf}}, \binits{T.R.}},
\bauthor{\bsnm{{Sato}}, \binits{J.}},
\bauthor{\bsnm{{Smith}}, \binits{D.M.}}:
\byear{2002},
\batitle{{The RHESSI Imaging Concept}}.
\bjtitle{\solphys}
\bvolume{210},
\bfpage{61}.
\doiurl{10.1023/A:1022436213688}.
\adsurl{2002SoPh..210...61H}.
\end{barticle}
\endbibitem

\bibitem[\protect\citeauthoryear{{Inglis} and
  {Dennis}}{2012}]{2012ApJ...748..139I}
\begin{barticle}
\bauthor{\bsnm{{Inglis}}, \binits{A.R.}},
\bauthor{\bsnm{{Dennis}}, \binits{B.R.}}:
\byear{2012},
\batitle{{The Relationship between Hard X-Ray Pulse Timings and the Locations
  of Footpoint Sources during Solar Flares}}.
\bjtitle{\apj}
\bvolume{748},
\bfpage{139}.
\doiurl{10.1088/0004-637X/748/2/139}.
\adsurl{2012ApJ...748..139I}.
\end{barticle}
\endbibitem

\bibitem[\protect\citeauthoryear{{Inglis} and
  {Gilbert}}{2013}]{2013ApJ...777...30I}
\begin{barticle}
\bauthor{\bsnm{{Inglis}}, \binits{A.R.}},
\bauthor{\bsnm{{Gilbert}}, \binits{H.R.}}:
\byear{2013},
\batitle{{Hard X-Ray and Ultraviolet Emission during the 2011 June 7 Solar
  Flare}}.
\bjtitle{\apj}
\bvolume{777},
\bfpage{30}.
\doiurl{10.1088/0004-637X/777/1/30}.
\adsurl{2013ApJ...777...30I}.
\end{barticle}
\endbibitem

\bibitem[\protect\citeauthoryear{{Inglis}, {Ireland}, and
  {Dominique}}{2015}]{2015ApJ...798..108I}
\begin{barticle}
\bauthor{\bsnm{{Inglis}}, \binits{A.R.}},
\bauthor{\bsnm{{Ireland}}, \binits{J.}},
\bauthor{\bsnm{{Dominique}}, \binits{M.}}:
\byear{2015},
\batitle{{Quasi-periodic Pulsations in Solar and Stellar Flares: Re-evaluating
  their Nature in the Context of Power-law Flare Fourier Spectra}}.
\bjtitle{\apj}
\bvolume{798},
\bfpage{108}.
\doiurl{10.1088/0004-637X/798/2/108}.
\adsurl{2015ApJ...798..108I}.
\end{barticle}
\endbibitem

\bibitem[\protect\citeauthoryear{{Jakimiec} and
  {Tomczak}}{2013}]{2013SoPh..286..427J}
\begin{barticle}
\bauthor{\bsnm{{Jakimiec}}, \binits{J.}},
\bauthor{\bsnm{{Tomczak}}, \binits{M.}}:
\byear{2013},
\batitle{{Quasi-periodic Variations in the Hard X-ray Emission of a Large
  Arcade Flare}}.
\bjtitle{\solphys}
\bvolume{286},
\bfpage{427}.
\doiurl{10.1007/s11207-013-0275-y}.
\adsurl{2013SoPh..286..427J}.
\end{barticle}
\endbibitem

\bibitem[\protect\citeauthoryear{{Jakimiec} and
  {Tomczak}}{2014}]{2014SoPh..289.2073J}
\begin{barticle}
\bauthor{\bsnm{{Jakimiec}}, \binits{J.}},
\bauthor{\bsnm{{Tomczak}}, \binits{M.}}:
\byear{2014},
\batitle{{Investigation of the X-Ray Emission of the Large Arcade Flare of 2
  March 1993}}.
\bjtitle{\solphys}
\bvolume{289},
\bfpage{2073}.
\doiurl{10.1007/s11207-013-0463-9}.
\adsurl{2014SoPh..289.2073J}.
\end{barticle}
\endbibitem

\bibitem[\protect\citeauthoryear{{Janvier}
  \textit{et~al.}}{2013}]{2013A&A...555A..77J}
\begin{barticle}
\bauthor{\bsnm{{Janvier}}, \binits{M.}},
\bauthor{\bsnm{{Aulanier}}, \binits{G.}},
\bauthor{\bsnm{{Pariat}}, \binits{E.}},
\bauthor{\bsnm{{D{\'e}moulin}}, \binits{P.}}:
\byear{2013},
\batitle{{The standard flare model in three dimensions. III. Slip-running
  reconnection properties}}.
\bjtitle{\aap}
\bvolume{555},
\bfpage{A77}.
\doiurl{10.1051/0004-6361/201321164}.
\adsurl{2013A\%26A...555A..77J}.
\end{barticle}
\endbibitem

\bibitem[\protect\citeauthoryear{{Ji}
  \textit{et~al.}}{2006}]{2006ApJ...636L.173J}
\begin{barticle}
\bauthor{\bsnm{{Ji}}, \binits{H.}},
\bauthor{\bsnm{{Huang}}, \binits{G.}},
\bauthor{\bsnm{{Wang}}, \binits{H.}},
\bauthor{\bsnm{{Zhou}}, \binits{T.}},
\bauthor{\bsnm{{Li}}, \binits{Y.}},
\bauthor{\bsnm{{Zhang}}, \binits{Y.}},
\bauthor{\bsnm{{Song}}, \binits{M.}}:
\byear{2006},
\batitle{{Converging Motion of H{$\alpha$} Conjugate Kernels: The Signature of
  Fast Relaxation of a Sheared Magnetic Field}}.
\bjtitle{\apjl}
\bvolume{636},
\bfpage{L173}.
\doiurl{10.1086/500203}.
\adsurl{2006ApJ...636L.173J}.
\end{barticle}
\endbibitem

\bibitem[\protect\citeauthoryear{{Kliem}, {Karlick{\'y}}, and
  {Benz}}{2000}]{2000A&A...360..715K}
\begin{barticle}
\bauthor{\bsnm{{Kliem}}, \binits{B.}},
\bauthor{\bsnm{{Karlick{\'y}}}, \binits{M.}},
\bauthor{\bsnm{{Benz}}, \binits{A.O.}}:
\byear{2000},
\batitle{{Solar flare radio pulsations as a signature of dynamic magnetic
  reconnection}}.
\bjtitle{\aap}
\bvolume{360},
\bfpage{715}.
\adsurl{2000A\%26A...360..715K}.
\end{barticle}
\endbibitem

\bibitem[\protect\citeauthoryear{{Kosugi}
  \textit{et~al.}}{1991}]{1991SoPh..136...17K}
\begin{barticle}
\bauthor{\bsnm{{Kosugi}}, \binits{T.}},
\bauthor{\bsnm{{Makishima}}, \binits{K.}},
\bauthor{\bsnm{{Murakami}}, \binits{T.}},
\bauthor{\bsnm{{Sakao}}, \binits{T.}},
\bauthor{\bsnm{{Dotani}}, \binits{T.}},
\bauthor{\bsnm{{Inda}}, \binits{M.}},
\bauthor{\bsnm{{Kai}}, \binits{K.}},
\bauthor{\bsnm{{Masuda}}, \binits{S.}},
\bauthor{\bsnm{{Nakajima}}, \binits{H.}},
\bauthor{\bsnm{{Ogawara}}, \binits{Y.}},
\bauthor{\bsnm{{Sawa}}, \binits{M.}},
\bauthor{\bsnm{{Shibasaki}}, \binits{K.}}:
\byear{1991},
\batitle{{The Hard X-ray Telescope (HXT) for the SOLAR-A Mission}}.
\bjtitle{\solphys}
\bvolume{136},
\bfpage{17}.
\doiurl{10.1007/BF00151693}.
\adsurl{1991SoPh..136...17K}.
\end{barticle}
\endbibitem

\bibitem[\protect\citeauthoryear{{Krucker}, {Hurford}, and
  {Lin}}{2003}]{2003ApJ...595L.103K}
\begin{barticle}
\bauthor{\bsnm{{Krucker}}, \binits{S.}},
\bauthor{\bsnm{{Hurford}}, \binits{G.J.}},
\bauthor{\bsnm{{Lin}}, \binits{R.P.}}:
\byear{2003},
\batitle{{Hard X-Ray Source Motions in the 2002 July 23 Gamma-Ray Flare}}.
\bjtitle{\apjl}
\bvolume{595},
\bfpage{L103}.
\doiurl{10.1086/378840}.
\adsurl{2003ApJ...595L.103K}.
\end{barticle}
\endbibitem

\bibitem[\protect\citeauthoryear{{Kupriyanova}
  \textit{et~al.}}{2010}]{2010SoPh..267..329K}
\begin{barticle}
\bauthor{\bsnm{{Kupriyanova}}, \binits{E.G.}},
\bauthor{\bsnm{{Melnikov}}, \binits{V.F.}},
\bauthor{\bsnm{{Nakariakov}}, \binits{V.M.}},
\bauthor{\bsnm{{Shibasaki}}, \binits{K.}}:
\byear{2010},
\batitle{{Types of Microwave Quasi-Periodic Pulsations in Single Flaring
  Loops}}.
\bjtitle{\solphys}
\bvolume{267},
\bfpage{329}.
\doiurl{10.1007/s11207-010-9642-0}.
\adsurl{2010SoPh..267..329K}.
\end{barticle}
\endbibitem

\bibitem[\protect\citeauthoryear{{Li}, {Ning}, and
  {Zhang}}{2015}]{2015ApJ...807...72L}
\begin{barticle}
\bauthor{\bsnm{{Li}}, \binits{D.}},
\bauthor{\bsnm{{Ning}}, \binits{Z.J.}},
\bauthor{\bsnm{{Zhang}}, \binits{Q.M.}}:
\byear{2015},
\batitle{{Imaging and Spectral Observations of Quasi-periodic Pulsations in a
  Solar Flare}}.
\bjtitle{\apj}
\bvolume{807},
\bfpage{72}.
\doiurl{10.1088/0004-637X/807/1/72}.
\adsurl{2015ApJ...807...72L}.
\end{barticle}
\endbibitem

\bibitem[\protect\citeauthoryear{{Li} and {Zhang}}{2015}]{2015ApJ...804L...8L}
\begin{barticle}
\bauthor{\bsnm{{Li}}, \binits{T.}},
\bauthor{\bsnm{{Zhang}}, \binits{J.}}:
\byear{2015},
\batitle{{Quasi-periodic Slipping Magnetic Reconnection During an X-class Solar
  Flare Observed by the Solar Dynamics Observatory and Interface Region Imaging
  Spectrograph}}.
\bjtitle{\apjl}
\bvolume{804},
\bfpage{L8}.
\doiurl{10.1088/2041-8205/804/1/L8}.
\adsurl{2015ApJ...804L...8L}.
\end{barticle}
\endbibitem

\bibitem[\protect\citeauthoryear{{Li} and {Gan}}{2008}]{2008SoPh..247...77L}
\begin{barticle}
\bauthor{\bsnm{{Li}}, \binits{Y.P.}},
\bauthor{\bsnm{{Gan}}, \binits{W.Q.}}:
\byear{2008},
\batitle{{Observational Studies of the X-Ray Quasi-Periodic Oscillations of a
  Solar Flare}}.
\bjtitle{\solphys}
\bvolume{247},
\bfpage{77}.
\doiurl{10.1007/s11207-007-9092-5}.
\adsurl{2008SoPh..247...77L}.
\end{barticle}
\endbibitem

\bibitem[\protect\citeauthoryear{{Lin}
  \textit{et~al.}}{2002}]{2002SoPh..210....3L}
\begin{barticle}
\bauthor{\bsnm{{Lin}}, \binits{R.P.}},
\bauthor{\bsnm{{Dennis}}, \binits{B.R.}},
\bauthor{\bsnm{{Hurford}}, \binits{G.J.}},
\bauthor{\bsnm{{Smith}}, \binits{D.M.}},
\bauthor{\bsnm{{Zehnder}}, \binits{A.}},
\bauthor{\bsnm{{Harvey}}, \binits{P.R.}},
\bauthor{\bsnm{{Curtis}}, \binits{D.W.}},
\bauthor{\bsnm{{Pankow}}, \binits{D.}},
\bauthor{\bsnm{{Turin}}, \binits{P.}},
\bauthor{\bsnm{{Bester}}, \binits{M.}},
\bauthor{\bsnm{{Csillaghy}}, \binits{A.}},
\bauthor{\bsnm{{Lewis}}, \binits{M.}},
\bauthor{\bsnm{{Madden}}, \binits{N.}},
\bauthor{\bsnm{{van Beek}}, \binits{H.F.}},
\bauthor{\bsnm{{Appleby}}, \binits{M.}},
\bauthor{\bsnm{{Raudorf}}, \binits{T.}},
\bauthor{\bsnm{{McTiernan}}, \binits{J.}},
\bauthor{\bsnm{{Ramaty}}, \binits{R.}},
\bauthor{\bsnm{{Schmahl}}, \binits{E.}},
\bauthor{\bsnm{{Schwartz}}, \binits{R.}},
\bauthor{\bsnm{{Krucker}}, \binits{S.}},
\bauthor{\bsnm{{Abiad}}, \binits{R.}},
\bauthor{\bsnm{{Quinn}}, \binits{T.}},
\bauthor{\bsnm{{Berg}}, \binits{P.}},
\bauthor{\bsnm{{Hashii}}, \binits{M.}},
\bauthor{\bsnm{{Sterling}}, \binits{R.}},
\bauthor{\bsnm{{Jackson}}, \binits{R.}},
\bauthor{\bsnm{{Pratt}}, \binits{R.}},
\bauthor{\bsnm{{Campbell}}, \binits{R.D.}},
\bauthor{\bsnm{{Malone}}, \binits{D.}},
\bauthor{\bsnm{{Landis}}, \binits{D.}},
\bauthor{\bsnm{{Barrington-Leigh}}, \binits{C.P.}},
\bauthor{\bsnm{{Slassi-Sennou}}, \binits{S.}},
\bauthor{\bsnm{{Cork}}, \binits{C.}},
\bauthor{\bsnm{{Clark}}, \binits{D.}},
\bauthor{\bsnm{{Amato}}, \binits{D.}},
\bauthor{\bsnm{{Orwig}}, \binits{L.}},
\bauthor{\bsnm{{Boyle}}, \binits{R.}},
\bauthor{\bsnm{{Banks}}, \binits{I.S.}},
\bauthor{\bsnm{{Shirey}}, \binits{K.}},
\bauthor{\bsnm{{Tolbert}}, \binits{A.K.}},
\bauthor{\bsnm{{Zarro}}, \binits{D.}},
\bauthor{\bsnm{{Snow}}, \binits{F.}},
\bauthor{\bsnm{{Thomsen}}, \binits{K.}},
\bauthor{\bsnm{{Henneck}}, \binits{R.}},
\bauthor{\bsnm{{McHedlishvili}}, \binits{A.}},
\bauthor{\bsnm{{Ming}}, \binits{P.}},
\bauthor{\bsnm{{Fivian}}, \binits{M.}},
\bauthor{\bsnm{{Jordan}}, \binits{J.}},
\bauthor{\bsnm{{Wanner}}, \binits{R.}},
\bauthor{\bsnm{{Crubb}}, \binits{J.}},
\bauthor{\bsnm{{Preble}}, \binits{J.}},
\bauthor{\bsnm{{Matranga}}, \binits{M.}},
\bauthor{\bsnm{{Benz}}, \binits{A.}},
\bauthor{\bsnm{{Hudson}}, \binits{H.}},
\bauthor{\bsnm{{Canfield}}, \binits{R.C.}},
\bauthor{\bsnm{{Holman}}, \binits{G.D.}},
\bauthor{\bsnm{{Crannell}}, \binits{C.}},
\bauthor{\bsnm{{Kosugi}}, \binits{T.}},
\bauthor{\bsnm{{Emslie}}, \binits{A.G.}},
\bauthor{\bsnm{{Vilmer}}, \binits{N.}},
\bauthor{\bsnm{{Brown}}, \binits{J.C.}},
\bauthor{\bsnm{{Johns-Krull}}, \binits{C.}},
\bauthor{\bsnm{{Aschwanden}}, \binits{M.}},
\bauthor{\bsnm{{Metcalf}}, \binits{T.}},
\bauthor{\bsnm{{Conway}}, \binits{A.}}:
\byear{2002},
\batitle{{The Reuven Ramaty High-Energy Solar Spectroscopic Imager (RHESSI)}}.
\bjtitle{\solphys}
\bvolume{210},
\bfpage{3}.
\doiurl{10.1023/A:1022428818870}.
\adsurl{2002SoPh..210....3L}.
\end{barticle}
\endbibitem

\bibitem[\protect\citeauthoryear{{Liu}, {Alexander}, and
  {Gilbert}}{2009}]{2009ApJ...691.1079L}
\begin{barticle}
\bauthor{\bsnm{{Liu}}, \binits{R.}},
\bauthor{\bsnm{{Alexander}}, \binits{D.}},
\bauthor{\bsnm{{Gilbert}}, \binits{H.R.}}:
\byear{2009},
\batitle{{Asymmetric Eruptive Filaments}}.
\bjtitle{\apj}
\bvolume{691},
\bfpage{1079}.
\doiurl{10.1088/0004-637X/691/2/1079}.
\adsurl{2009ApJ...691.1079L}.
\end{barticle}
\endbibitem

\bibitem[\protect\citeauthoryear{{McAteer}
  \textit{et~al.}}{2007}]{2007ApJ...662..691M}
\begin{barticle}
\bauthor{\bsnm{{McAteer}}, \binits{R.T.J.}},
\bauthor{\bsnm{{Young}}, \binits{C.A.}},
\bauthor{\bsnm{{Ireland}}, \binits{J.}},
\bauthor{\bsnm{{Gallagher}}, \binits{P.T.}}:
\byear{2007},
\batitle{{The Bursty Nature of Solar Flare X-Ray Emission}}.
\bjtitle{\apj}
\bvolume{662},
\bfpage{691}.
\doiurl{10.1086/518086}.
\adsurl{2007ApJ...662..691M}.
\end{barticle}
\endbibitem

\bibitem[\protect\citeauthoryear{{McLaughlin}, {Thurgood}, and
  {MacTaggart}}{2012}]{2012A&A...548A..98M}
\begin{barticle}
\bauthor{\bsnm{{McLaughlin}}, \binits{J.A.}},
\bauthor{\bsnm{{Thurgood}}, \binits{J.O.}},
\bauthor{\bsnm{{MacTaggart}}, \binits{D.}}:
\byear{2012},
\batitle{{On the periodicity of oscillatory reconnection}}.
\bjtitle{\aap}
\bvolume{548},
\bfpage{A98}.
\doiurl{10.1051/0004-6361/201220234}.
\adsurl{2012A\%26A...548A..98M}.
\end{barticle}
\endbibitem

\bibitem[\protect\citeauthoryear{{Nakariakov} and
  {Melnikov}}{2009}]{2009SSRv..149..119N}
\begin{barticle}
\bauthor{\bsnm{{Nakariakov}}, \binits{V.M.}},
\bauthor{\bsnm{{Melnikov}}, \binits{V.F.}}:
\byear{2009},
\batitle{{Quasi-Periodic Pulsations in Solar Flares}}.
\bjtitle{\ssr}
\bvolume{149},
\bfpage{119}.
\doiurl{10.1007/s11214-009-9536-3}.
\adsurl{2009SSRv..149..119N}.
\end{barticle}
\endbibitem

\bibitem[\protect\citeauthoryear{{Nakariakov} and
  {Verwichte}}{2005}]{2005LRSP....2....3N}
\begin{barticle}
\bauthor{\bsnm{{Nakariakov}}, \binits{V.M.}},
\bauthor{\bsnm{{Verwichte}}, \binits{E.}}:
\byear{2005},
\batitle{{Coronal Waves and Oscillations}}.
\bjtitle{Living Reviews in Solar Physics}
\bvolume{2}.
\doiurl{10.12942/lrsp-2005-3}.
\adsurl{2005LRSP....2....3N}.
\end{barticle}
\endbibitem

\bibitem[\protect\citeauthoryear{{Nakariakov} and
  {Zimovets}}{2011}]{2011ApJ...730L..27N}
\begin{barticle}
\bauthor{\bsnm{{Nakariakov}}, \binits{V.M.}},
\bauthor{\bsnm{{Zimovets}}, \binits{I.V.}}:
\byear{2011},
\batitle{{Slow Magnetoacoustic Waves in Two-ribbon Flares}}.
\bjtitle{\apjl}
\bvolume{730},
\bfpage{L27}.
\doiurl{10.1088/2041-8205/730/2/L27}.
\adsurl{2011ApJ...730L..27N}.
\end{barticle}
\endbibitem

\bibitem[\protect\citeauthoryear{{Nakariakov}
  \textit{et~al.}}{2006}]{2006A&A...452..343N}
\begin{barticle}
\bauthor{\bsnm{{Nakariakov}}, \binits{V.M.}},
\bauthor{\bsnm{{Foullon}}, \binits{C.}},
\bauthor{\bsnm{{Verwichte}}, \binits{E.}},
\bauthor{\bsnm{{Young}}, \binits{N.P.}}:
\byear{2006},
\batitle{{Quasi-periodic modulation of solar and stellar flaring emission by
  magnetohydrodynamic oscillations in a nearby loop}}.
\bjtitle{\aap}
\bvolume{452},
\bfpage{343}.
\doiurl{10.1051/0004-6361:20054608}.
\adsurl{2006A\%26A...452..343N}.
\end{barticle}
\endbibitem

\bibitem[\protect\citeauthoryear{{Nakariakov}
  \textit{et~al.}}{2010}]{2010PPCF...52l4009N}
\begin{barticle}
\bauthor{\bsnm{{Nakariakov}}, \binits{V.M.}},
\bauthor{\bsnm{{Inglis}}, \binits{A.R.}},
\bauthor{\bsnm{{Zimovets}}, \binits{I.V.}},
\bauthor{\bsnm{{Foullon}}, \binits{C.}},
\bauthor{\bsnm{{Verwichte}}, \binits{E.}},
\bauthor{\bsnm{{Sych}}, \binits{R.}},
\bauthor{\bsnm{{Myagkova}}, \binits{I.N.}}:
\byear{2010},
\batitle{{Oscillatory processes in solar flares}}.
\bjtitle{Plasma Physics and Controlled Fusion}
\bvolume{52}(\bissue{12}),
\bfpage{124009}.
\doiurl{10.1088/0741-3335/52/12/124009}.
\adsurl{2010PPCF...52l4009N}.
\end{barticle}
\endbibitem

\bibitem[\protect\citeauthoryear{{Nakariakov}
  \textit{et~al.}}{2016}]{2016SSRv..XXX..YYYZ}
\begin{botherref}
\oauthor{\bsnm{{Nakariakov}}, \binits{V.M.}},
\oauthor{\bsnm{{Pilipenko}}, \binits{V.}},
\oauthor{\bsnm{{Heilig}}, \binits{B.}},
\oauthor{\bsnm{{Jelinek}}, \binits{P.}},
\oauthor{\bsnm{{Karlicky}}, \binits{M.}},
\oauthor{\bsnm{{Klimushkin}}, \binits{D.Y.}},
\oauthor{\bsnm{{Kolotkov}}, \binits{D.Y.}},
\oauthor{\bsnm{{Lee}}, \binits{D.-H.}},
\oauthor{\bsnm{{Nistico}}, \binits{G.}},
\oauthor{\bsnm{{Van Doorsselaere}}, \binits{T.}},
\oauthor{\bsnm{{Verth}}, \binits{G.}},
\oauthor{\bsnm{{Zimovets}}, \binits{I.V.}}:
2016,
{Magnetohydrodynamic Oscillations in the Solar Corona and Earth�s
  Magnetosphere: Towards Consolidated Understanding}.
\textit{\ssr},
1.
\doiurl{10.1007/s11214-015-0233-0}.
\end{botherref}
\endbibitem

\bibitem[\protect\citeauthoryear{{Nordlund}, {Stein}, and
  {Asplund}}{2009}]{2009LRSP....6....2N}
\begin{barticle}
\bauthor{\bsnm{{Nordlund}}, \binits{{\AA}.}},
\bauthor{\bsnm{{Stein}}, \binits{R.F.}},
\bauthor{\bsnm{{Asplund}}, \binits{M.}}:
\byear{2009},
\batitle{{Solar Surface Convection}}.
\bjtitle{Living Reviews in Solar Physics}
\bvolume{6}.
\doiurl{10.12942/lrsp-2009-2}.
\adsurl{2009LRSP....6....2N}.
\end{barticle}
\endbibitem

\bibitem[\protect\citeauthoryear{{Ofman} and {Sui}}{2006}]{2006ApJ...644L.149O}
\begin{barticle}
\bauthor{\bsnm{{Ofman}}, \binits{L.}},
\bauthor{\bsnm{{Sui}}, \binits{L.}}:
\byear{2006},
\batitle{{Oscillations of Hard X-Ray Flare Emission Observed by RHESSI: Effects
  of Super-Alfv{\'e}nic Beams?}}
\bjtitle{\apjl}
\bvolume{644},
\bfpage{L149}.
\doiurl{10.1086/505622}.
\adsurl{2006ApJ...644L.149O}.
\end{barticle}
\endbibitem

\bibitem[\protect\citeauthoryear{{Priest} and
  {D{\'e}moulin}}{1995}]{1995JGR...10023443P}
\begin{barticle}
\bauthor{\bsnm{{Priest}}, \binits{E.R.}},
\bauthor{\bsnm{{D{\'e}moulin}}, \binits{P.}}:
\byear{1995},
\batitle{{Three-dimensional magnetic reconnection without null points. 1. Basic
  theory of magnetic flipping}}.
\bjtitle{\jgr}
\bvolume{100},
\bfpage{23443}.
\doiurl{10.1029/95JA02740}.
\adsurl{1995JGR...10023443P}.
\end{barticle}
\endbibitem

\bibitem[\protect\citeauthoryear{{Priest} and
  {Forbes}}{2002}]{2002A&ARv..10..313P}
\begin{barticle}
\bauthor{\bsnm{{Priest}}, \binits{E.R.}},
\bauthor{\bsnm{{Forbes}}, \binits{T.G.}}:
\byear{2002},
\batitle{{The magnetic nature of solar flares}}.
\bjtitle{\aapr}
\bvolume{10},
\bfpage{313}.
\doiurl{10.1007/s001590100013}.
\adsurl{2002A\%26ARv..10..313P}.
\end{barticle}
\endbibitem

\bibitem[\protect\citeauthoryear{{Sakao}, {Kosugi}, and
  {Masuda}}{1998}]{1998ASSL..229..273S}
\begin{bchapter}
\bauthor{\bsnm{{Sakao}}, \binits{T.}},
\bauthor{\bsnm{{Kosugi}}, \binits{T.}},
\bauthor{\bsnm{{Masuda}}, \binits{S.}}:
\byear{1998},
\bctitle{{Energy Release and Particle Acceleration in Solar Flares with Respect
  to Flaring Magnetic Loops}}.
In: \beditor{\bsnm{{Watanabe}}, \binits{T.}},
\beditor{\bsnm{{Kosugi}}, \binits{T.}} (eds.)
\bbtitle{Observational Plasma Astrophysics : Five Years of YOHKOH and Beyond},
\bsertitle{Astrophysics and Space Science Library}
\bseriesno{229},
\bfpage{273}.
\doiurl{10.1007/978-94-011-5220-4_44}.
\adsurl{1998ASSL..229..273S}.
\end{bchapter}
\endbibitem

\bibitem[\protect\citeauthoryear{{Scherrer}
  \textit{et~al.}}{1995}]{1995SoPh..162..129S}
\begin{barticle}
\bauthor{\bsnm{{Scherrer}}, \binits{P.H.}},
\bauthor{\bsnm{{Bogart}}, \binits{R.S.}},
\bauthor{\bsnm{{Bush}}, \binits{R.I.}},
\bauthor{\bsnm{{Hoeksema}}, \binits{J.T.}},
\bauthor{\bsnm{{Kosovichev}}, \binits{A.G.}},
\bauthor{\bsnm{{Schou}}, \binits{J.}},
\bauthor{\bsnm{{Rosenberg}}, \binits{W.}},
\bauthor{\bsnm{{Springer}}, \binits{L.}},
\bauthor{\bsnm{{Tarbell}}, \binits{T.D.}},
\bauthor{\bsnm{{Title}}, \binits{A.}},
\bauthor{\bsnm{{Wolfson}}, \binits{C.J.}},
\bauthor{\bsnm{{Zayer}}, \binits{I.}},
\bauthor{\bsnm{{MDI Engineering Team}}}:
\byear{1995},
\batitle{{The Solar Oscillations Investigation - Michelson Doppler Imager}}.
\bjtitle{\solphys}
\bvolume{162},
\bfpage{129}.
\doiurl{10.1007/BF00733429}.
\adsurl{1995SoPh..162..129S}.
\end{barticle}
\endbibitem

\bibitem[\protect\citeauthoryear{{Scherrer}
  \textit{et~al.}}{2012}]{2012SoPh..275..207S}
\begin{barticle}
\bauthor{\bsnm{{Scherrer}}, \binits{P.H.}},
\bauthor{\bsnm{{Schou}}, \binits{J.}},
\bauthor{\bsnm{{Bush}}, \binits{R.I.}},
\bauthor{\bsnm{{Kosovichev}}, \binits{A.G.}},
\bauthor{\bsnm{{Bogart}}, \binits{R.S.}},
\bauthor{\bsnm{{Hoeksema}}, \binits{J.T.}},
\bauthor{\bsnm{{Liu}}, \binits{Y.}},
\bauthor{\bsnm{{Duvall}}, \binits{T.L.}},
\bauthor{\bsnm{{Zhao}}, \binits{J.}},
\bauthor{\bsnm{{Title}}, \binits{A.M.}},
\bauthor{\bsnm{{Schrijver}}, \binits{C.J.}},
\bauthor{\bsnm{{Tarbell}}, \binits{T.D.}},
\bauthor{\bsnm{{Tomczyk}}, \binits{S.}}:
\byear{2012},
\batitle{{The Helioseismic and Magnetic Imager (HMI) Investigation for the
  Solar Dynamics Observatory (SDO)}}.
\bjtitle{\solphys}
\bvolume{275},
\bfpage{207}.
\doiurl{10.1007/s11207-011-9834-2}.
\adsurl{2012SoPh..275..207S}.
\end{barticle}
\endbibitem

\bibitem[\protect\citeauthoryear{{Schmieder}, {D{\'e}moulin}, and
  {Aulanier}}{2013}]{2013AdSpR..51.1967S}
\begin{barticle}
\bauthor{\bsnm{{Schmieder}}, \binits{B.}},
\bauthor{\bsnm{{D{\'e}moulin}}, \binits{P.}},
\bauthor{\bsnm{{Aulanier}}, \binits{G.}}:
\byear{2013},
\batitle{{Solar filament eruptions and their physical role in triggering
  coronal mass ejections}}.
\bjtitle{Advances in Space Research}
\bvolume{51},
\bfpage{1967}.
\doiurl{10.1016/j.asr.2012.12.026}.
\adsurl{2013AdSpR..51.1967S}.
\end{barticle}
\endbibitem

\bibitem[\protect\citeauthoryear{{Schrijver}}{2009}]{2009AdSpR..43..739S}
\begin{barticle}
\bauthor{\bsnm{{Schrijver}}, \binits{C.J.}}:
\byear{2009},
\batitle{{Driving major solar flares and eruptions: A review}}.
\bjtitle{Advances in Space Research}
\bvolume{43},
\bfpage{739}.
\doiurl{10.1016/j.asr.2008.11.004}.
\adsurl{2009AdSpR..43..739S}.
\end{barticle}
\endbibitem

\bibitem[\protect\citeauthoryear{{Shibata} and
  {Magara}}{2011}]{2011LRSP....8....6S}
\begin{barticle}
\bauthor{\bsnm{{Shibata}}, \binits{K.}},
\bauthor{\bsnm{{Magara}}, \binits{T.}}:
\byear{2011},
\batitle{{Solar Flares: Magnetohydrodynamic Processes}}.
\bjtitle{Living Reviews in Solar Physics}
\bvolume{8}.
\doiurl{10.12942/lrsp-2011-6}.
\adsurl{2011LRSP....8....6S}.
\end{barticle}
\endbibitem

\bibitem[\protect\citeauthoryear{{Sim{\~o}es}, {Hudson}, and
  {Fletcher}}{2015}]{2015SoPh..290.3625S}
\begin{barticle}
\bauthor{\bsnm{{Sim{\~o}es}}, \binits{P.J.A.}},
\bauthor{\bsnm{{Hudson}}, \binits{H.S.}},
\bauthor{\bsnm{{Fletcher}}, \binits{L.}}:
\byear{2015},
\batitle{{Soft X-Ray Pulsations in Solar Flares}}.
\bjtitle{\solphys}
\bvolume{290},
\bfpage{3625}.
\doiurl{10.1007/s11207-015-0691-2}.
\adsurl{2015SoPh..290.3625S}.
\end{barticle}
\endbibitem

\bibitem[\protect\citeauthoryear{{Sim{\~o}es}
  \textit{et~al.}}{2013}]{2013ApJ...777..152S}
\begin{barticle}
\bauthor{\bsnm{{Sim{\~o}es}}, \binits{P.J.A.}},
\bauthor{\bsnm{{Fletcher}}, \binits{L.}},
\bauthor{\bsnm{{Hudson}}, \binits{H.S.}},
\bauthor{\bsnm{{Russell}}, \binits{A.J.B.}}:
\byear{2013},
\batitle{{Implosion of Coronal Loops during the Impulsive Phase of a Solar
  Flare}}.
\bjtitle{\apj}
\bvolume{777},
\bfpage{152}.
\doiurl{10.1088/0004-637X/777/2/152}.
\adsurl{2013ApJ...777..152S}.
\end{barticle}
\endbibitem

\bibitem[\protect\citeauthoryear{{Somov}}{2007}]{2007plas.book.....S}
\begin{bbook}
\bauthor{\bsnm{{Somov}}, \binits{B.V.}}:
\byear{2007},
\bbtitle{{Plasma Astrophysics, Part II: Reconnection and Flares}}.
\adsurl{2007plas.book.....S}.
\end{bbook}
\endbibitem

\bibitem[\protect\citeauthoryear{{Sych}
  \textit{et~al.}}{2009}]{2009A&A...505..791S}
\begin{barticle}
\bauthor{\bsnm{{Sych}}, \binits{R.}},
\bauthor{\bsnm{{Nakariakov}}, \binits{V.M.}},
\bauthor{\bsnm{{Karlicky}}, \binits{M.}},
\bauthor{\bsnm{{Anfinogentov}}, \binits{S.}}:
\byear{2009},
\batitle{{Relationship between wave processes in sunspots and quasi-periodic
  pulsations in active region flares}}.
\bjtitle{\aap}
\bvolume{505},
\bfpage{791}.
\doiurl{10.1051/0004-6361/200912132}.
\adsurl{2009A\%26A...505..791S}.
\end{barticle}
\endbibitem

\bibitem[\protect\citeauthoryear{{Tajima}
  \textit{et~al.}}{1987}]{1987ApJ...321.1031T}
\begin{barticle}
\bauthor{\bsnm{{Tajima}}, \binits{T.}},
\bauthor{\bsnm{{Sakai}}, \binits{J.}},
\bauthor{\bsnm{{Nakajima}}, \binits{H.}},
\bauthor{\bsnm{{Kosugi}}, \binits{T.}},
\bauthor{\bsnm{{Brunel}}, \binits{F.}},
\bauthor{\bsnm{{Kundu}}, \binits{M.R.}}:
\byear{1987},
\batitle{{Current loop coalescence model of solar flares}}.
\bjtitle{\apj}
\bvolume{321},
\bfpage{1031}.
\doiurl{10.1086/165694}.
\adsurl{1987ApJ...321.1031T}.
\end{barticle}
\endbibitem

\bibitem[\protect\citeauthoryear{{Torrence} and
  {Compo}}{1998}]{1998BAMS...79...61T}
\begin{barticle}
\bauthor{\bsnm{{Torrence}}, \binits{C.}},
\bauthor{\bsnm{{Compo}}, \binits{G.P.}}:
\byear{1998},
\batitle{{A Practical Guide to Wavelet Analysis.}}
\bjtitle{Bulletin of the American Meteorological Society}
\bvolume{79},
\bfpage{61}.
\doiurl{10.1175/1520-0477(1998)079<0061:APGTWA>2.0.CO;2}.
\adsurl{1998BAMS...79...61T}.
\end{barticle}
\endbibitem

\bibitem[\protect\citeauthoryear{{van Beek}, {de Feiter}, and {de
  Jager}}{1974a}]{1974spre.conf..447V}
\begin{bchapter}
\bauthor{\bsnm{{van Beek}}, \binits{H.F.}},
\bauthor{\bsnm{{de Feiter}}, \binits{L.D.}},
\bauthor{\bsnm{{de Jager}}, \binits{C.}}:
\byear{1974}a,
\bctitle{{Hard X-ray observations of elementary flare bursts, and their
  interpretation}}.
In: \beditor{\bsnm{{Rycroft}}, \binits{M.J.}},
\beditor{\bsnm{{Reasenberg}}, \binits{R.D.}} (eds.)
\bbtitle{Space Research XIV},
\bfpage{447}.
\adsurl{1974spre.conf..447V}.
\end{bchapter}
\endbibitem

\bibitem[\protect\citeauthoryear{{van Beek}, {de Feiter}, and {de
  Jager}}{1974b}]{1974ASSL...42..533V}
\begin{bchapter}
\bauthor{\bsnm{{van Beek}}, \binits{H.F.}},
\bauthor{\bsnm{{de Feiter}}, \binits{L.D.}},
\bauthor{\bsnm{{de Jager}}, \binits{C.}}:
\byear{1974}b,
\bctitle{{Time profiles and photon spectra of solar hard X-rays}}.
In: \beditor{\bsnm{{Page}}, \binits{D.E.}} (ed.)
\bbtitle{Correlated Interplanetary and Magnetospheric Observations},
\bsertitle{Astrophysics and Space Science Library}
\bseriesno{42},
\bfpage{533}.
\doiurl{10.1007/978-94-010-2172-2_34}.
\adsurl{1974ASSL...42..533V}.
\end{bchapter}
\endbibitem

\bibitem[\protect\citeauthoryear{{Vaughan}}{2005}]{2005A&A...431..391V}
\begin{barticle}
\bauthor{\bsnm{{Vaughan}}, \binits{S.}}:
\byear{2005},
\batitle{{A simple test for periodic signals in red noise}}.
\bjtitle{\aap}
\bvolume{431},
\bfpage{391}.
\doiurl{10.1051/0004-6361:20041453}.
\adsurl{2005A\%26A...431..391V}.
\end{barticle}
\endbibitem

\bibitem[\protect\citeauthoryear{{Veronig} and
  {Brown}}{2004}]{2004ApJ...603L.117V}
\begin{barticle}
\bauthor{\bsnm{{Veronig}}, \binits{A.M.}},
\bauthor{\bsnm{{Brown}}, \binits{J.C.}}:
\byear{2004},
\batitle{{A Coronal Thick-Target Interpretation of Two Hard X-Ray Loop
  Events}}.
\bjtitle{\apjl}
\bvolume{603},
\bfpage{L117}.
\doiurl{10.1086/383199}.
\adsurl{2004ApJ...603L.117V}.
\end{barticle}
\endbibitem

\bibitem[\protect\citeauthoryear{{White}
  \textit{et~al.}}{2011}]{2011SSRv..159..225W}
\begin{barticle}
\bauthor{\bsnm{{White}}, \binits{S.M.}},
\bauthor{\bsnm{{Benz}}, \binits{A.O.}},
\bauthor{\bsnm{{Christe}}, \binits{S.}},
\bauthor{\bsnm{{F{\'a}rn{\'{\i}}k}}, \binits{F.}},
\bauthor{\bsnm{{Kundu}}, \binits{M.R.}},
\bauthor{\bsnm{{Mann}}, \binits{G.}},
\bauthor{\bsnm{{Ning}}, \binits{Z.}},
\bauthor{\bsnm{{Raulin}}, \binits{J.-P.}},
\bauthor{\bsnm{{Silva-V{\'a}lio}}, \binits{A.V.R.}},
\bauthor{\bsnm{{Saint-Hilaire}}, \binits{P.}},
\bauthor{\bsnm{{Vilmer}}, \binits{N.}},
\bauthor{\bsnm{{Warmuth}}, \binits{A.}}:
\byear{2011},
\batitle{{The Relationship Between Solar Radio and Hard X-ray Emission}}.
\bjtitle{\ssr}
\bvolume{159},
\bfpage{225}.
\doiurl{10.1007/s11214-010-9708-1}.
\adsurl{2011SSRv..159..225W}.
\end{barticle}
\endbibitem

\bibitem[\protect\citeauthoryear{{Yang}
  \textit{et~al.}}{2009}]{2009ApJ...693..132Y}
\begin{barticle}
\bauthor{\bsnm{{Yang}}, \binits{Y.-H.}},
\bauthor{\bsnm{{Cheng}}, \binits{C.Z.}},
\bauthor{\bsnm{{Krucker}}, \binits{S.}},
\bauthor{\bsnm{{Lin}}, \binits{R.P.}},
\bauthor{\bsnm{{Ip}}, \binits{W.H.}}:
\byear{2009},
\batitle{{A Statistical Study of Hard X-Ray Footpoint Motions in Large Solar
  Flares}}.
\bjtitle{\apj}
\bvolume{693},
\bfpage{132}.
\doiurl{10.1088/0004-637X/693/1/132}.
\adsurl{2009ApJ...693..132Y}.
\end{barticle}
\endbibitem

\bibitem[\protect\citeauthoryear{{Zaitsev} and
  {Stepanov}}{1982}]{1982SvAL....8..132Z}
\begin{barticle}
\bauthor{\bsnm{{Zaitsev}}, \binits{V.V.}},
\bauthor{\bsnm{{Stepanov}}, \binits{A.V.}}:
\byear{1982},
\batitle{{On the Origin of the Hard X-Ray Pulsations during Solar Flares}}.
\bjtitle{Soviet Astronomy Letters}
\bvolume{8},
\bfpage{132}.
\adsurl{1982SvAL....8..132Z}.
\end{barticle}
\endbibitem

\bibitem[\protect\citeauthoryear{{Zaitsev} and
  {Stepanov}}{2008}]{2008PhyU...51.1123Z}
\begin{barticle}
\bauthor{\bsnm{{Zaitsev}}, \binits{V.V.}},
\bauthor{\bsnm{{Stepanov}}, \binits{A.V.}}:
\byear{2008},
\batitle{{REVIEWS OF TOPICAL PROBLEMS: Coronal magnetic loops}}.
\bjtitle{Physics Uspekhi}
\bvolume{51},
\bfpage{1123}.
\doiurl{10.1070/PU2008v051n11ABEH006657}.
\adsurl{2008PhyU...51.1123Z}.
\end{barticle}
\endbibitem

\bibitem[\protect\citeauthoryear{{Zimovets} and
  {Struminsky}}{2009}]{2009SoPh..258...69Z}
\begin{barticle}
\bauthor{\bsnm{{Zimovets}}, \binits{I.V.}},
\bauthor{\bsnm{{Struminsky}}, \binits{A.B.}}:
\byear{2009},
\batitle{{Imaging Observations of Quasi-Periodic Pulsatory Nonthermal Emission
  in Two-Ribbon Solar Flares}}.
\bjtitle{\solphys}
\bvolume{258},
\bfpage{69}.
\doiurl{10.1007/s11207-009-9394-x}.
\adsurl{2009SoPh..258...69Z}.
\end{barticle}
\endbibitem

\bibitem[\protect\citeauthoryear{{Zimovets} and
  {Struminsky}}{2010}]{2010SoPh..263..163Z}
\begin{barticle}
\bauthor{\bsnm{{Zimovets}}, \binits{I.V.}},
\bauthor{\bsnm{{Struminsky}}, \binits{A.B.}}:
\byear{2010},
\batitle{{Observations of Double-Periodic X-Ray Emission in Interacting Systems
  of Solar Flare Loops}}.
\bjtitle{\solphys}
\bvolume{263},
\bfpage{163}.
\doiurl{10.1007/s11207-010-9518-3}.
\adsurl{2010SoPh..263..163Z}.
\end{barticle}
\endbibitem

\bibitem[\protect\citeauthoryear{{Zimovets}, {Kuznetsov}, and
  {Struminsky}}{2013}]{2013AstL...39..267Z}
\begin{barticle}
\bauthor{\bsnm{{Zimovets}}, \binits{I.V.}},
\bauthor{\bsnm{{Kuznetsov}}, \binits{S.A.}},
\bauthor{\bsnm{{Struminsky}}, \binits{A.B.}}:
\byear{2013},
\batitle{{Fine structure of the sources of quasi-periodic pulsations in
  ''single-loop'' solar flares}}.
\bjtitle{Astronomy Letters}
\bvolume{39},
\bfpage{267}.
\doiurl{10.1134/S1063773713040063}.
\adsurl{2013AstL...39..267Z}.
\end{barticle}
\endbibitem

\end{thebibliography}

\newpage

\begin{figure} 
\centerline{\includegraphics[width=0.99\textwidth,clip=]{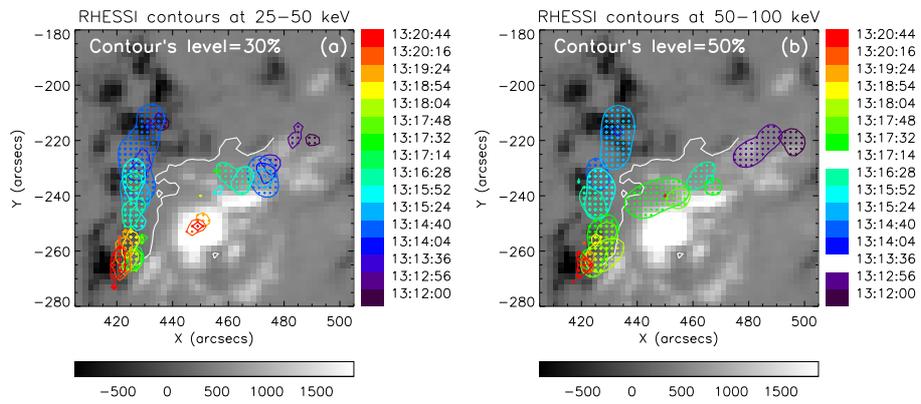}}

\caption{Comparison of positions of HXR sources in the 25--50 keV (a) and 50--100 keV (b) energy ranges for significant HXR peaks of flare No 6 (see Figure~\ref{F-1}). The PIXON algorithm was used for reconstruction of the HXR sources shown. The background image is the photospheric line-of-sight magnetogram obtained with the SOHO/MDI. The magnetic field colorbar (in gauss) is presented below the images. The MPIL is shown by the white curves. Colors of the HXR sources correspond to the times of the HXR peaks shown near the colorbars to the right of the images. White color in the colorbar (right) indicates that there are no good HXR images for these peaks. Contour levels of the HXR sources (in percentage of the maximum brightness) are written on the figures.} 
\label{F-5}
\end{figure}

\begin{figure}
\centerline{\includegraphics[width=0.968\textwidth,clip=]{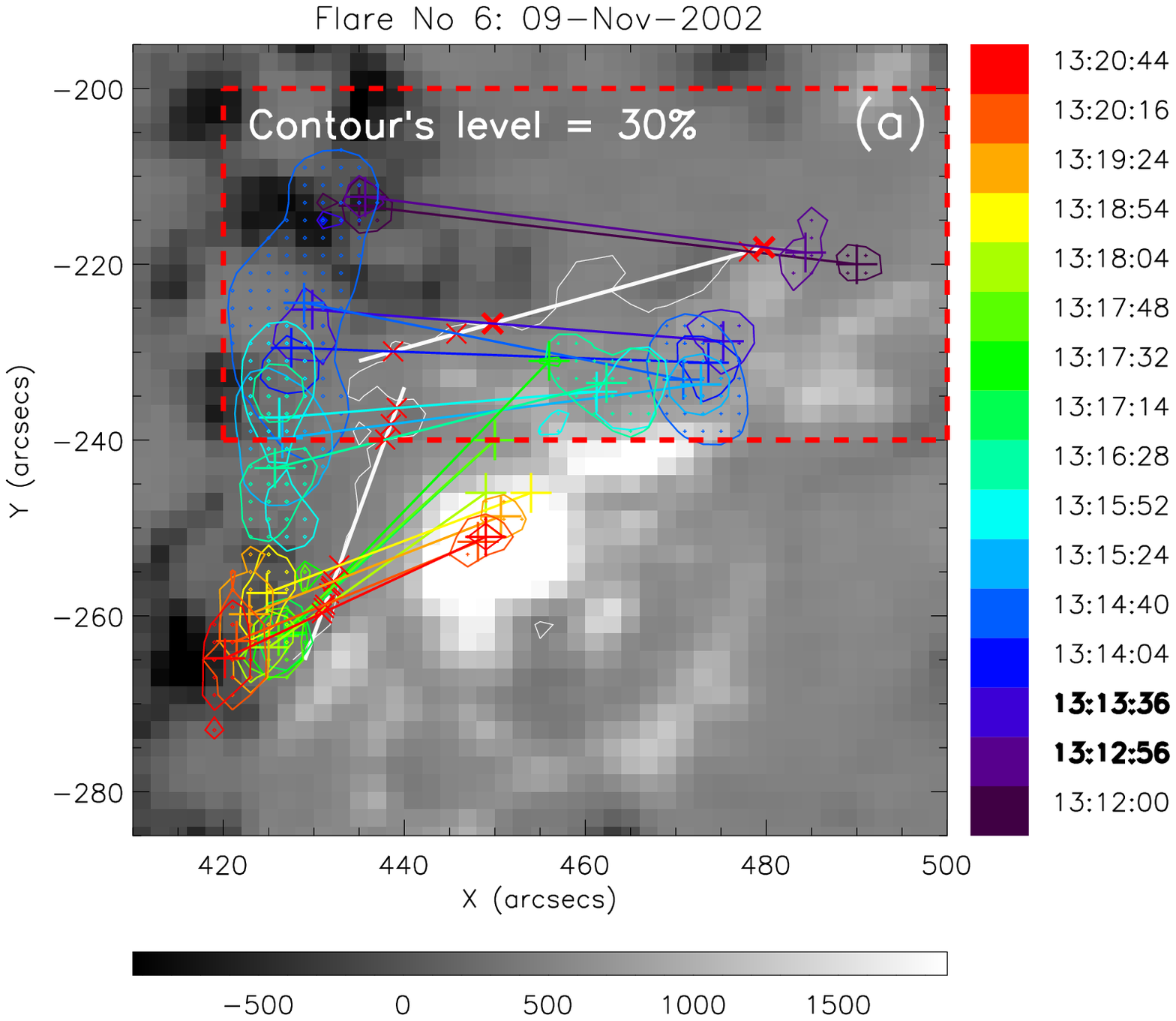}}
\hspace{-0.089\textwidth}
\centerline{\includegraphics[width=0.8\textwidth,clip=]{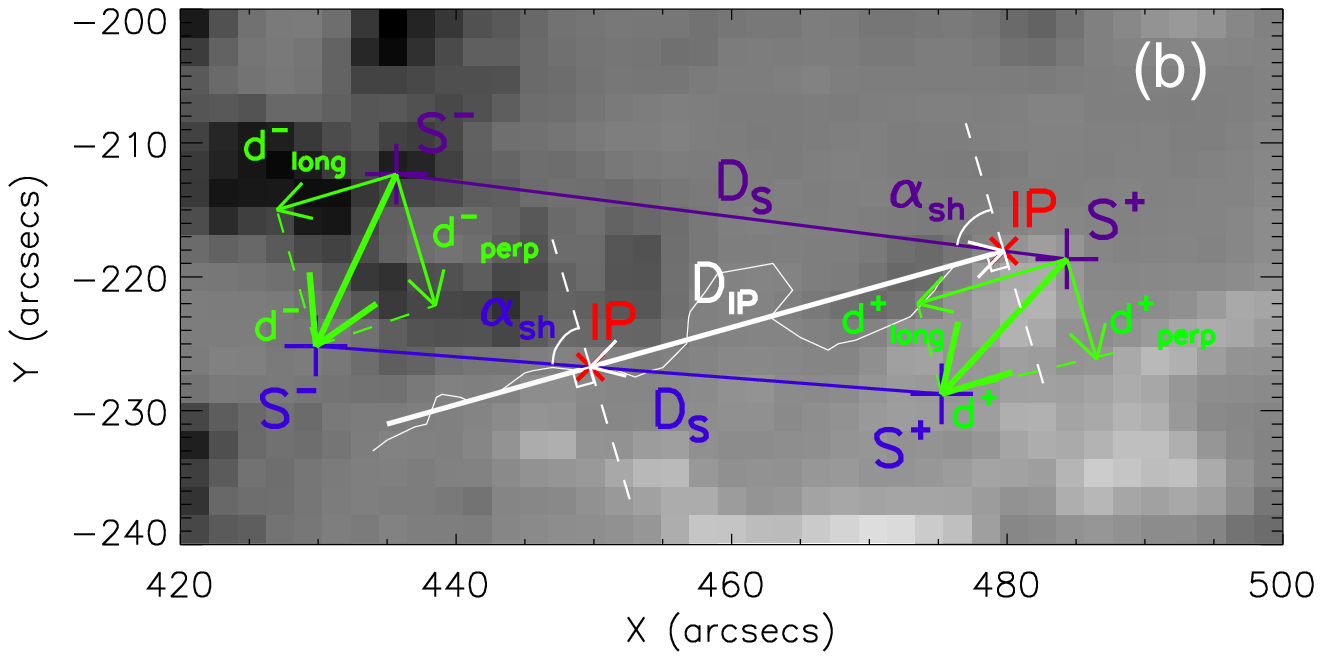}}

\caption{(a) Spatio-temporal evolution of the sources of the HXR pulsations in flare No 6 of the group-1. The date and time of the flare are shown above. The PIXON algorithm was used for reconstruction of the HXR sources shown. The background image is the photospheric line-of-sight magnetogram obtained with the SOHO/MDI. The magnetic field colorbar (in gauss) is presented below. Colors of the HXR sources correspond to the times of the HXR peaks shown near the colorbar to the right. Contour levels of the HXR sources are written on the images. Large color crosses show average positions (centroids) of the HXR sources at the corresponding times. Linear sizes of these crosses are equal to the doubled FWHM of the RHESSI's collimator number 1. Straight color lines connect paired HXR sources located in opposite magnetic polarities. Small red crosses indicate intersection points (IPs) of these lines with the approximate MPIL (thick white lines). The real MPIL derived from the magnetogram is shown by the thin white curve. (b) Zoom of the parental active region within the red dashed boundaries indicated on the panel (a) with designations of the quantitative characteristics of the HXR sources dynamics calculated in Section~\ref{SS-hxrcha}. Paired HXR sources $S^{+}$ and $S^{-}$ in positive and negative magnetic polarities for the two successive significant HXR peaks are shown by different colors. Corresponding times of these peaks are marked by bold font near the colorbar to the right of the panel (a). Thick green arrows $d^{+}$ and $d^{-}$ indicate the total displacements of the sources $S^{+}$ and $S^{-}$ between two successive HXR peaks respectively. White dashed lines are the perpendiculars to the approximate MPIL at the intersection points (IPs).}
\label{F-6}
\end{figure}

\begin{figure} 
\centerline{\includegraphics[width=0.99\textwidth,clip=]{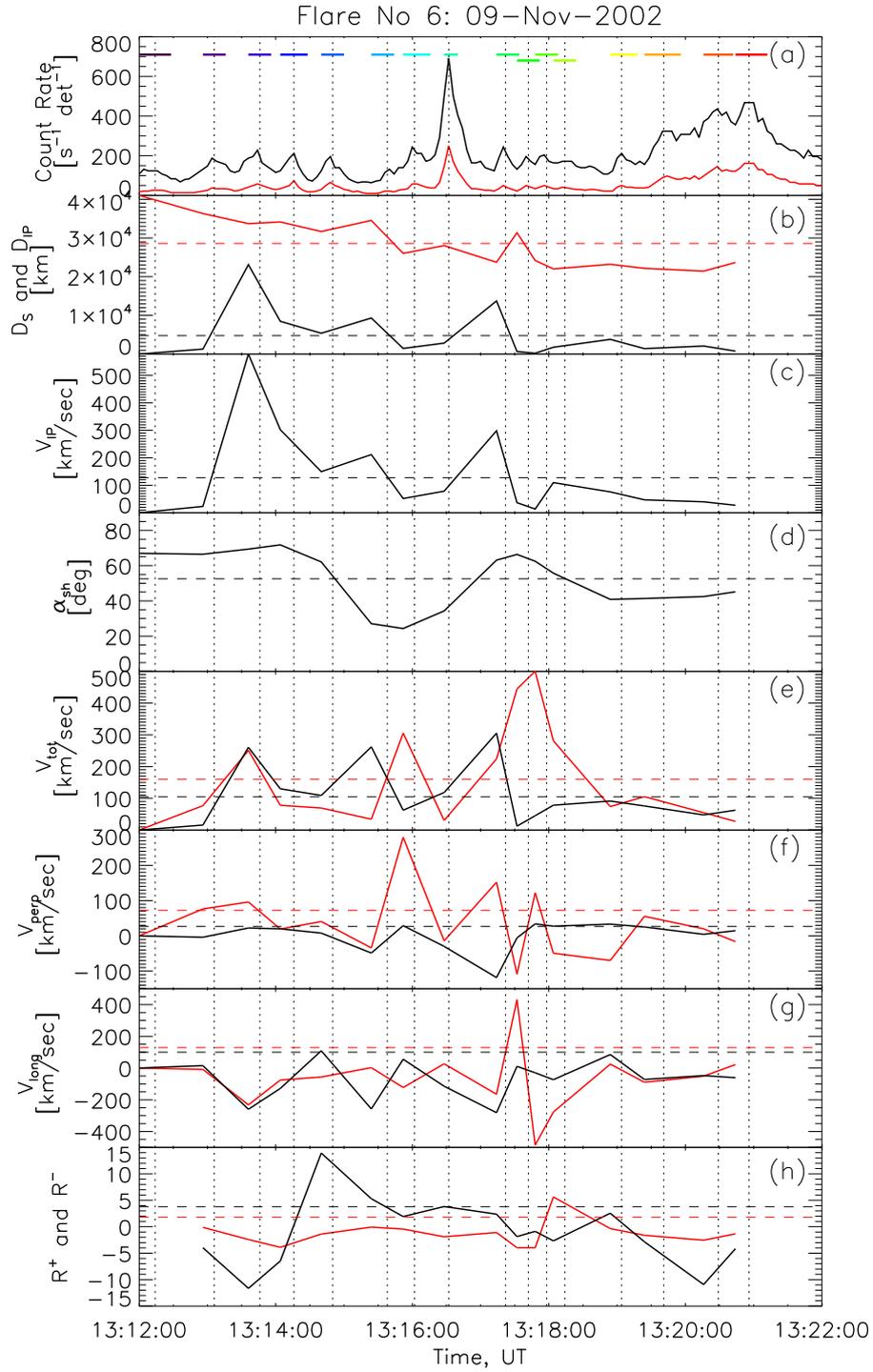}}

\caption{Time variations of quantitative characteristics of paired HXR sources calculated for flare No 6. (a) RHESSI's corrected count rates in the 25--50 keV (black) and 50--100 keV (red) channels. Black vertical dotted lines show identified significant peaks. Color horizontal segments at the top mark time intervals for which HXR images were synthesized (see Figure~\ref{F-5}(b) and Figure~\ref{F-6}(a)). (b) $D_{S}$ (red) and $D_{IP}$ (black); (c) $V_{IP}$; (d) $\alpha_{sh}$; (e) $V^{+}$ (red) and $V^{-}$ (black); (f) $V_{\bot}^{+}$ (red) and $V_{\bot}^{-}$ (black); (g) $V_{\parallel}^{+}$ (red) and $V_{\parallel}^{-}$ (black); (h) $R^{+}$ (red) and $R^{-}$ (black). See Section~\ref{SS-hxrcha} for explanation of these notations. Horizontal dashed lines show means of the presented characteristics on (b)--(e), while means of absolute values of corresponding characteristics are shown on (f)--(h).} 
 \label{F-7}
 \end{figure}

\begin{landscape}
\newpage
\begin{table}
\caption{General information about the studied events.}
\begin{tabular}{|c|c|c|c|c|c|c|c|c|c|c|c|c|} 
\hline
1 & 2 & 3 & 4 & 5 & 6 & 7 & 8 & 9 & 10 & 11 & 12 & 13 \\
\hline
Flare & Date & RHESSI  & GOES  & GOES & GOES & GOES  & NOAA & Hale  & RHESS & Flare & CME & Radio \\
No    &      & Flare & Start & Peak & End  & Class & AR   & Class & Flare & Group   &  First  & Burst \\
 & & No & Time & Time & Time & & No & & (arcsec) & No & Appear. & Types \\
\hline
1 &  14--Mar--02 & 2031402  & 01:38 & 01:50  & 02:02 & M5.7 & 9866  & bdg & -377, -76 & 1 & No & III, IV\\
2 &  19--Aug--02 & 20819155 & 20:56 & 21:02  & 21:06 & M3.1 & 10069 & bdg & +487, -272 & 2 & No & III\\
3 &  20--Aug--02 & 20820159 & 01:33 & 01:40  & 01:43 & M5.0 & 10069 & bdg & +533, -265 & 2 & 01:54 & III\\
4 &  20--Aug--02 & 20820140 & 08:22 & 08:26  & 08:30 & M3.4 & 10069 & bdg & +569, -264 & 2 & 08:55 & III, V\\
5 &  21--Aug--02 & 208211203& 01:35 & 01:41  & 01:45 & M1.4 & 10069 & bdg & +693, -250 & 2 & 02:06 & II, III\\
6 &  09--Nov--02 & 2110912  & 13:08 & 13:23  & 13:36 & M4.6 & 10180& bdg & +432, -255 & 1 & 13:32 & II, III, IV\\
7 &  29--May--03 & 30529100 & 00:51 & 01:05  & 01:12 & X1.2 & 10365 & bdg & +497, -106 & 1 & 01:27 & II, III, IV\\
8 &  13--Jul--04 & 40713103 & 00:09 & 00:17  & 00:23 & M6.7 & 10646 & b & +656, +184 & 1 & 00:54 & II, III\\
9 &  15--Jul--04 & 4071514  & 18:15 & 18:24  & 18:28 & X1.6 & 10649 & bdg & -648, -233 & 2 & No & II\\
10 & 30--Oct--04 & 4103006  & 03:23 & 03:33  & 03:37 & M3.3 & 10691 & b & +319, +143 & 1 & 03:54 & II, III\\
11 & 03--Nov--04 & 4110305 & 03:23 & 03:35  & 03:57 & M1.6 & 10696 & b & -679, +101 & 1 & 03:54 & II, III, IV, V\\
12 & 04--Nov--04 & 4110460 & 22:53 & 23:09  & 23:26 & M5.4 & 10696 & bd & -283, +71 & 1 & 23:20 & III, IV\\
13 & 06--Nov--04 & 4110630 & 00:11 & 00:34  & 00:42 & M9.3 & 10696 & bdg & -76, +88 & 1 & 01:32 & II, III, IV\\
14 & 10--Nov--04 & 4111002 & 01:59 & 02:13  & 02:20 & X2.5 & 10696 & bdg & +710, +102 & 2 & 02:26 & II, III, IV\\
15 & 01--Jan--05 & 5010148 & 00:01 & 00:31  & 00:39 & X1.7 & 10715 & bdg & -535, +125 & 1 & 00:54 & II, III, IV\\
16 & 15--Jan--05 & 5011586 & 00:22 & 00:43  & 01:02 & X1.2 & 10720 & bd & -124, +294 & 2 & No & IIIN\\
17 & 15--Jan--05 & 50115152& 22:25 & 23:02  & 23:31 & X2.6 & 10720 & bd & +20, +310 & 1 & 23:07 & II, III, IV\\
18 & 17--Jan--05 & 5011710 & 06:59 & 09:52  & 10:07 & X3.8 & 10720 & bd & +432, +305 & 1 & 09:54 & II, III, IV\\
19 & 19--Jan--05 & 5011911 & 08:03 & 08:22  & 08:40 & X1.3 & 10720 & bg & +696, +314 & 1 & 08:30 & II, III, IV\\
20 & 13--May--05 & 5051310 & 16:13 & 16:57  & 17:28 & M8.0 & 10759 & b & -174, +238 & 2 & 17:12 & II, III, IV\\
21 & 10--Sep--05 & 5091026 & 21:30 & 22:11  & 22:43 & X2.1 & 10808 & bgd & -665, -254 & 2 & 21:52 & II, III, IV\\
22 & 13--Sep--05 & 5091367 & 23:15 & 23:22  & 23:30 & X1.7 & 10808 & bgd & -24, -297 & 2 & 23:36 & III, IV\\
23 & 15--Feb--11 & 11021565& 01:44 & 01:56  & 02:06 & X2.2 & 11158 & bg & +205, -222 & 2 & 02:24 & II, III, IV, V\\
24 & 07--Jun--11 & 11060702& 06:16 & 06:41  & 06:59 & M2.5 & 11226 & b & +705, -353 & 1 & 06:49 & II, III, IV\\
25 & 06--Sep--11 & 11090678& 22:12 & 22:20  & 22:24 & X2.1 & 11283 & bd & +284, +131 & 1 & 23:06 & II, III\\
26 & 18--Apr--14 & 14041818& 12:31 & 13:03  & 13:20 & M7.3 & 12036 & bd & +499, -229 & 2 & 13:26 & II, III, IV\\
27 & 22--Oct--14 & 14102243& 14:02 & 14:28  & 14:50 & X1.6 & 12192 & bdg & -225, -317 & 1 & 16:16 & III\\
28 & 24--Oct--14 & 14102478& 21:07 & 21:41  & 22:13 & X3.1 & 12192 & bdg & +287, -329 & 2 & 21:48 & No\\
29 & 09--Nov--14 & 14110936& 15:24 & 15:32  & 15:38 & M2.3 & 12205 & bgd & -184, +188 & 1 & No & III\\
\hline
\end{tabular}
\label{T-1}
\end{table}
\end{landscape}

\begin{landscape}
\newpage
\begin{table}
\caption{Results of analysis of the RHESSI's background-subtracted corrected count rates in the 25--50 and 50--100 keV channels.}
\begin{tabular}{|c|c|c|c|c|c|c|c|c|c|c|c|c|c|c|}
\hline
 1 & 2 & 3 & 4 & 5 & 6 & 7 & 8 & 9 & 10 & 11 & 12 & 13 & 14 & 15 \\
\hline
Flare & $n_{p}$ & $\left\langle P \right\rangle$, & $\sigma_{P}$, & N of peaks in & N of peaks in & $ n_{p}$ & $\left\langle\ P \right\rangle$,  & $\sigma_{P}$,  &  N of peaks in & N of peaks in & $k_{p}$ & $\left\langle P^{\prime} \right\rangle$, & $\sigma(P^{\prime})$, & $r_{p}$  \\
No & & [s] & [s] & $\left\langle P \right\rangle \pm 0.5 \left\langle P \right\rangle$ & $\left\langle P \right\rangle \pm 3.0 \sigma_{P}$ & & [s] & [s] & $\left\langle P \right\rangle \pm 0.5 \left\langle P \right\rangle$ & $\left\langle P \right\rangle \pm 3.0 \sigma_{P}$ & & [s] & [s] &  \\   
\hline
 & \multicolumn{5}{c|}{25--50 keV channel} & \multicolumn{5}{c|}{50--100 keV channel} & \multicolumn{4}{c|}{50--100 keV channel}\\ 
\hline
1 &  27 & 22 & 13 & 19 (70\%) & 26 (96\%)& 25 & 28 & 25 & 10 (40\%) & 24 (96\%) & 10 & 61 & 35 & 2.5 \\
2 &  5  & 19 & 12 & 2 (40\%) & 5 (100\%)& 5 & 23 & 10 & 3 (60\%) & 5 (100\%) & 5 & 23 & 10 & 1.0 \\
3 &  15 & 20 & 11 & 8 (53\%) & 15 (100\%)& 9 & 26 & 18 & 5 (56\%) & 9 (100\%) & 7& 25 & 13 & 1.3 \\
4 &  9  & 21 & 12 & 6 (67\%) & 9 (100\%)& 9 & 17 & 6 & 7 (78\%) & 9 (100\%) & 5& 22& 4 & 1.8 \\
5 &  10 & 21 & 7 & 8 (80\%) & 10 (100\%)& 9 & 26 & 21& 6 (67\%) & 9 (100\%) & 5& 18 & 4 & 1.8 \\
6 &  21 & 27 & 13 & 16 (76\%) & 21 (100\%)& 21&26& 12& 14 (67\%) & 21 (100\%) & 16& 36 & 13 & 1.3 \\
7 &  18 & 28 & 15 & 7 (39\%) & 18 (100\%)& 13 & 31& 17 & 7 (54\%) & 13 (100\%) & 6& 70 & 7 & 2.2 \\
8 &  15 & 16 & 5  & 13 (87\%) & 15 (100\%)& 13 & 19& 5 & 13 (100\%) & 13 (100\%) & 11 & 21 & 11 & 1.2 \\
9 &  7  & 17 & 7  & 7 (100\%) & 7 (100\%)& 7 & 20& 7 & 7 (100\%) & 7 (100\%) & 5 & 23 & 2 & 1.4 \\
10 & 14 & 22 & 12 & 8 (57\%) & 14 (100\%)& 13 & 22& 11 & 8 (62\%) & 13 (100\%) & 8 & 34 & 13 & 1.6 \\
11 & 15 & 19 & 7  & 13 (87\%) & 15 (100\%)& 12 & 20& 9 & 8 (67\%) & 12 (100\%) & 6 & 30 & 12 & 2.0 \\
12 & 36 & 24 & 26 & 19 (53\%) & 36 (100\%)& 19 & 40& 31 & 12 (63\%) & 18 (95\%) & 8 & 46 & 28 & 2.4 \\
13 & 6  & 43 & 23 & 3 (50\%) & 6 (100\%)& 9 & 30& 22 & 6 (67\%) & 9 (100\%) & 7 & 39 & 22 & 1.3 \\
14 & 31 & 19 & 10 & 17 (55\%) & 31 (100\%)& 13 & 44 & 22 & 10 (77\%) & 12 (92\%) & 8 & 47 & 10 & 1.6 \\
15 & 12 & 22 & 14 & 9 (75\%) & 12 (100\%)& 7 & 19 & 9 & 3 (43\%) & 7 (100\%) & 5 & 22 & 10 & 1.4 \\
16 & 10 & 19 & 11 & 5 (50\%) & 10 (100\%)& 9 & 23 & 12 & 9 (56\%) & 9 (100\%) & 8 & 30 & 6 & 1.1 \\
17 & 54 & 39 & 44 & 27 (50\%) & 54 (100\%)& 103 & 22 & 19 & 87 (84\%) & 100 (97\%) & 15 & 86 & 25 & 6.9 \\
18 & 36 & 22 & 15 & 27 (75\%) & 35 (97\%)& 32 & 25 & 18 & 18 (56\%) & 32 (100\%) & 13 & 55 & 37 & 2.5 \\
19 & 55 & 20 & 15 & 46 (84\%) & 54 (98\%)& 18 & 60 & 46 & 8 (44\%) & 18 (100\%) & 11 & 87 & 47 & 1.6 \\
20 & 20 & 21 & 9 & 16 (80\%) & 20 (100\%)& 14 & 33 & 34 & 6 (43\%) & 14 (100\%) & 6 & 86 & 25 & 2.3 \\
21 & 103 & 23 & 25 & 65 (63\%) & 102 (99\%)& 89 & 31 & 34 & 44 (49\%) & 87 (98\%) & 12 & 133 & 30 & 7.4 \\
22 & 6  & 27 & 30 & 1 (17\%) & 6 (100\%)& 13 & 33 & 34 & 6 (46\%) & 13 (100\%) & 6 & 22 & 12 & 2.2\\
23 & 35 & 16 & 7 & 31 (89\%) & 35 (100\%)& 29 & 20 & 9 & 20 (69\%) & 29 (100\%) & 12 & 33 & 7 & 2.4 \\
24 & 36 & 26 & 14 & 21 (58\%) & 36 (100\%)& 41 & 23 & 14 & 29 (71\%) & 40 (98\%) & 13 & 51 & 11 & 3.2 \\
25 & 20 & 21 & 14 & 13 (65\%) & 20 (100\%)& 10 & 28 & 21 & 3 (33\%) & 10 (100\%) & 8 & 23 & 8 & 1.3 \\
26 & 23 & 24 & 16 & 15 (65\%) & 23 (100\%)& 30 & 20 & 13 & 18 (60\%) & 29 (97\%) & 11 & 48 & 35 & 2.7 \\
27 & 10 & 21 & 10 & 5 (50\%) & 10 (100\%)& 6 & 35 & 10 & 6 (100\%) & 6 (100\%) & 5 & 30 & 6 & 1.2 \\
28 & 34 & 17 & 8 & 22 (65\%) & 34 (100\%)& 15 & 27 & 18 & 11 (73\%) & 15 (100\%) & 9 & 44 & 17 & 1.7 \\
29 & 8  & 20 & 12& 4 (50\%) & 8 (100\%)& 7 & 19 & 5 & 7 (100\%) & 7 (100\%) & 7 & 26 & 8 & 1.0 \\
\hline
\end{tabular}
\label{T-2}
\end{table}
\end{landscape}

\newpage
\begin{landscape}
\newpage
\begin{table}
\caption{Results of the search for quasi-periodicities of HXR emission in the studied flares. The characteristic time scales found in the RHESSI's background-subtracted, normalized-to-maximum, corrected count rates in the 50--100 keV channel are shown above the red-noise level for the significance levels of 95.0\% and 99.7\%.}
\begin{tabular}{|c|c|c|c|c|c|c|}
\hline
1     &       2      &      3     &   4    &   5	  & 	 6     & 7  \\
\hline
Flare& Start time  & End time & \multicolumn{2}{c|}{Periods from wavelet analysis, [s]}&\multicolumn{2}{c|}{Periods from periodogram analysis, [s]}\\
No   &    UT   &        UT        &  95.0\%  & 99.7\% &  95.0\%  &99.7\%\\
\hline
1    &  	01:41:00   &	01:53:00  &  no      &   no   & 8.1--9.1 & 8.6\\
2    &  	21:00:00	 &	21:02:00  &	 no      &   no   &    no    & no \\
3    &  	01:38:00	 &	01:43:00  &  no      &   no   &    no    & no \\
4    &  	08:24:00	 &	08:27:00  &	 no      &   no   &    no    & no \\
5    &  	01:38:00   &	01:43:00  &	26--28   &   no   &    no    & no \\
6    &  	13:11:40	 &	13:21:40  &	25--40 &   no   &    no    & no \\
7    &  	01:00:00	 &	01:08:00  &	  no   &   no   &    no    & no \\
8    &  	00:13:00	 &	00:18:00  &	  no   &   no   &    no    & no \\
9    &  	18:21:30	 &	18:24:00  &	  no   &   no   &    no    & no \\
10   &  	03:29:00	 &	03:35:00  &	  no   &   no   &    no    & no \\
11   &  	03:28:00	 &	03:33:00  &	  no   &   no   &    no    & no \\
12   &  	22:53:00	 &	23:08:00  &	40--148& 50--130&    no    & no \\
13   &  	00:28:00	 &	00:33:00  &	  no   &   no   &    no    & no \\
14   &  	02:05:00	 &	02:15:00  &	  no   &   no   &    no    & no \\
15   &  	00:27:00	 &	00:32:00  &	  no   &   no   &    no    & no \\
16   &  	00:39:00	 &	00:43:00  &	  no   &   no   &    no    & no \\
17   &  	22:30:00	 &	23:10:00  &	  no   &   no   &400--800  & 400\\
18   &  	09:41:00	 &	09:55:00  &	50--130& 80--110&    no    & no \\
19   &  	08:11:00	 &	08:31:00  &	90--290& 90--235&    no    & no \\
20   &  	16:37:00	 &	16:45:00  &	  no   &   no   &    no    & no \\
21   &  	21:30:00	 &	22:20:00  &	  no   &   no   &   8.2    & no \\
22   &  	23:18:00   &	23:22:00  &	  no   &   no   &    no    & no \\
23   &  	01:48:00	 &	01:58:00  &	  no   &   no   &    no    & no \\
24   &  	06:25:00	 &	06:41:00  &	  no   &   no   &    no    & no \\
25   &  	22:12:00	 &	22:24:00  &	  no   &   no   &    no    & no \\
26   &  	12:50:00	 &	13:00:00  &	  no   &   no   &    no    & no \\
27   &  	14:05:00	 &	14:09:00  &	  no   &   no   &   8.6    & no \\
28   &  	21:09:00	 &	21:19:00  &	  no   &   no   &    no    & no \\
29   &  	15:28:00	 &	15:31:00  &	  no   &   no   &    no    & no \\
\hline
\end{tabular}
\label{T-3}
\end{table}
\end{landscape}

\begin{landscape}
\newpage
\begin{table}
\caption{Results of analysis of quantitative characteristics of dynamics of the sources of HXR pulsations in the group-1 flares.}
\tiny
\begin{tabular}{|c|c|c|c|c|c|c|c|c|c|c|c|c|c|c|c|}
\hline
1 & 2 & 3 & 4 & 5 & 6 & 7 & 8 & 9 & 10 & 11 & 12 & 13 & 14 & 15 & 16\\
\hline
Flare & $\left\langle D_{S}\right\rangle$, & $\sigma{(D_{S})}$, & $\left\langle \alpha_{sh}\right\rangle$, & $\sigma{(\alpha_{sh})}$, & $\left\langle D_{IP}\right\rangle$, & $\sigma{(D_{IP})}$, & $\left\langle v_{IP}\right\rangle$, & $\sigma{(v_{IP})}$, & $\left\langle v^{+}\right\rangle$, & $\sigma{(v^{+})}$, & $\left\langle v^{-}\right\rangle$, & $\sigma{(v^{-})}$, & $R^{+}$ & $R^{-}$ & Types of \\
No & [Mm] & [Mm] & [Deg] & [Deg] & [km] & [km] & [km s$^{-1}$] & [km s$^{-1}$] & [km s$^{-1}$] & [km s$^{-1}$] & [km s$^{-1}$] & [km s$^{-1}$] & & & Dynamics \\
\hline
1 & 24 & 8 & 40 & 17 & 1600 & 1460 & 36 & 36 & 57& 61& 57& 63& 2.4& 0.6& A+D+C\\
6 & 29 & 6 & 53 & 16 & 4767 & 6236 & 128 & 154 & 160 & 156& 105& 93& 1.8& 3.8& A+D\\
7 & 16 & 4 & 32 & 15 & 965 & 860 & 14& 12& 56& 93& 90& 66& 1.3& 1.8& A+C\\
8 & 10& 4 & 28 & 18 & 2932 & 7876 & 120& 301& 71& 52& 61& 91& 1.3& 1.1& A+B+C\\
10 & 13 & 4 & 48 & 16 & 1612 & 1440 & 51& 50& 53& 48& 37& 49& 1.4& 2.1& A\\
11 & 22 & 4 & 40 & 6 & 1570 & 1485 & 61& 51& 110& 87& 81& 89& 3.5& 1.0& A\\
12 & 40 & 1 & 61 & 3 & 1931 & 1630 & 23& 17& 28& 31& 18& 20& 0.9& 1.4& A\\
13 & 22 & 1 & 20 & 5 & 1177 & 1024 & 40& 42& 48& 42& 50& 65& 3.2& 3.2& A+C\\
15 & 17 & 2 & 5 & 7 & 723 & 989 & 27& 41& 72& 50& 46& 45& 0.8& 1.8& A+D+C\\
17 & 30 & 6 & 52 & 21 & 6680 & 8550 & 50& 51& 64& 57& 50& 94& 3.1& 1.7& A+C\\
18 & 26 & 10 & 31 & 14 & 1146 & 1113 & 44& 64& 107& 132& 73& 76& 1.0& 1.6& A+C\\
19 & 36 & 10 & 1 & 8 & 5046 & 3890 & 72& 63& 70& 59& 35& 26& 4.9& 0.3& A+B+C\\
24 & 18 & 4 & 31 & 10 & 1395 & 1170 & 24& 23& 48& 54& 32& 23& 4.4& 6.0& A+C\\
25 & 9 & 4 & 37 & 17 & 1843 & 2100 & 61& 60& 86& 106& 103& 155& 1.6& 0.8& A+C\\
27 & 42 & 6 & 34 & 5 & 1546 & 2212 & 38& 50& 67& 68& 88& 101& 3.5& 1.1& A+D\\
29 & 11 & 3 & 72 & 4 & 1060 & 1161 & 59& 66& 63& 46& 97& 71& 5.3& 4.9& A\\
\hline
\end{tabular}
\normalsize
\label{T-4}
\end{table}
\end{landscape}

\newpage
\appendix   
\section{HXR light curves, pulsations, and time differences between successive pulsations of the studied flares} 
\label{S-appendix-A}

\begin{figure}
\centerline{\includegraphics[width=0.4\textwidth,clip=]{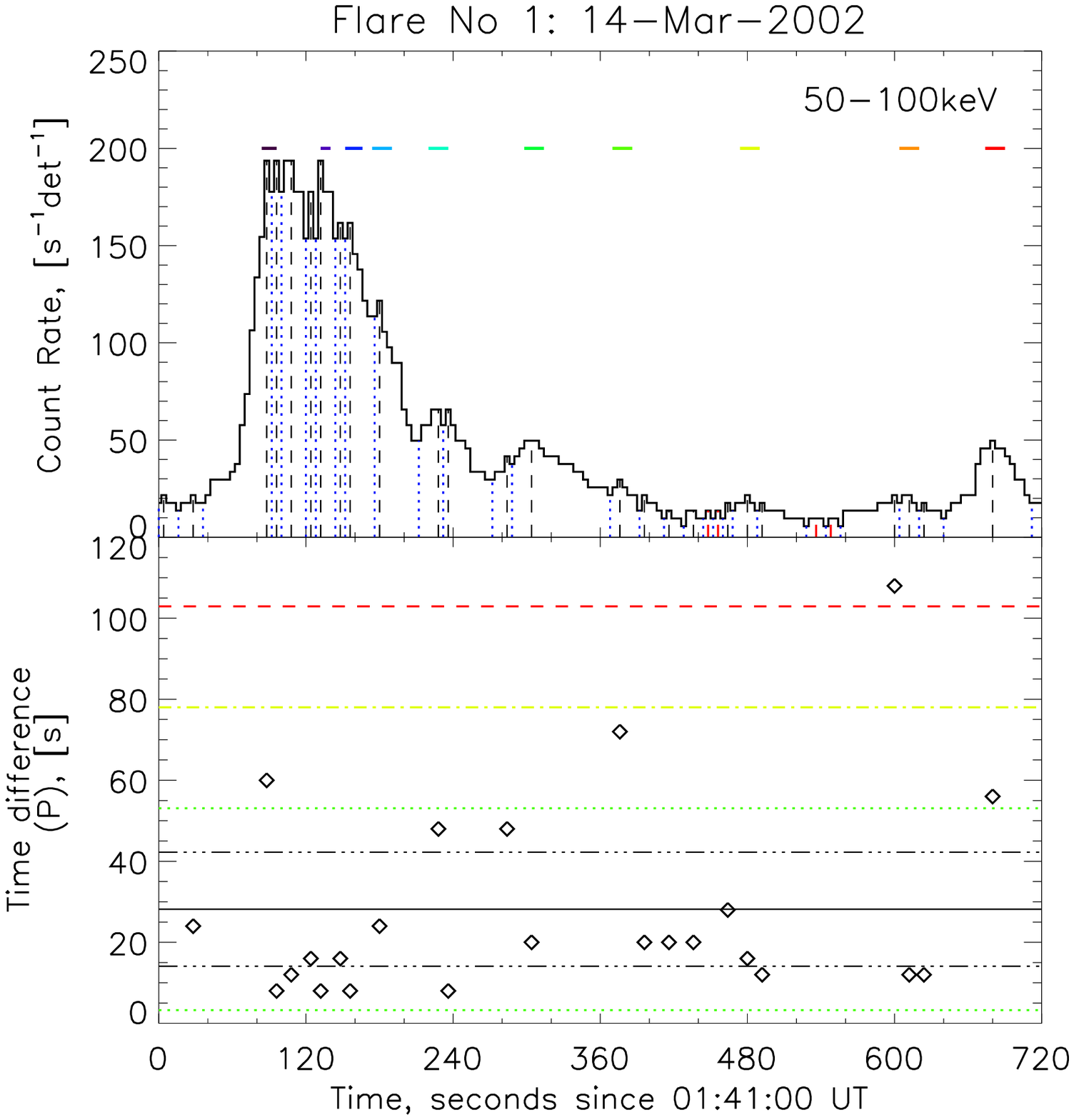}
\includegraphics[width=0.4\textwidth,clip=]{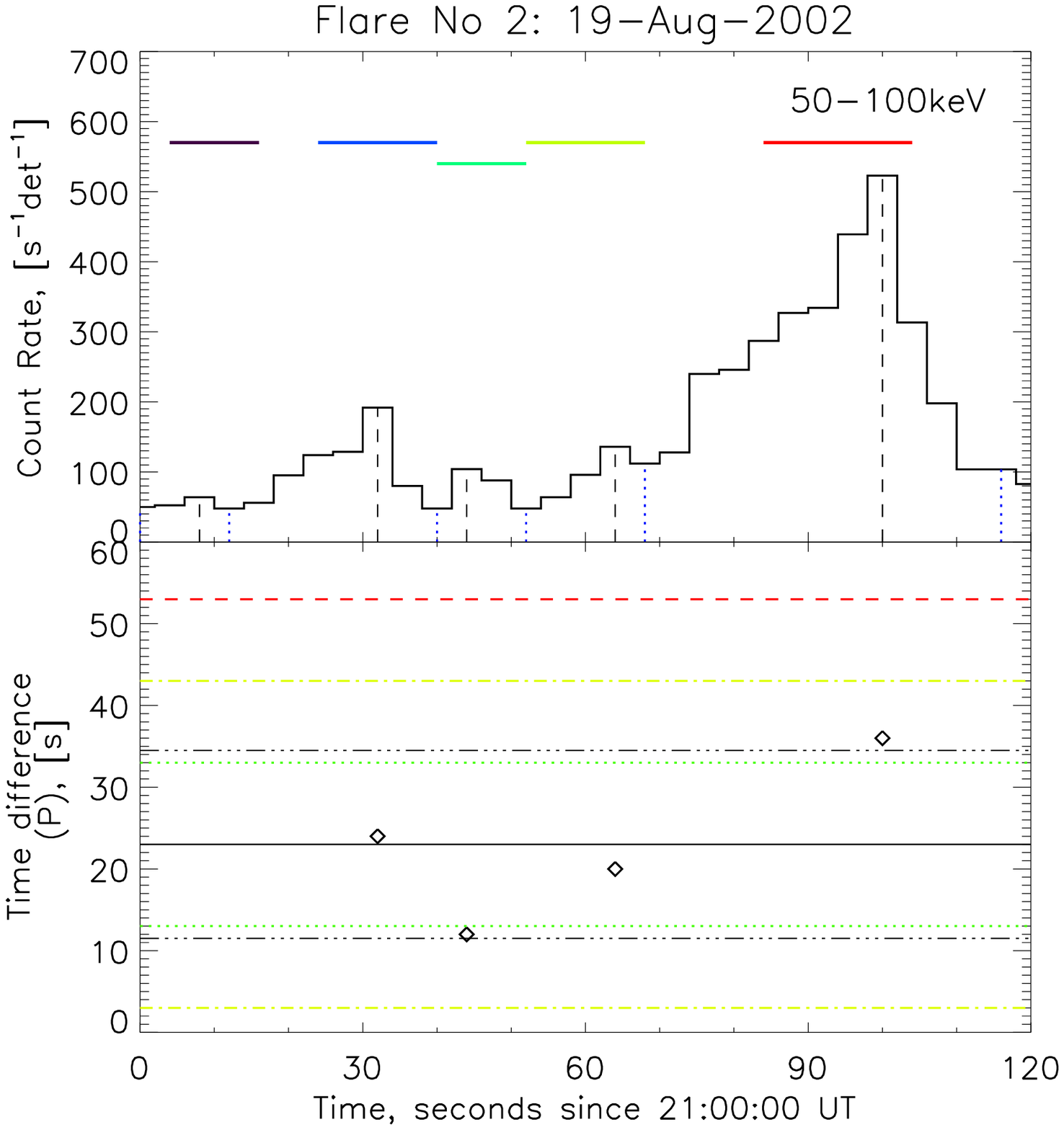}}
\vspace{-0.4\textwidth}   
\centerline{\small \bf    
\hspace{0.02\textwidth}  \color{black}{(a)}
\hspace{0.82\textwidth}  \color{black}{(b)}
\hfill}
\vspace{0.37\textwidth}   

\centerline{\includegraphics[width=0.4\textwidth,clip=]{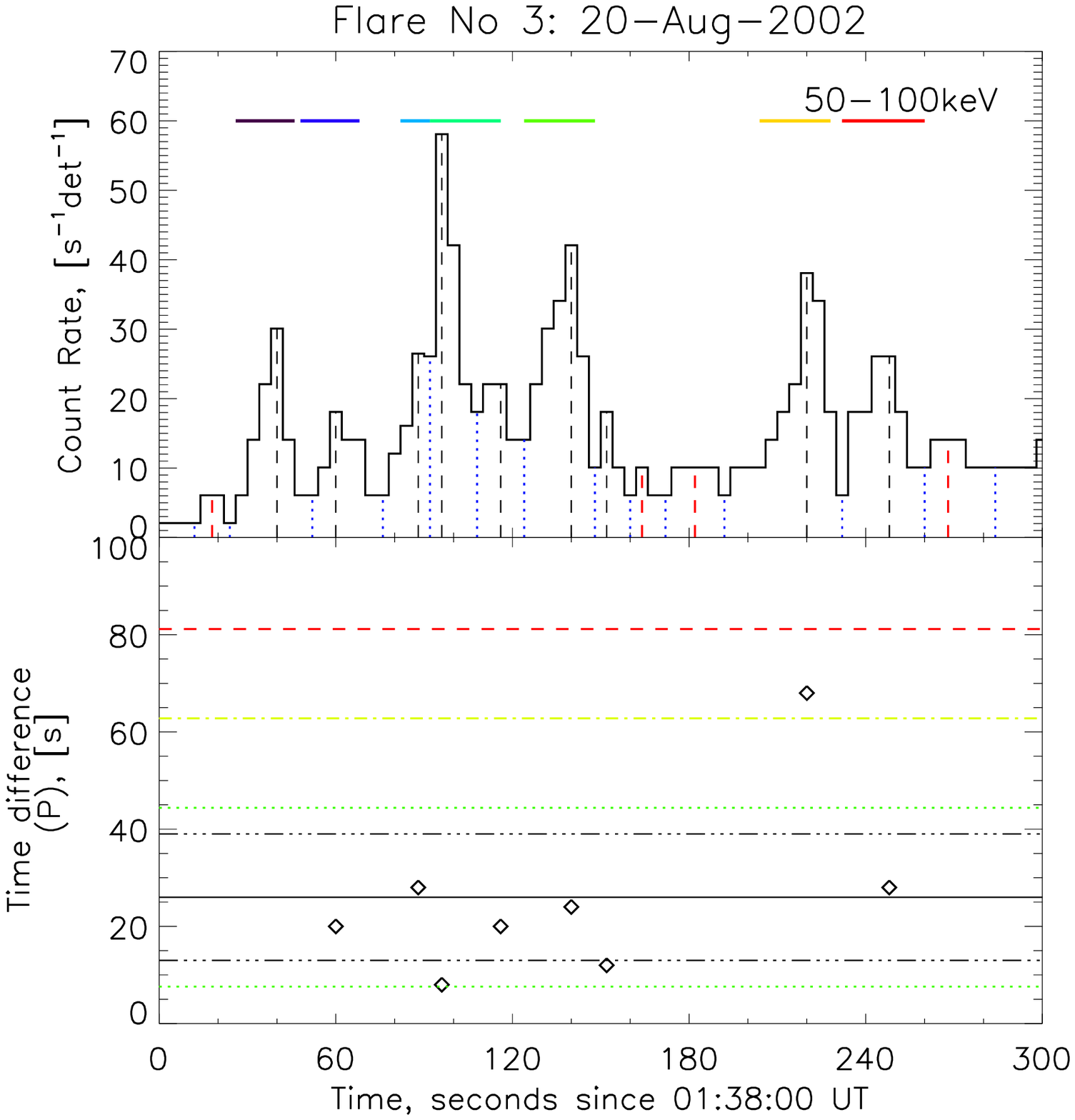}
\includegraphics[width=0.4\textwidth,clip=]{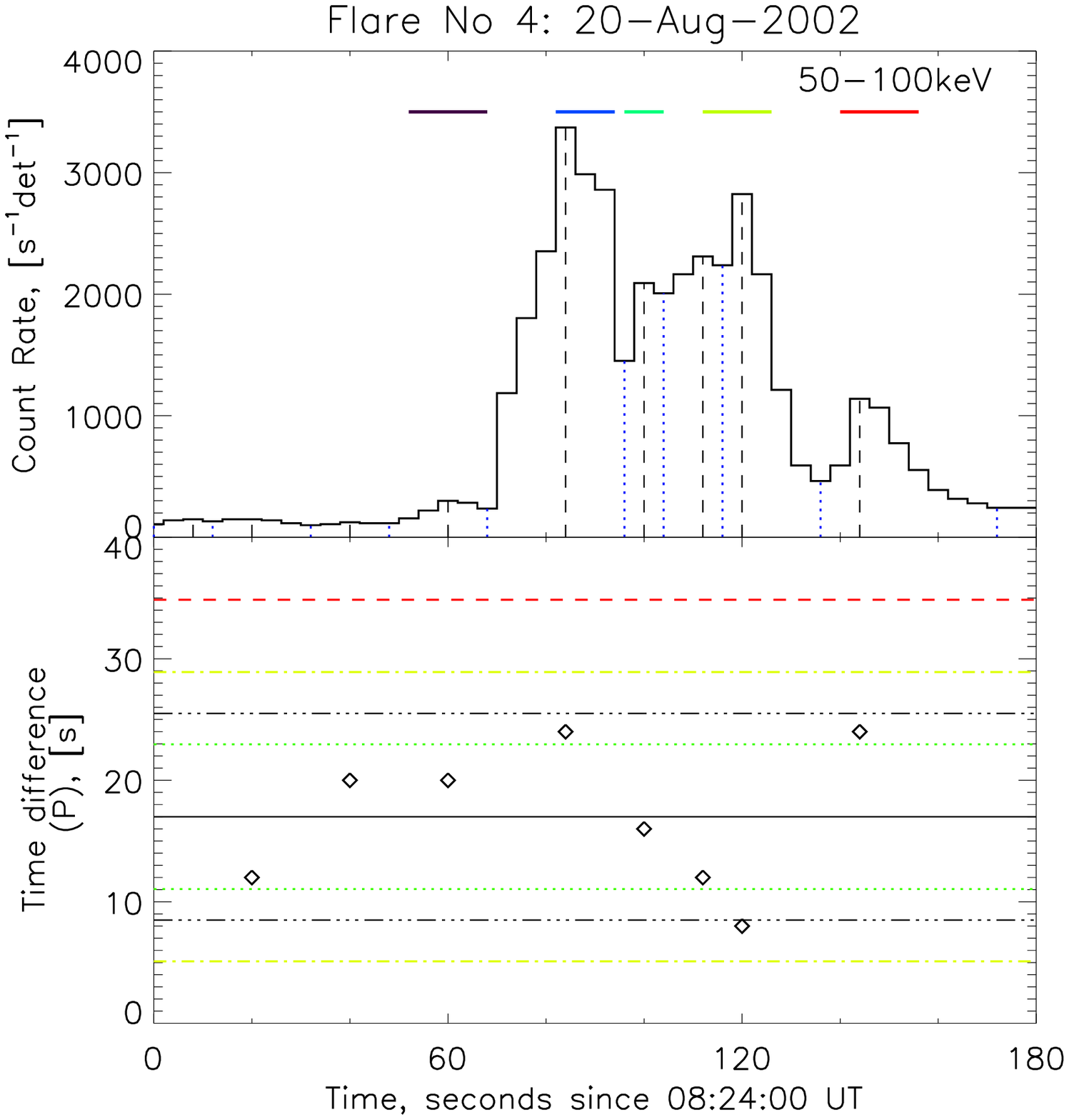}}
\vspace{-0.4\textwidth}   
\centerline{\small \bf   
\hspace{0.02\textwidth}  \color{black}{(c)}
\hspace{0.82\textwidth}  \color{black}{(d)}
\hfill}
\vspace{0.37\textwidth}   

\centerline{\includegraphics[width=0.4\textwidth,clip=]{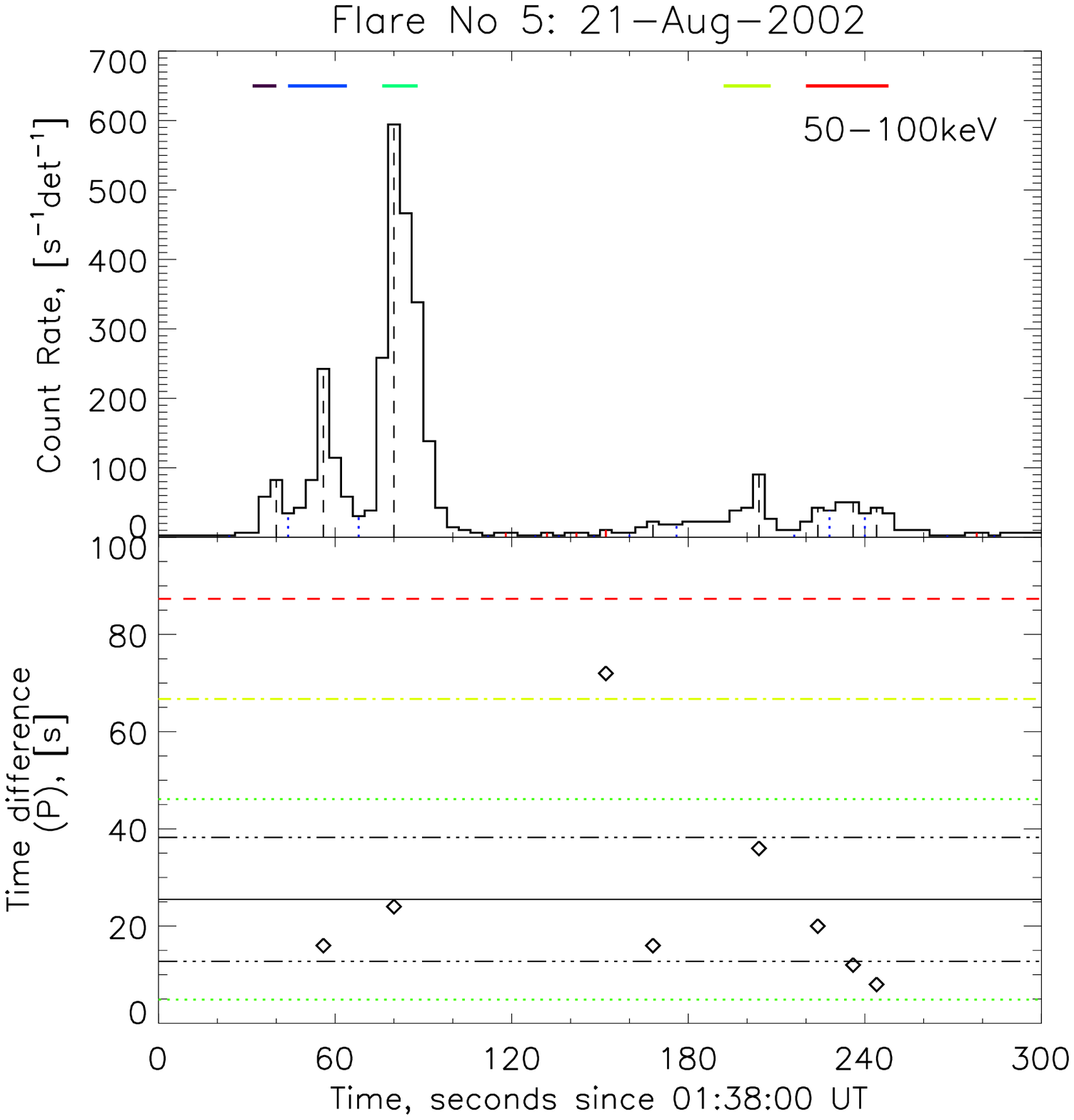}
\includegraphics[width=0.4\textwidth,clip=]{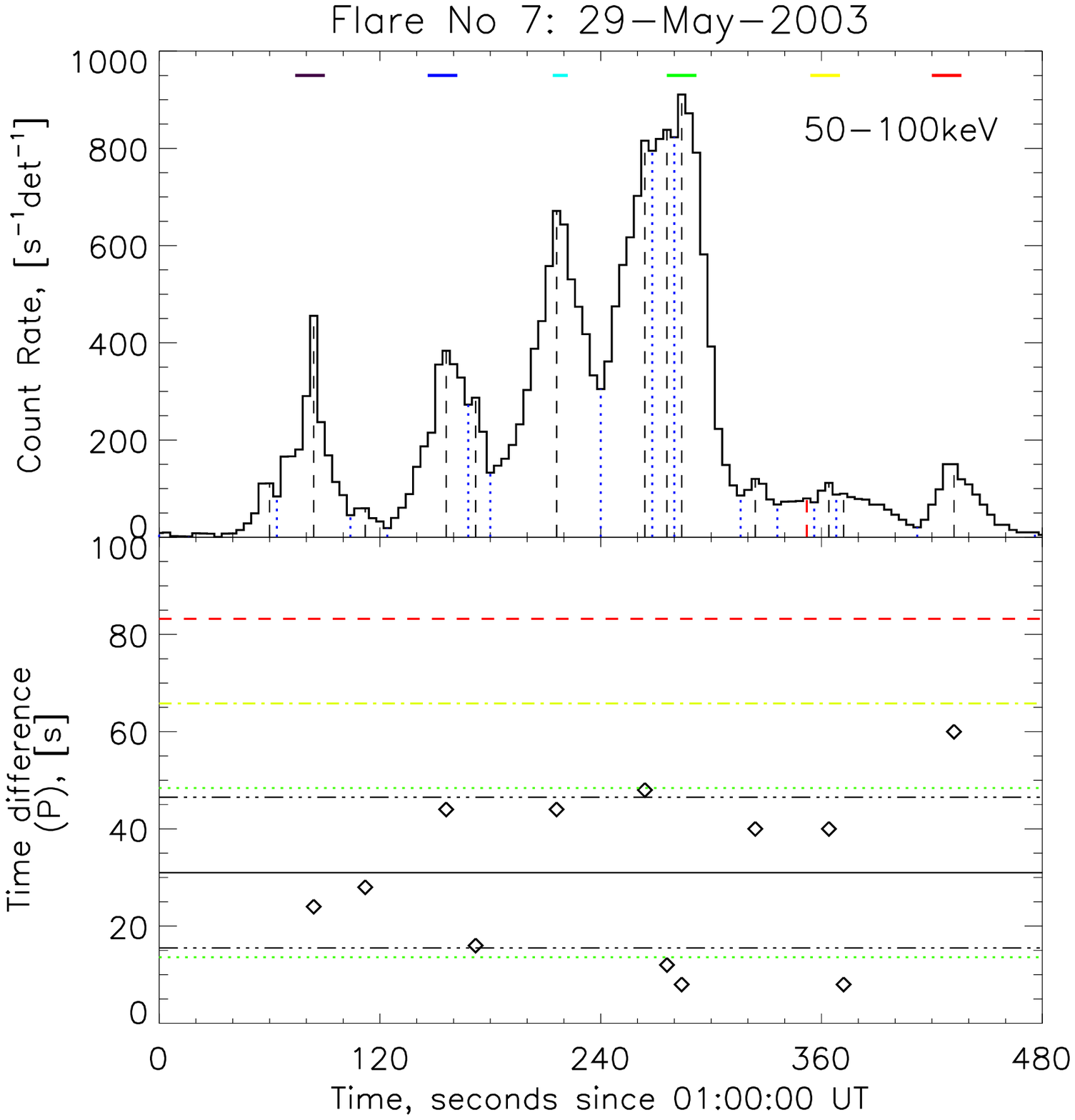}}
\vspace{-0.4\textwidth}   
\centerline{\small \bf   
\hspace{0.02\textwidth}  \color{black}{(e)}
\hspace{0.82\textwidth}  \color{black}{(f)}
\hfill}
\vspace{0.37\textwidth}   

\caption{Peaks and time differences in the RHESSI 50--100 keV corrected count rates in the studied flares No 1--5, 7 (see Table~\ref{T-1}). (\textit{Top}) RHESSI 4-second background-subtracted corrected count rate in the 50--100 keV channel is shown by black curve. Black/red vertical dashed lines indicate significant/insignificant local maxima. Blue vertical dotted lines indicate local minima. Color horizontal segments at the top mark time intervals for which HXR images were synthesized for further analysis (see Figures~\ref{F-2}--\ref{F-8}). (\textit{Bottom}) Time differences between successive significant peaks are shown by diamonds. Average value of all time differences $\left\langle P\right\rangle$ is shown by black horizontal solid line. Black, green, yellow and red horizontal dashed and dotted lines indicate levels: $\left\langle P\right\rangle \pm 0.5 \left\langle P\right\rangle$, $\left\langle P\right\rangle \pm 1.0 \sigma_{\left\langle P\right\rangle}$, $\left\langle P\right\rangle \pm 2.0 \sigma_{\left\langle P\right\rangle}$ and $\left\langle P\right\rangle \pm 3.0 \sigma_{\left\langle P\right\rangle}$ respectively, where $\sigma_{\left\langle P\right\rangle}$ is the standard deviation of the time differences.}
\label{AF-1}
\end{figure}

\begin{figure}
\centerline{\includegraphics[width=0.4\textwidth,clip=]{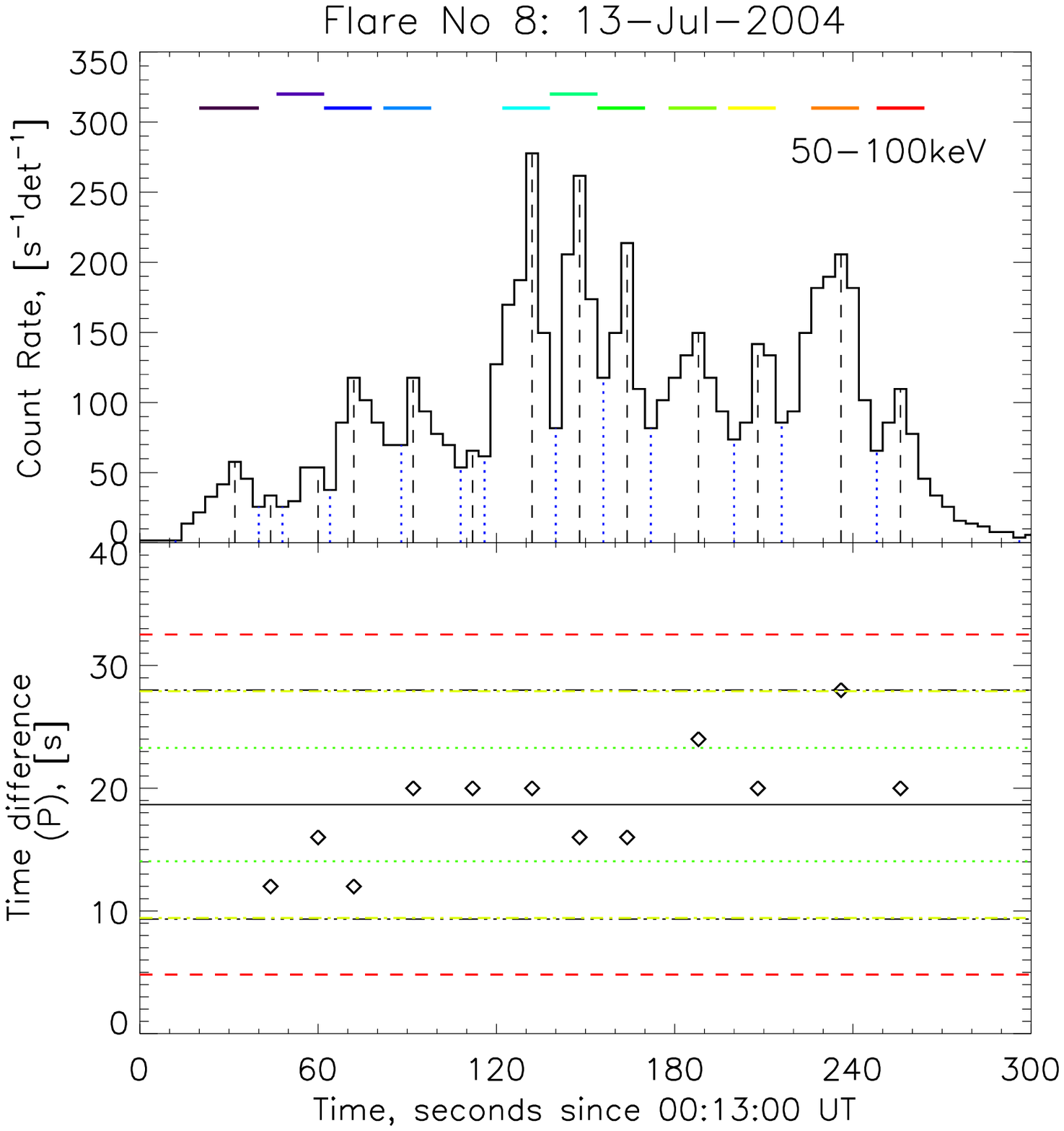}
\includegraphics[width=0.4\textwidth,clip=]{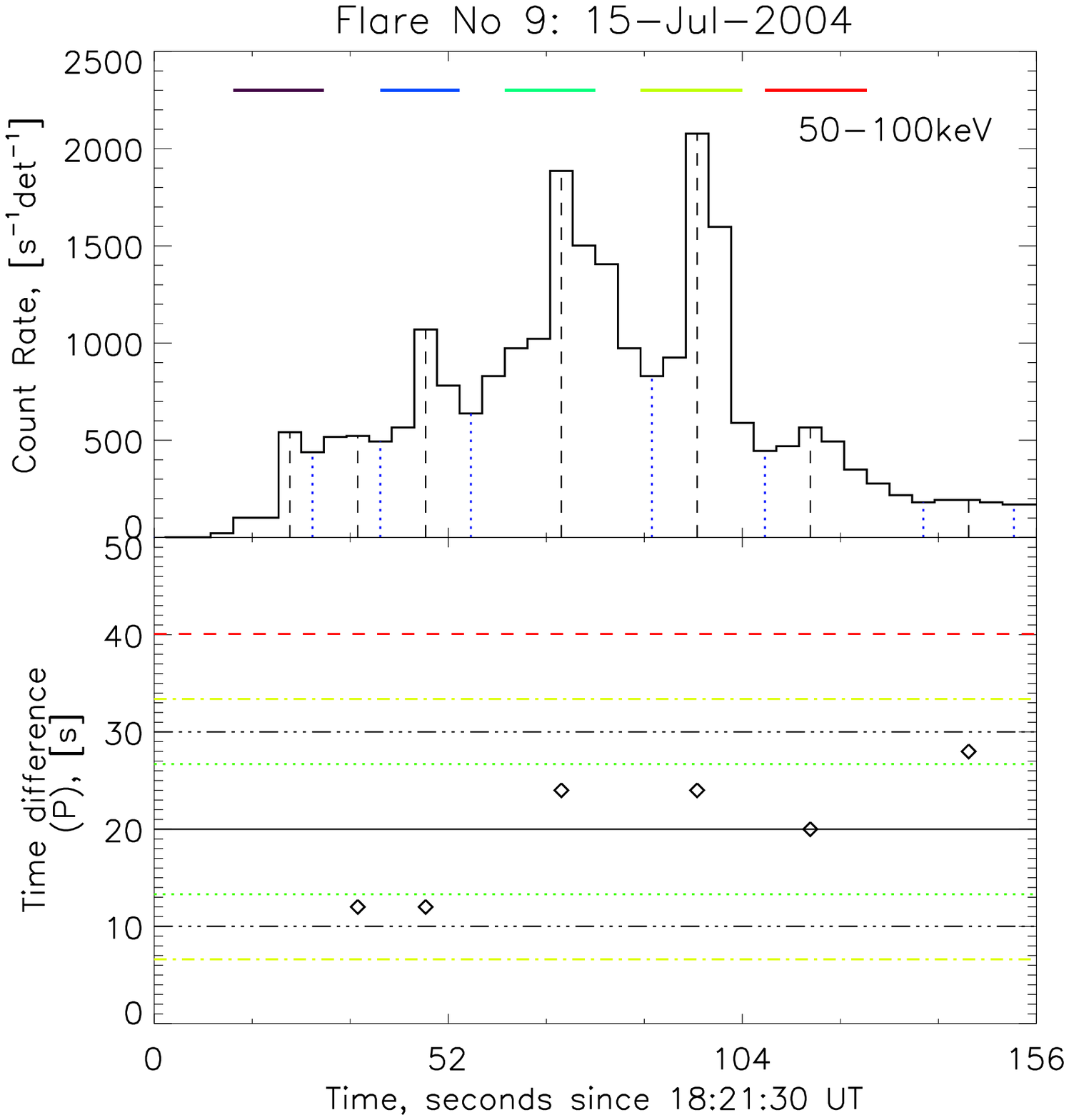}}
\vspace{-0.4\textwidth}   
\centerline{\small \bf    
\hspace{0.02\textwidth}  \color{black}{(a)}
\hspace{0.82\textwidth}  \color{black}{(b)}
\hfill}
\vspace{0.37\textwidth}   

\centerline{\includegraphics[width=0.4\textwidth,clip=]{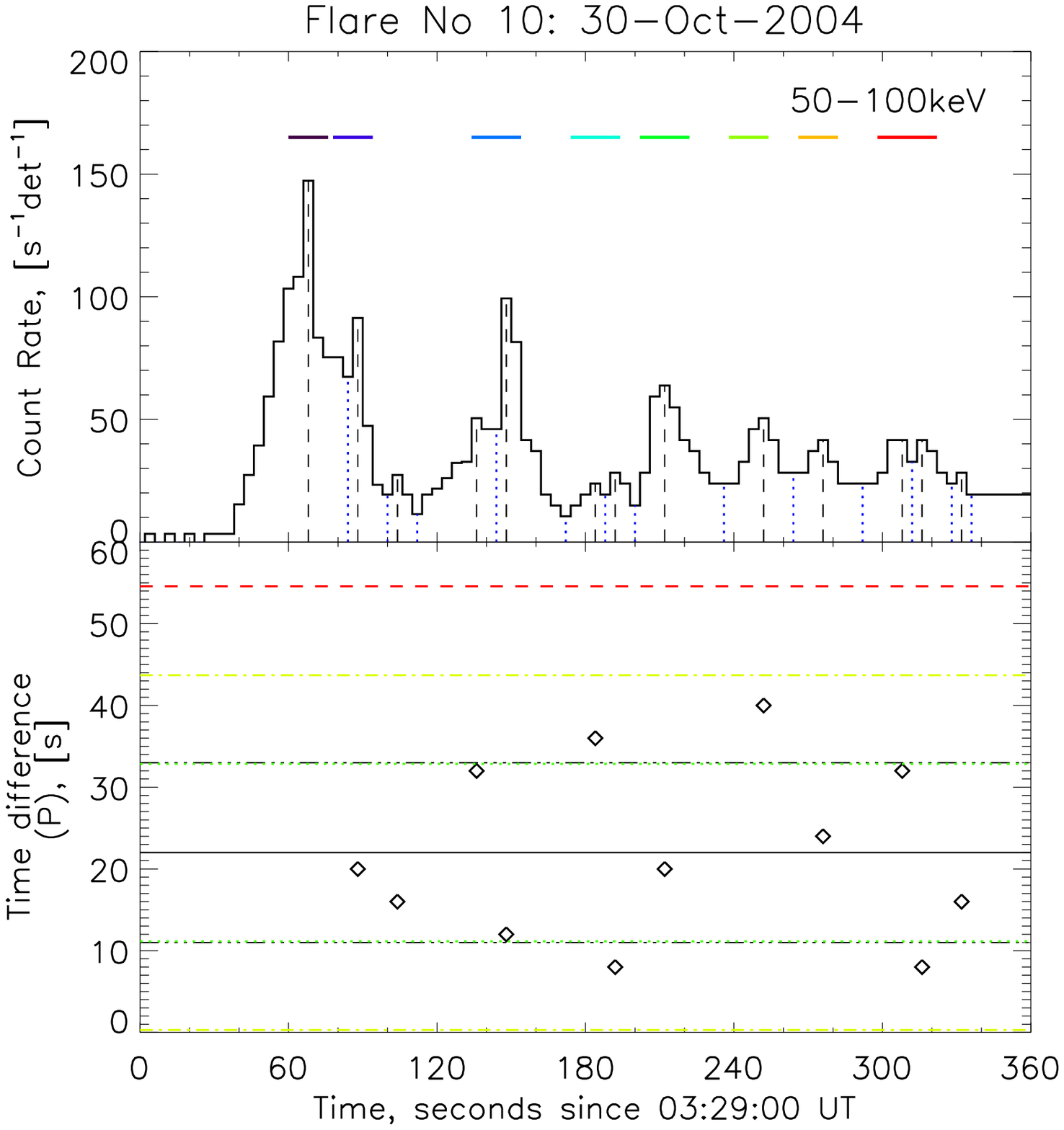}
\includegraphics[width=0.4\textwidth,clip=]{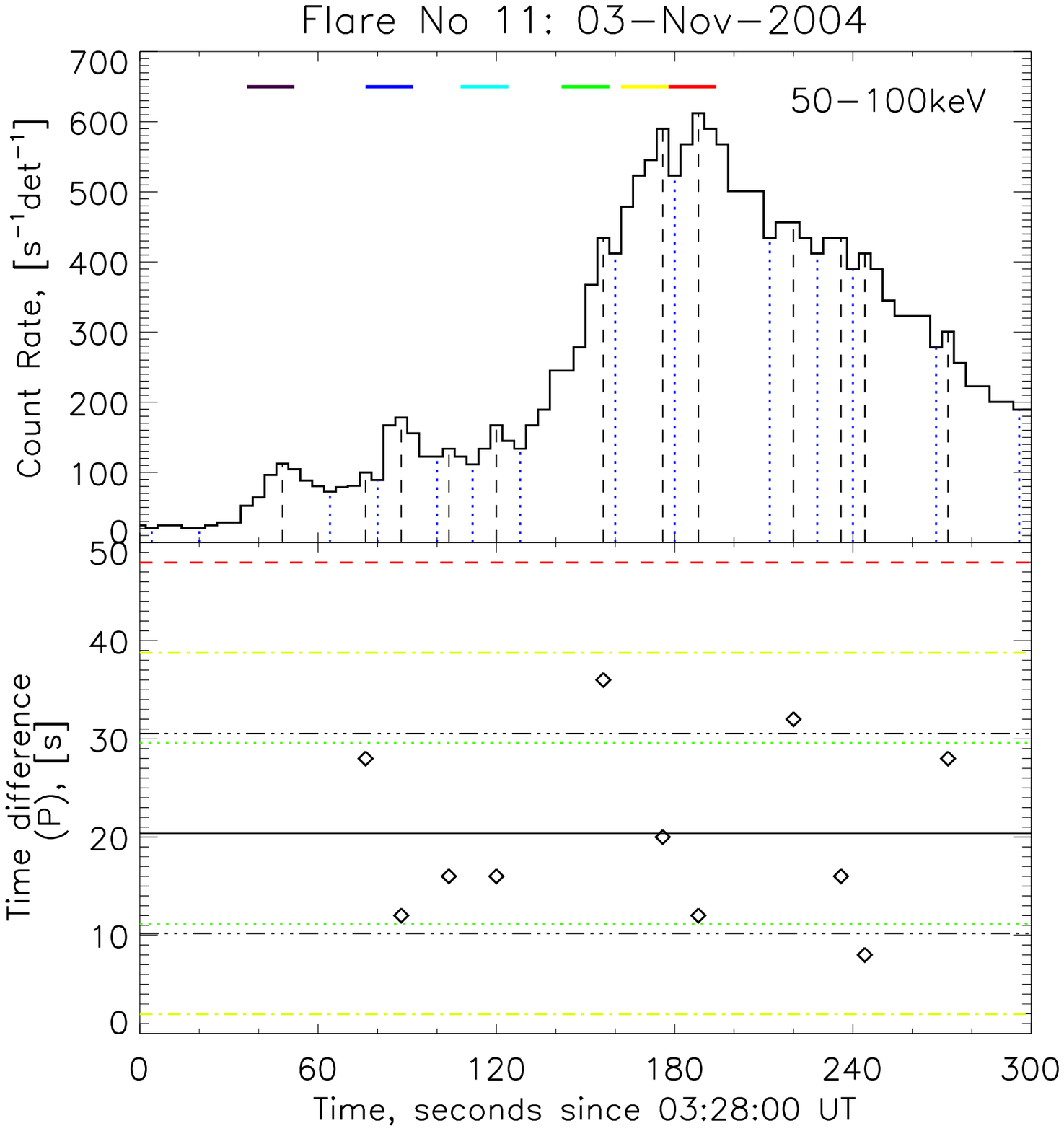}}
\vspace{-0.4\textwidth}   
\centerline{\small \bf    
\hspace{0.02\textwidth}  \color{black}{(c)}
\hspace{0.82\textwidth}  \color{black}{(d)}
\hfill}
\vspace{0.37\textwidth}   

\centerline{\includegraphics[width=0.4\textwidth,clip=]{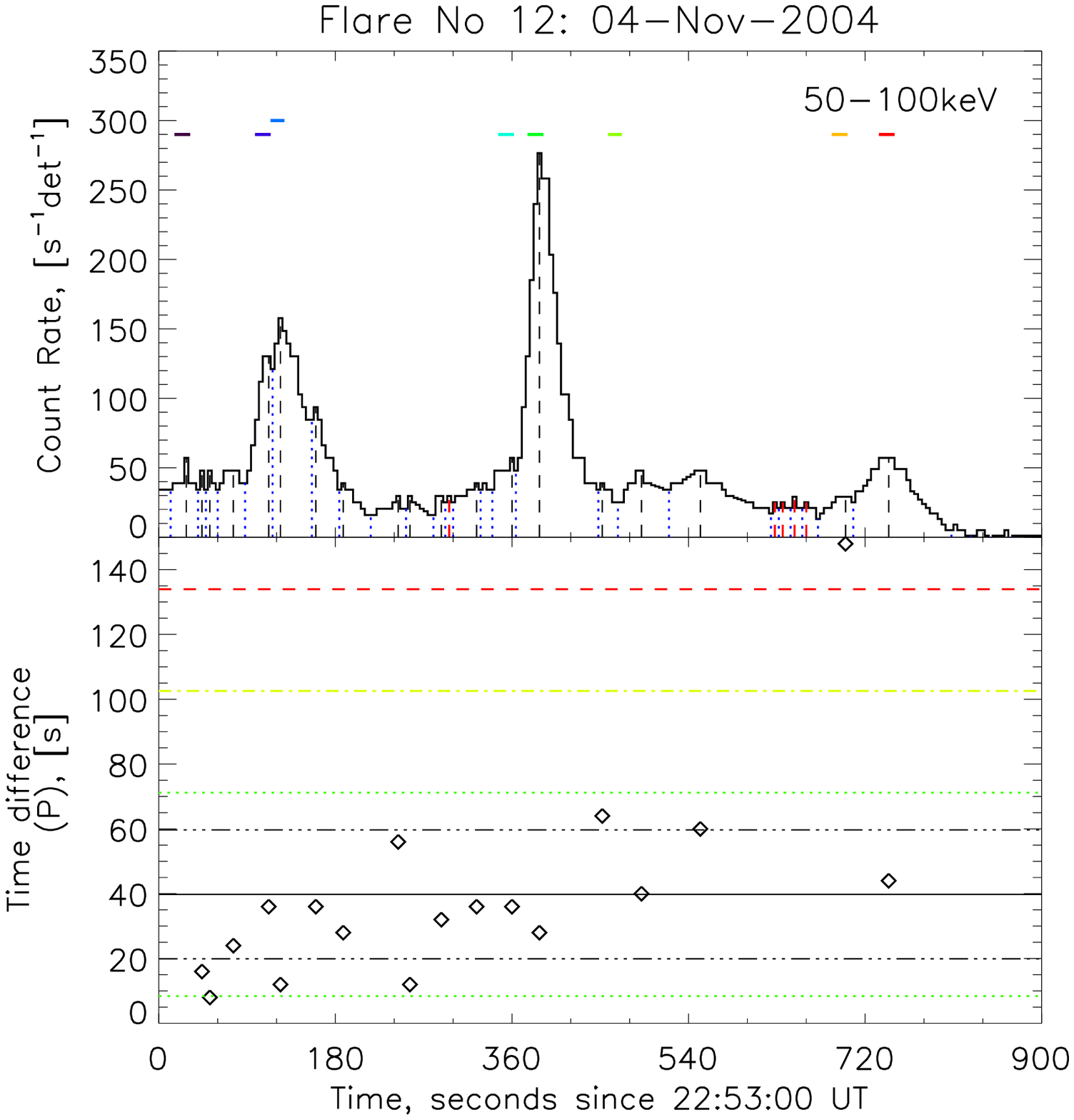}
\includegraphics[width=0.4\textwidth,clip=]{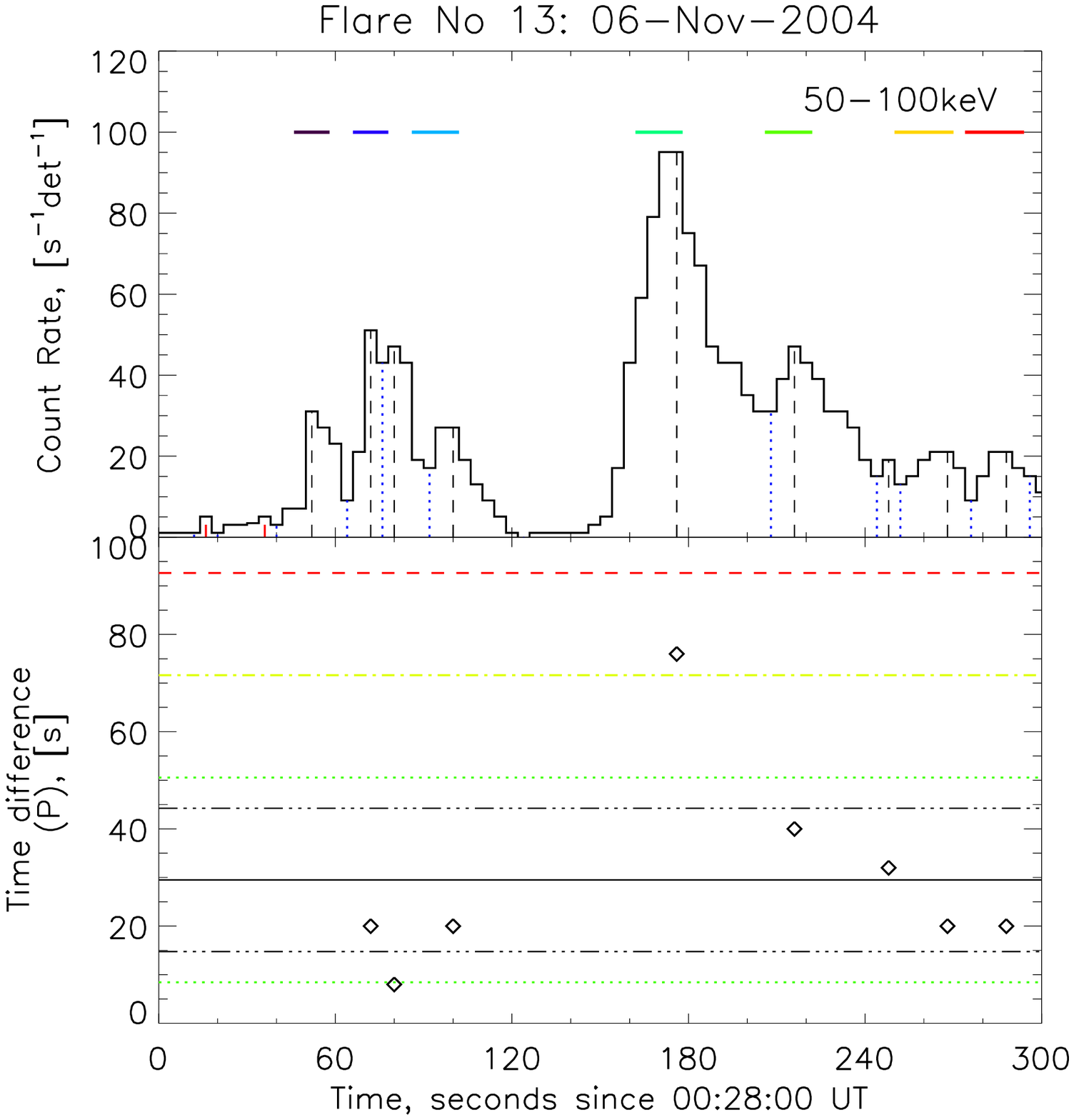}}
\vspace{-0.4\textwidth}   
\centerline{\small \bf    
\hspace{0.02\textwidth}  \color{black}{(e)}
\hspace{0.82\textwidth}  \color{black}{(f)}
\hfill}
\vspace{0.37\textwidth}   

\centerline{\includegraphics[width=0.4\textwidth,clip=]{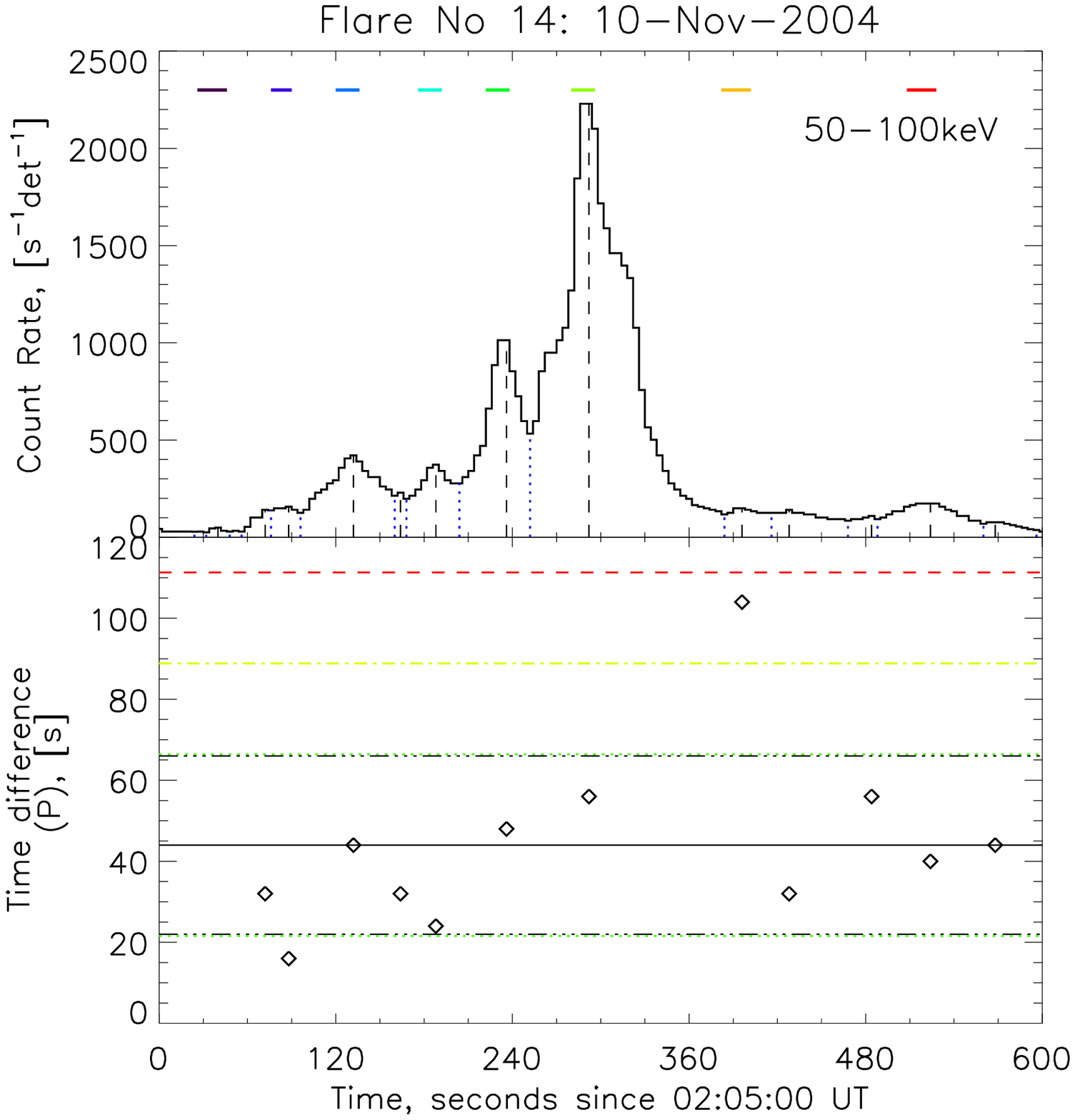}
\includegraphics[width=0.4\textwidth,clip=]{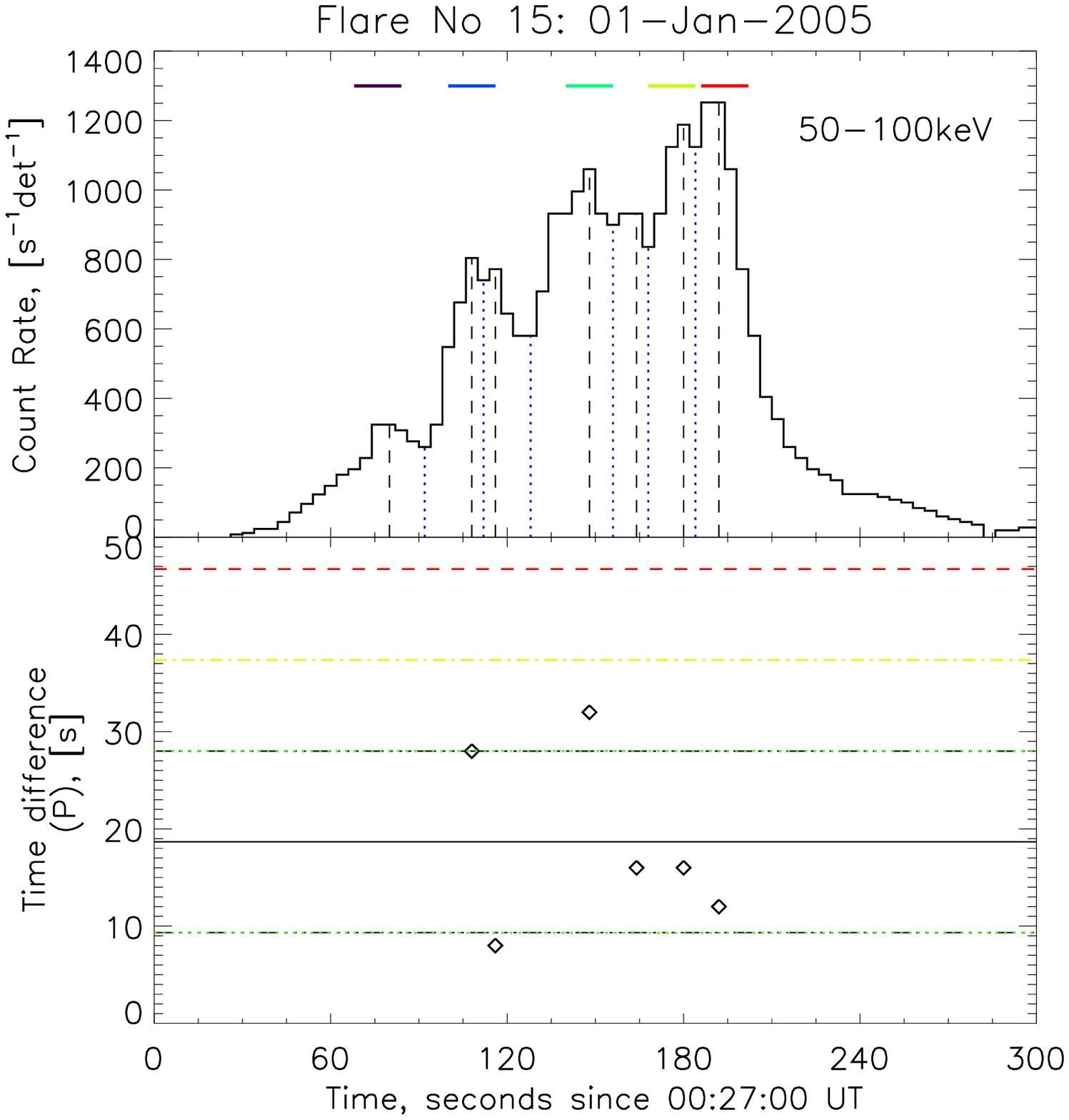}}
\vspace{-0.4\textwidth}   
\centerline{\small \bf    
\hspace{0.02\textwidth}  \color{black}{(g)}
\hspace{0.82\textwidth}  \color{black}{(h)}
\hfill}
\vspace{0.37\textwidth}   

\caption{Peaks and time differences in the RHESSI 50--100 keV corrected count rates in the studied flares No 8--15 (see Table~\ref{T-1}). The same designations as in Figure~\ref{AF-1}.}
\label{AF-2}
\end{figure}

\begin{figure}
\centerline{\includegraphics[width=0.4\textwidth,clip=]{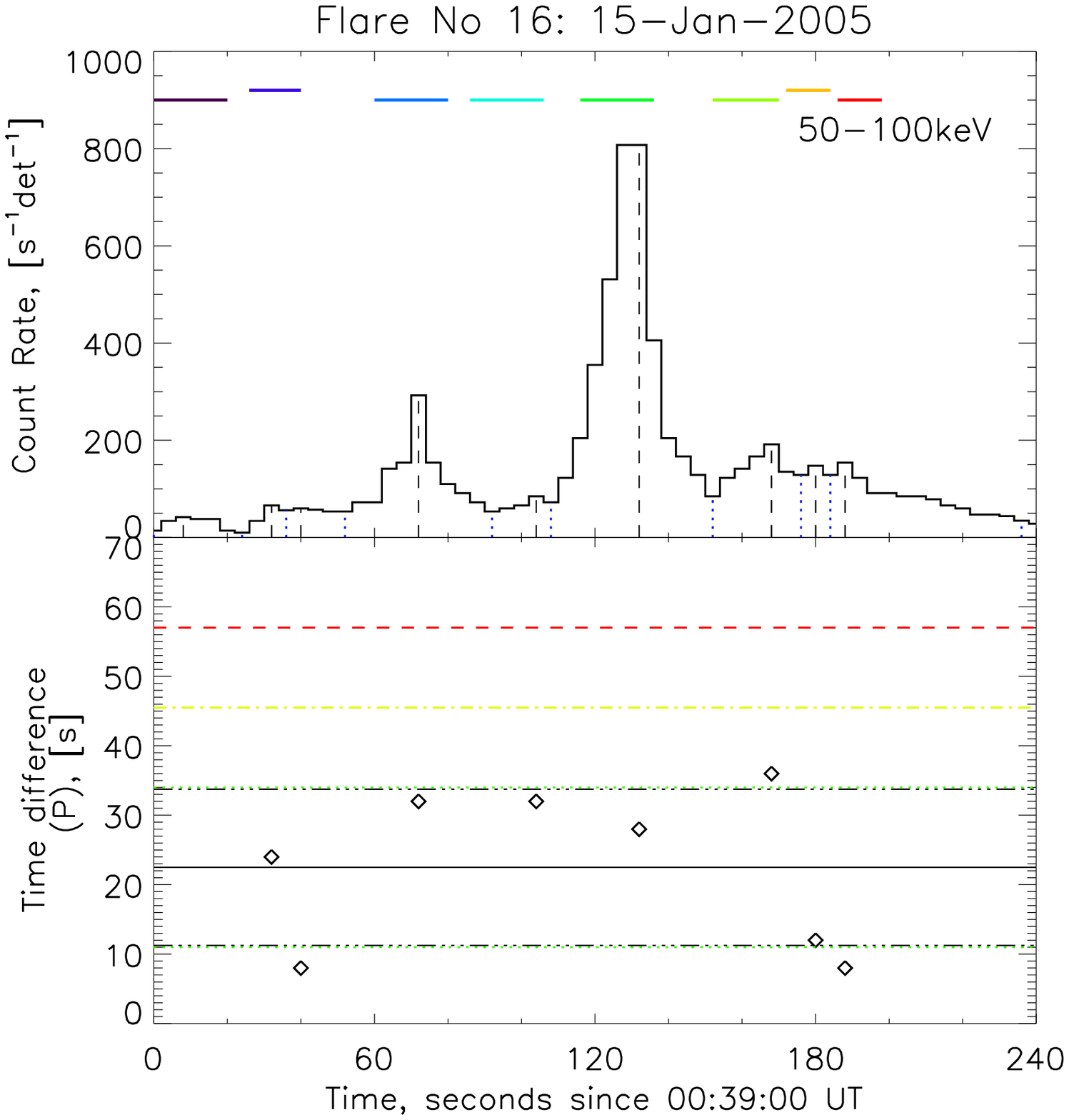}
\includegraphics[width=0.4\textwidth,clip=]{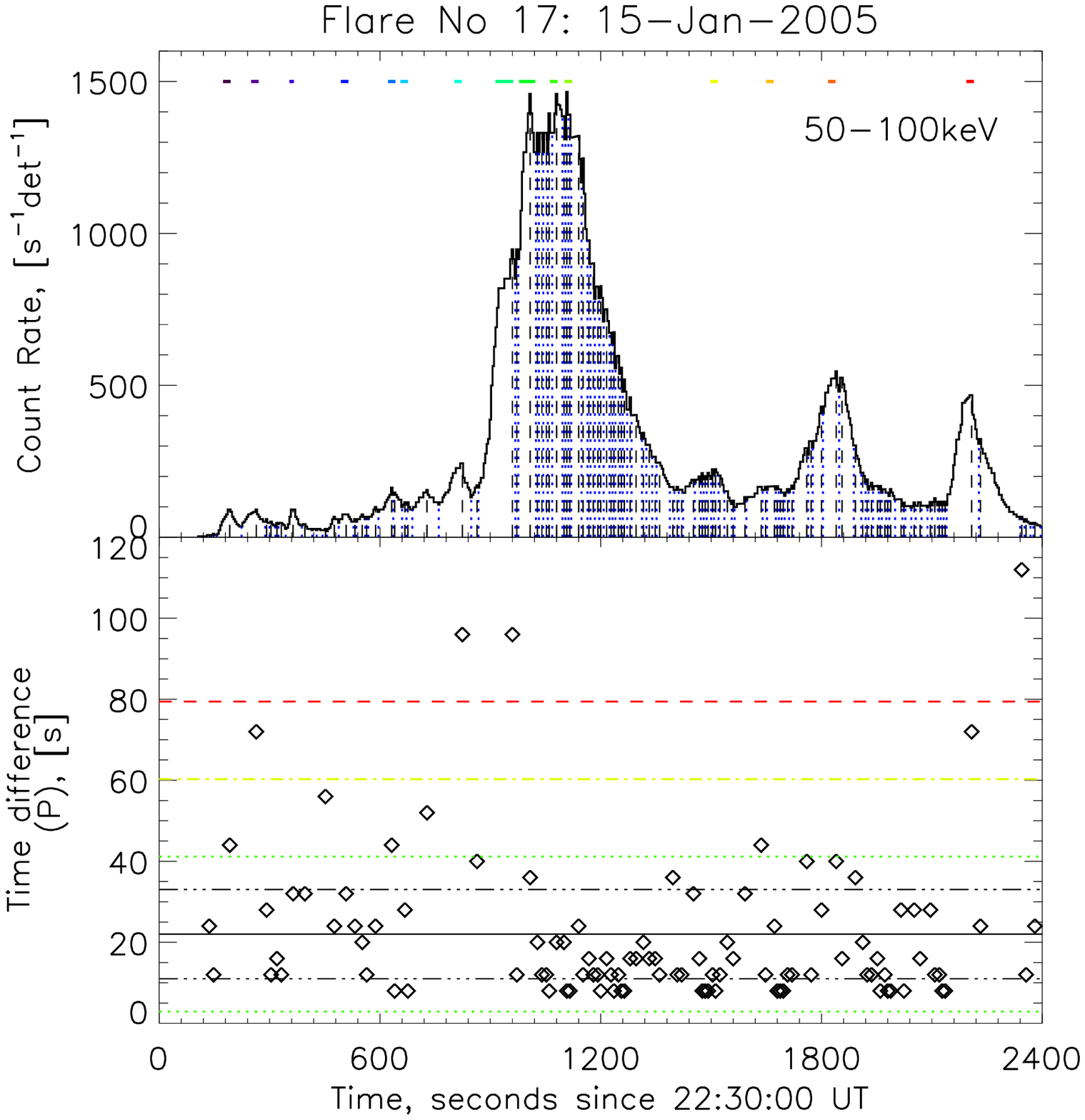}}
\vspace{-0.4\textwidth}   
\centerline{\small \bf    
\hspace{0.02\textwidth}  \color{black}{(a)}
\hspace{0.82\textwidth}  \color{black}{(b)}
\hfill}
\vspace{0.37\textwidth}   

\centerline{\includegraphics[width=0.4\textwidth,clip=]{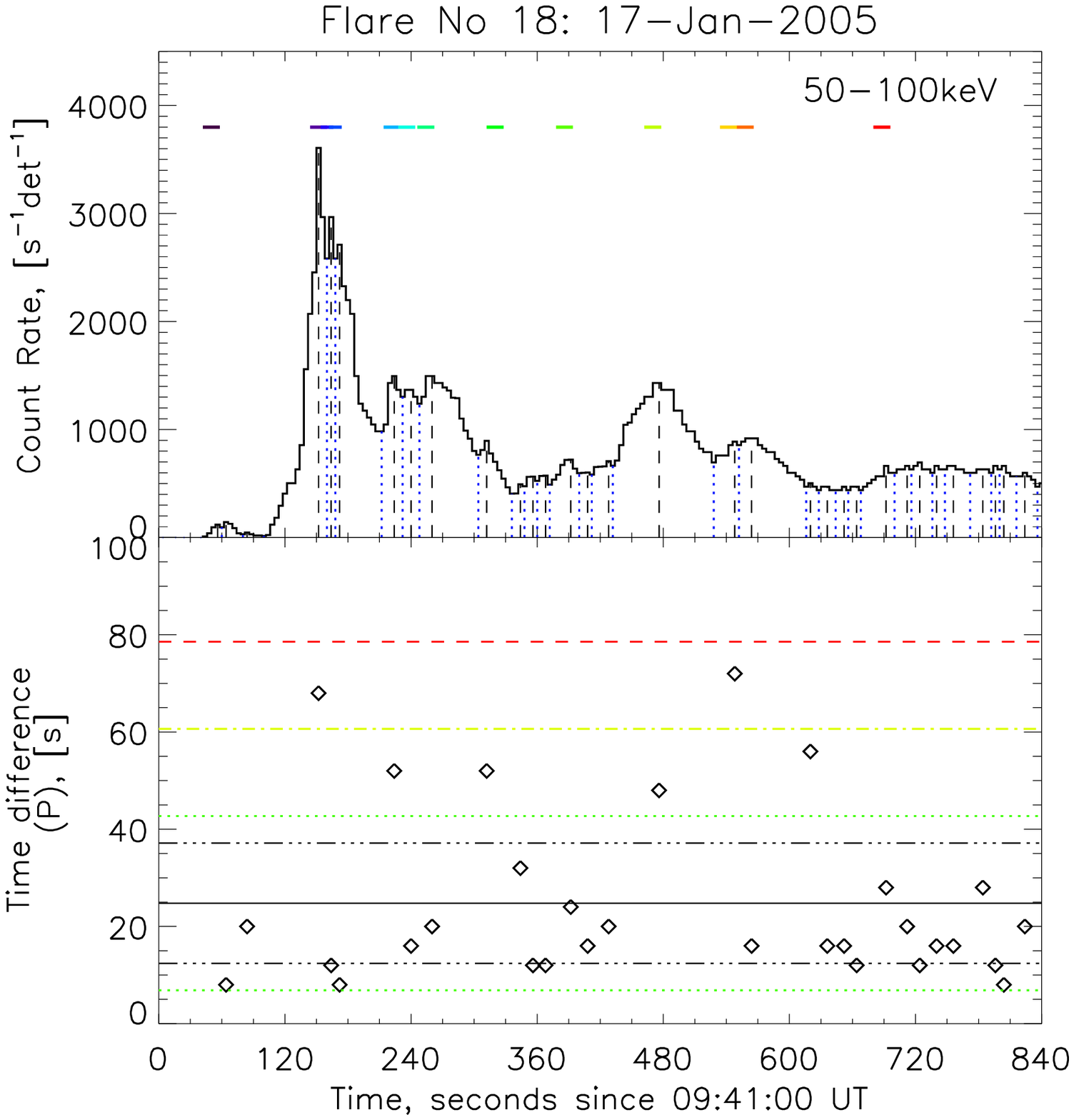}
\includegraphics[width=0.4\textwidth,clip=]{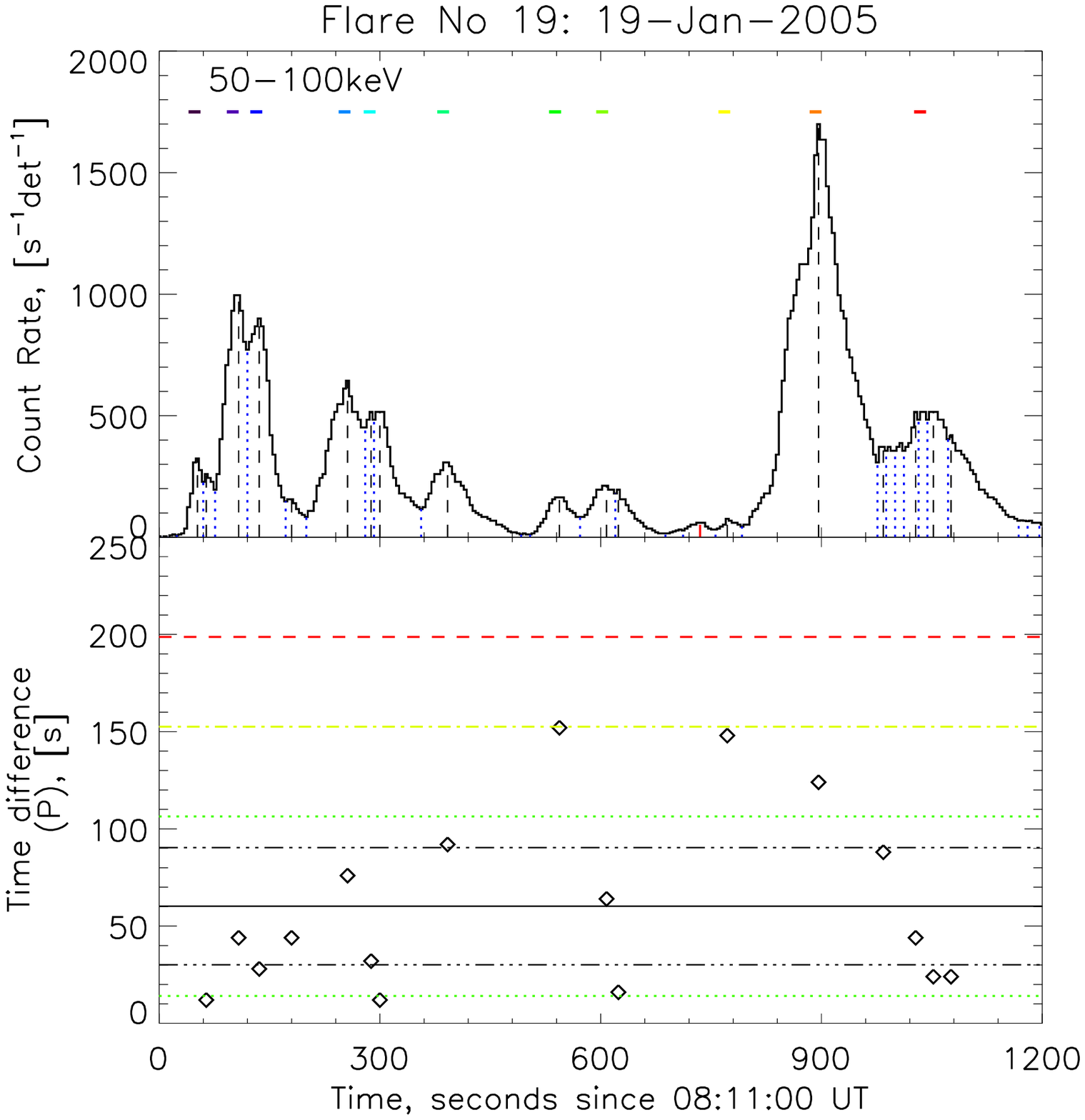}}
\vspace{-0.4\textwidth}   
\centerline{\small \bf    
\hspace{0.02\textwidth}  \color{black}{(c)}
\hspace{0.82\textwidth}  \color{black}{(d)}
\hfill}
\vspace{0.37\textwidth}   

\centerline{\includegraphics[width=0.4\textwidth,clip=]{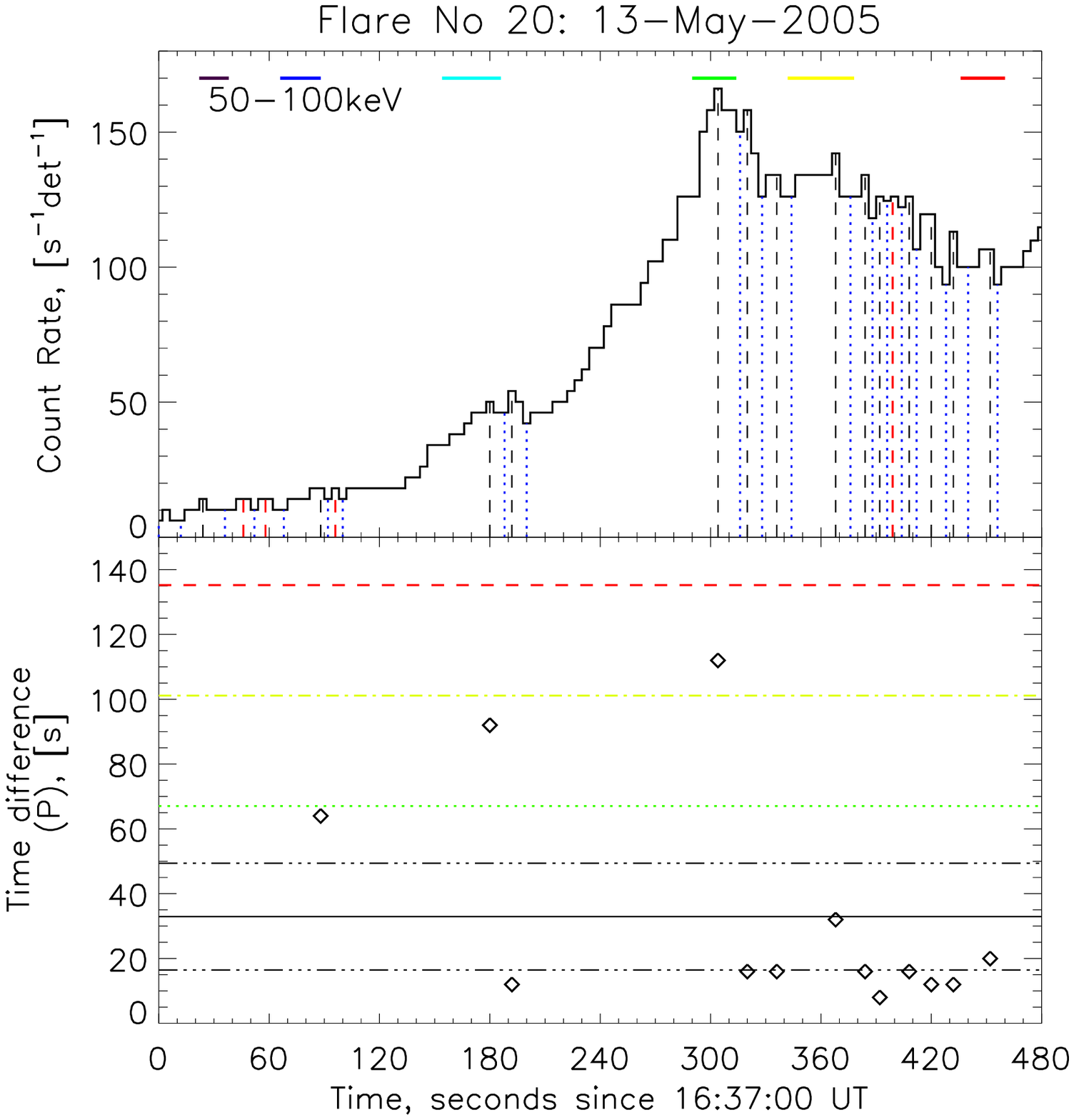}
\includegraphics[width=0.4\textwidth,clip=]{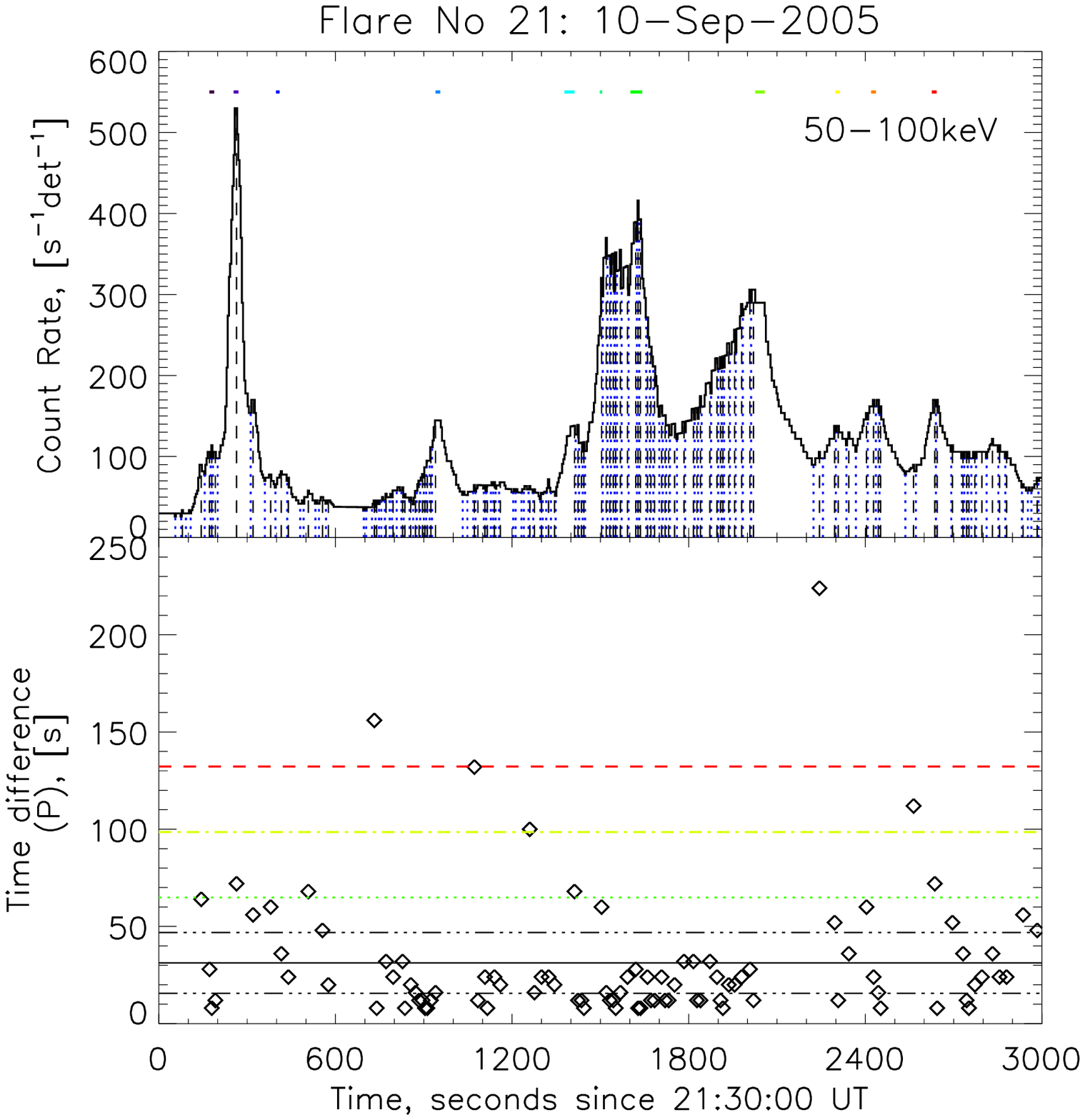}}
\vspace{-0.4\textwidth}   
\centerline{\small \bf    
\hspace{0.02\textwidth}  \color{black}{(e)}
\hspace{0.82\textwidth}  \color{black}{(f)}
\hfill}
\vspace{0.37\textwidth}   

\centerline{\includegraphics[width=0.4\textwidth,clip=]{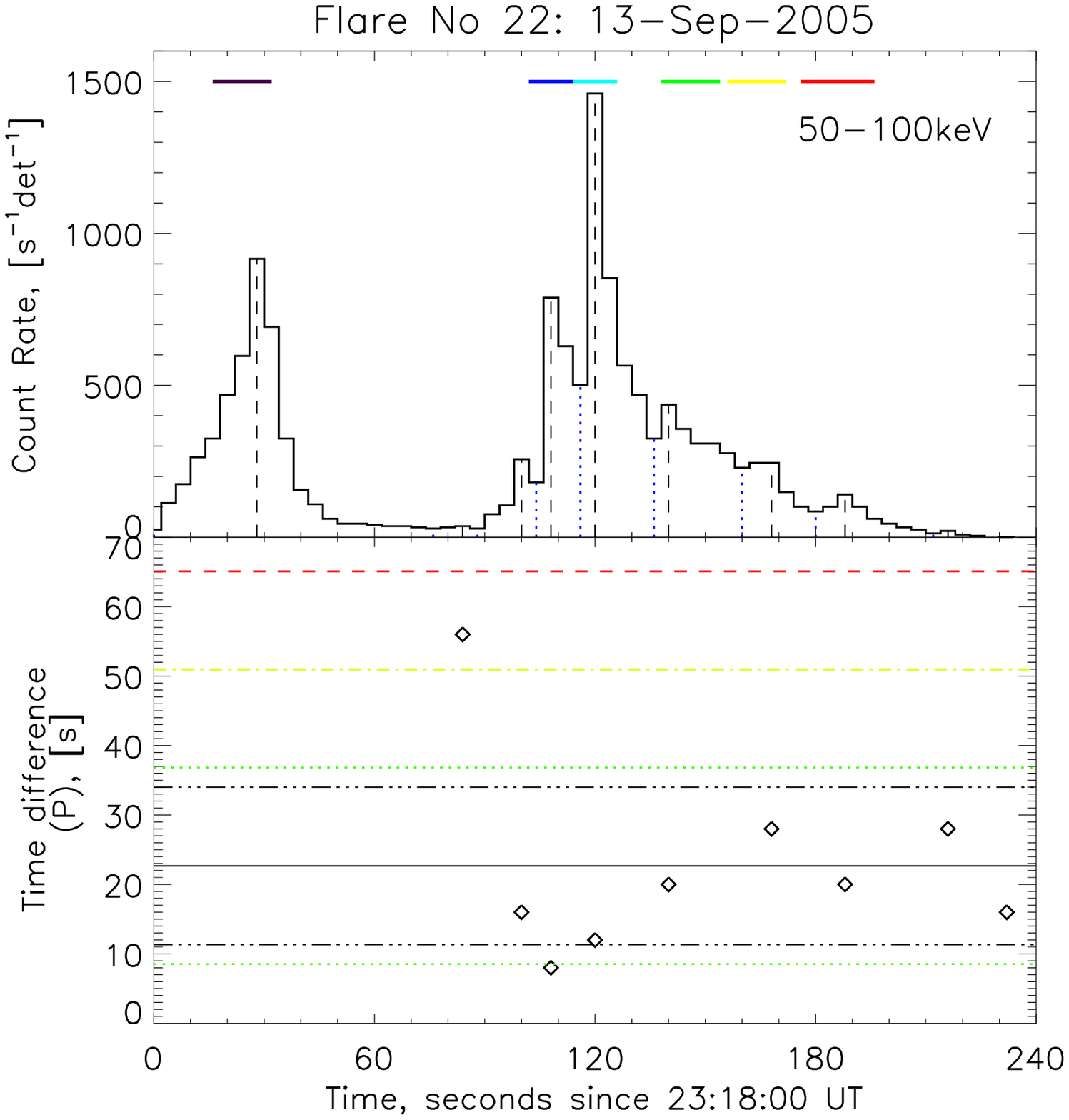}
\includegraphics[width=0.4\textwidth,clip=]{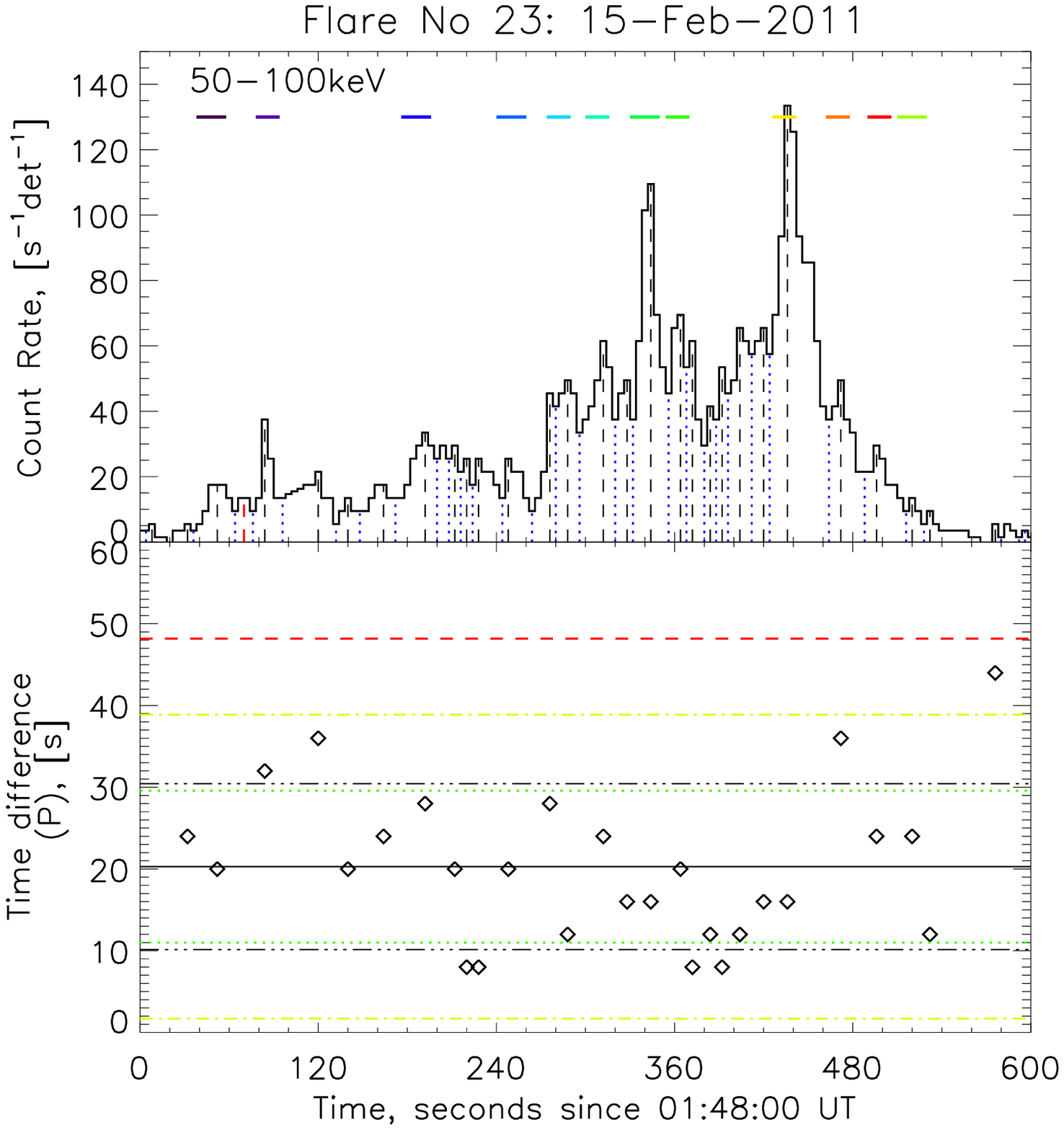}}
\vspace{-0.4\textwidth}   
\centerline{\small \bf    
\hspace{0.02\textwidth}  \color{black}{(g)}
\hspace{0.82\textwidth}  \color{black}{(h)}
\hfill}
\vspace{0.37\textwidth}   

\caption{Peaks and time differences in the RHESSI 50--100 keV corrected count rates in the studied flares No 16--23 (see Table~\ref{T-1}). The same designations as in Figure~\ref{AF-1}.}
\label{AF-3}
\end{figure}

\begin{figure}
\centerline{\includegraphics[width=0.4\textwidth,clip=]{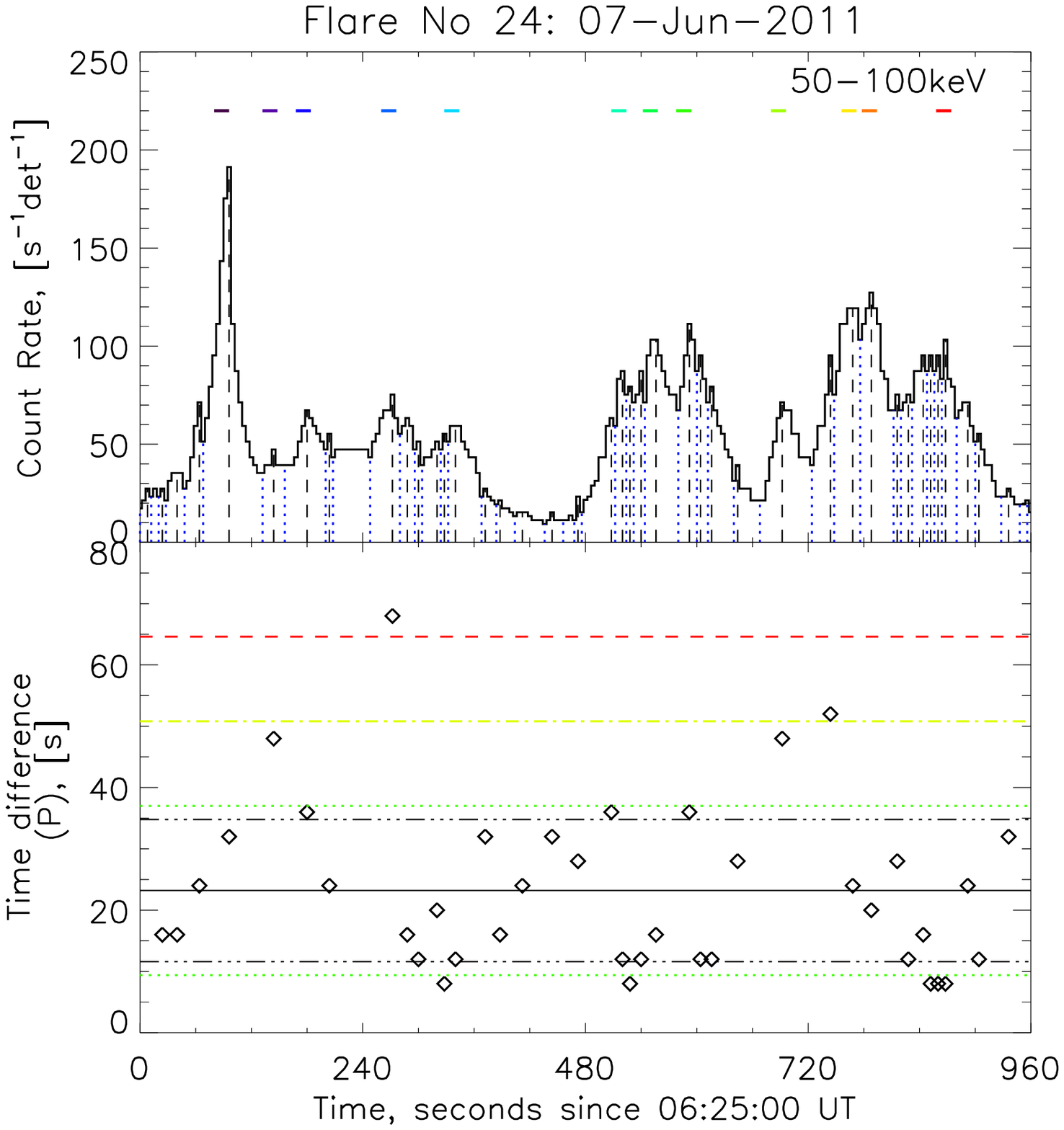}
\includegraphics[width=0.4\textwidth,clip=]{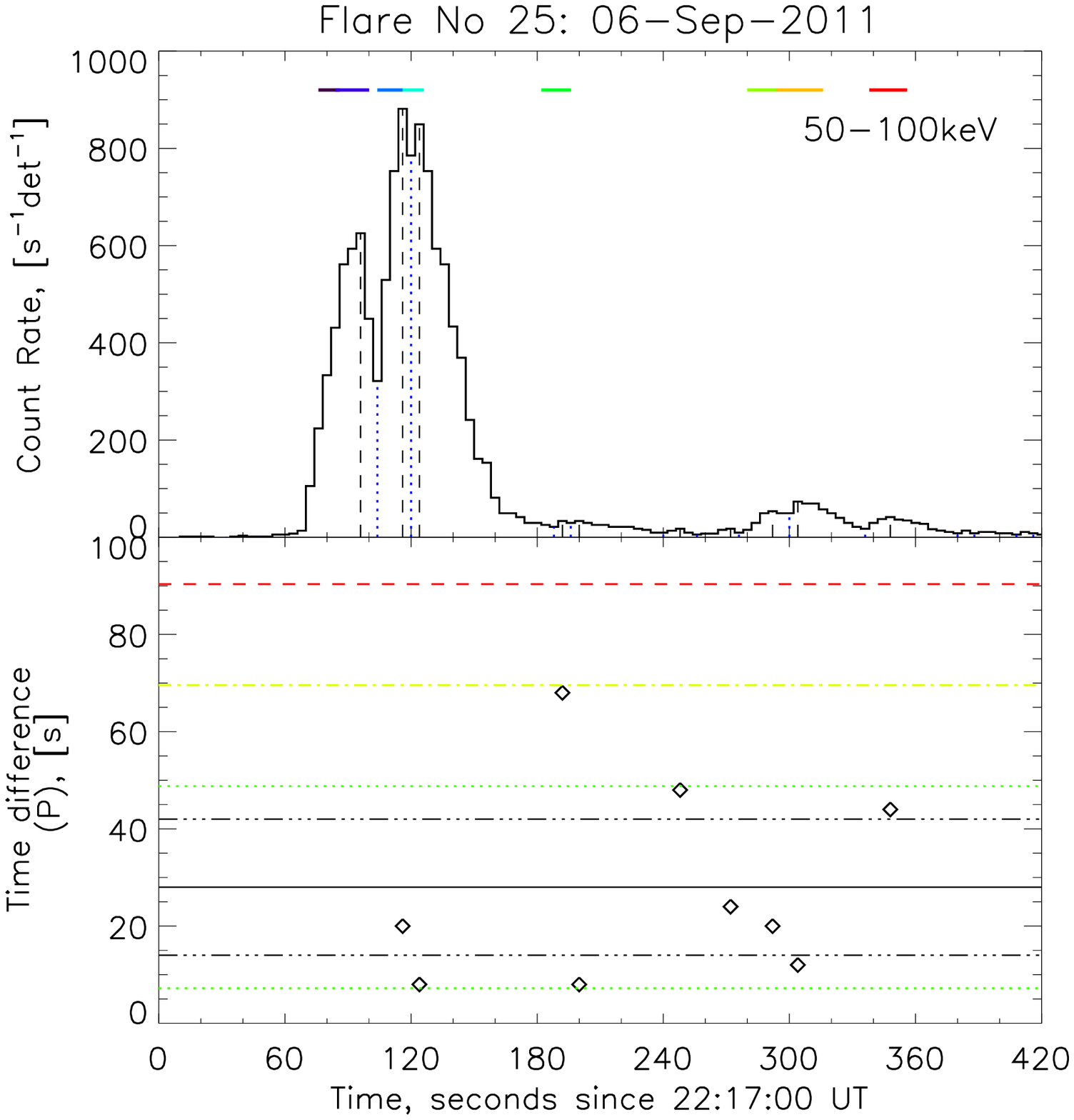}}
\vspace{-0.4\textwidth}   
\centerline{\small \bf    
\hspace{0.02\textwidth}  \color{black}{(a)}
\hspace{0.82\textwidth}  \color{black}{(b)}
\hfill}
\vspace{0.37\textwidth}   

\centerline{\includegraphics[width=0.4\textwidth,clip=]{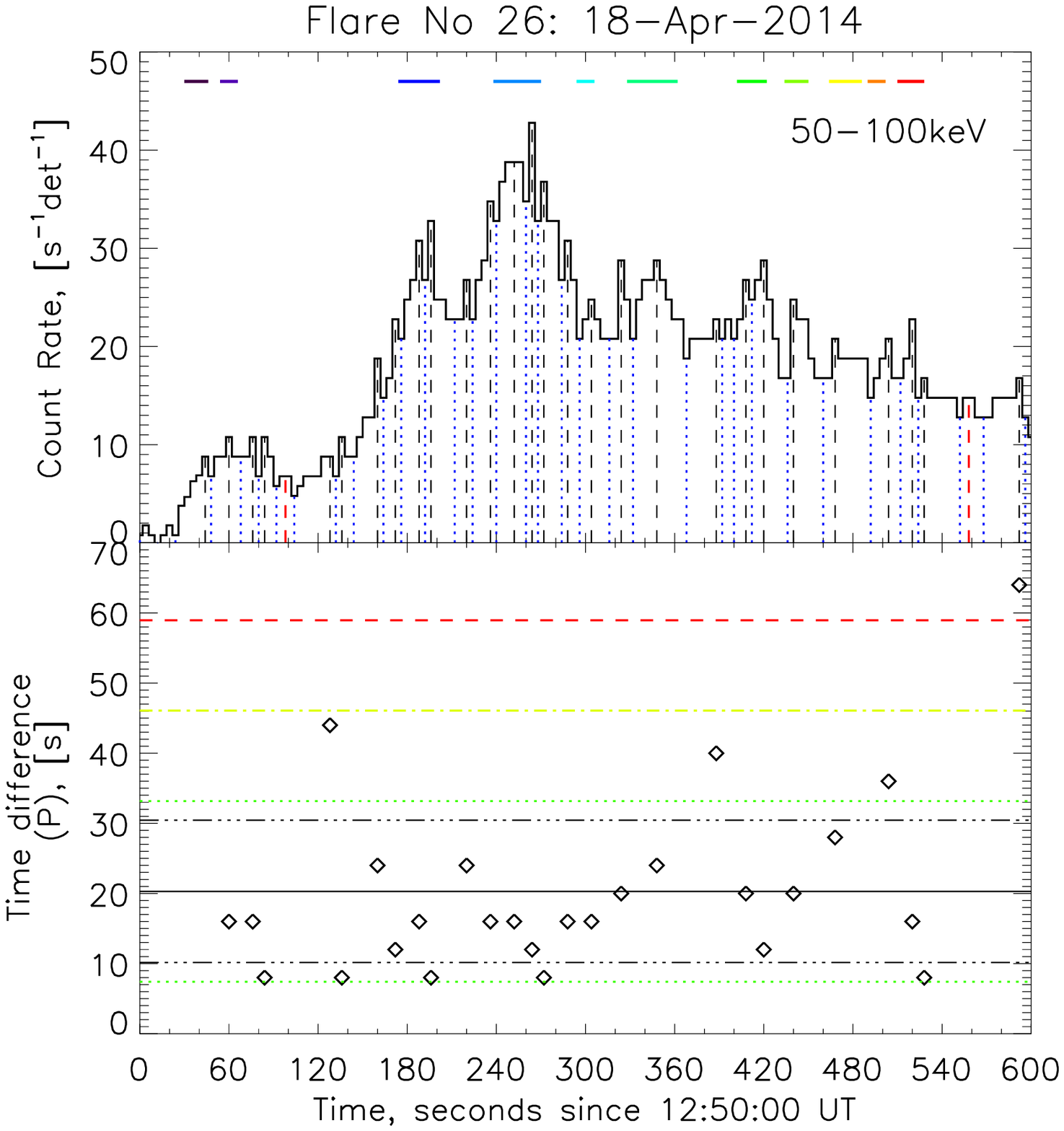}
\includegraphics[width=0.4\textwidth,clip=]{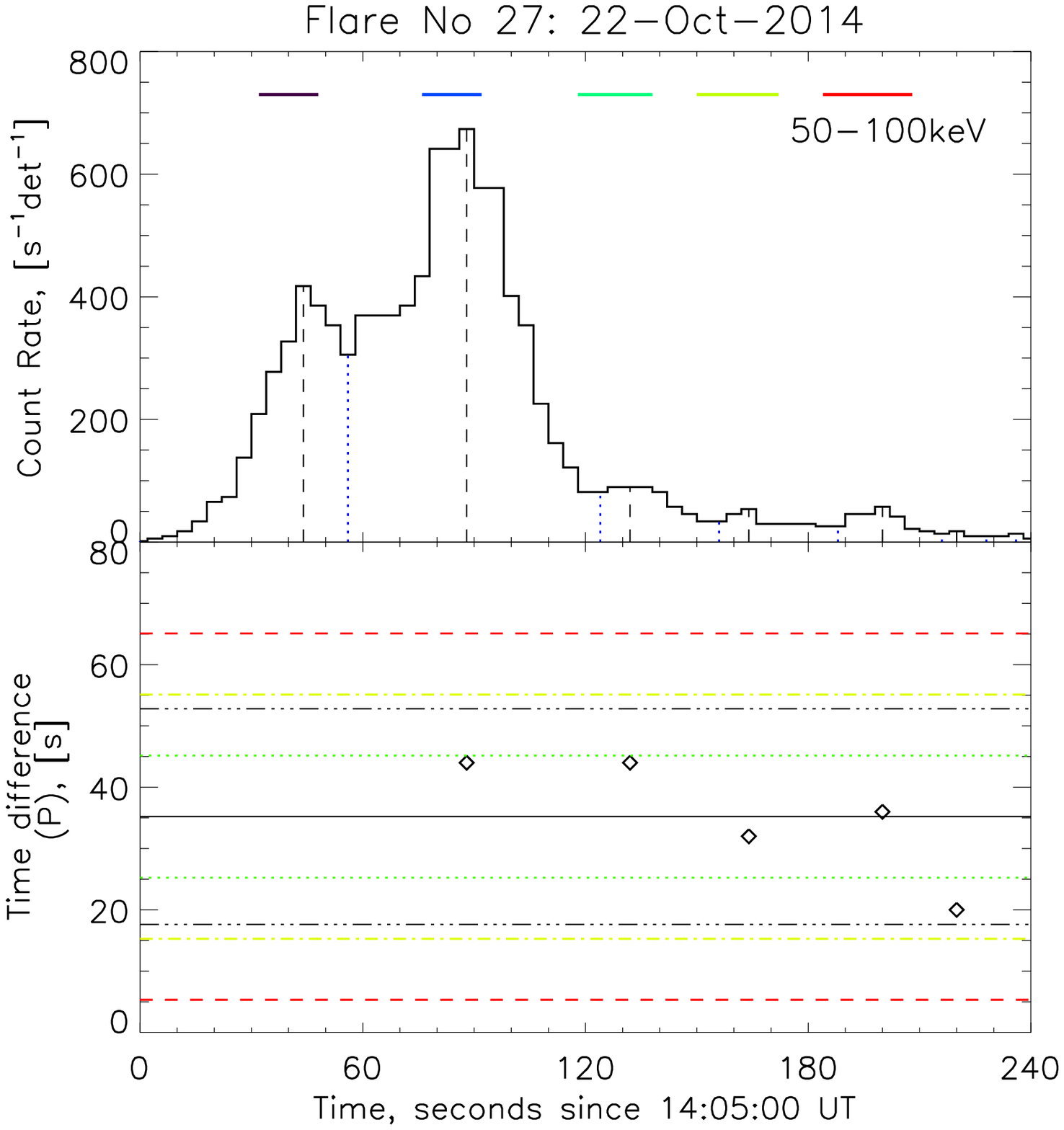}}
\vspace{-0.4\textwidth}   
\centerline{\small \bf    
\hspace{0.02\textwidth}  \color{black}{(c)}
\hspace{0.82\textwidth}  \color{black}{(d)}
\hfill}
\vspace{0.37\textwidth}   

\centerline{\includegraphics[width=0.4\textwidth,clip=]{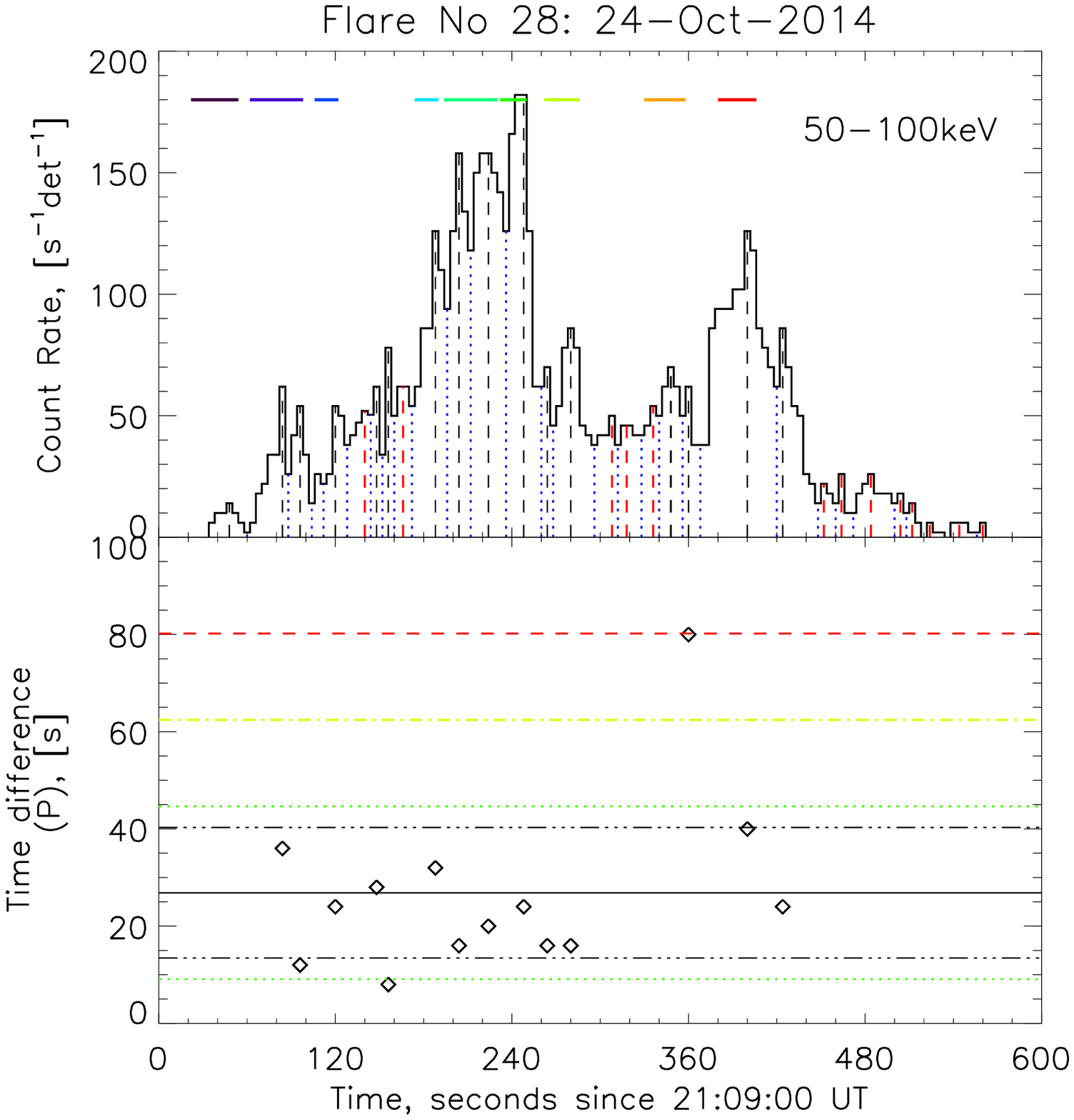}
\includegraphics[width=0.4\textwidth,clip=]{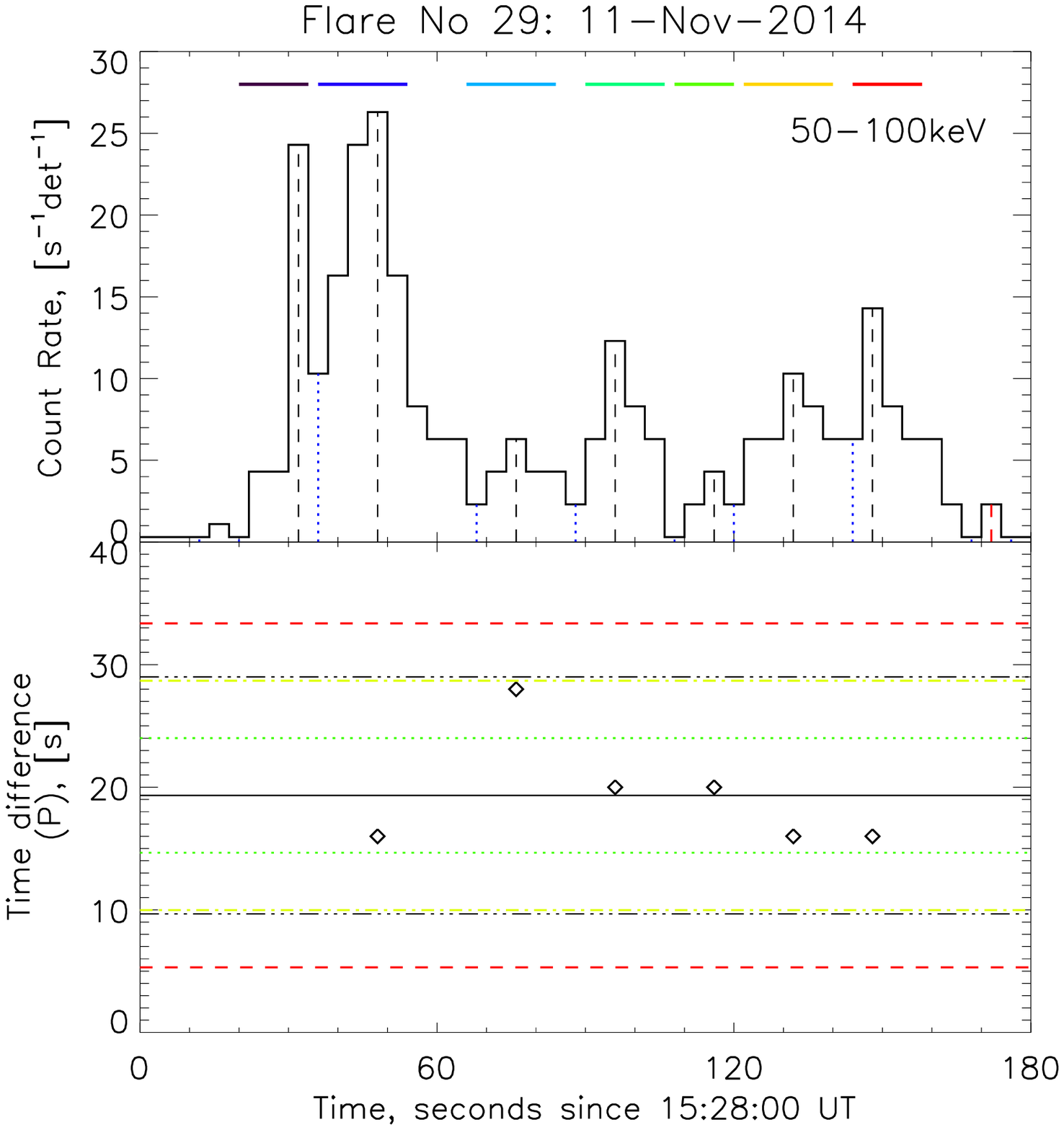}}
\vspace{-0.4\textwidth}   
\centerline{\small \bf    
\hspace{0.02\textwidth}  \color{black}{(e)}
\hspace{0.82\textwidth}  \color{black}{(f)}
\hfill}
\vspace{0.37\textwidth}   

\caption{Peaks and time differences in the RHESSI 50--100 keV corrected count rates in the studied flares No 24--29 (see Table~\ref{T-1}). The same designations as in Figure~\ref{AF-1}.}
\label{AF-4}
\end{figure}

\newpage
\section{Spatio-temporal evolution of the sources of the HXR pulsations in the studied flares}
\label{S-appendix-B}
\begin{figure}
\centerline{\includegraphics[width=0.45\textwidth,clip=]{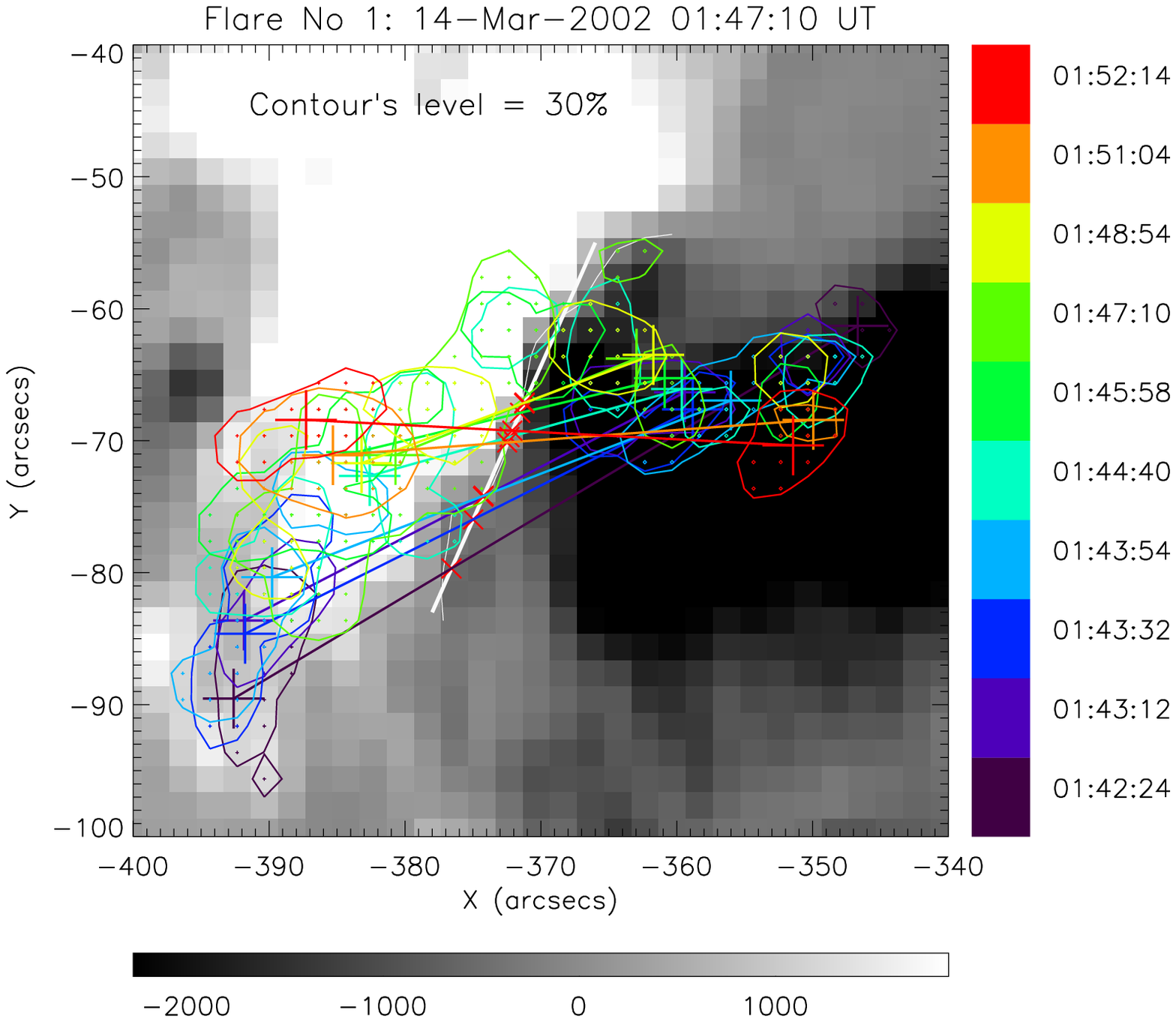}
\includegraphics[width=0.45\textwidth,clip=]{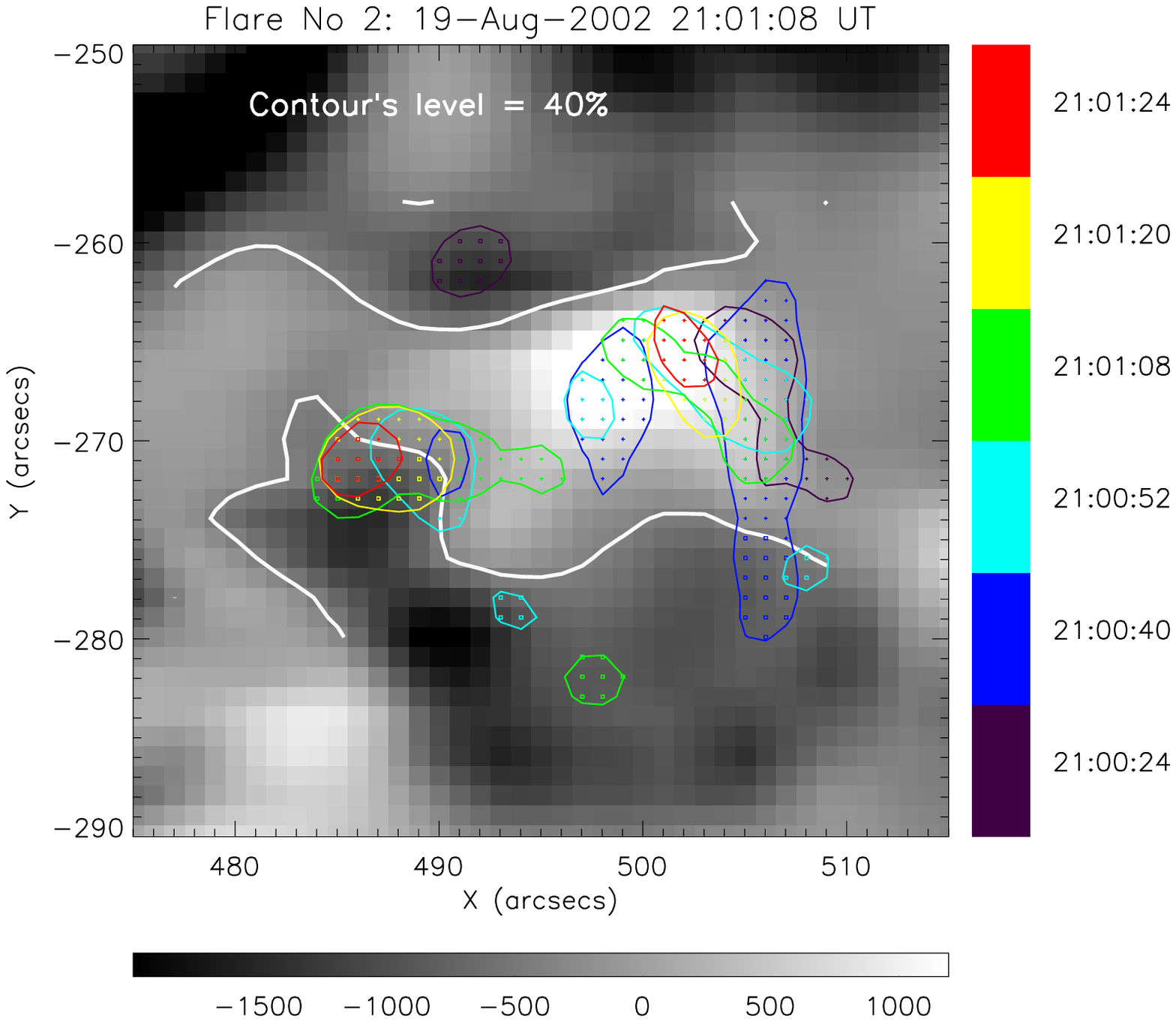}}
\vspace{-0.13\textwidth}   
\centerline{\Large \bf     
\hspace{0.32 \textwidth} \color{white}{(a)}
\hspace{0.37\textwidth}  \color{white}{(b)}
\hfill}
\vspace{0.09\textwidth}    

\centerline{\includegraphics[width=0.45\textwidth,clip=]{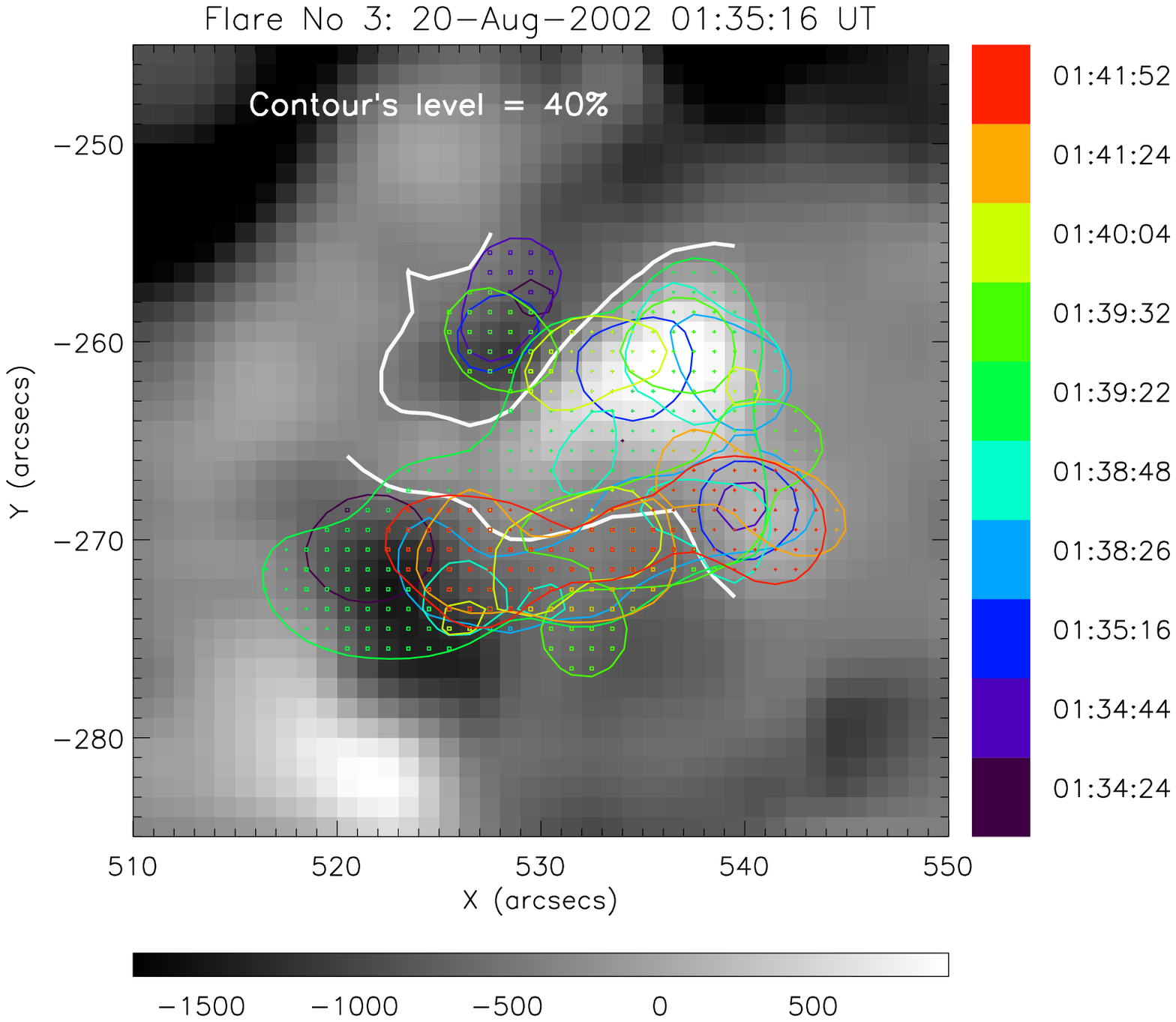}
\includegraphics[width=0.45\textwidth,clip=]{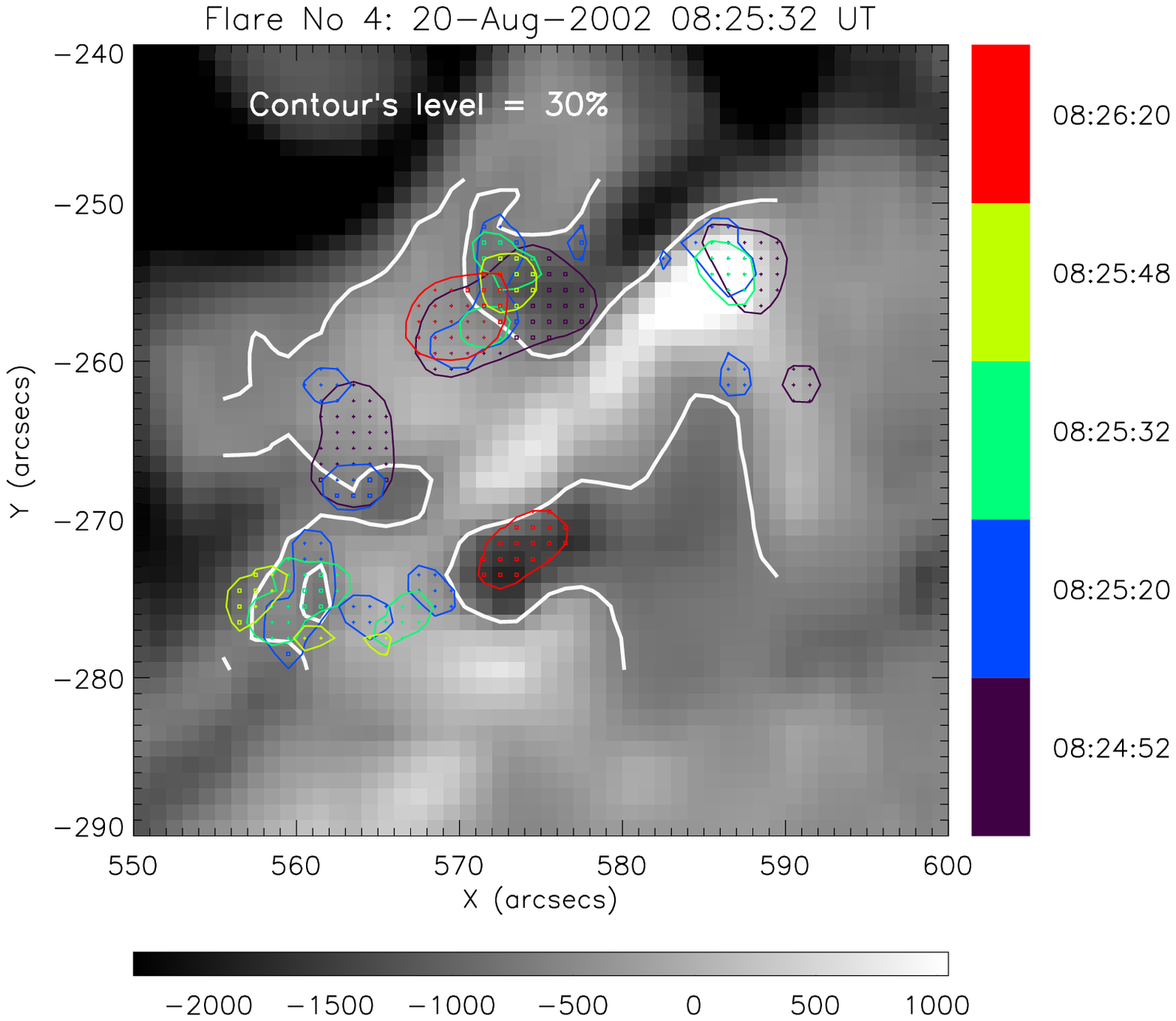}}
\vspace{-0.13\textwidth}   
\centerline{\Large \bf     
\hspace{0.32 \textwidth} \color{white}{(c)}
\hspace{0.37\textwidth}  \color{white}{(d)}
\hfill}
\vspace{0.09\textwidth}    
              
\centerline{\includegraphics[width=0.45\textwidth,clip=]{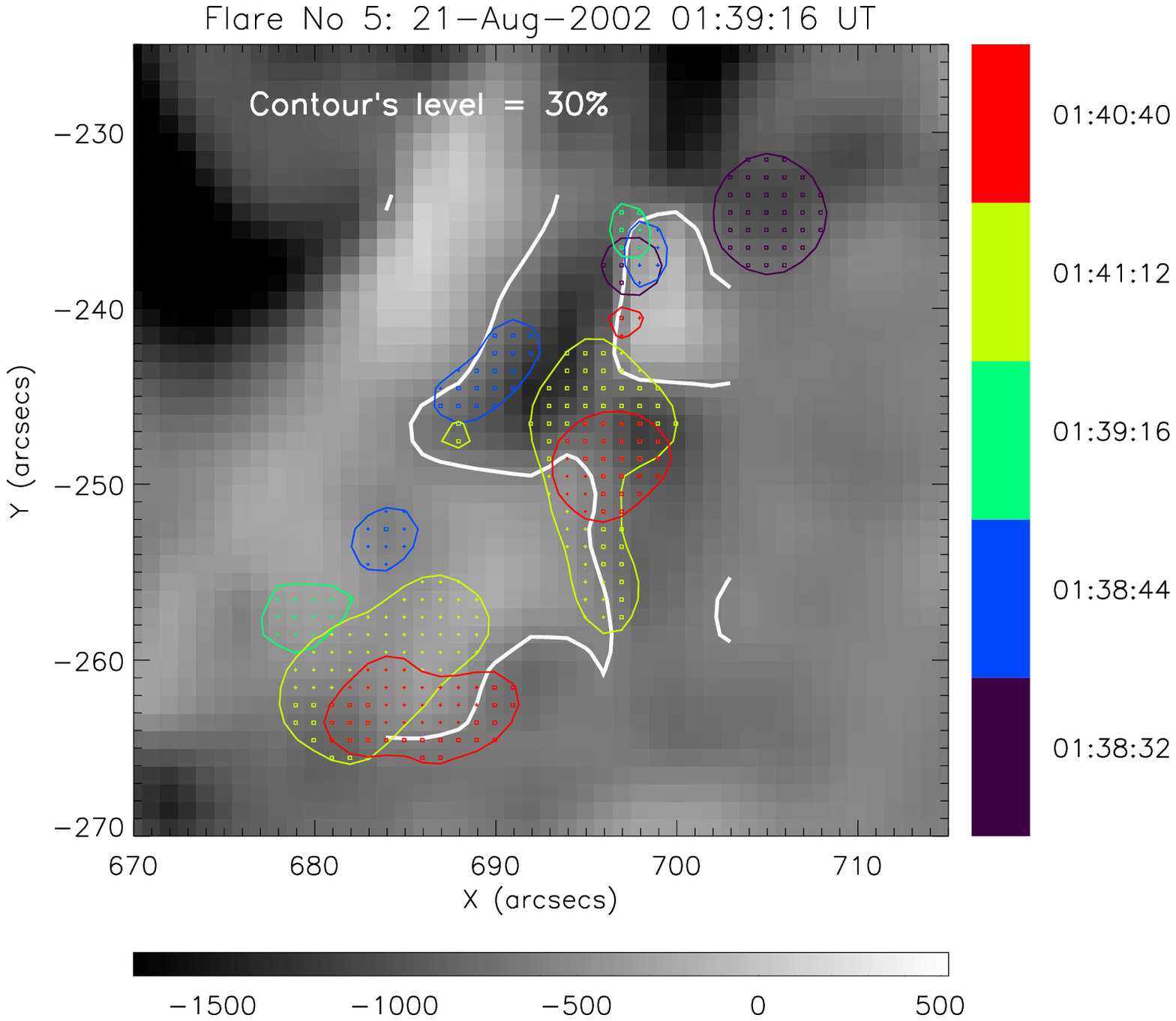}
\includegraphics[width=0.45\textwidth,clip=]{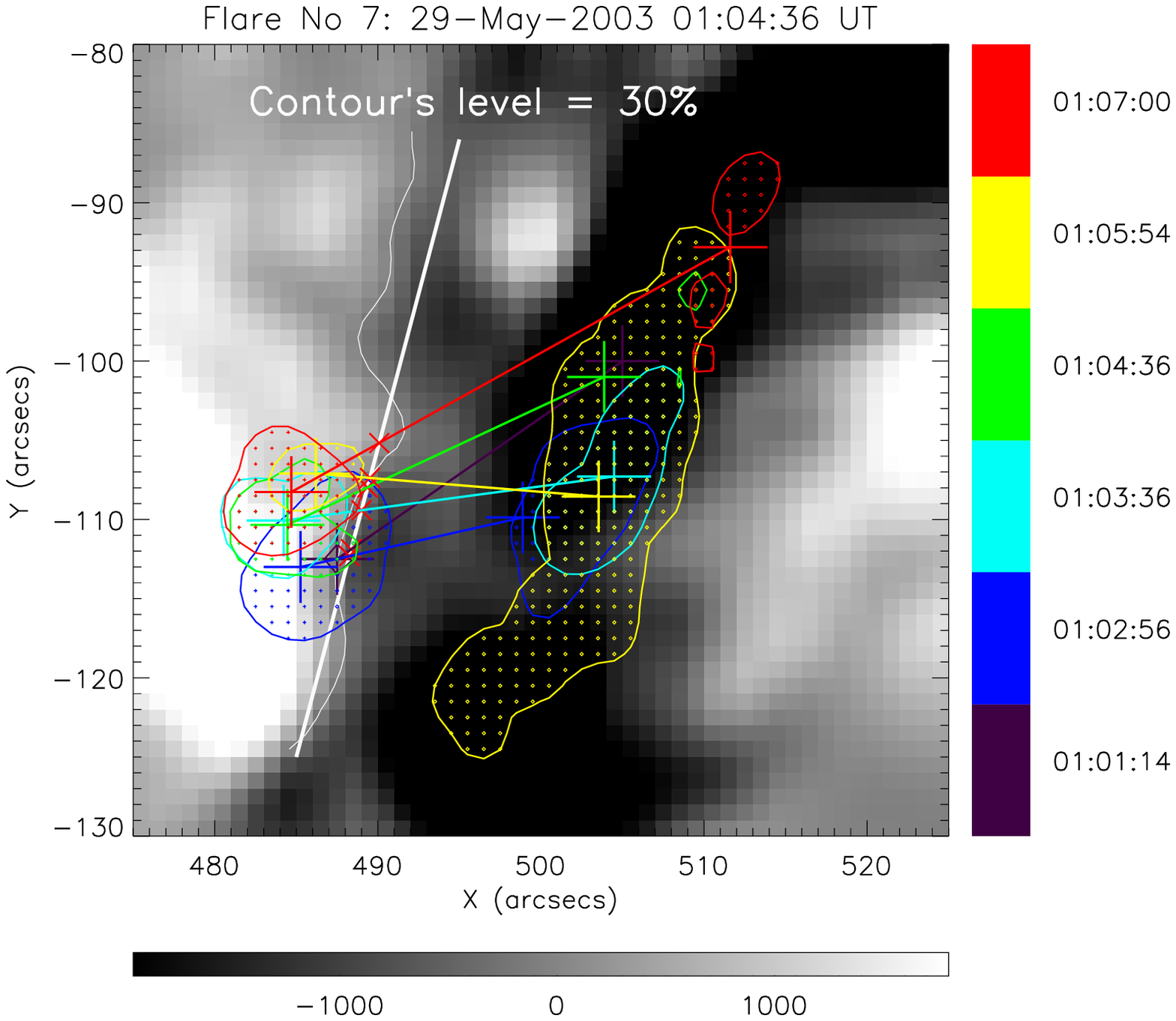}}
\vspace{-0.13\textwidth}   
\centerline{\Large \bf     
\hspace{0.32 \textwidth} \color{white}{(e)}
\hspace{0.37\textwidth}  \color{white}{(f)}
\hfill}
\vspace{0.09\textwidth}    
									
\caption{Spatio-temporal evolution of the sources of the HXR pulsations in the flares No 1--5, 7. The PIXON algorithm was used for reconstruction of the HXR sources shown. The date and time of flare are shown above. The background images are the photospheric line-of-sight magnetograms obtained with the SOHO/MDI or SDO/HMI. The magnetic field colorbars (in gauss) are presented below. Colors of the HXR sources correspond to the times of the HXR peaks shown near the colorbars to the right. Contour levels of the HXR sources are written on the images. Large color crosses show average positions (centroids) of the HXR sources at the corresponding times. Linear sizes of these crosses are equal to the doubled FWHM of the RHESSI collimator number 1. Straight color lines connect paired HXR sources located in opposite magnetic polarities. Small red crosses indicate intersection points of these lines with the model MPIL (thick white lines). The real MPILs found from the magnetograms are shown by the thin white curves.}
\label{AF-5}
\end{figure}

\begin{figure}
\centerline{\includegraphics[width=0.45\textwidth,clip=]{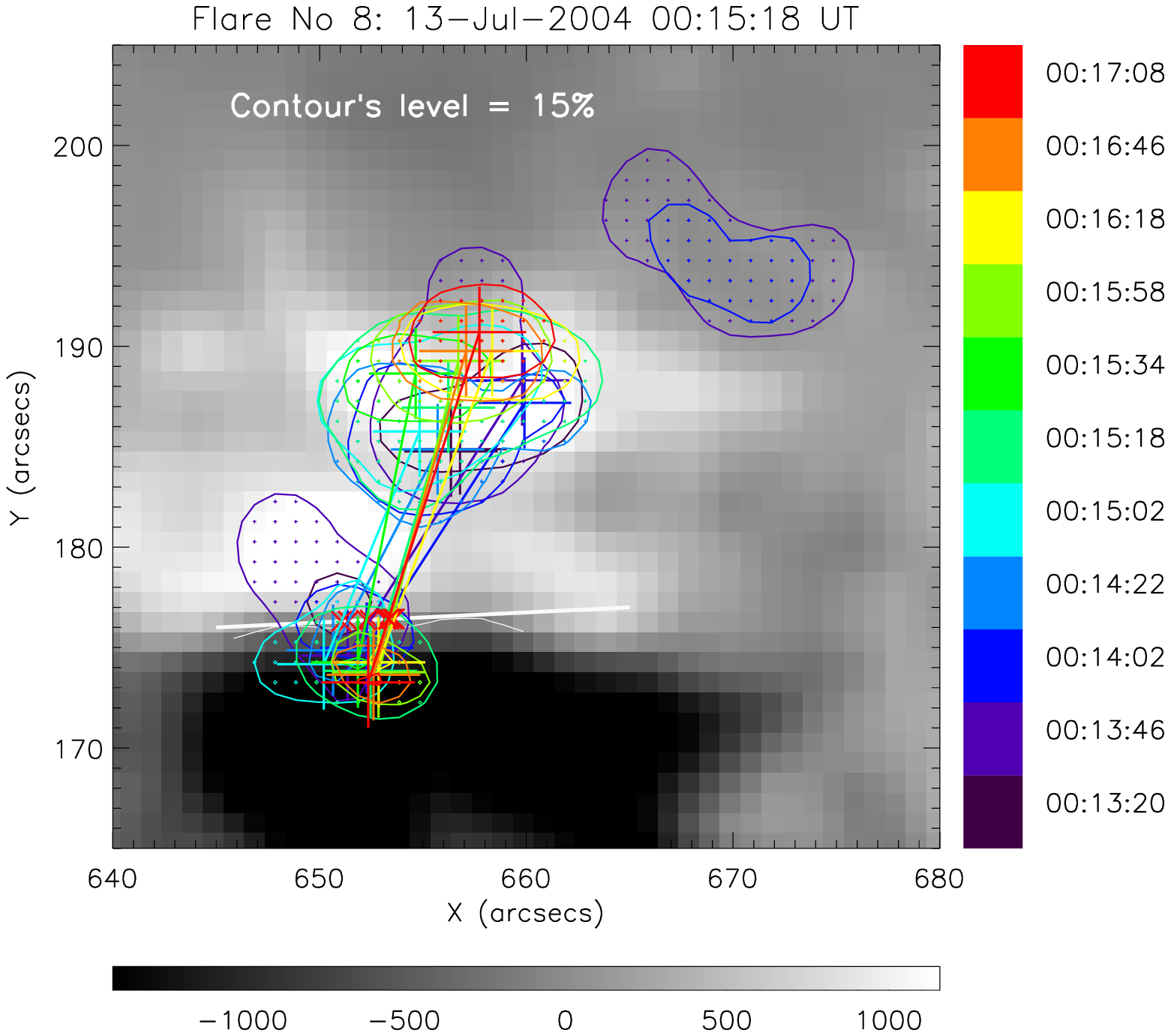}
\includegraphics[width=0.45\textwidth,clip=]{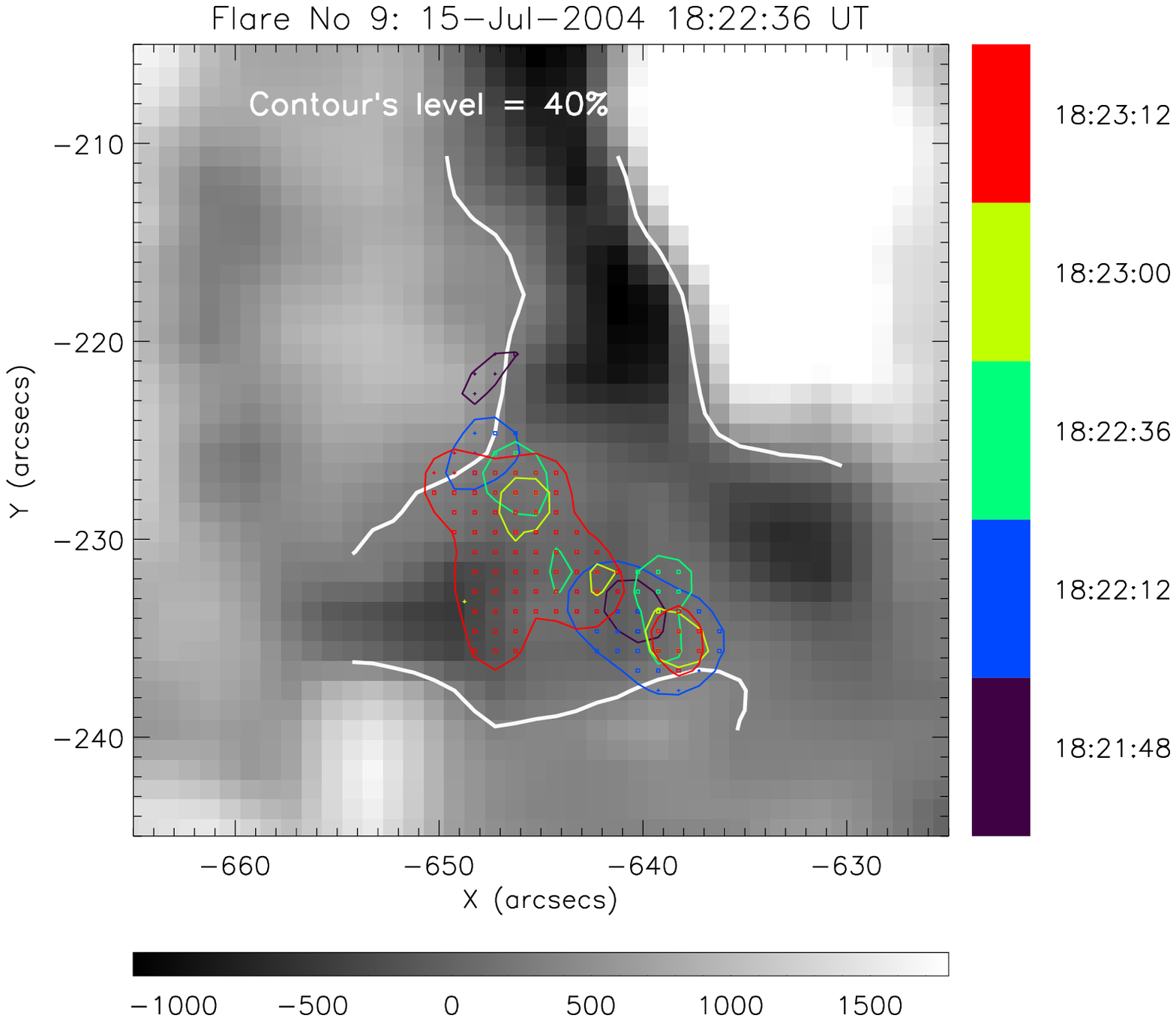}}
\vspace{-0.13\textwidth}   
\centerline{\Large \bf     
\hspace{0.32 \textwidth} \color{white}{(a)}
\hspace{0.37\textwidth}  \color{white}{(b)}
\hfill}
\vspace{0.09\textwidth}    

\centerline{\includegraphics[width=0.45\textwidth,clip=]{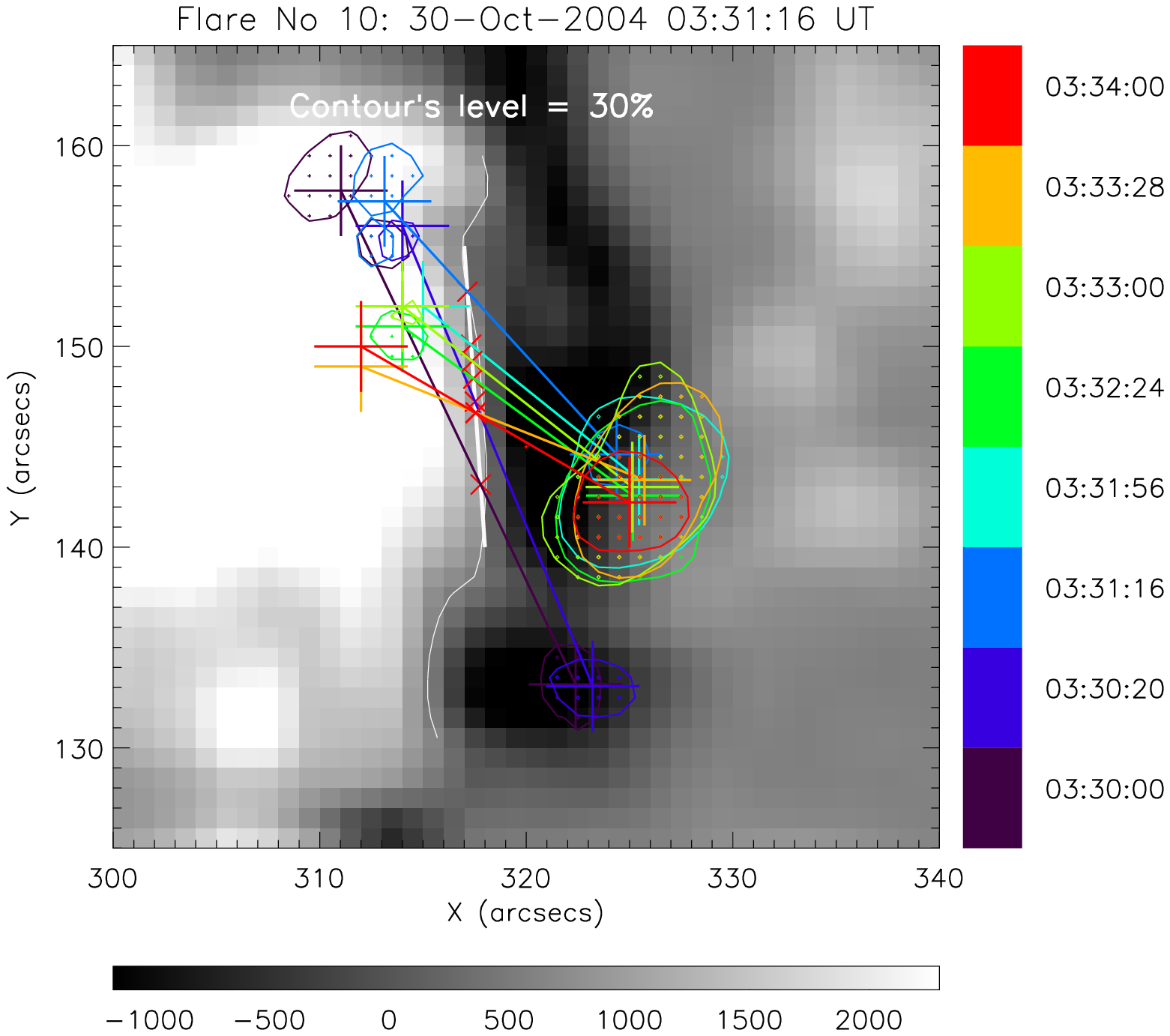}
\includegraphics[width=0.45\textwidth,clip=]{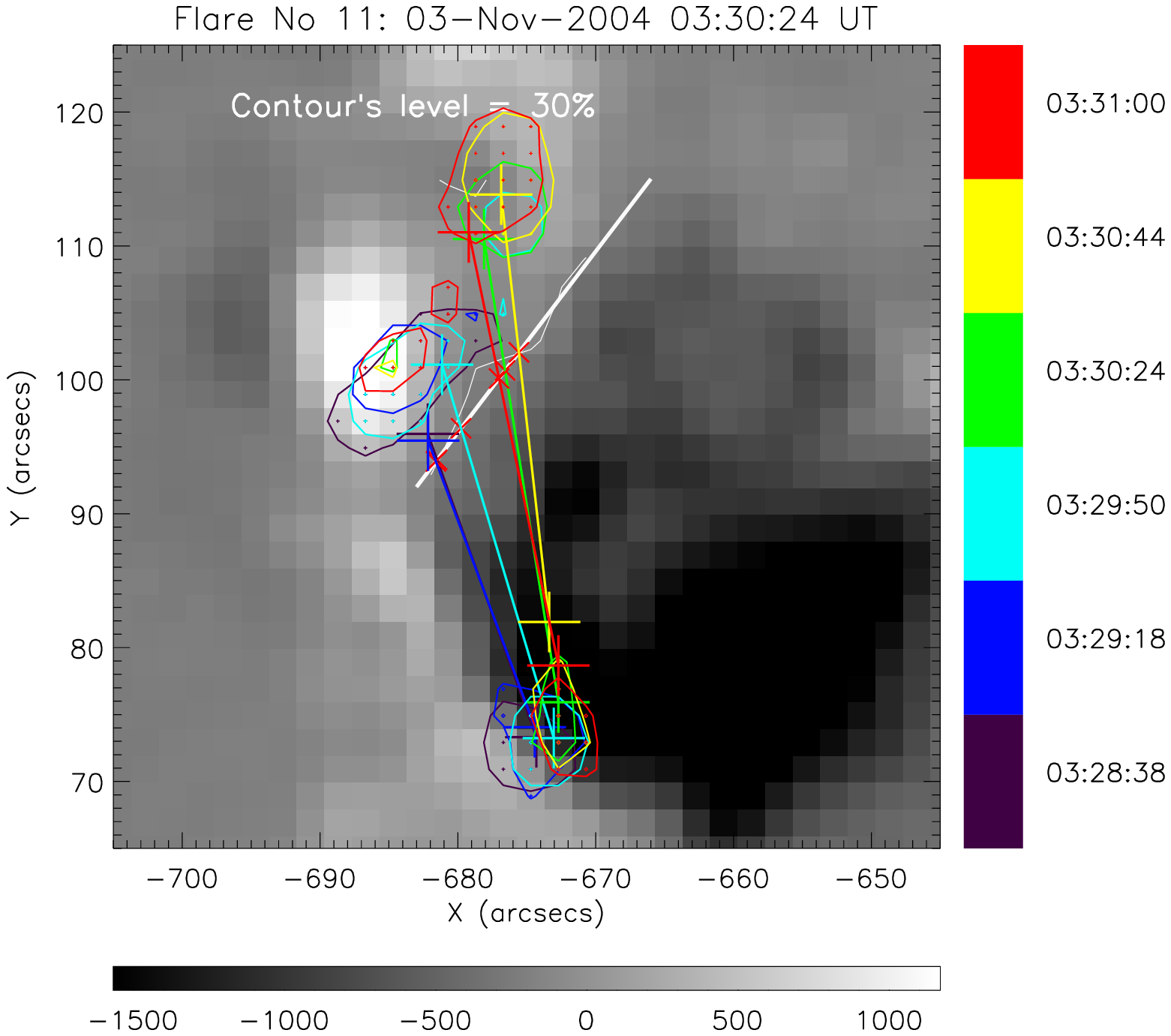}}
\vspace{-0.13\textwidth}   
\centerline{\Large \bf     
\hspace{0.32 \textwidth} \color{white}{(c)}
\hspace{0.37\textwidth}  \color{white}{(d)}
\hfill}
\vspace{0.09\textwidth}    
              
\centerline{\includegraphics[width=0.45\textwidth,clip=]{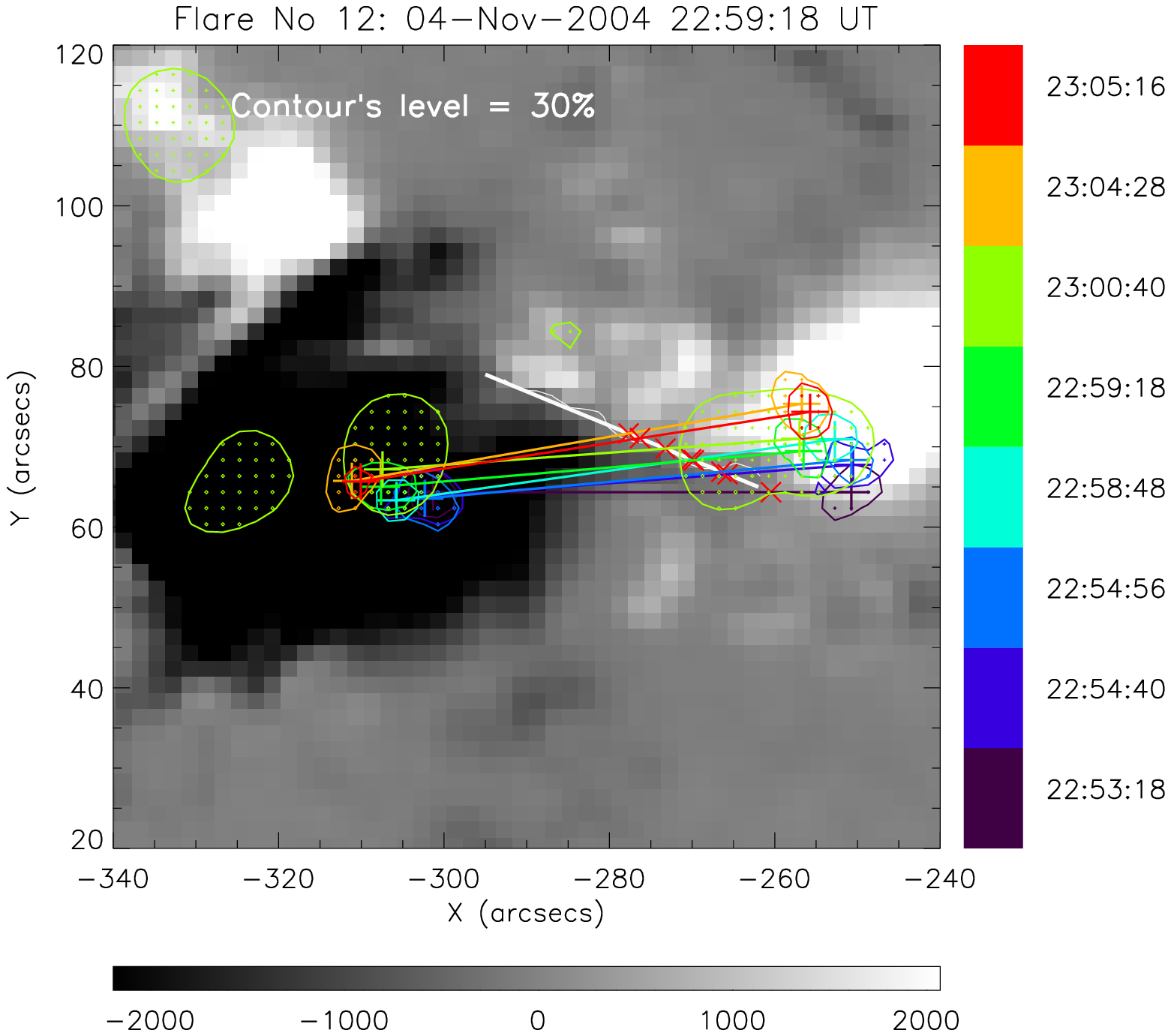}
\includegraphics[width=0.45\textwidth,clip=]{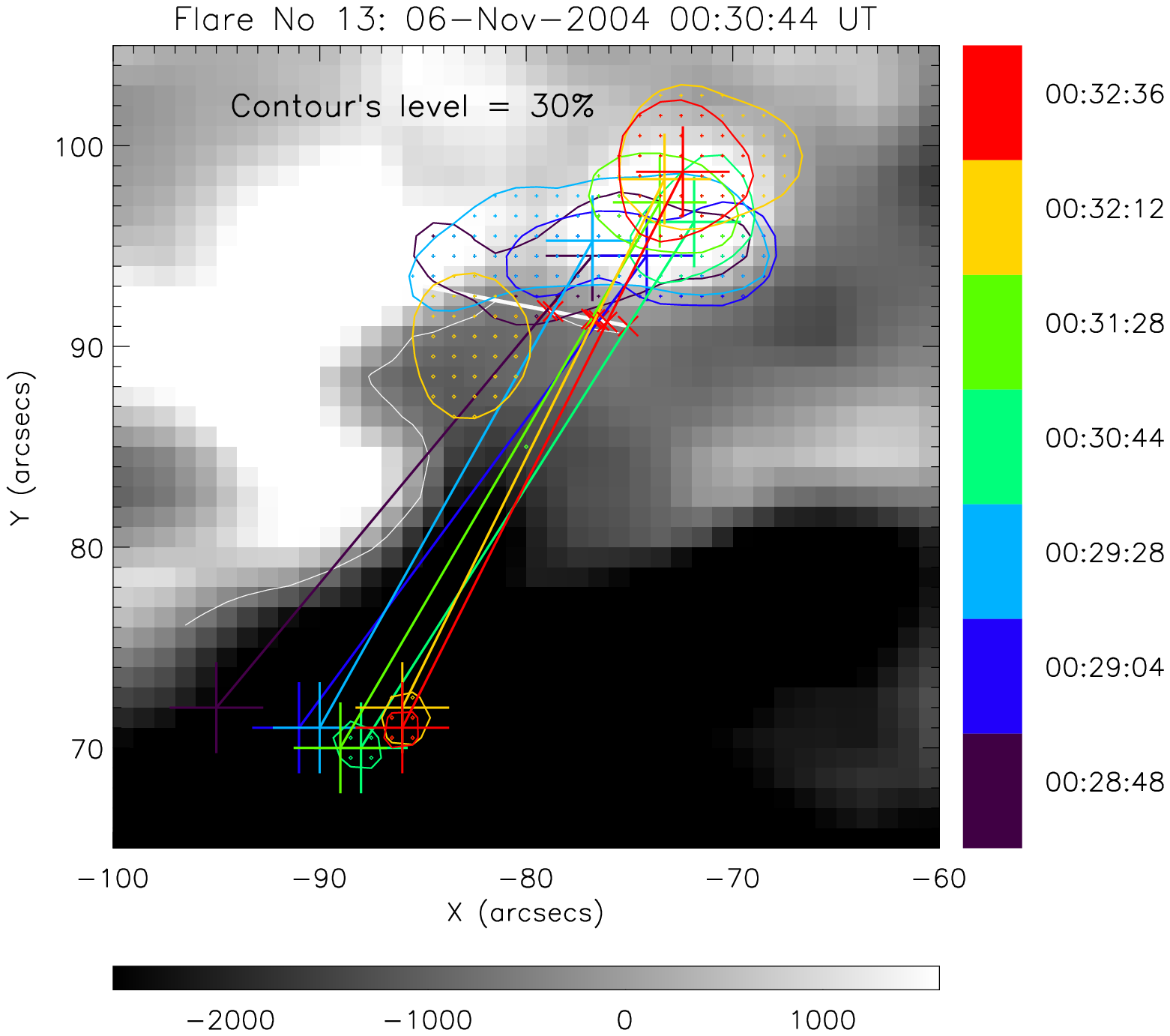}}
\vspace{-0.13\textwidth}   
\centerline{\Large \bf     
\hspace{0.32 \textwidth} \color{white}{(e)}
\hspace{0.37\textwidth}  \color{white}{(f)}
\hfill}
\vspace{0.09\textwidth}    

\centerline{\includegraphics[width=0.45\textwidth,clip=]{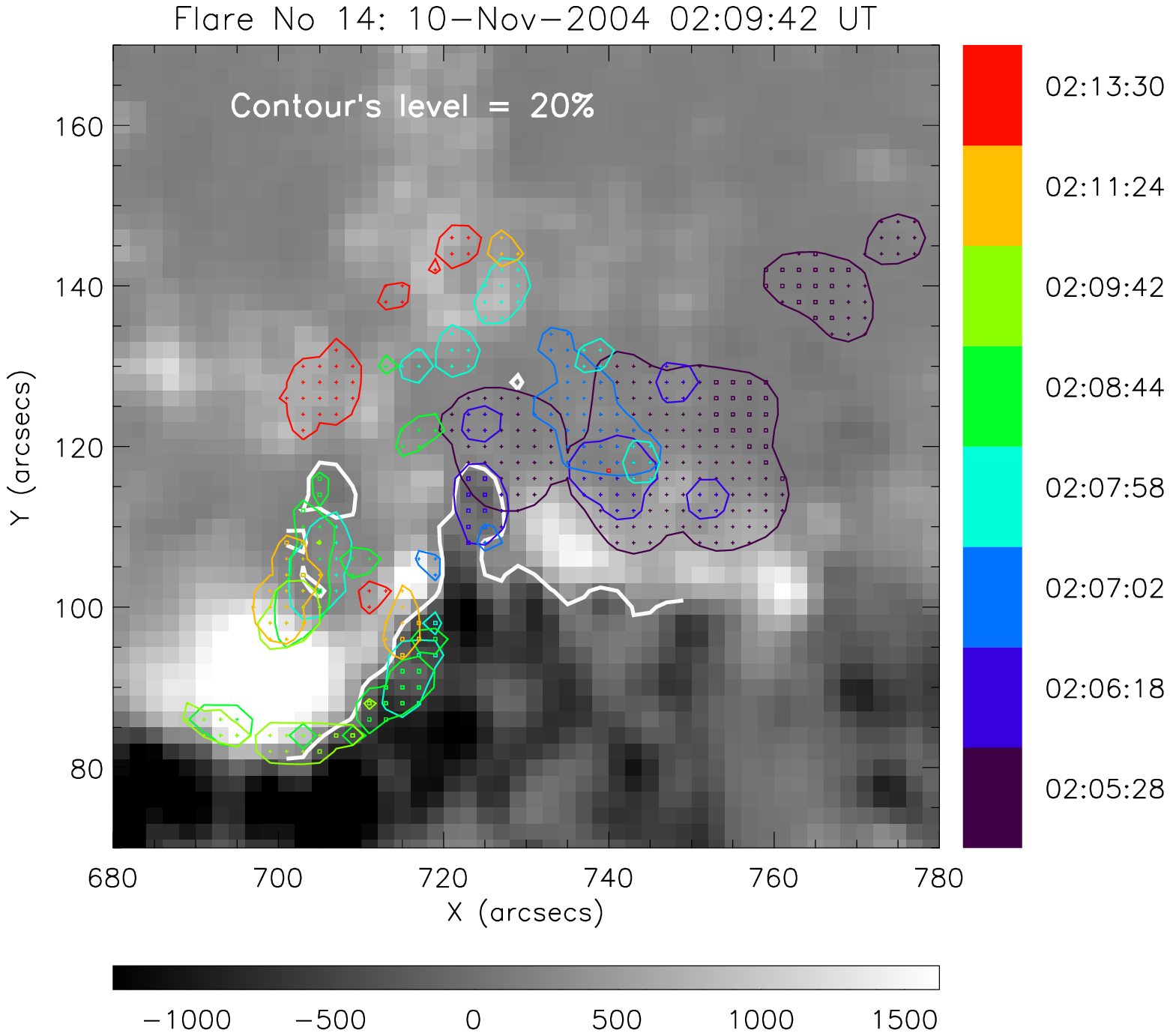}
\includegraphics[width=0.45\textwidth,clip=]{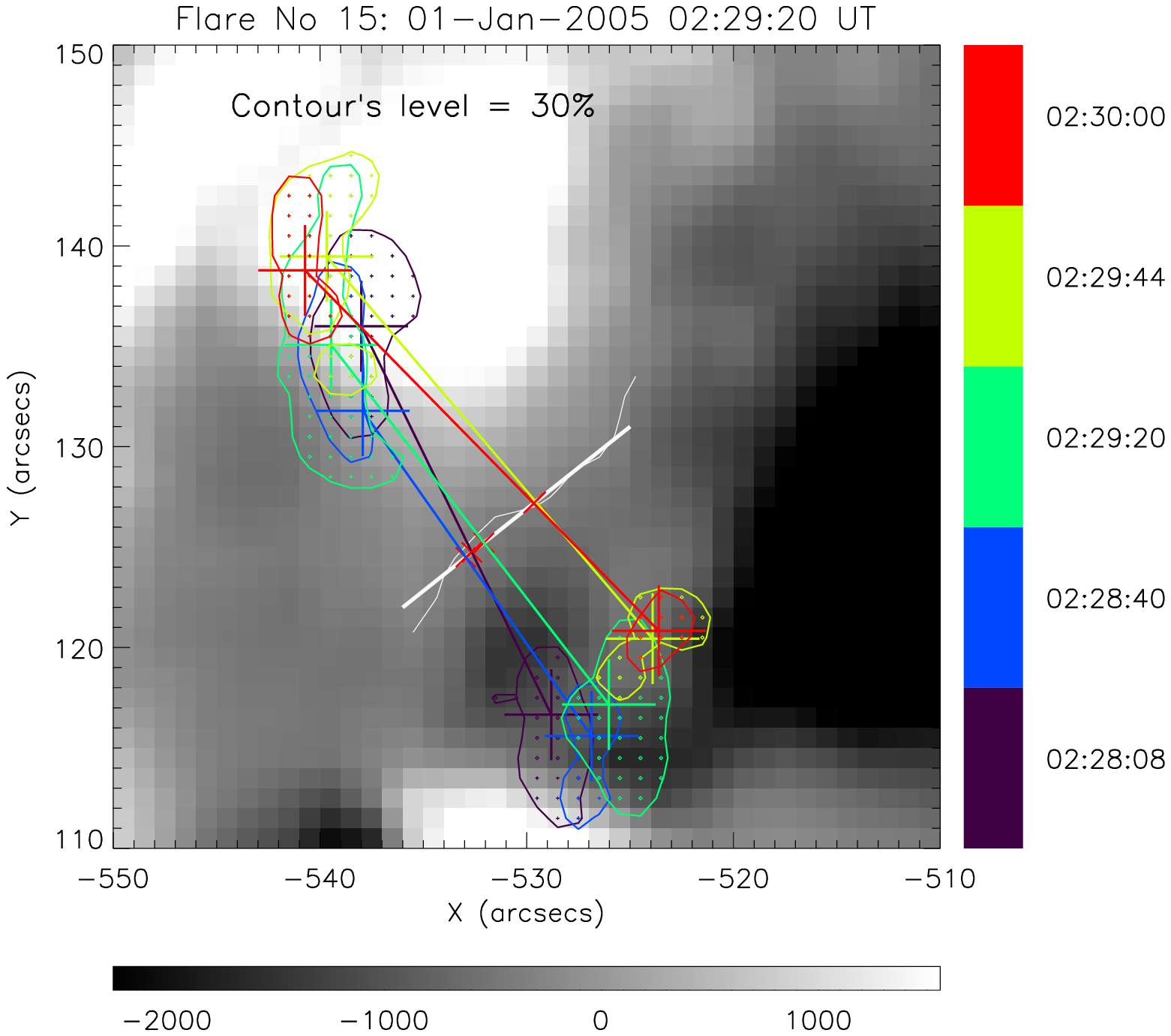}}
\vspace{-0.13\textwidth}   
\centerline{\Large \bf     
\hspace{0.32 \textwidth} \color{white}{(g)}
\hspace{0.37\textwidth}  \color{white}{(h)}
\hfill}
\vspace{0.09\textwidth}    

\caption{Spatio-temporal evolution of the sources of the HXR pulsations in the flares No 8--15. The method and convention are the same as in Figure~\ref{AF-5}.}
\label{AF-6}
\end{figure}

\begin{figure}
\centerline{\includegraphics[width=0.45\textwidth,clip=]{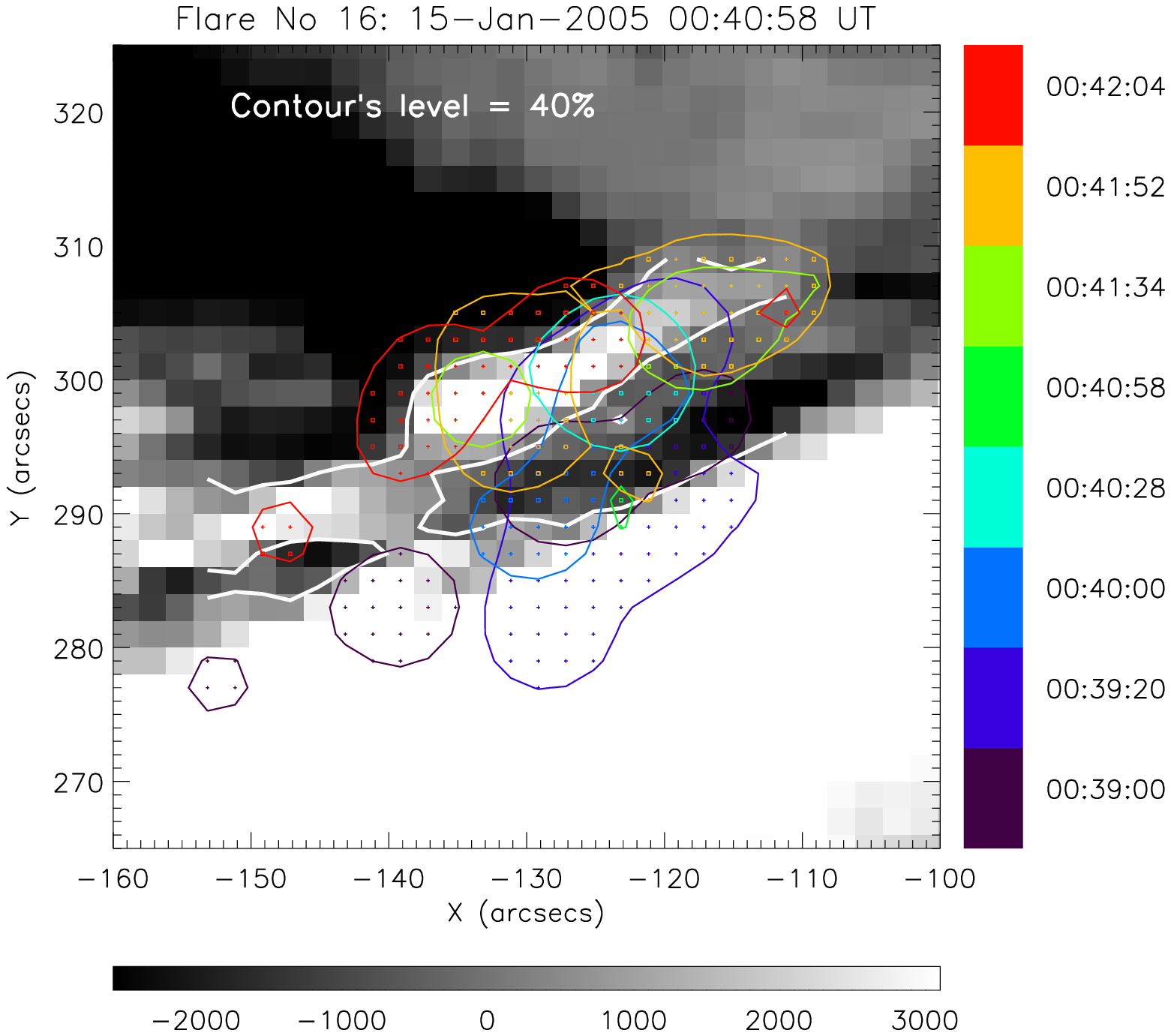}
\includegraphics[width=0.45\textwidth,clip=]{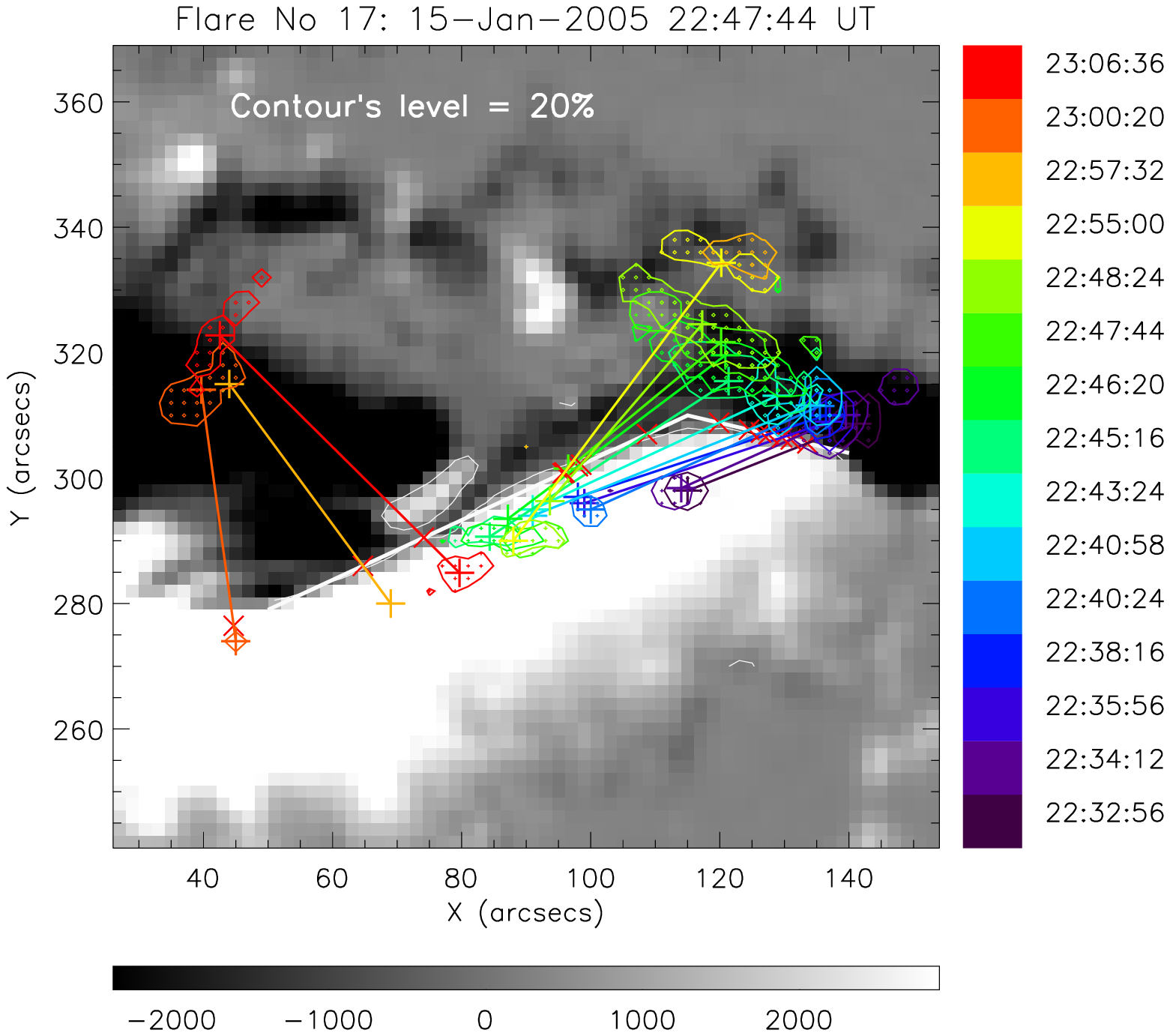}}
\vspace{-0.13\textwidth}   
\centerline{\Large \bf     
\hspace{0.32 \textwidth} \color{black}{(a)}
\hspace{0.37\textwidth}  \color{white}{(b)}
\hfill}
\vspace{0.09\textwidth}    

\centerline{\includegraphics[width=0.45\textwidth,clip=]{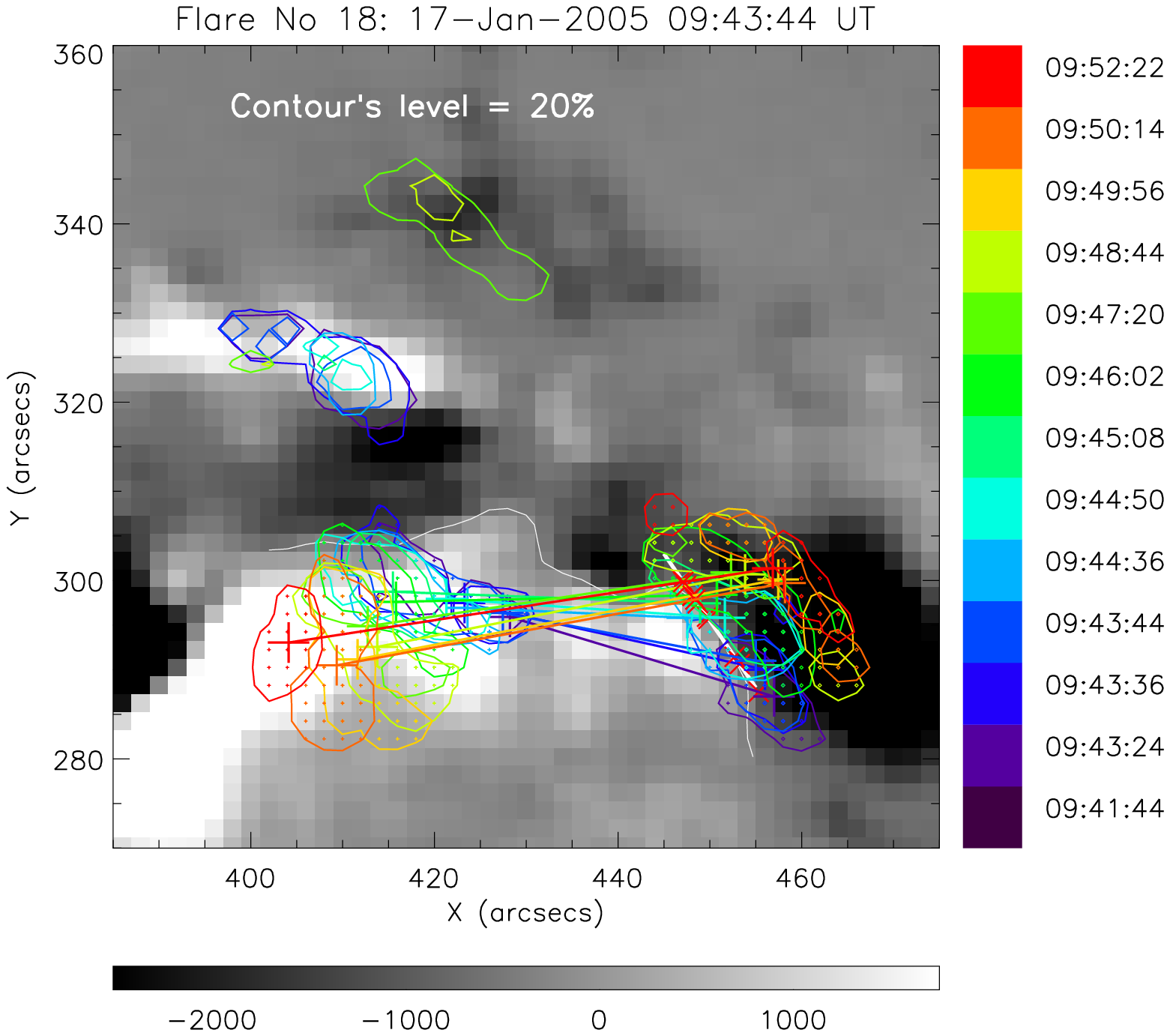}
\includegraphics[width=0.45\textwidth,clip=]{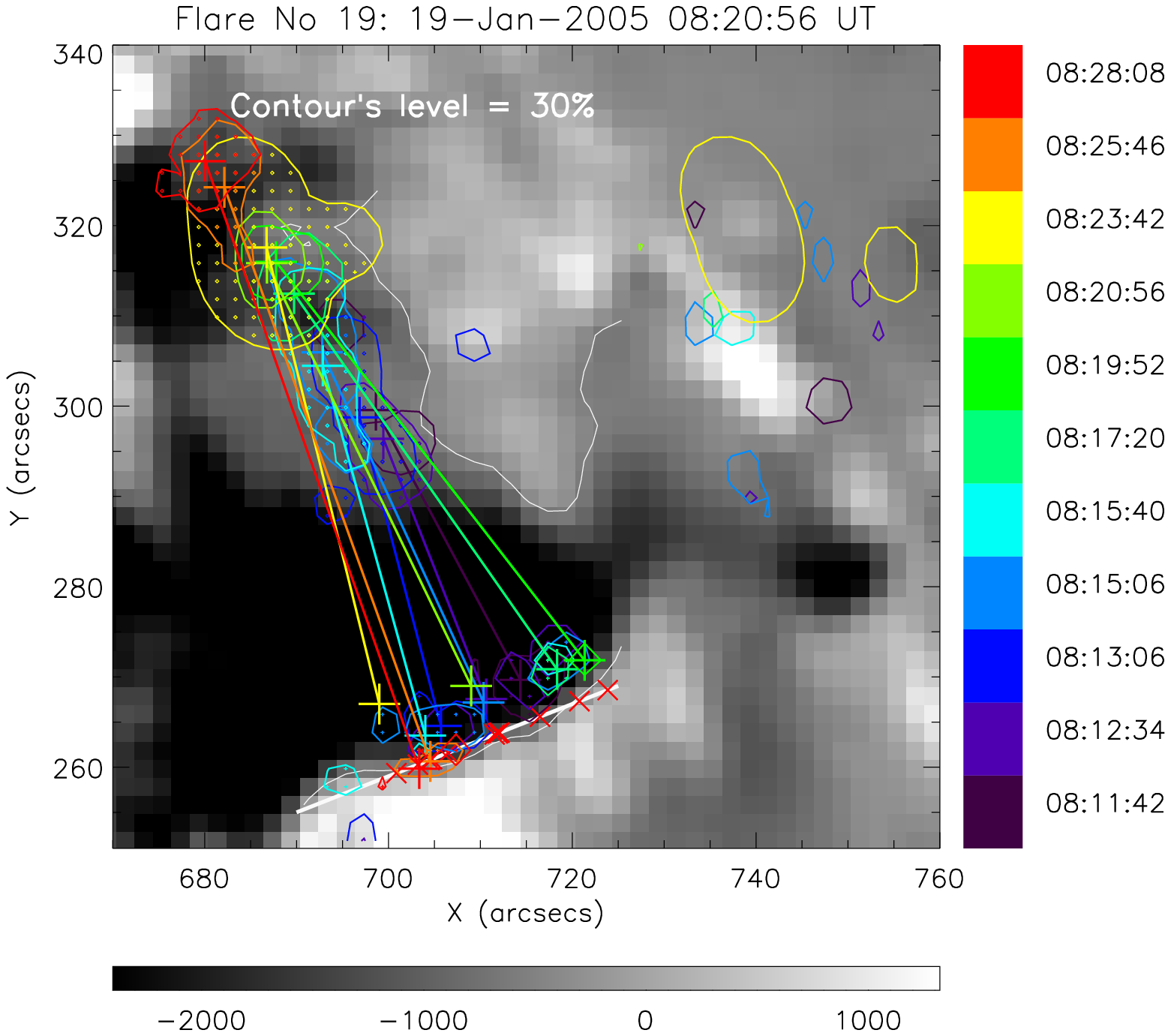}}
\vspace{-0.13\textwidth}   
\centerline{\Large \bf     
\hspace{0.32 \textwidth} \color{white}{(c)}
\hspace{0.37\textwidth}  \color{white}{(d)}
\hfill}
\vspace{0.09\textwidth}    
              
\centerline{\includegraphics[width=0.45\textwidth,clip=]{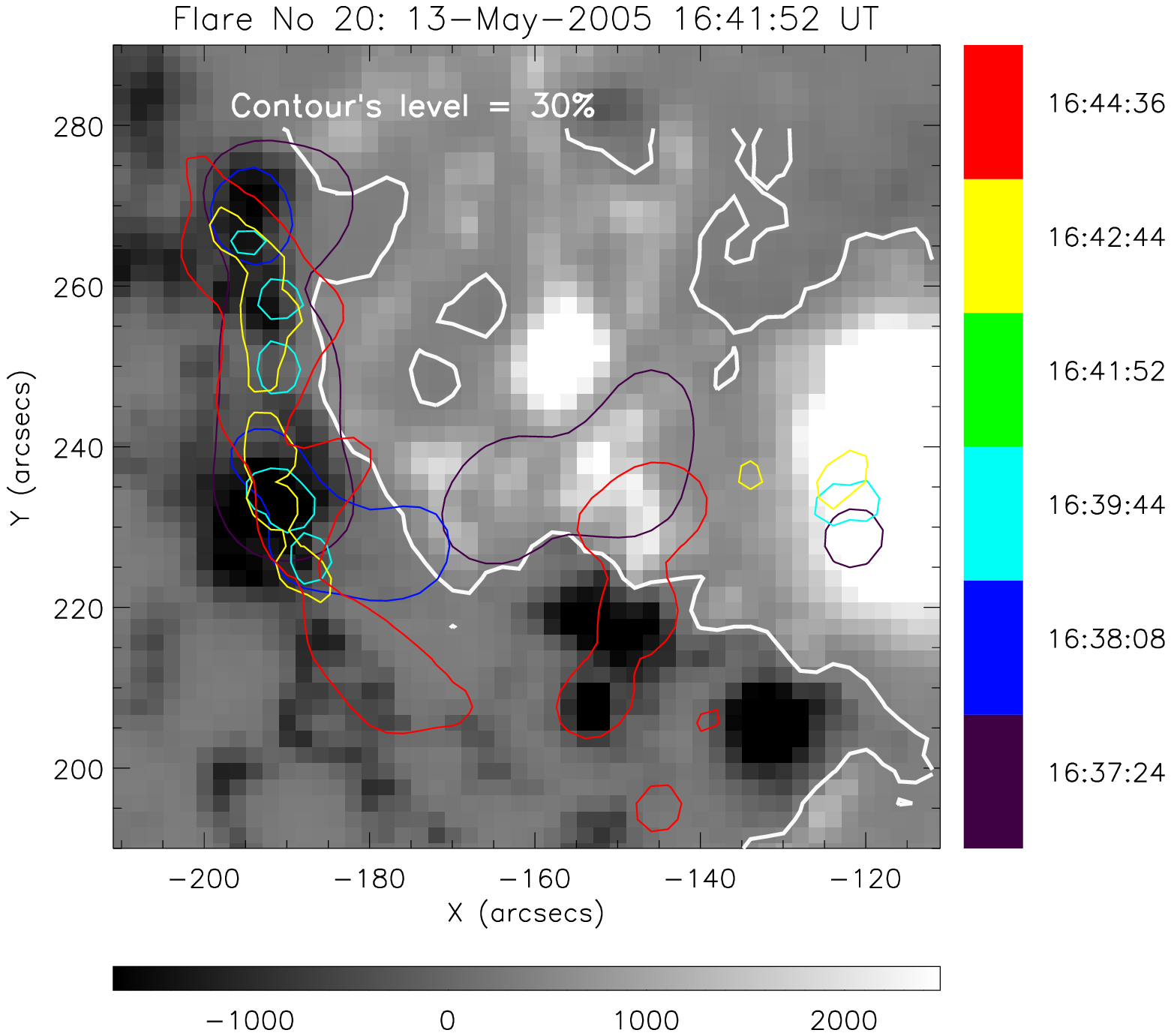}
\includegraphics[width=0.45\textwidth,clip=]{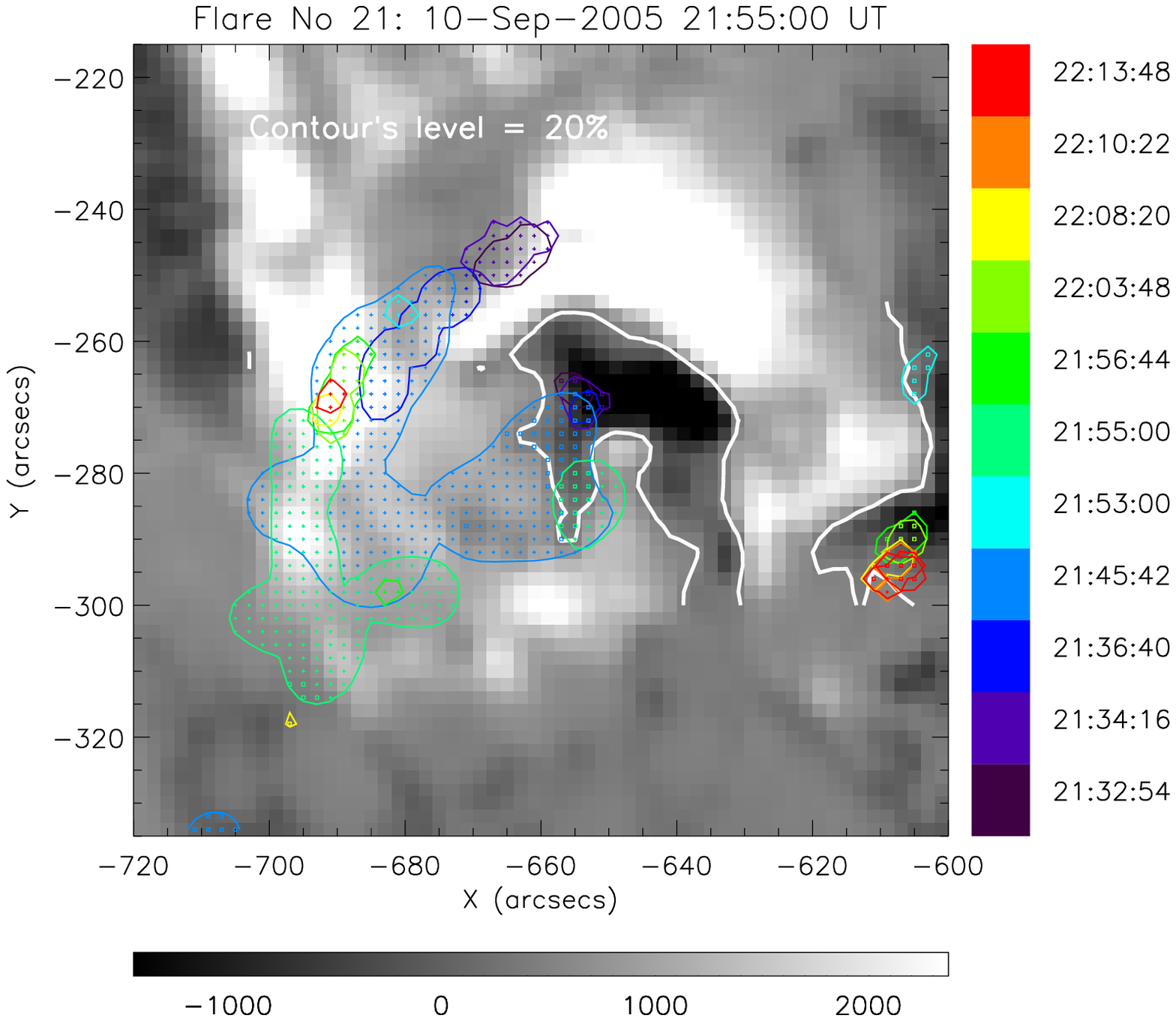}}
\vspace{-0.13\textwidth}   
\centerline{\Large \bf     
\hspace{0.32 \textwidth} \color{white}{(e)}
\hspace{0.37\textwidth}  \color{white}{(f)}
\hfill}
\vspace{0.09\textwidth}    

\centerline{\includegraphics[width=0.45\textwidth,clip=]{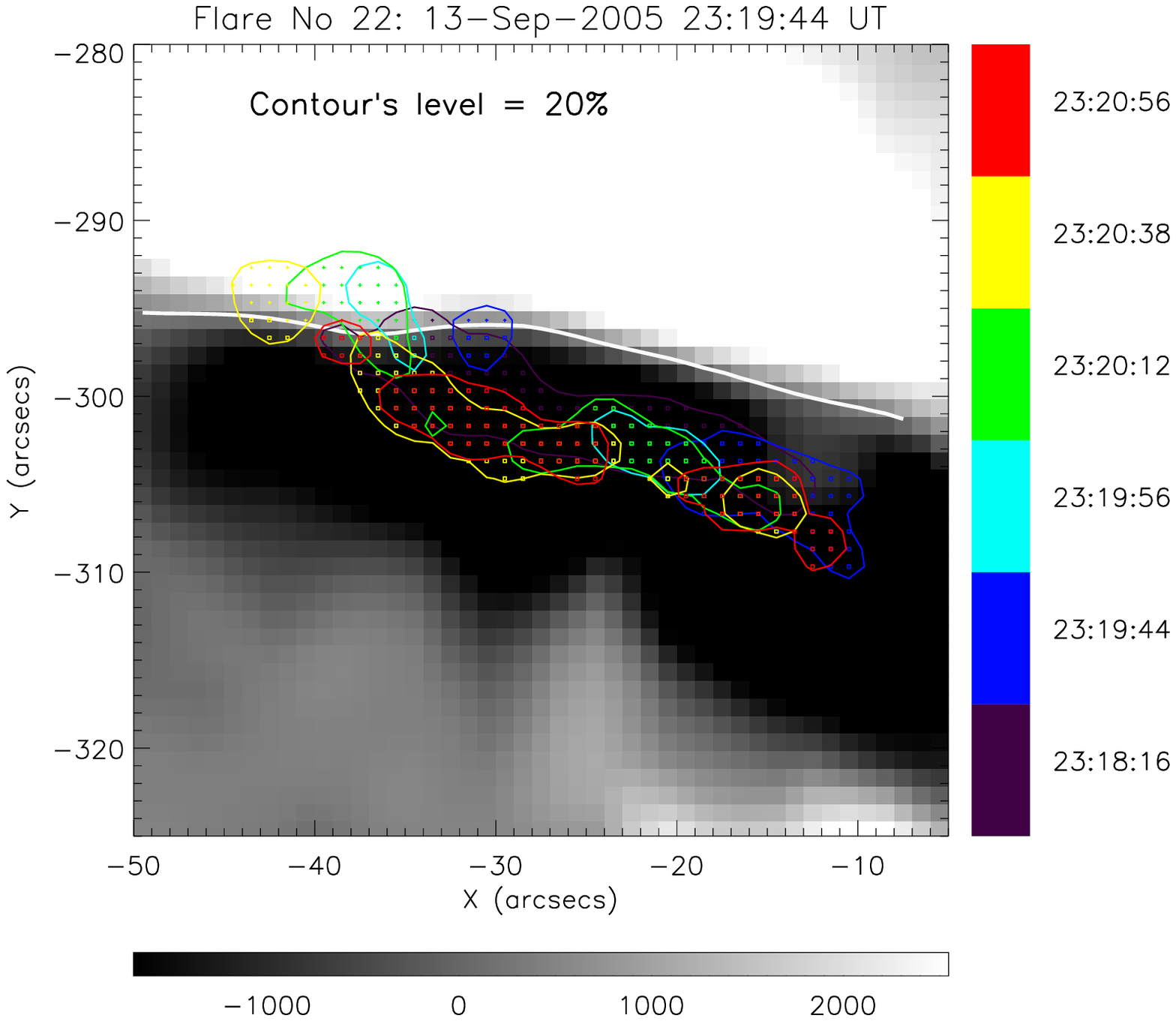}
\includegraphics[width=0.45\textwidth,clip=]{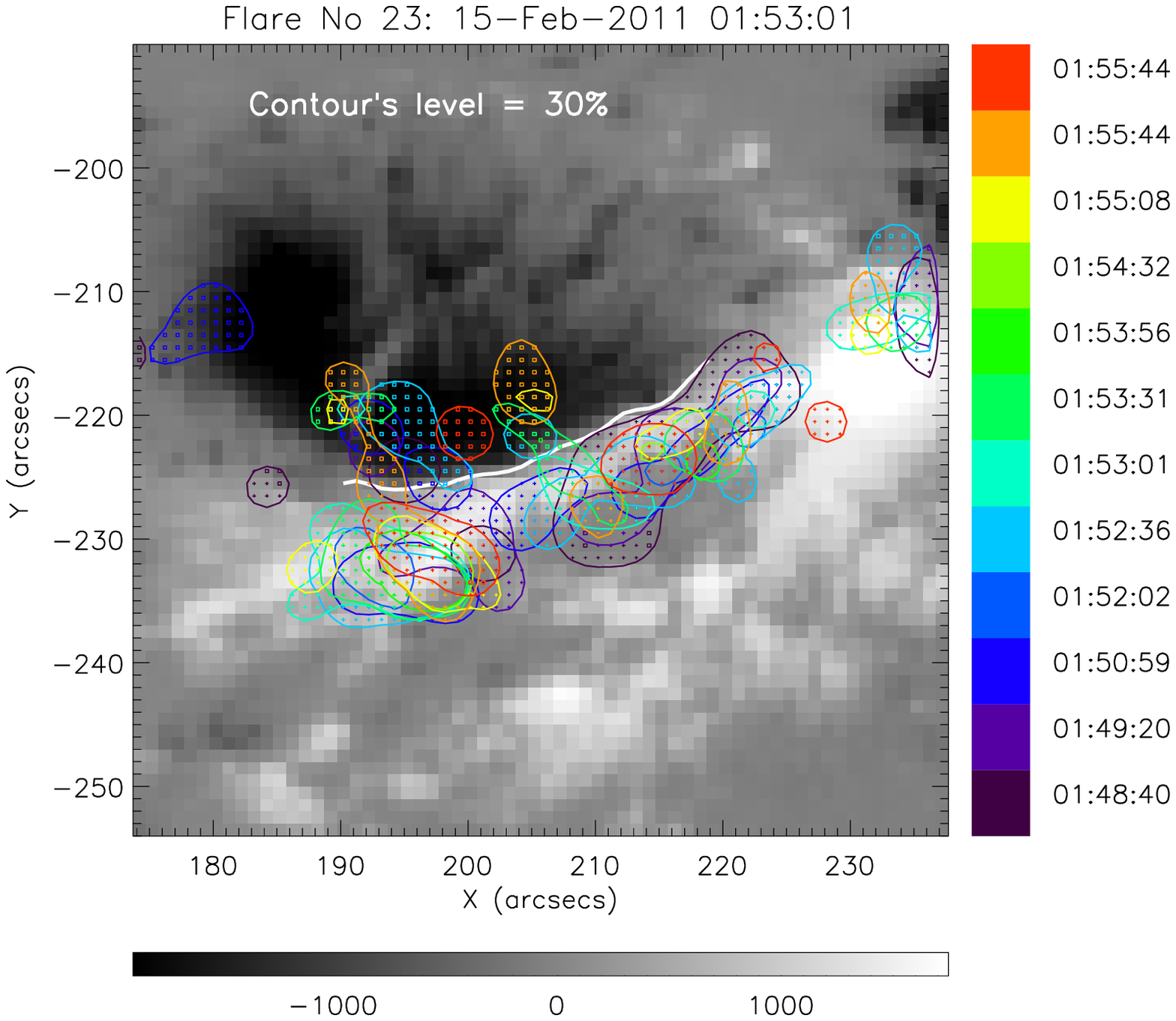}}
\vspace{-0.13\textwidth}   
\centerline{\Large \bf     
\hspace{0.32 \textwidth} \color{white}{(g)}
\hspace{0.37\textwidth}  \color{white}{(h)}
\hfill}
\vspace{0.09\textwidth}    

\caption{Spatio-temporal evolution of the sources of the HXR pulsations in the flares No 16--23. The method and convention are the same as in Figure~\ref{AF-5}.}
\label{AF-7}
\end{figure}

\begin{figure}
\centerline{\includegraphics[width=0.45\textwidth,clip=]{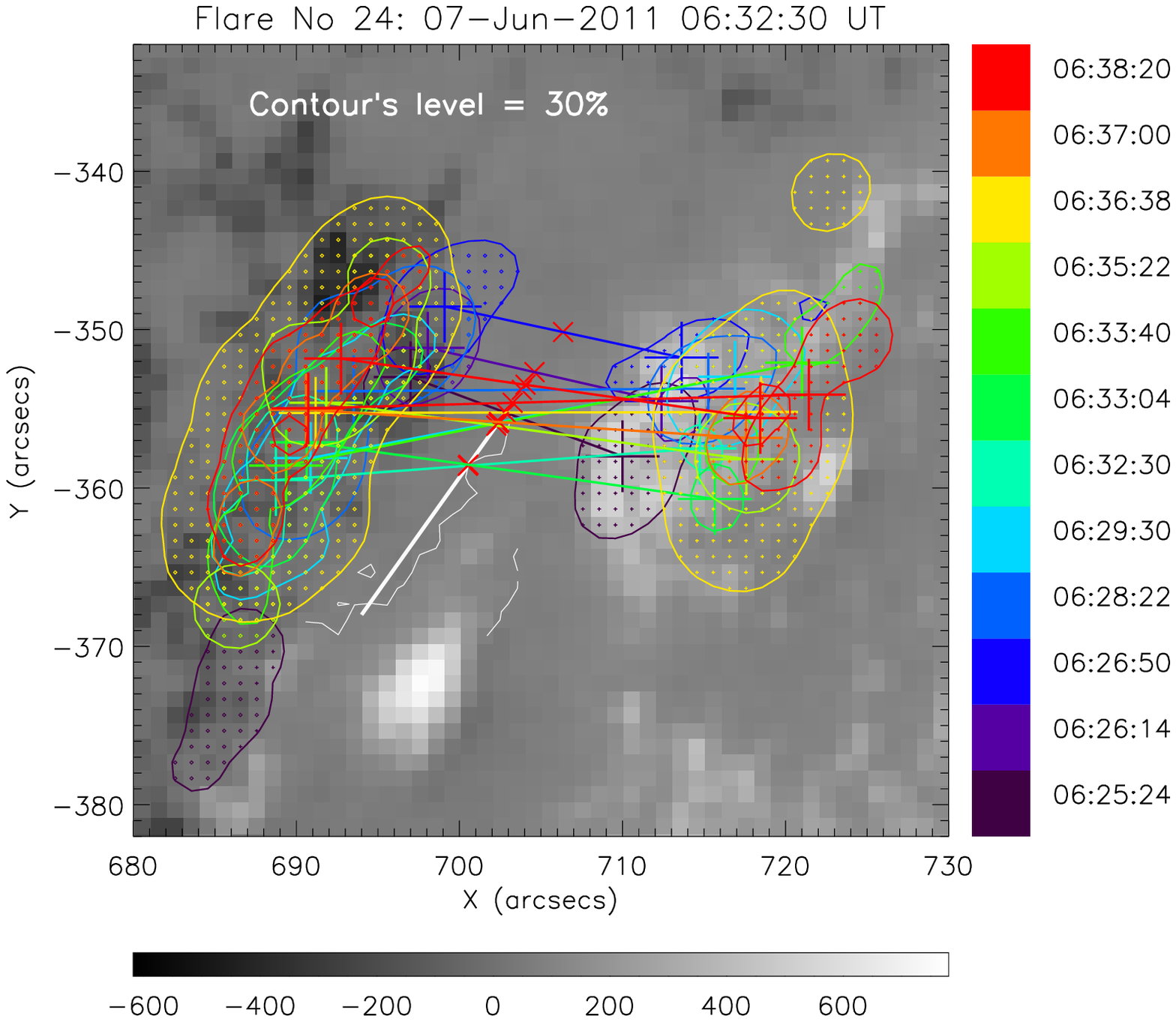}
\includegraphics[width=0.45\textwidth,clip=]{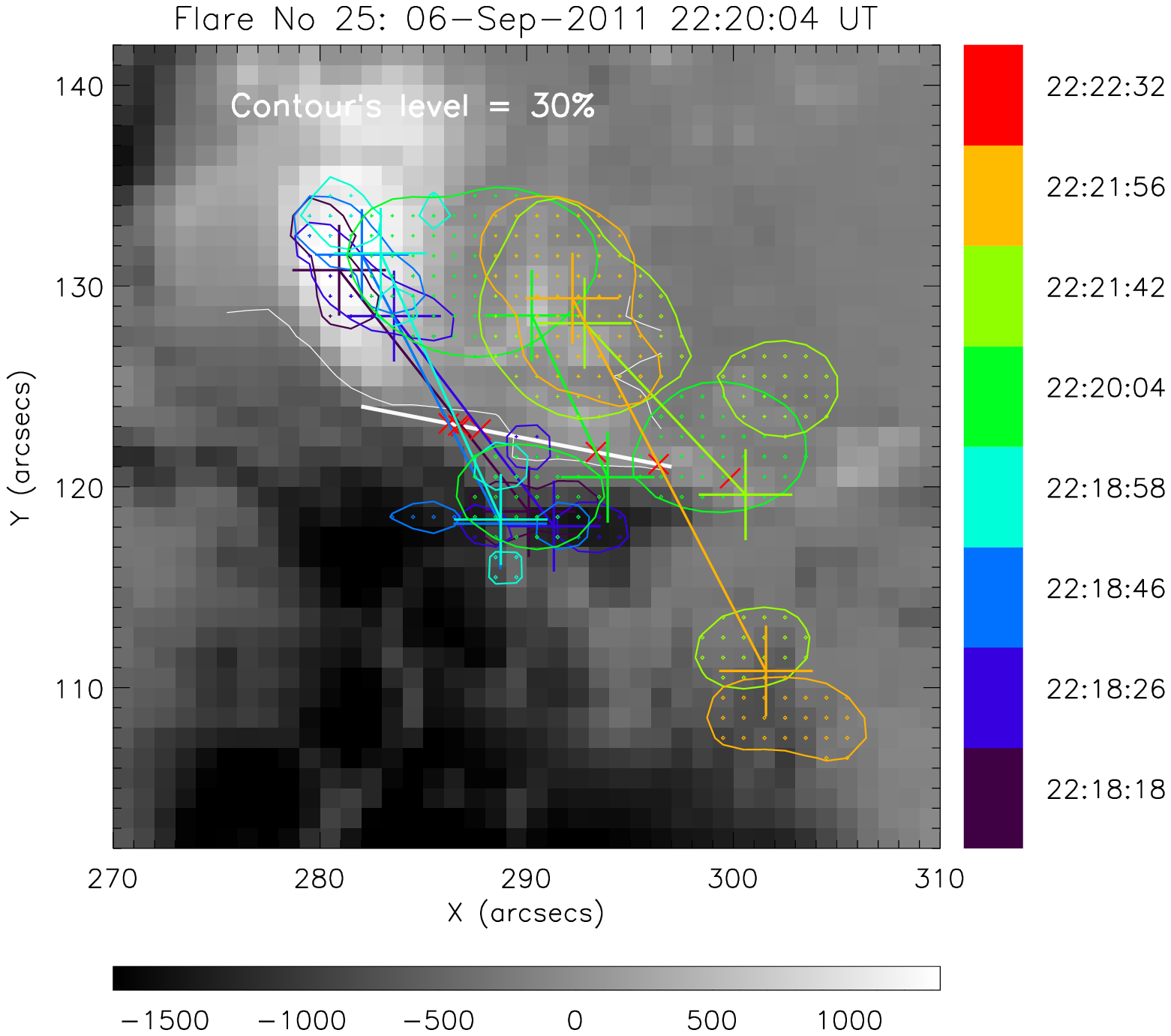}}
\vspace{-0.13\textwidth}   
\centerline{\Large \bf     
\hspace{0.32 \textwidth} \color{white}{(a)}
\hspace{0.37\textwidth}  \color{white}{(b)}
\hfill}
\vspace{0.09\textwidth}    

\centerline{\includegraphics[width=0.45\textwidth,clip=]{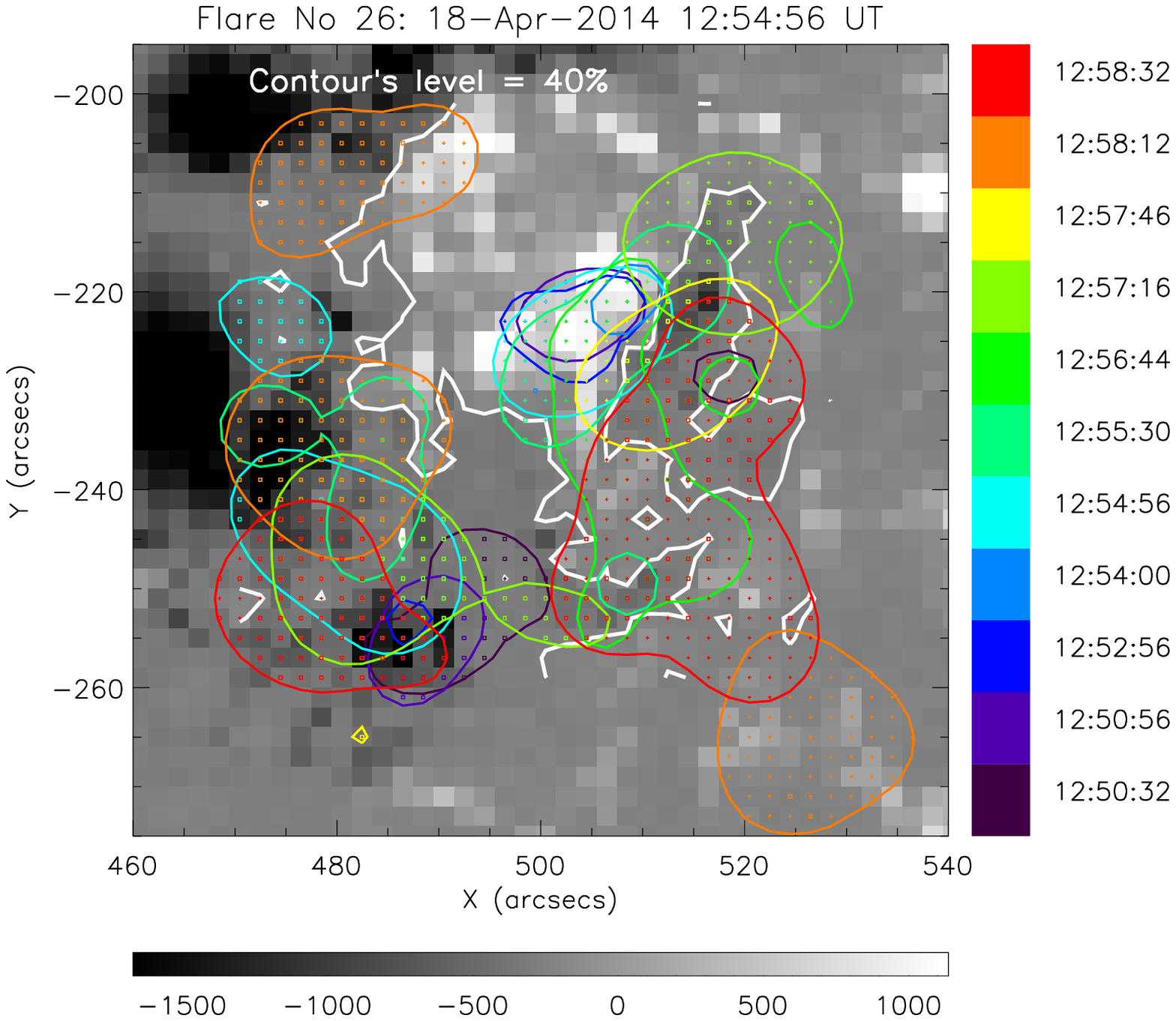}
\includegraphics[width=0.45\textwidth,clip=]{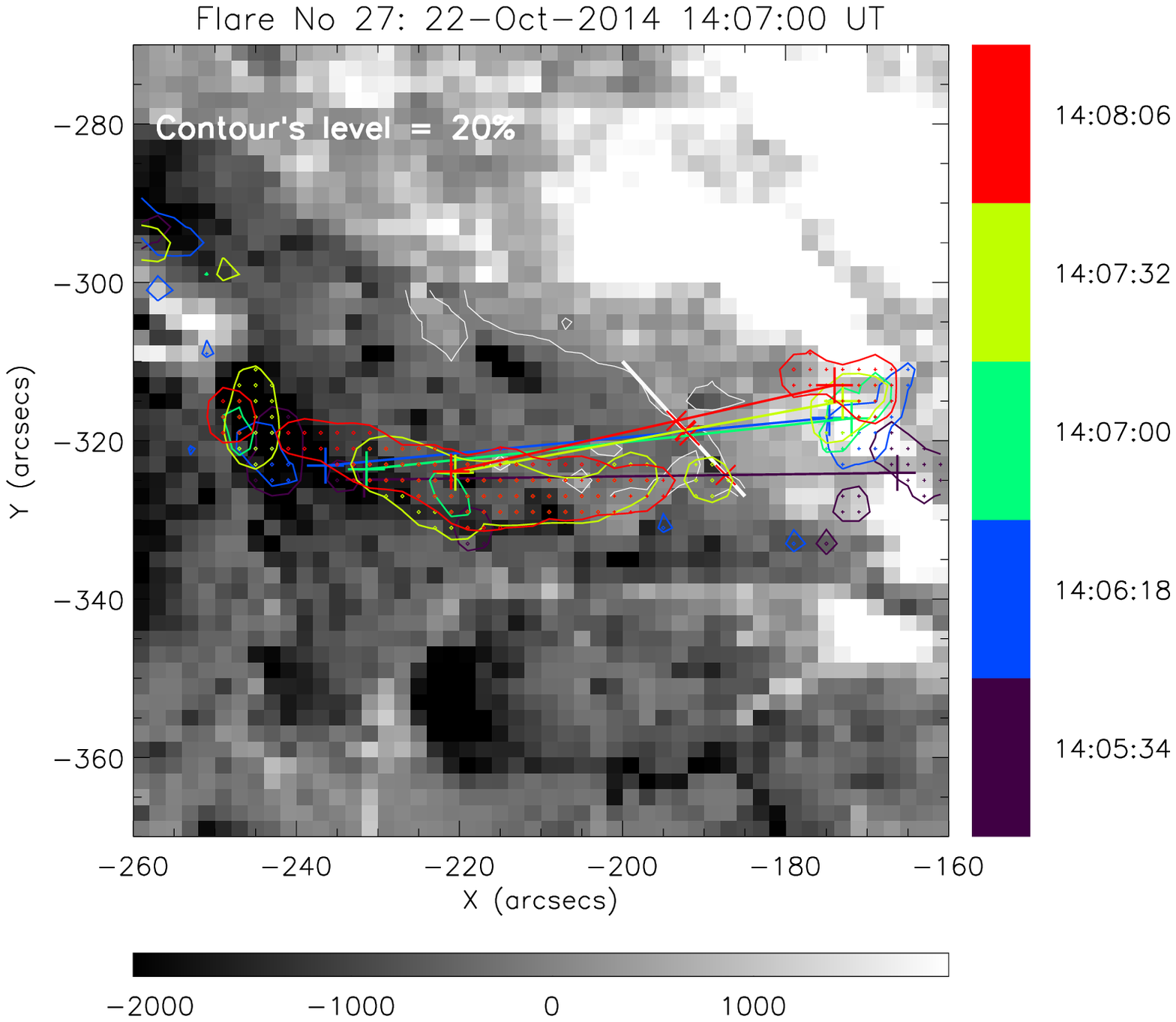}}
\vspace{-0.13\textwidth}   
\centerline{\Large \bf     
\hspace{0.32 \textwidth} \color{white}{(c)}
\hspace{0.37\textwidth}  \color{white}{(d)}
\hfill}
\vspace{0.09\textwidth}    
              
\centerline{\includegraphics[width=0.45\textwidth,clip=]{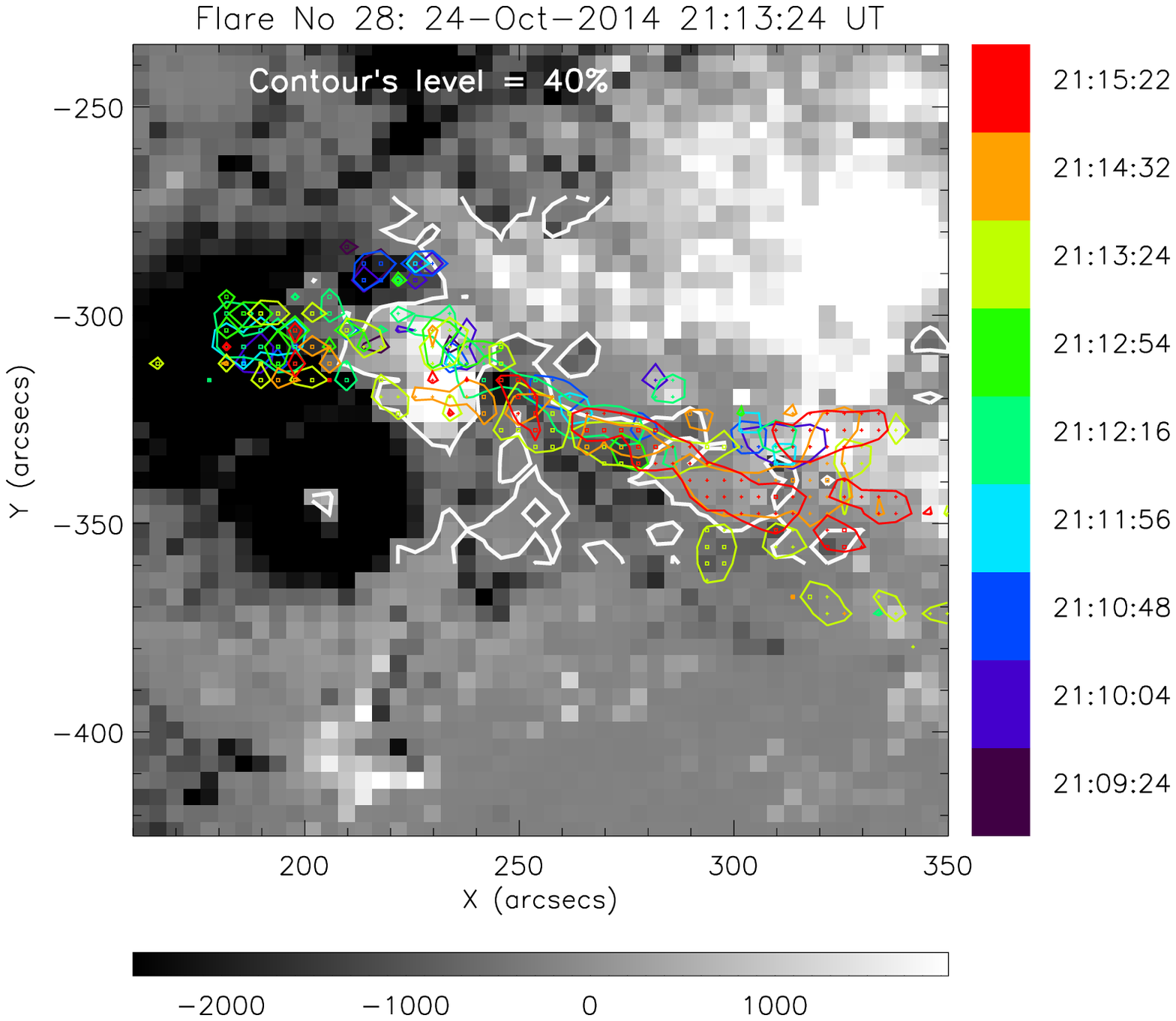}
\includegraphics[width=0.45\textwidth,clip=]{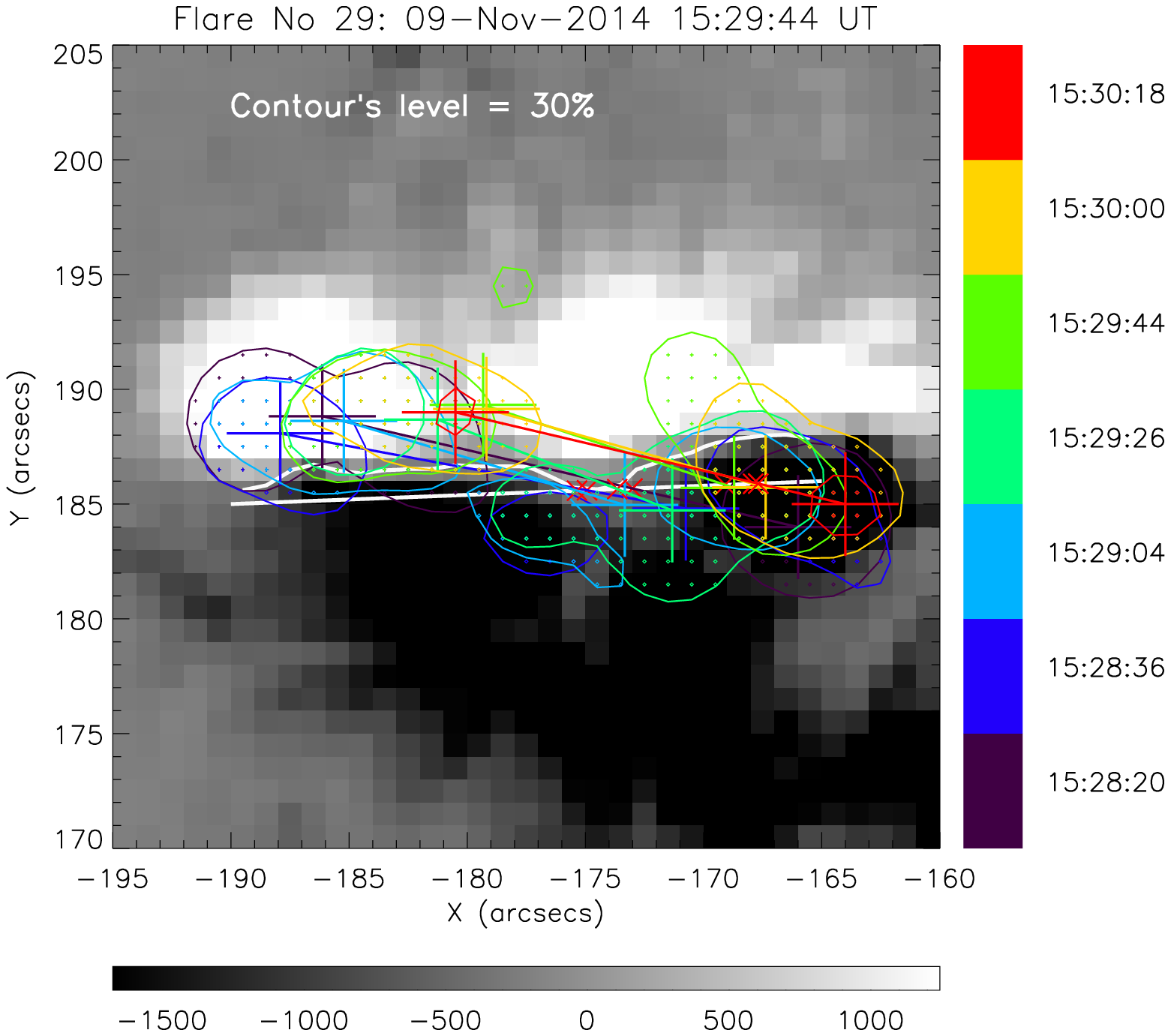}}
\vspace{-0.13\textwidth}   
\centerline{\Large \bf     
\hspace{0.32 \textwidth} \color{white}{(e)}
\hspace{0.37\textwidth}  \color{white}{(f)}
\hfill}
\vspace{0.09\textwidth}    

\caption{Spatio-temporal evolution of the sources of the HXR pulsations in the flares No 24--29. The method and convention are the same as in Figure~\ref{AF-5}.}
\label{AF-8}
\end{figure}

\end{article} 
\end{document}